\documentclass[preprint]{elsarticle}
\usepackage[margin=15mm]{geometry}
\usepackage{times}
\usepackage{natbib}
\usepackage{pdfcomment}                        
\usepackage{graphicx}\graphicspath{{figures/}} 
\usepackage[justification=centering]{caption}  
\usepackage{grffile}                           
\usepackage{placeins}                          
\usepackage[export]{adjustbox}                 
\usepackage{setspace}                          
\usepackage{subcaption}                        
\usepackage{booktabs}                          
\usepackage{flafter}                           
\usepackage{multirow}                          
\usepackage{makecell}                          
\usepackage{tabularx}                          
\usepackage[fleqn]{mathtools}                  
\usepackage{amsfonts,amsthm,amssymb}           
\usepackage{mathrsfs,bm,siunitx}               
\usepackage{pifont}                            
\usepackage{xpatch}                            
\usepackage{enumitem}                          
\usepackage{scrextend}
\usepackage{color,soul}                        
\usepackage{pgfplots}                          
\usepackage [autostyle]{csquotes}              
\MakeOuterQuote{"}
\newcommand{\vu}{\bm{u}}

\newcommand{\dvu}{\dot{\bm{u}}}

\newcommand{\vw}{\bm{w}}
\newcommand{\vn}{\bm{n}}
\newcommand{\dt}{\Delta t}
\newcommand{\intom}{\int_{\Omega}}
\newcommand{\intome}{\int_{\Omega_e}}
\newcommand{\intgam}{\int_{\Gamma}}
\newcommand{\dom}{\mathop{} \! {\mathrm d} \Omega}
\newcommand{\dgam}{\mathop{} \! {\mathrm d} \Gamma}

\newcommand{\rhoi}{\rho_\infty}

\newcommand*{\sdr}[3]{\ensuremath{\frac{\mathrm{d}^{#1} #2}{\mathrm{d}#3 ^{#1}}}}

\newlist{todolist}{itemize}{2}
\setlist[todolist]{label=\textcolor{blue}{$\square$}}

\newlist{todolistb}{itemize}{2}
\setlist[todolistb]{label=\textcolor{blue}{$\circ$}}

\usepackage{newfloat}

\DeclareFloatingEnvironment[fileext=frm,placement={!ht},name=Frame]{mytable}

\begin{document}


\setlength{\jot}{5.5pt}
\setlength{\abovedisplayskip}{11pt}
\setlength{\belowdisplayskip}{11pt}
\setlength{\abovedisplayshortskip}{11pt}
\setlength{\belowdisplayshortskip}{11pt}

\numberwithin{equation}{section}
\setcounter{page}{1}\pagenumbering{arabic}


\definecolor{mygreen}{rgb}{0.0, 0.5, 0.0}

\hfuzz=1pt
\vfuzz=1pt

	\title{Low Order Finite Element Methods for the Navier-Stokes-Cahn-Hilliard Equations}

\author{Aleksander Lovri\'c\corref{cor1}} 
\cortext[cor1]{Corresponding author.}
\ead{alexlovric@gmail.com}

\author{Wulf G. Dettmer\corref{}}
\author{Djordje Peri\'c\corref{}}

\address{Zienkiewicz Centre for Computational Engineering, College of Engineering, Swansea University, Fabian Way, Swansea SA1 8EN, Wales, UK}

	\begin{abstract}	
	A computationally efficient, low order finite element formulation is developed for modelling the Navier-Stokes-Cahn-Hilliard equations, which have been established as a promising phase field modelling approach for simulation of immiscible multiphase flows. The present study suggests that traditional Navier-Stokes-Cahn-Hilliard models do not allow for surface tension effects to be neglected due to the presence of the surface tension parameter in the Cahn-Hilliard equation. This motivates the proposed formulation, which allows surface tension effects to be changed without affecting the behaviour of the phase transport. Two methods are proposed: The first uses stabilised SUPG/PSPG linear elements, while the second is based on mixed Taylor-Hood elements. The proposed models are applied to a number of benchmark and example problems, including both capillary regime in which surface tension effects are dominant, and inertial regime in which surface tension effects are negligibly small. All results obtained agree very well with reference solutions.
\end{abstract}	


	\maketitle

	\section{Introduction}
Modelling multiphase fluid flow in the presence of surface tension has for a long period been the subject of scientific investigation. These types of flows can be observed in numerous industrial and natural applications, for instance in emulsification, fluidized beds, combustion reactors, gas-liquid pipeline flows, etc. Typically multiphase problems are subject to a high degree of topological change, making interface tracking difficult, if not impossible. Strategies for simulating such flows include volume-of-fluid \cite{Hirt1981}, level-set \cite{Osher1988} and diffuse interface methods \cite{Tryggvason1988,Abels2011,Boyer2002}, as well as arbitrary Lagrangian-Eulerian \cite{Hughes1981,Benson1992,Donea2004,Dettmer2006} or purely Lagrangian finite element formulations \cite{Bach1985,Ramaswamy1987,Saksono2006,Saksono2006a,Javili2009}.
\par Diffuse-interface methods have become a popular choice for the simulation of multiphase problems involving complex topological changes. The fundamental concepts behind diffuse-interface methods for immiscible multiphase fluids were pioneered by Van der Waals \cite{VDW1979} and Korteweg \cite{Korteweg1901}. For an in depth explanation of the types of phase-field models, refer to \cite{Anderson1998,Gomez2015}. The fundamental principle is that instead of having a sharp interface partitioning two phases, the interface is modelled as a region within which physical quantities (density, viscosity etc.) and interfacial forces vary smoothly. In order to achieve this an auxiliary field variable, typically called the \textit{phase-field variable} or \textit{order parameter}, is introduced to smoothly transition quantities across the interface between two distinct values in the bulk phases. The dynamics of the phase-field variable are typically governed by the Cahn-Hilliard equation \cite{Cahn1959}, or its non-conserving counterpart the Allen-Cahn equation \cite{Allen1972}. Various numerical strategies for solving the Cahn-Hilliard and Allen-Cahn equations are explored in for instance \cite{Wells2006,Gomez2008,Gomez2011,Guillen-gonzalez2013}.
\par Coupling the Cahn-Hilliard equation with hydrodynamic Navier-Stokes equations results in a set of equations known as the Navier-Stokes-Cahn-Hilliard equations. These equations have been extensively used in the simulation of immiscible two-component fluid flows, and have been widely researched from a mathematical and computational perspective, see for instance \cite{Antanovskii1995,Gurtin1996,Jacqmin1999,Lowengrub1998,Shen2010}. These types of equations were utilised as early as the 1970's \cite{Hohenberg1977}, with the so-called Model H. This model combines the phase dynamics of the Cahn-Hilliard equation with the hydrodynamics of the Navier-Stokes equation for the purpose of modelling immiscible, incompressible two-component flows. The drawback of the Model H is that it assumes that the mixture is comprised of constant matching densities, and is hence severely restricted in its application.
The difficulty with considering non-matching densities is that the macroscopic description of the density differs from the direct average of the microscopic descriptions, meaning that even if the individual components are incompressible, the mixture may not be. Typically models proposed for mixtures of non-matching densities are characterised as either: incompressible - where the volume averaged velocity field is divergence-free (\textit{i.e.} satisfies incompressibility), and  quasi-incompressible - where the mass-averaged velocity field satisfies mass conservation only, resulting in a slight compressibility in the interface region.
A quasi-incompressible modification of the Model H by Lowengrub and Truskinovsky \cite{Lowengrub1998} allows for the the use of non-matching densities in the bulk regions. However, as mentioned the definition of the average velocity results in the loss of the divergence-free velocity field. Furthermore, a stronger coupling is present since the pressure from the momentum equation of the Navier-Stokes equations is present in the Cahn-Hilliard part of the equations. 
More recently, another quasi-incompressible form of the equations is proposed in \cite{Roudbari2016}, where notably a linear method is used which satisfies discrete energy dissipation unconditionally.
On the other hand, models by Boyer \cite{Boyer2002} and Ding et al. \cite{Ding2007} present appealing formulations with divergence-free velocity fields; however the energy inequalities are unknown for the model.
Shen et al. \cite{Shen2010} propose a physically consistent model for incompressible mixtures with non-matching densities, implementing energy stable, accurate time integration schemes. More recently, Abels et al. \cite{Abels2011} derived a variable density variation of the Model H which is thermodynamically consistent. The model is shown to recover sharp interfaces when the interface thickness tends to zero. The Abels et al. model has been effectively used in numerous numerical studies, and is chosen as the base model of this paper. The Allen-Cahn equation has also been successfully coupled with the Navier-Stokes equations in \cite{Xu2010,Joshi2018,Favre2019}.
\par Numerous numerical strategies have been developed to solve the types of Navier-Stokes-Cahn-Hilliard equations discussed above. 
The use of a spectral-element type discretisation has been shown effective in \cite{Dong2012a,Lee2002,Liu2003}.
A finite difference discretisation is effectively demonstrated by Kim et al. \cite{Kim2004}.
Approaches based on finite element methods, including those with adaptive re-meshing, are also frequently employed (see, for instance \cite{Feng2006,Du2008,Kay2008,Zhang2010,Bao2012}).
Guo et al. \cite{Guo2014} present an adaptive mesh strategy, using an energy preserving $C_0$ finite element method to solve problems of high topological complexity. A discontinuous Galerkin finite element approach is used by Giesselmann and Pryer \cite{Giesselmann2015} to model the quasi-incompressible Navier-Stokes-Cahn-Hilliard equations.
One of the more recent approaches is based on using isogeometric analysis, which is successfully demonstrated in \cite{Gomez2015,Espath2016,Hosseini2017}.
There have also been recent advancements in fractional step type schemes for the mentioned incompressible phase field models. Shen et al. \cite{Shen2010} present a three stage decoupling where the computations of the Allen-Cahn equation, i.e. the phase-field variable and chemical potential, are decoupled from the Navier-Stokes equations, in which the pressure and velocity fields are also decoupled. Similar strategies for the Navier-Stokes-Chan-Hilliard equations is proposed in \cite{Minjeaud2013,Liu2014,Chen2016}, and \cite{Guillen-Gonzalez2014a}, where the latter only decouples the Cahn-Hilliard equation from the Navier-Stokes equations.
Similar numerical strategies have been demonstrated using the Navier-Stokes-Allen-Cahn equations. For instance Yang et al. \cite{Yang2013} and more recently Chiu \cite{Chiu2019} successfully implemented a Navier-Stokes-Allen-Cahn type model to simulate surface tension dominated drop dynamics problems.
\par The primary objectives of this work are:
\begin{enumerate}
	\item To present a new Navier-Stokes-Cahn-Hilliard model capable of simulating problems  where the choice of the surface tension coefficient is based on the physics of the problem at hand. 
	This addresses the drawback of traditional forms of the Navier-Stokes-Cahn-Hilliard equations, wherein the surface tension coefficient cannot be set to zero without removing all stabilising effects of the surface and bulk energies.	
	\item To present computationally efficient, low order finite element formulations of the Navier-Stokes-Cahn-Hilliard equations. As mentioned previously, the simulation of these equations generally involves the use of higher order b-splines or adaptive re-meshing, 
	which are generally associated with high computational cost. In this work emphasis is placed on computational efficiency, hence an equal order SUPG/PSPG stabilisation strategy as well as a mixed Taylor-Hood type methodology based on standard finite element discretisation with Lagrange polynomials are presented.
	\item To verify the presented methods with benchmark problems and realistic examples of two-phase flows. A focus is placed on reproducing a number of surface tension dominated benchmark problems. Additionally, sloshing and breaking dam problems are considered where surface tension effects are negligible and have therefore been deactivated during simulation.
\end{enumerate}
\par The remainder of this work is structured as follows: In Section \ref{sec: Governing equations}, the governing equations associated with the Navier-Stokes-Cahn-Hilliard equations are presented. In Section \ref{sec: Numerical formulation} these equations are formulated for mixed and stabilised finite elements. In Section \ref{sec: Numerical examples} various numerical examples are demonstrated using the proposed formulations. In Section \ref{sec: Conclusions} conclusions are drawn.
%
\section{Governing equations} \label{sec: Governing equations}
%
\subsection{Cahn-Hilliard equation}
Phase-field models are characterised by the introduction of an auxiliary function $\varphi$, a phase-field variable which localises the individual phases. $\varphi$ is represented by distinct values outside the interface region (i.e. the bulk region), for example,
\begin{align}
	\varphi(x,t) =
	\begin{cases}
		1,  &\quad \textrm{phase $a$}\\
		-1, &\quad \textrm{phase $b$} .
	\end{cases}
	\label{eq:order_paramater_def}
\end{align}
The Cahn-Hilliard equation governs the transport and decomposition of $\varphi$ through dissipation of the energy functional,
\begin{equation}
	\mathcal{F}(\varphi) = \intom \Psi \dom  ,
	\label{eq:GL_functional}
\end{equation}
where,
\begin{align}
	 \Psi = \left( F(\varphi) + \frac{\epsilon^2}{2}\lvert \nabla \varphi\rvert^2 \right).
\end{align}
$\Psi$ is the Ginzburg-Landau free energy composed of a hydrophobic bulk component $F(\varphi)$, and a hydrophilic surface component $\frac{\epsilon^2}{2}\lvert \nabla \varphi\rvert^2$. Here $\epsilon$ is a length scale related to the interface thickness. In the context of this work we consider the bulk free energy component as a double well potential,
\begin{align}
	F(\varphi) = W \left( 1 - \varphi^2 \right)^2,
\end{align}
where $W$ is the height of the well.
An in depth mathematical analysis and derivation of the Cahn-Hilliard equation can be found for instance in \cite{Cahn1958,Elliott1986,Blowey1992,Gomez2008}.
\par The Cahn-Hilliard equation is presented in a two equation form, similarly to \cite{Gomez2011}: Consider a domain $\Omega \subset \mathbb{R} ^d$ ($d \leq 3$) with boundary $\Gamma$ which is separable into Dirichlet and Neumann subsets, $\Gamma_g$ and $\Gamma_q$. The solution variables $\varphi$ and $\eta$ are described by the following governing equations:
\begin{subequations}
	\begin{alignat}{2}
		\frac{\partial \varphi}{\partial t} 
		+ \bm{a} \cdot \nabla\varphi 
		- \nabla \cdot \left( M(\varphi) \nabla \eta \right) 
		&= 0 &&\qquad \text{in} \ \Omega_{\phantom{N}}
		\label{eq:CH_problem_a}
		\\\eta - f(\varphi) + \epsilon^2 \Delta \varphi &= 0 &&\qquad \text{in} \ \Omega_{\phantom{N}}
		\label{eq:CH_problem_b}
		\\\nabla \eta\cdot\vn &= 0 &&\qquad \text{on} \ \Gamma_{q}
		\label{eq:CH_bc1}	
		\\\nabla \varphi\cdot\vn + \lVert\nabla\varphi\rVert \cos{(\alpha)} &= 0 &&\qquad \text{on} \ \Gamma_{q}.
		\label{eq:CH_bc2}	
	\end{alignat}
\label{eq:CHGovEq}
\end{subequations}
Here $\eta$ is the chemical potential, which is the variational derivative of the free energy given by Equation \eqref{eq:GL_functional}, with $f(\varphi)=4W(\varphi^3-\varphi)$. $M(\varphi)$ is the mobility function, which is described in Section \ref{sec: Mobility}. The angle $\alpha$ is the three-phase contact angle at the boundary.
\subsection{Navier-Stokes-Cahn-Hilliard equations}
Consider again the domain $\Omega \subset \mathbb{R} ^d$ ($d \leq 3$) bounded by $\Gamma$ and containing a mixture of two immiscible incompressible fluids with different densities $\rho_a$ and $\rho_b$. In this work, the following NSCH model is used:
\begin{subequations}
	\begin{alignat}{2}
		\rho(\varphi) \left(\frac{\partial \vu}{\partial t} 
		+ \left(\vu\cdot\nabla\right)\vu - \bm{b} \right)
		+ \left(\bm{J}(\varphi,\eta)\cdot\nabla\right)\vu 
		+ \nabla p 
		- \nabla\cdot\left( 2\mu(\varphi) \nabla^s \vu \right) 
	    - \kappa \eta \nabla \varphi
		&= \bm{0} &&\qquad \text{in} \ \Omega
		\label{eq:NSCHGovEqA}
		\\\nabla\cdot\vu &= 0 &&\qquad \text{in} \ \Omega
		\label{eq:NSCHGovEqB}
		\\\frac{\partial \varphi}{\partial t} 
		+ \bm{u} \cdot \nabla\varphi 
		- \nabla \cdot \left( M(\varphi) \nabla \eta \right) 
		&= 0 &&\qquad \text{in} \ \Omega
		\label{eq:NSCHGovEqC}
		\\\eta - f(\varphi) + \epsilon^2 \Delta \varphi &= 0 &&\qquad \text{in} \ \Omega
		\label{eq:NSCHGovEqD}
		\\\nabla \eta\cdot\vn &= 0 &&\qquad \text{on} \ \Gamma_{q}	
		\\\nabla \varphi\cdot\vn + \lVert\nabla\varphi\rVert \cos{(\alpha)} &= 0 &&\qquad \text{on} \ \Gamma_{q}		
	\end{alignat}
\label{eq:NSCHGovEq}
\end{subequations}
where $\vu$ and $p$ represent the velocity and pressure, respectively, and $\bm{b}$ is the body force. The symmetric gradient operator is defined as $\nabla^s := \frac{1}{2}\left(\nabla + \nabla^T\right)$.
\par The model presented in Equation \eqref{eq:NSCHGovEq} is based on the Abels et al. \cite{Abels2011} and is thermodynamically consistent. It allows for variable densities, and agrees with sharp interface models for $\epsilon \to 0$. 
\par Cahn and Hilliard \cite{Cahn1959} describe surface tension as the excess free energy per unit surface area. It follows that in the case of a plane phase-field interface at equilibrium, the surface tension coefficient, $\gamma$, is related to the phase-field variable $\varphi$ by
\begin{align}
	\gamma = \kappa \int^{\infty}_{-\infty} \left( \sdr{}{\varphi}{x} \right) \mathrm{d} x.
	\label{eq:gamma}
\end{align}
In order to achieve consistency of the surface tension term in \eqref{eq:NSCHGovEqA} with the Young-Laplace sharp interface surface tension model, the factor $\kappa$ must be chosen as
\begin{align}
	\kappa = \frac{3}{4\sqrt{2W}} \frac{\gamma}{\epsilon},
	\label{eq:kappa}
\end{align}
Equation \eqref{eq:gamma} is successfully implemented in numerous studies \cite{Jacqmin1999,Villanueva2006,Ding2007,Hosseini2017}. An NSCH model using Equation \eqref{eq:kappa} is shown to approach the sharp interface model as $\epsilon \to 0$ in \cite{Xu2018}.
\par In the present work, the surface tension stress term $\eta\nabla\varphi$ (see also Boyer \cite{Boyer2002}) is used in Equation \eqref{eq:NSCHGovEqA}, instead of $\gamma\epsilon\nabla\cdot(\nabla\varphi\otimes \nabla\varphi)$ (see Abels et al. \cite{Abels2011}). The former term is easily derived from the latter, and the resulting pressure term in Equation \eqref{eq:NSCHGovEqA} is now a modified pressure term $\nabla \hat{p}$, where $\hat{p}=(p+\Psi)$. This modified pressure is similar to the original pressure, in fact, it is identical in the bulk regions where $\Psi$ vanishes. Advantageously the modified pressure varies generally more smoothly across interface regions than the original pressure.
\par In Equation \eqref{eq:NSCHGovEqA}, the relative diffusive flux is expressed as
\begin{align}
	\bm{J} = -\rho_{dif} M(\varphi) \nabla \eta .
\end{align}
The density and viscosity are described by the linear approximations
\begin{align}
	\begin{alignedat}{2}
		\rho(\varphi) &= \rho_{dif}\varphi &&+ \rho_{avg}
		\\\mu(\varphi) &= \mu_{dif}\varphi &&+  \mu_{avg},
	\end{alignedat}
	\label{eq:density viscosity interp}	
\end{align}
where,
\begin{align}
	 \rho_{dif} = \frac{\rho_a-\rho_b}{2}, \quad \rho_{avg} = \frac{\rho_a+\rho_b}{2} \quad \textrm{and} \quad
	 \mu_{dif} = \frac{\mu_a-\mu_b}{2}, \quad \mu_{avg} = \frac{\mu_a+\mu_b}{2}.
	\label{eq:density viscosity interp2}
\end{align}
If necessary, a cut-off function $\bar{\varphi}$ can be incorporated in Equation \eqref{eq:density viscosity interp}, where
\begin{align}
	 	\bar{\varphi} = 
	 	\begin{cases}
	 	\varphi \quad&\text{if} \lvert\varphi\rvert \leq 1 \\
	 	\mathrm{sign}(\varphi) \quad&\text{if} \lvert\varphi\rvert > 1.
	 	\end{cases}
	\label{eq:cut-off}
\end{align}
This would ensure that $\rho$ and $\mu$ remain within the physical bounds of the specified bulk phases, and has been effectively used in \cite{Yu2017}.
\\
\\\textbf{Remark 1. }
In the model given by Equation \eqref{eq:NSCHGovEq} the physical quantities and the Young-Laplace surface tension term occur only in the momentum equation. The CH Equation \eqref{eq:CHGovEq} merely governs the transport of the order parameter $\varphi$. Traditional representations of the NSCH equations such as those used in \cite{Lowengrub1998,Ding2007,Abels2011} can easily be recovered by introducing appropriate scalar factors in Equations \eqref{eq:NSCHGovEqC} or \eqref{eq:NSCHGovEqD}.
%
%
\subsection{Mobility function} \label{sec: Mobility}
\begin{figure}[bt]
	\captionsetup[subfigure]{labelformat=empty}
	\centering
	\begin{subfigure}{.4\textwidth}
		\centering
		\includegraphics[width = 0.85\textwidth]{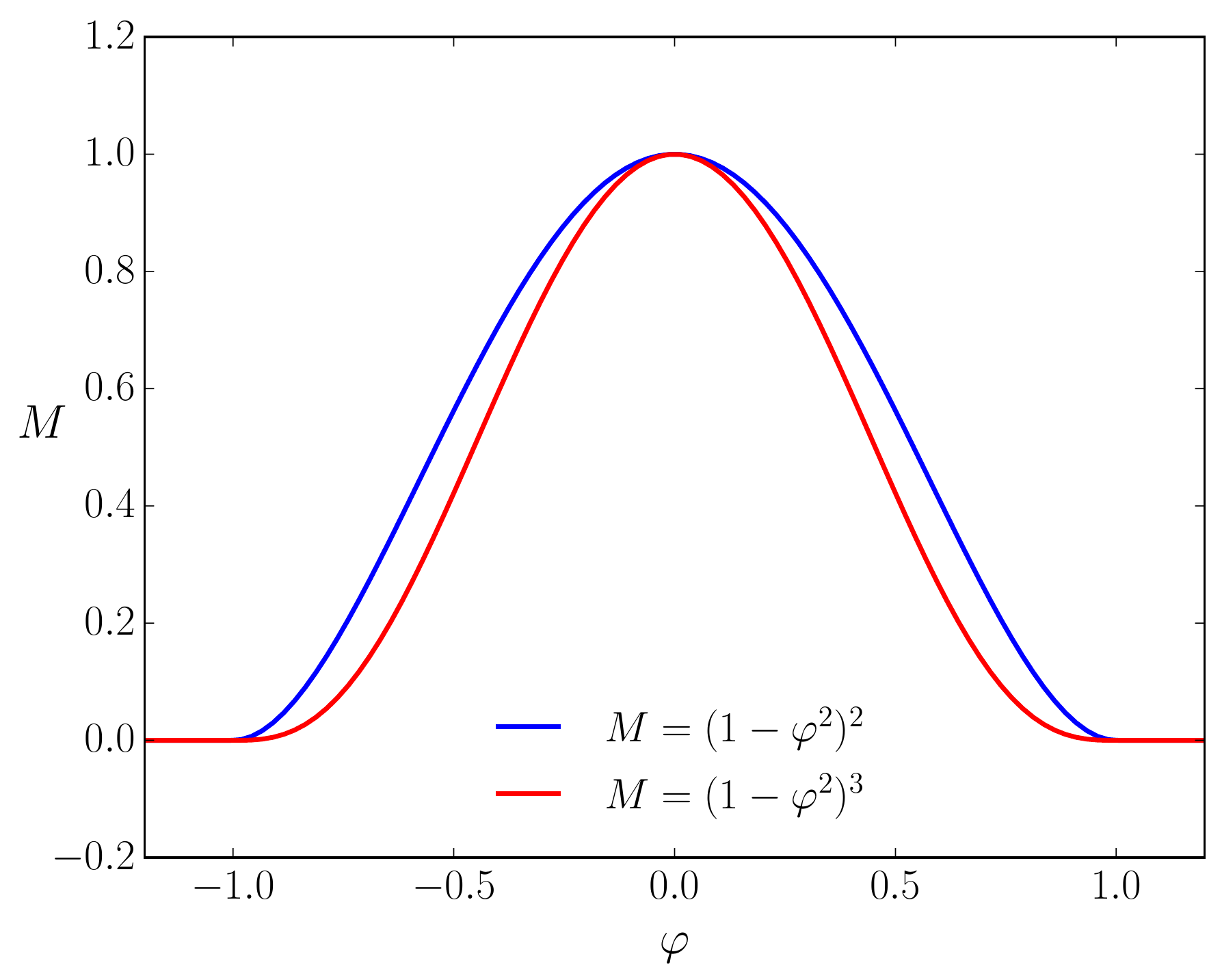}
	\end{subfigure}
	\begin{subfigure}{.4\textwidth}
		\centering
		\includegraphics[width = 0.85\textwidth]{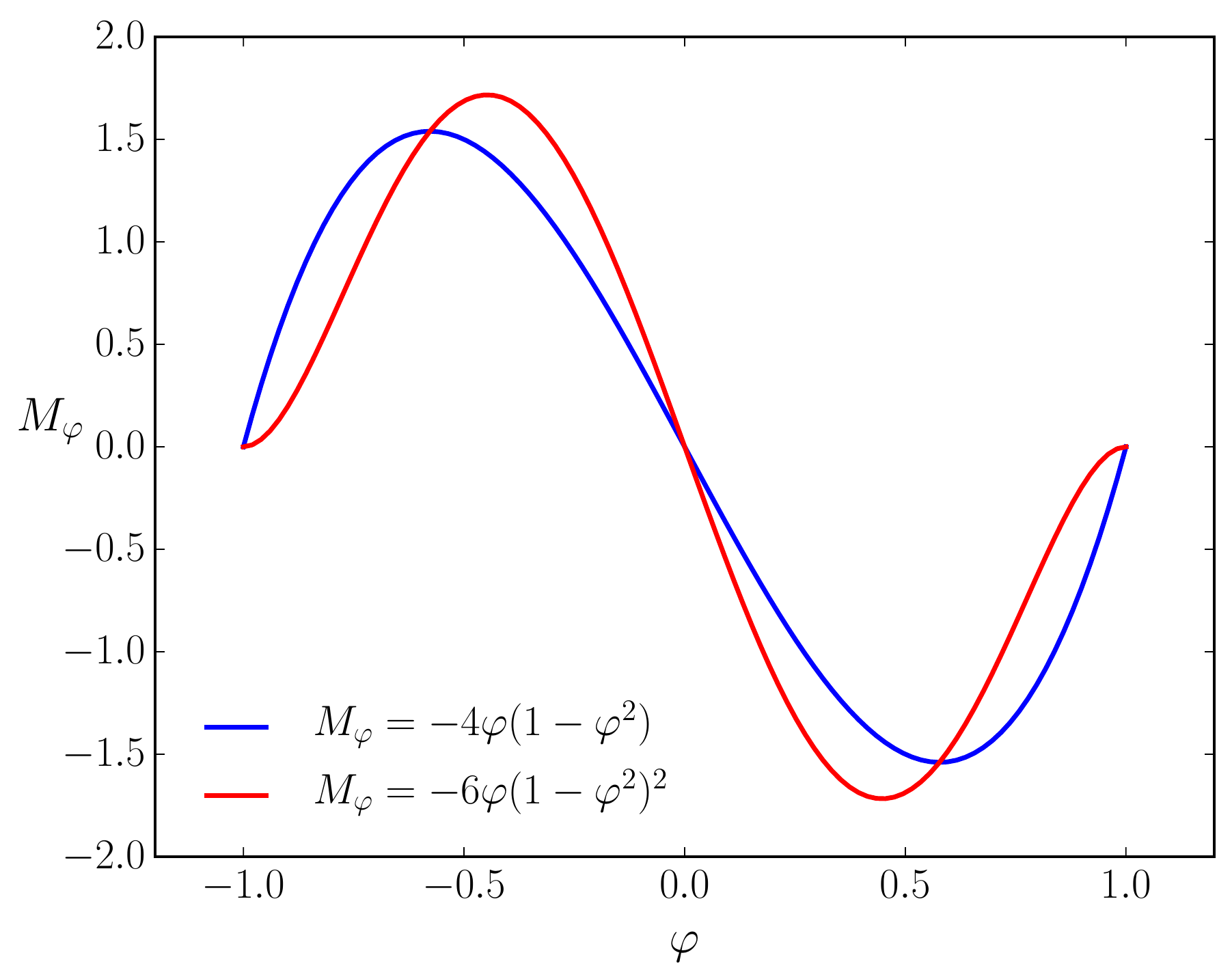}
	\end{subfigure}	
	\caption{Degenerate mobility functions, $M_{2,3}$ (left) and the respective first derivatives (right), with $D=1$ in both cases.}
	\label{fig:mobility}	
\end{figure}

%
In this work three options are considered for the mobility, i.e.
\begin{align}
	M_0 = D,
\end{align}
\begin{align}
	M_2(\varphi) =
	\begin{cases}
		D(1 - \varphi^2), \quad &\text{if} \ \lvert \varphi \rvert \leq 1 \\
		0, \quad &\text{elsewhere},
	\end{cases}
\end{align}
and
\begin{align}
	M_3(\varphi) = 
	\begin{cases}
	D(-2\varphi^3 - 3\varphi^2 + 1) \quad&\text{if} \ \varphi \geq -1 \\
	D(2\varphi^3 - 3\varphi^2 + 1) \quad&\text{if} \ \varphi \leq  1 \\
	0 \quad&\text{elsewhere},
	\end{cases}	
\end{align}
where $D$ is a constant. The latter two mobility functions are degenerate, \textit{i.e.} they are non-zero only in the interface region. This is illustrated in Figure \ref{fig:mobility} for $D=1$.
Notably, $M_3(\varphi)$, transitions smoother in the derivative.
On coarse meshes, the choice of mobility can drastically change the behaviour of the model. Choosing a large constant mobility can result in the acceleration of the \textit{Ostwald ripening} or \textit{coarsening} effect. In essence this means the total interfacial area will reduce with time in an effort to reach the lowest energy state and hence thermodynamic equilibrium. For this reason $D$ must be chosen small enough to ensure that the associated timescale is far larger than the time domain of interest. The reader is referred to \cite{Voorhees1992} for a detailed explanation of Ostwald ripening.
%
\section{Numerical formulation} \label{sec: Numerical formulation}
%
\subsection{Navier-Stokes-Cahn-Hilliard mixed Taylor-Hood formulation}
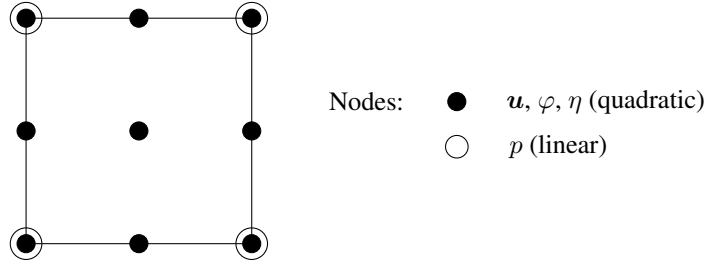
\begin{figure}[tb!]
	\captionsetup[subfigure]{labelformat=empty}
	\centering
	\begin{subfigure}[b]{.25\textwidth}
		\centering
		\begin{tikzpicture}[scale=3.0]
			\draw[] (0,0) -- (1,0) -- (1,1) -- (0,1) -- cycle;
			\draw[black,fill] (0,0) circle(0.04);
			\draw[black,fill] (0.5,0) circle(0.04);
			\draw[black,fill] (1,0) circle(0.04);
			\draw[black,fill] (1,0.5) circle(0.04);
			\draw[black,fill] (1,1) circle(0.04);
			\draw[black,fill] (0.5,1) circle(0.04);
			\draw[black,fill] (0,1) circle(0.04);
			\draw[black,fill] (0,0.5) circle(0.04);
			\draw[black,fill] (0.5,0.5) circle(0.04);
			\draw[black] (0,0) circle(0.07);
			\draw[black] (1,0) circle(0.07);
			\draw[black] (1,1) circle(0.07);				
			\draw[black] (0,1) circle(0.07);	
		\end{tikzpicture}
	\end{subfigure}
	\begin{subfigure}[b]{.25\textwidth}
		\centering
		\begin{tikzpicture}[scale=3.0]
			\draw[white] (0,0) -- (1,0) -- (1,1) -- (0,1) -- cycle;
			\draw[white,fill] (0,0) circle(0.05);
			\draw[white,fill] (0.5,0) circle(0.05);
			\draw[white,fill] (1,0) circle(0.05);
			\draw[white,fill] (1,0.5) circle(0.05);
			\draw[white,fill] (1,1) circle(0.05);
			\draw[white,fill] (0.5,1) circle(0.05);
			\draw[white,fill] (0,1) circle(0.05);
			\draw[white,fill] (0,0.5) circle(0.05);
			\draw[white,fill] (0.5,0.5) circle(0.05);
			\draw[white] (0,0) circle(0.1);
			\draw[white] (1,0) circle(0.1);
			\draw[white] (1,1) circle(0.1);				
			\draw[white] (0,1) circle(0.1);				
			\node[] at (0.05,0.6) (A) {Nodes:};
			\draw[black,fill] (0.45,0.6) circle(0.04);
			\node[] at (1.11,0.6) (A) {$\vu$, $\varphi$, $\eta$ (quadratic)};
			\draw[black] (0.45,0.4) circle(0.05);
			\node[] at (0.90,0.4) (A) {$p$ (linear)};			
		\end{tikzpicture}
	\end{subfigure}	
	\caption{Two-dimensional Taylor-Hood interpolations.}
	\label{fig:Taylor-Hood interpolations}	
\end{figure}

%
In order to satisfy the LBB stability condition, a Taylor-Hood element is chosen for the spatial discretisation (see Figure \ref{fig:Taylor-Hood interpolations}). Thus, the velocity, phase-field variable and chemical potential interpolations are piecewise quadratic, while the pressure interpolation is piecewise linear.
\par The weak form of the NSCH problem described in Equation \eqref{eq:NSCHGovEq} reads: Find $(\vu^h$, $p^h$, $\varphi^h$, $\eta^h) \in (\mathcal{S}^h,\mathcal{P}^h,\mathcal{X}^h,\mathcal{Z}^h)$, such that, for all $(\vw^h$, $q^h$, $s^h$, $v^h) \in (\mathcal{V}^h,\mathcal{P}^h,\mathcal{Y}^h,\mathcal{W}^h)$,
\begin{subequations}
	\begin{align}
		\begin{split}
		\int_{\Omega}
		\vw^h &\cdot \left( \rho(\varphi^h) \left(  \frac{\partial\vu^h}{\partial t}
		+ \vu^h \cdot \nabla \vu^h - \bm{b} \right) - \bm{J}(\varphi^h,\eta^h) \cdot \nabla \vu^h - \kappa\eta^h \nabla \varphi^h \right)
		-\left(\nabla \cdot \vw^h \right) p^h
		\\&+\nabla \vw^h : \left( 2 \mu(\varphi^h) \nabla^s \vu^h \right) \dom = 0
		\end{split}
		\label{eq:NSCHweakA}
	\end{align}
	\vspace{-22pt}
	\begin{equation}
		\int_{\Omega} \nabla q^h \cdot \vu^h \dom = 0
		\label{eq:NSCHweakB}
	\end{equation}
	\vspace{-22pt}
	\begin{equation}
		\intom v^h \left( \frac{\partial \varphi^h}{\partial t} + \vu^h\cdot \nabla \varphi^h \right)
		+ \nabla v^h \cdot \left( M(\varphi^h) \nabla \eta^h \right) \dom = 0
		\label{eq:NSCHweakC}		
	\end{equation}
	\vspace{-22pt}
	\begin{equation}
		\intom s^h \ \left( \eta^h - f(\varphi^h) \right) - \nabla s^h \cdot \epsilon^2 \nabla \varphi^h \dom 
		- \intgam s^h \ \epsilon^2 \lVert\nabla\varphi^h\rVert \cos{(\alpha)} \dgam = 0,
		\label{eq:NSCHweakD}		
	\end{equation}
	\label{eq:NSCHweak}
\end{subequations}
where $\mathcal{S}^h$, $\mathcal{P}^h$, $\mathcal{X}^h$, $\mathcal{Z}^h$, $\mathcal{V}^h$, $\mathcal{Y}^h$ and $\mathcal{W}^h$ are the appropriate finite element spaces of piecewise continuous quadratic and linear basis functions.
%
\subsection{Navier-Stokes-Cahn-Hilliard stabilised formulation}
%
The standard SUPG/PSPG stabilisation strategy (see, for instance \cite{Tezduyar1992,Tezduyar2003,Dettmer2003,Scovazzi2007a,Hughes2017} ) is applied to the Navier-Stokes momentum equation. This allows for the use of piecewise linear equal order interpolations for $u$, $p$, $\varphi$ and $\eta$.
\par The SUPG/PSPG stabilised weak form of \eqref{eq:NSCHGovEq} reads: Find $(\vu^h$, $p^h$, $\varphi^h$, $\eta^h) \in (\mathcal{S}^h,\mathcal{P}^h,\mathcal{X}^h,\mathcal{Z}^h)$, such that, for all $(\vw^h$, $q^h$, $s^h$, $v^h) \in (\mathcal{V}^h,\mathcal{P}^h,\mathcal{Y}^h,\mathcal{W}^h)$,
\begin{subequations}
	\begin{align}
		\begin{split}
		&\int_{\Omega}
		\vw^h \cdot \left( \rho(\varphi^h)\left(  \frac{\partial\vu^h}{\partial t}
		+ \vu^h \cdot \nabla \vu^h - \bm{b} \right)
		- \bm{J}(\varphi^h,\eta^h) \cdot \nabla \vu^h - \kappa\eta^h \nabla \varphi^h \right)
		-\left(\nabla \cdot \vw^h\right) p^h
		\\&\quad\qquad\qquad+\nabla \vw^h : \left( 2 \mu(\varphi^h) \nabla^s \vu^h \right) \dom
		\\&+\sum_{e=1}^{n_{el}} \intome \Big[ \tau_{\vu} \rho(\varphi^h) \left(\vu^h \cdot \nabla \vw^h \right) + \tau_p \nabla q^h \Big] 
		\\&\quad\qquad\qquad\cdot \Big[ \rho(\varphi^h)\left( \frac{\partial\vu^h}{\partial t} + \vu^h \cdot \nabla \vu^h - \bm{b} \right) - \bm{J}(\varphi^h,\eta^h) \cdot \nabla \vu^h - \kappa\eta^h \nabla \varphi^h + \nabla p^h \Big] \dom = 0
		\end{split}
		\label{eq:NSCHstabWeakA}
	\end{align}
	\vspace{-22pt}
	\begin{equation}
		\int_{\Omega} \nabla q^h \cdot \vu^h \dom = 0
		\label{eq:NSCHstabWeakB}
	\end{equation}
	\vspace{-22pt}
	\begin{equation}
		\intom v^h \left( \frac{\partial \varphi^h}{\partial t} + \vu^h\cdot \nabla \varphi^h \right)
		+ \nabla v^h \cdot \left(M(\varphi^h)\nabla\eta^h\right) \dom = 0
		\label{eq:NSCHstabWeakC}		
	\end{equation}
	\vspace{-22pt}
	\begin{equation}
		\intom s^h \left( \eta^h - f \right) - \nabla s^h \cdot \epsilon^2 \nabla \varphi^h \dom 
		- \intgam s^h \ \epsilon^2 \lVert\nabla\varphi^h\rVert \cos{(\alpha)} \dgam= 0.
		\label{eq:NSCHstabWeakD}		
	\end{equation}
	\label{eq:NSCHstabWeak}
\end{subequations}
where $\mathcal{S}^h$, $\mathcal{P}^h$, $\mathcal{X}^h$, $\mathcal{Z}^h$, $\mathcal{V}^h$, $\mathcal{Y}^h$ and $\mathcal{W}^h$ are the appropriate finite element spaces of piecewise continuous linear basis functions.
\par Following \cite{Dettmer2016}, the stabilisation parameters, $\tau_{\vu}$ and $\tau_p$ are defined as
\begin{align}
	\tau_{\vu} = \left( \frac{1}{\tau_p^2} + \left(\frac{2 \rho \lVert \vu_e \rVert}{h_e}\right)^2 \right)^{-\frac{1}{2}},
	\quad
	\tau_{p} = \frac{h_e^2}{4\mu},
	\label{eq:stab parameters}
\end{align}
where $h_e$ is the characteristic size of the element, evaluated as $h_e=V_e^{1/d}$, with $d$ representing the number of spatial dimensions and $V_e$, the element volume or area. The vector $\vu^e$ is the velocity in the element centroid.
%
%
\subsection{Temporal discretisation}
%
The generalised-$\alpha$ method is employed for the temporal discretisation.
This method is an unconditionally stable, implicit single-step time integration scheme (refer to \cite{Chung2008,Jansen2000}). The scheme allows for high frequency damping to be controlled by without compromising the second order accuracy (see \cite{Dettmer2003,Lovric2018}). Applying the generalised-$\alpha$ method to a generic first order problem gives
\begin{equation}
	\dvu^{n + \alpha_m} = \bm{f} \left(t^{n + \alpha_f},\bm{u}^{n + \alpha_f} \right),
	\label{eq:AMgeneral}
\end{equation}		
with
\begin{align}
	\dvu^{n+\alpha_m} &= \left(1 - \alpha_m \right) \dvu^n + \alpha_m \dvu^{n+1}
	\label{eq:AMacc}
	\\t^{n+\alpha_f} &= \left(1 - \alpha_f \right)t^n + \alpha_f t^{n+1}
	\label{eq:AMtime}	
	\\\vu^{n+\alpha_f} &= \left(1 - \alpha_f \right)\vu^n + \alpha_f \vu^{n+1}
	\label{eq:AMvelf}
	\\\frac{\vu^{n+1} - \vu^{n}}{\dt} &= \left(1 - \gamma \right) \dvu^n + \gamma \dvu^{n+1},
	\label{eq:AMvel}
\end{align}
and,
\begin{equation}
	\alpha_m = \frac{1}{2} \ \frac{3 - \rho_\infty}{1 + \rho_\infty}
	\text{,}\qquad
	\alpha_f = \frac{1}{1 + \rho_\infty}	
	\text{,}\qquad
	\gamma = \frac{1}{2} + \alpha_m - \alpha_f,
	\label{eq:AMpar}
\end{equation}
where $\gamma$ must not be confused with the surface tension parameter. Note that $\alpha_m$, $\alpha_f$ and $\gamma$ are expressed in terms of the spectral radius $\rhoi$ for an infinitely large time step size.
\par This scheme has been successfully employed in solution of a number of challenging coupled field problems \cite{Dettmer2006,Dettmer2016,Kadapa2017,Kadapa2018}.
%
\subsection{Solver}
%
The problems described by Equations \eqref{eq:NSCHweak} and \eqref{eq:NSCHstabWeak} are highly nonlinear and thus a Newton-Raphson procedure is employed based on the consistent linearisation of all nonlinear terms. For smaller scale problems, a direct linear solver (PARDISO \cite{pardiso-6.0a,pardiso-6.0b,pardiso-6.0c}) is used, whereas for larger scale problems, an iterative parallel solver (PETSC \cite{petsc-efficient}) with a block Jacobi pre-conditioner is employed.
%
\subsection{Adaptive time stepping}
%
Several of the numerical examples in Section \ref{sec: Numerical examples} implement an adaptive time stepping procedure defined by 
\begin{align}
	\Delta t = \left( t^{n+1}-t^{n} \right)\theta^{\left(n_{\mathrm{iter}}-n_{\mathrm{opt}}\right)},
\end{align}
where $n_{\mathrm{iter}}$ and $n_{\mathrm{opt}}$ are respectively the number of Newton iteration steps required in the previous time step and the desired number of iteration steps. The constant $\theta$ is typically chosen between $0.5$ and $1$.
%
%
\section{Numerical examples} \label{sec: Numerical examples}
%
In the following examples the proposed methodology is applied to a number of two and three-dimensional problems. Sections \ref{sec: numerical examples: static bubble}-\ref{sec: numerical examples: filling drop_3D} present surface tension dominated problems, while the simulations described in Sections \ref{sec: numerical examples: broken dam}-\ref{sec: numerical examples: rayleigh-taylor instability} do not feature any surface tension effects.
\par The properties of water and air used in the following examples are:
\begin{alignat*}{2}
	\textrm{water:} \quad
	&\rho = 0.998 \ &&\si[per-mode=symbol]{\gram\per\centi\metre^3}\\[-5pt]
	&\mu  = 0.0101 \ &&\si[per-mode=symbol]{\gram\per\centi\metre\per\second}\\[-5pt]
	\textrm{air:} \quad
	&\rho = 0.0012 \ &&\si[per-mode=symbol]{\gram\per\centi\metre^3}\\[-5pt]
	&\mu  = 0.000182 \ &&\si[per-mode=symbol]{\gram\per\centi\metre\per\second}\\[-5pt]
	\textrm{water/air:} \quad 
	&\gamma = 73.0 \ &&\si[per-mode=symbol]{\gram\per\second^2}\\[-5pt]
	\textrm{gravity:} \quad 
	&g = 980.0 \ &&\si[per-mode=symbol]{\centi\metre\per\second^2}	
\end{alignat*}
Unless otherwise stated, the computations are based on the stabilised formulation given by Equation \eqref{eq:NSCHstabWeak} and the Cahn-Hilliard parameters are set as follows: 
\begin{alignat*}{2}
	\textrm{mobility function:} &\quad
	M_0 &&= 10^{-3} \ \si[per-mode=symbol]{\centi\metre^2\per\second}\\[-5pt]
	\textrm{well height:} &\quad
	W &&= 0.25\\[-5pt]
	\textrm{interface thickness:} &\quad 
	\epsilon  &&= 2h
\end{alignat*}
where $h$ corresponds to the characteristic size of the element used, which is taken as the largest element size of the respective mesh.
\par The relative diffusive flux term $\left(\bm{J}\cdot\nabla\right) \vu$ is neglected in the following examples as its effect is observably negligible. 
%
%
\subsection{Static bubble} \label{sec: numerical examples: static bubble}
\begin{figure}[tb!]
	\centering
	\captionsetup[subfigure]{labelformat=empty}
	\includegraphics[width=0.35\textwidth]{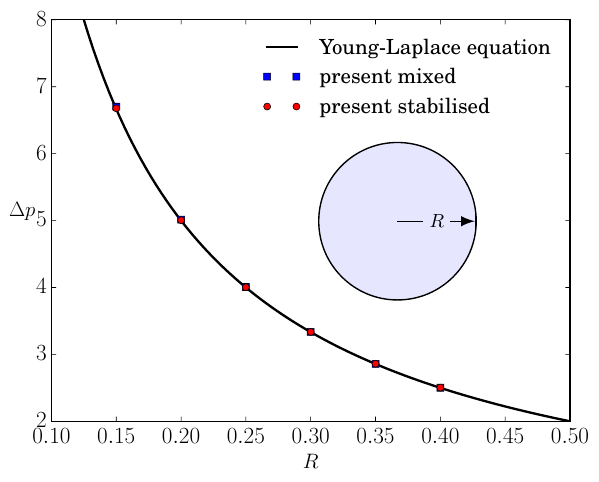}
	\caption{Static bubble: Comparison between numerical and analytical (Young-Laplace) pressure drop.}
	\label{fig:static_bubble_comp2}
\end{figure}

%
A bubble of radius $R$ is placed at coordinate $[0.5,0.5]$ in the centre of a square domain of dimensions $[0,1]\times[0,1]$. The densities and viscosities, as well as the surface tension coefficient, are set to $1$. The initial condition is taken as
\begin{align}
	\varphi(\bm{x}) = \tanh\left( \frac{R - d(\bm{x})}{\sqrt{2}\epsilon} \right).
\end{align}
where $d(\bm{x})$ is the Euclidean distance between the bubble centre and $\bm{x}$.
The pressure difference between the centre of the bubble and a point outside the bubble at coordinate $[1.0,0.5]$, are compared to the Young-Laplace equation,
\begin{align}
	\Delta p = \frac{\gamma}{R}.
	\label{eq: Young-Laplace}
\end{align}
Two meshes are considered: a mesh with $256\times256$ mixed Taylor-Hood elements, and a mesh with $512\times512$ stabilised linear quadrilateral elements. For both meshes, the interface thickness is set as $\epsilon = 1/128$. All simulations are run until $T=10$.
Figure \ref{fig:static_bubble_comp2} shows the excellent agreement between the analytical expression and the numerical experiment, for a range of values of $R$.
%
\subsection{Small amplitude oscillations of a two-dimensional drop} \label{sec: numerical examples: small oscillations}
%
A small two-dimensional drop of water with radius $R=0.0125 \ \si{cm}$ is placed in a square domain filled with air. The initial configuration of the drop is described by the equation 
\begin{align}
	r_n(\theta) = R + A \cos(n\theta),
\end{align}
where $A$ is set to $0.02 R$ for all simulations, and $n$ is the mode order (see \cite{Dettmer2003a,Dettmer2006,Saksono2006a,BrantFoote1973}). The effects of gravity are neglected, and the mobility is set as $M_3(\varphi)$, with $D=10^{-2}$. The mesh considered has $512\times512$ linear quadrilateral elements, and a fixed time step size of $\Delta t =1\times10^{-6} \ \si{\second}$ is used.
\par
\begin{figure}[tb!]
	\captionsetup[subfigure]{labelformat=empty}
	\centering
	\begin{subfigure}{.32\textwidth}
		\centering
		\includegraphics[width = \textwidth]{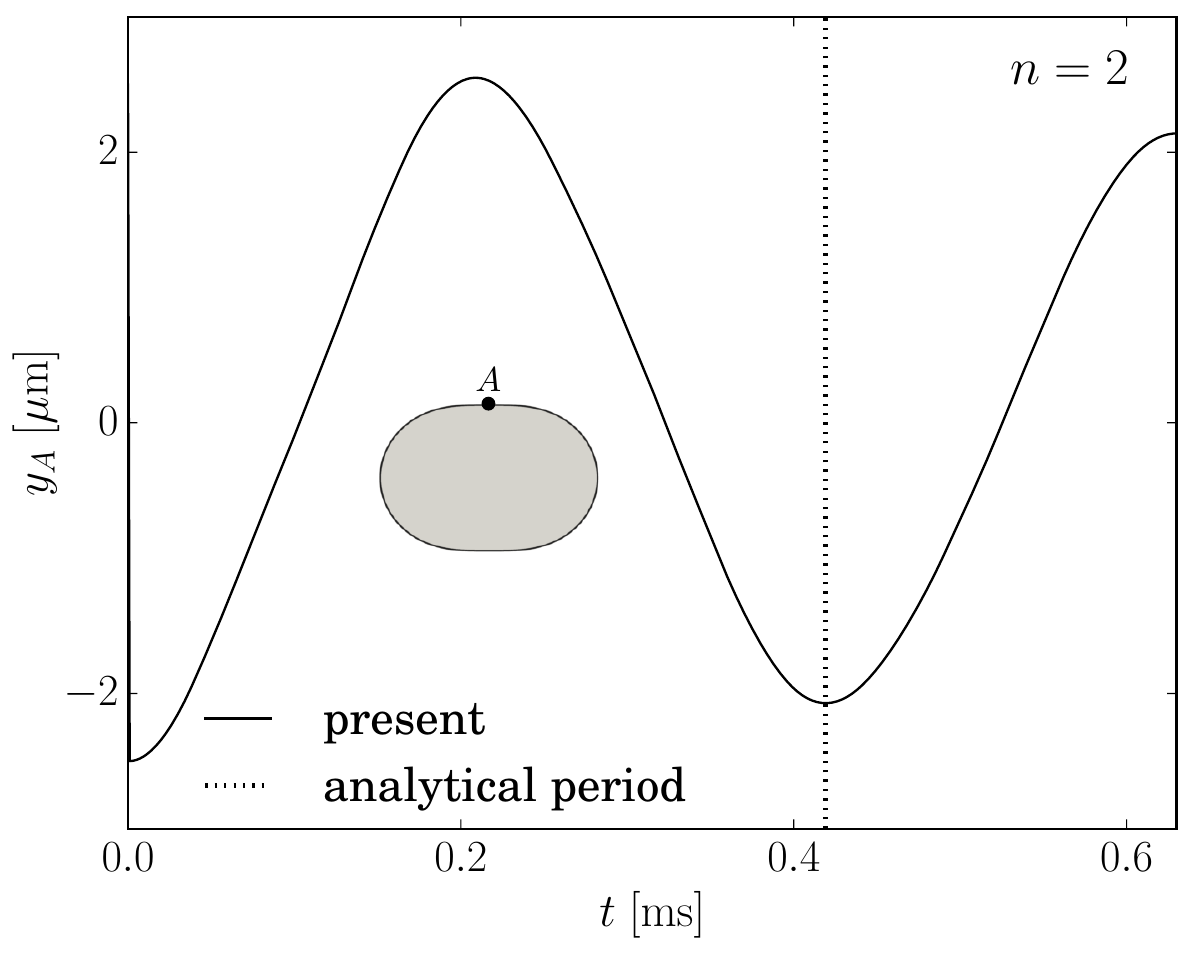}
	\end{subfigure}
	\begin{subfigure}{.32\textwidth}
		\centering
		\includegraphics[width = \textwidth]{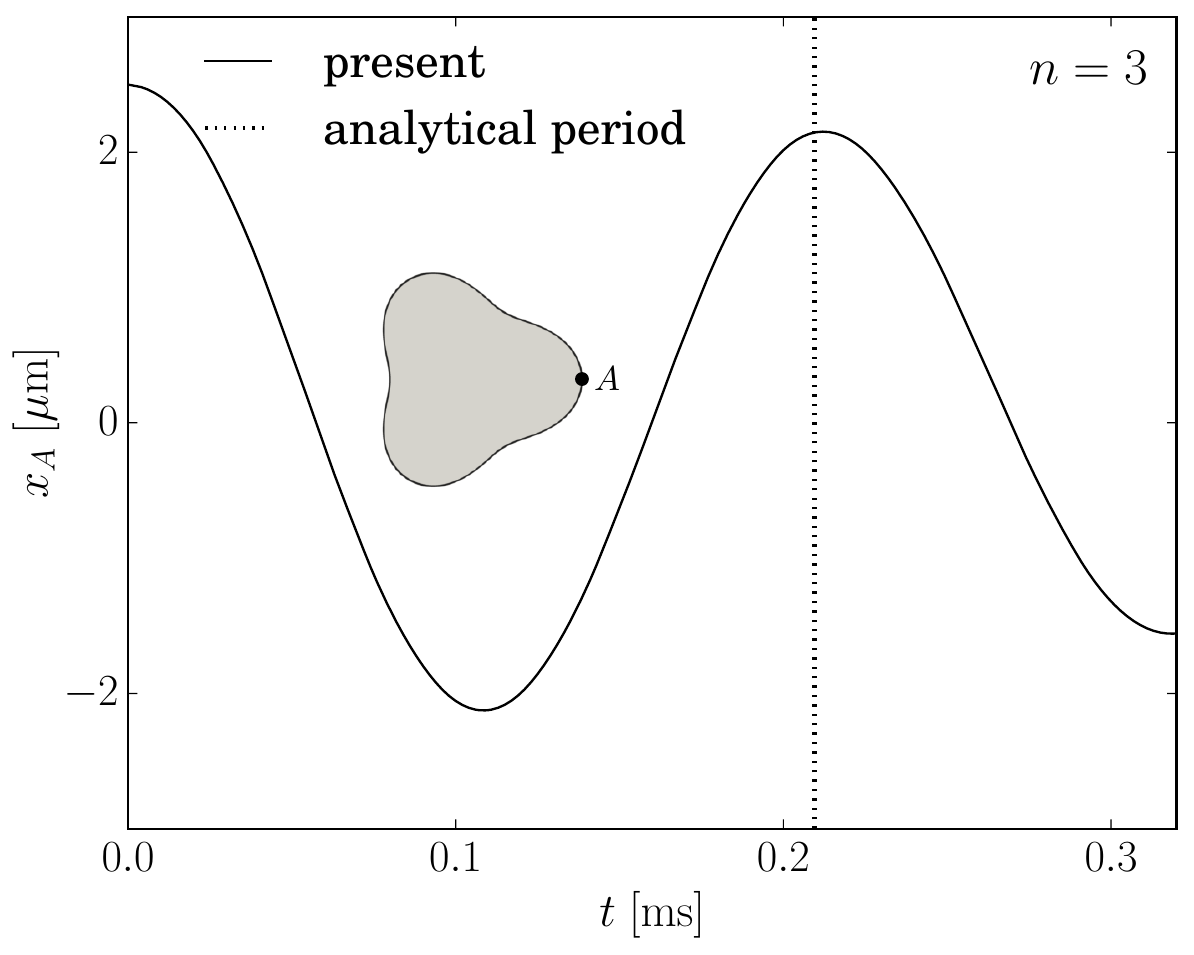}
	\end{subfigure}	
	\begin{subfigure}{.32\textwidth}
		\centering
		\includegraphics[width = \textwidth]{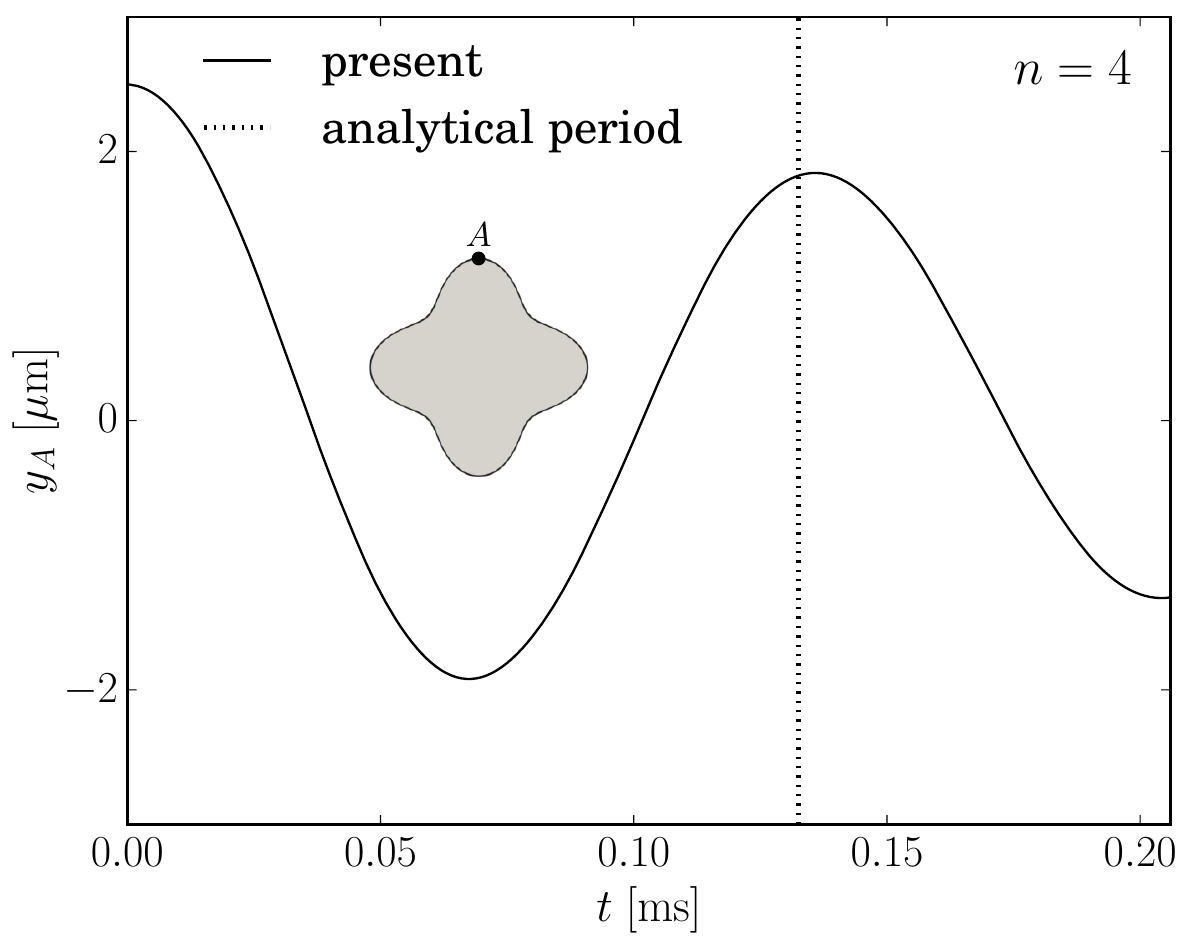}
	\end{subfigure}
	\caption{Small amplitude oscillations of a two-dimensional drop: Amplitude evolution for $n=2,3,4$, with $512\times512$ linear elements.}
	\label{fig:small_oscillations_amplitude}	
\end{figure}

 Figure \ref{fig:small_oscillations_amplitude} shows the evolution of the oscillation amplitude for modes $n=2,3,4$. The observed period matches very well with the analytical period,
\begin{align}
	\tau = \frac{2\pi}{\omega_n},
	\label{eq: oscillations analytical period}
\end{align}
where following \cite{Saksono2006a},
\begin{align}
	\omega_n^2 = \left( n^3 - n \right) \frac{\gamma}{\rho R^3},
\end{align}
A comparison of the numerical and analytical periods is given in Table \ref{tab:small oscillations}.
\begin{table}[bt!]
	\centering
	\caption{Small amplitude oscillations of a two-dimensional drop: Numerical and analytical periods.}
	
	\begin{tabular}{ ccc }
		\toprule
		$n$ & $\tau$ [numerical] & $\tau$ [analytical \eqref{eq: oscillations analytical period}] \\
		\midrule
		$2$ & $4.190 \cdot 10^{-4}$ & $4.192 \cdot 10^{-4}$ \\
		$3$ & $2.110 \cdot 10^{-4}$ & $2.096 \cdot 10^{-4}$ \\
		$4$ & $1.350 \cdot 10^{-4}$ & $1.326 \cdot 10^{-4}$ \\	
		\bottomrule				
	\end{tabular}
	\label{tab:small oscillations}
\end{table}
%
\subsection{Large amplitude oscillation of a two-dimensional drop} \label{sec: numerical examples: large oscillations}
%
A two-dimensional drop is set up similarly to Section \ref{sec: numerical examples: small oscillations}, with a larger initial amplitude $A=0.2R$. 
\par
\begin{figure}[tb!]
	\captionsetup[subfigure]{labelformat=empty}
	\centering
	\begin{subfigure}{.32\textwidth}
		\centering	
		\includegraphics[width = \textwidth]{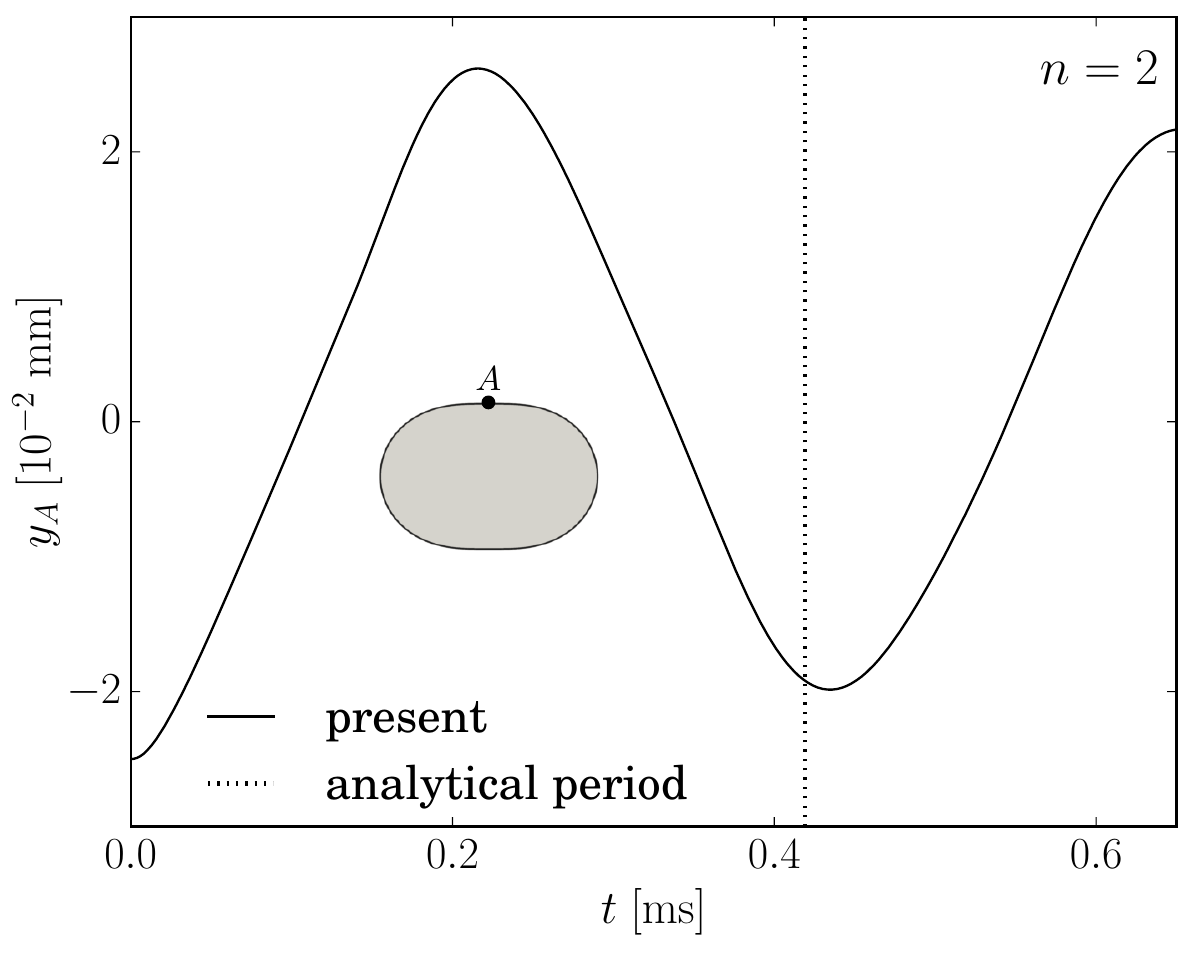}	
	\end{subfigure}
	\begin{subfigure}{.32\textwidth}
		\centering	
		\includegraphics[width = \textwidth]{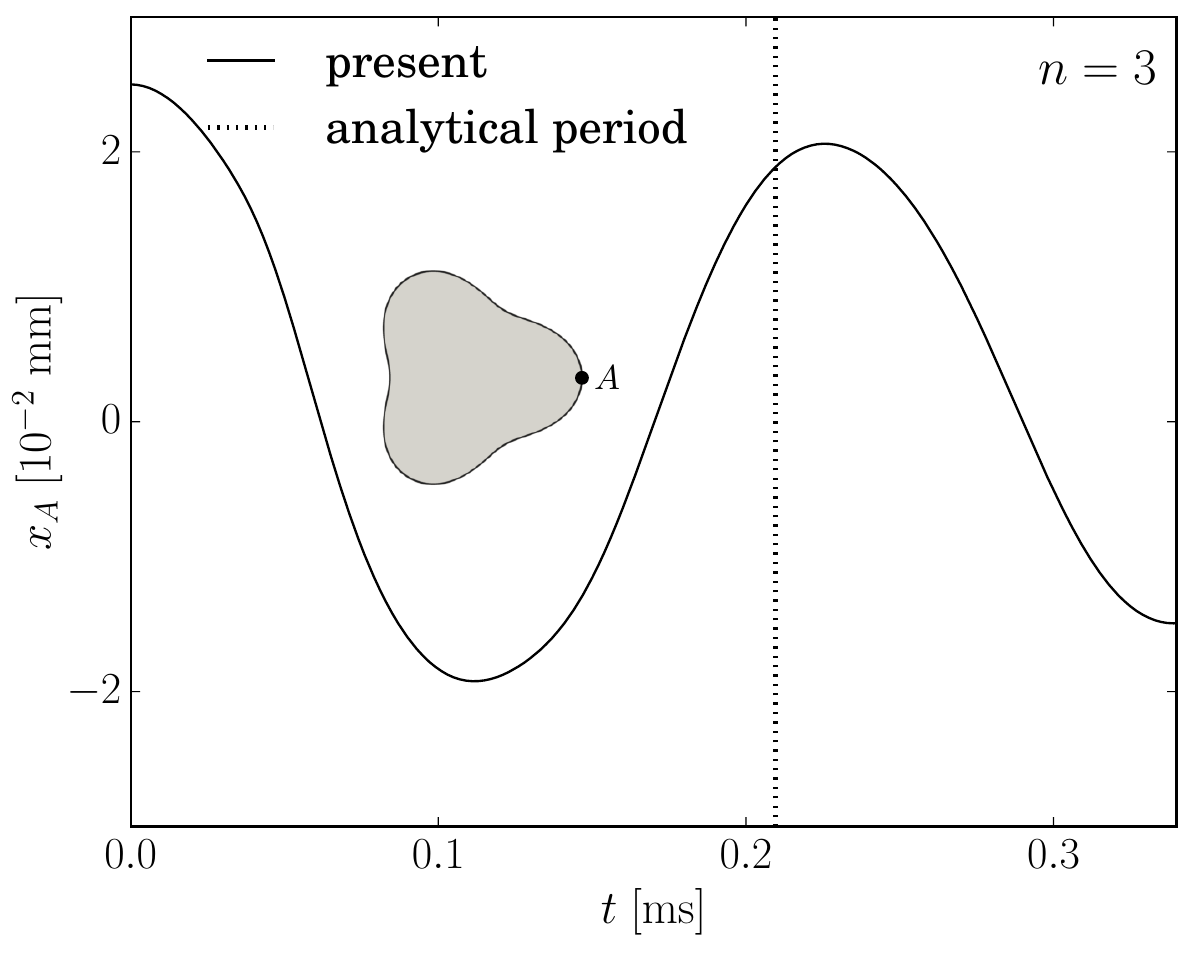}	
	\end{subfigure}
	\begin{subfigure}{.32\textwidth}
		\centering	
		\includegraphics[width = \textwidth]{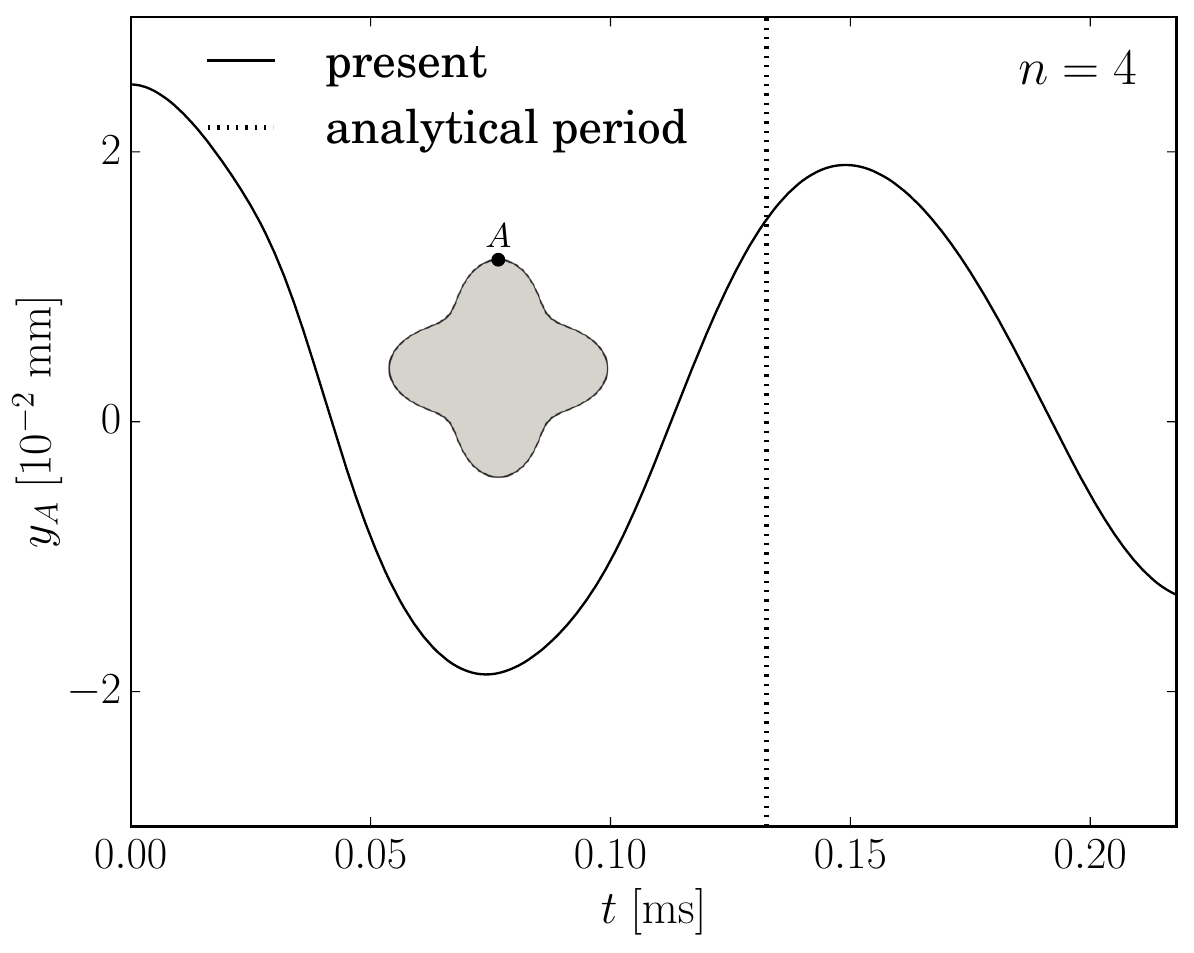}	
	\end{subfigure}	
	\caption{Large amplitude oscillation of a two-dimensional drop: Amplitude evolution for $n=2,3,4$, with $512\times512$ linear elements.}
	\label{fig:large_oscillations_amplitude}
\end{figure}

\begin{figure}[tb!]
	\captionsetup[subfigure]{labelformat=empty}
	\centering
	\begin{subfigure}{.163\textwidth}
		\centering
		\includegraphics[width = 0.95\textwidth]{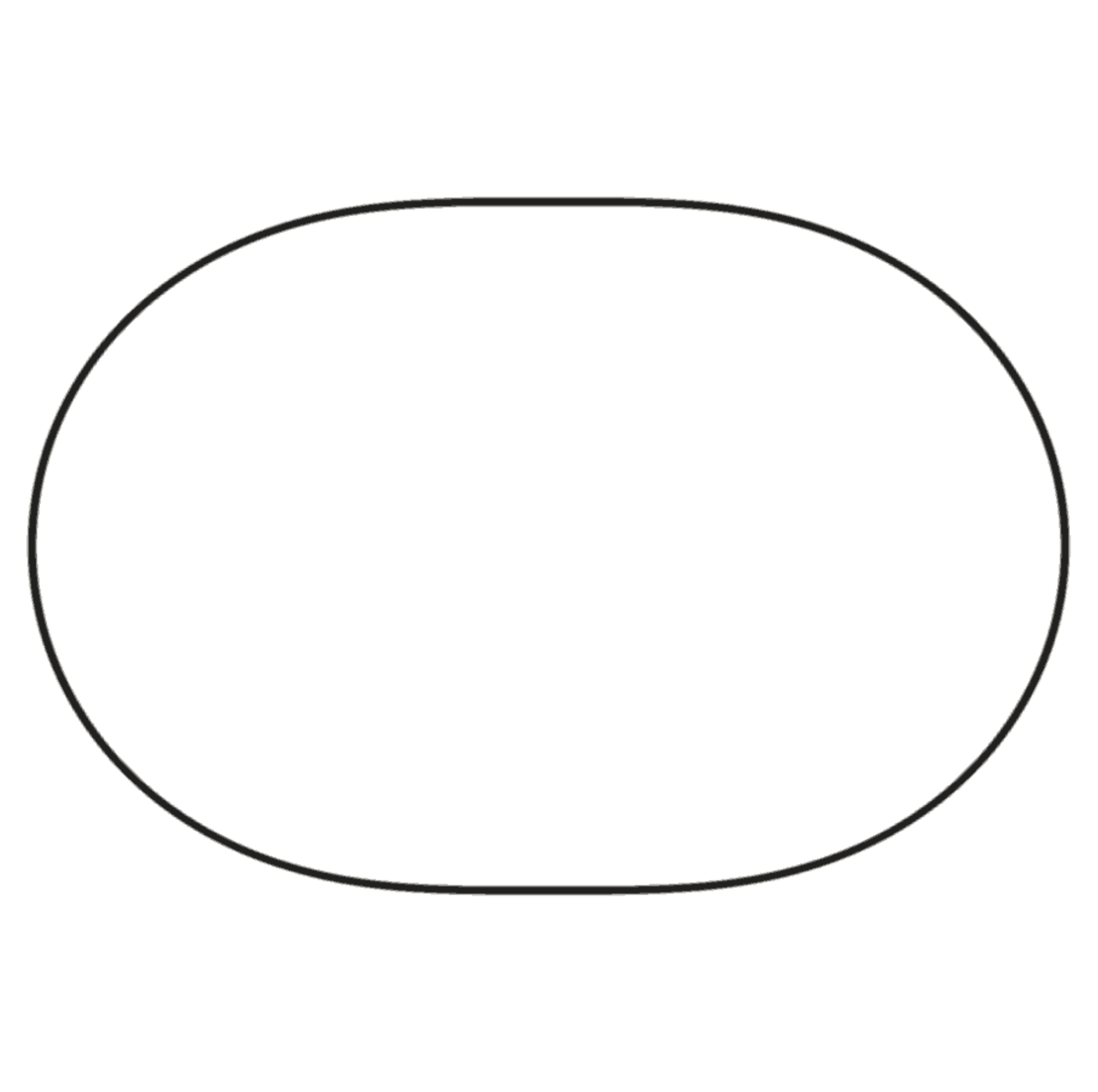}
		\subcaption{$t=0 \ \si{\second}$}
	\end{subfigure}
	\begin{subfigure}{.163\textwidth}
		\centering	
		\includegraphics[width = 0.95\textwidth]{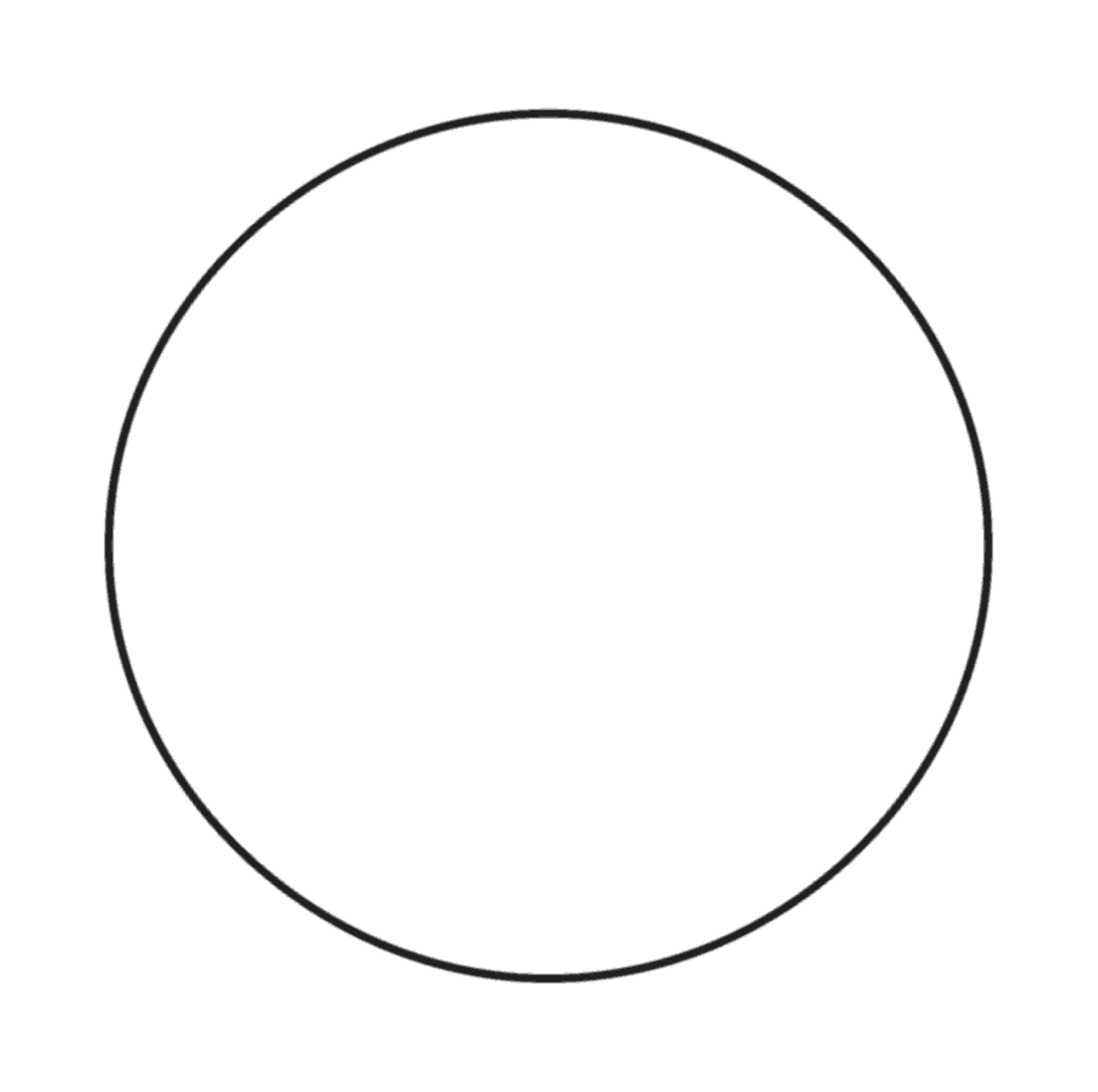}
		\subcaption{$t=1.08\cdot 10^{-4} \ \si{\second}$}	
	\end{subfigure}
	\begin{subfigure}{.163\textwidth}
		\centering		
		\includegraphics[width = 0.95\textwidth]{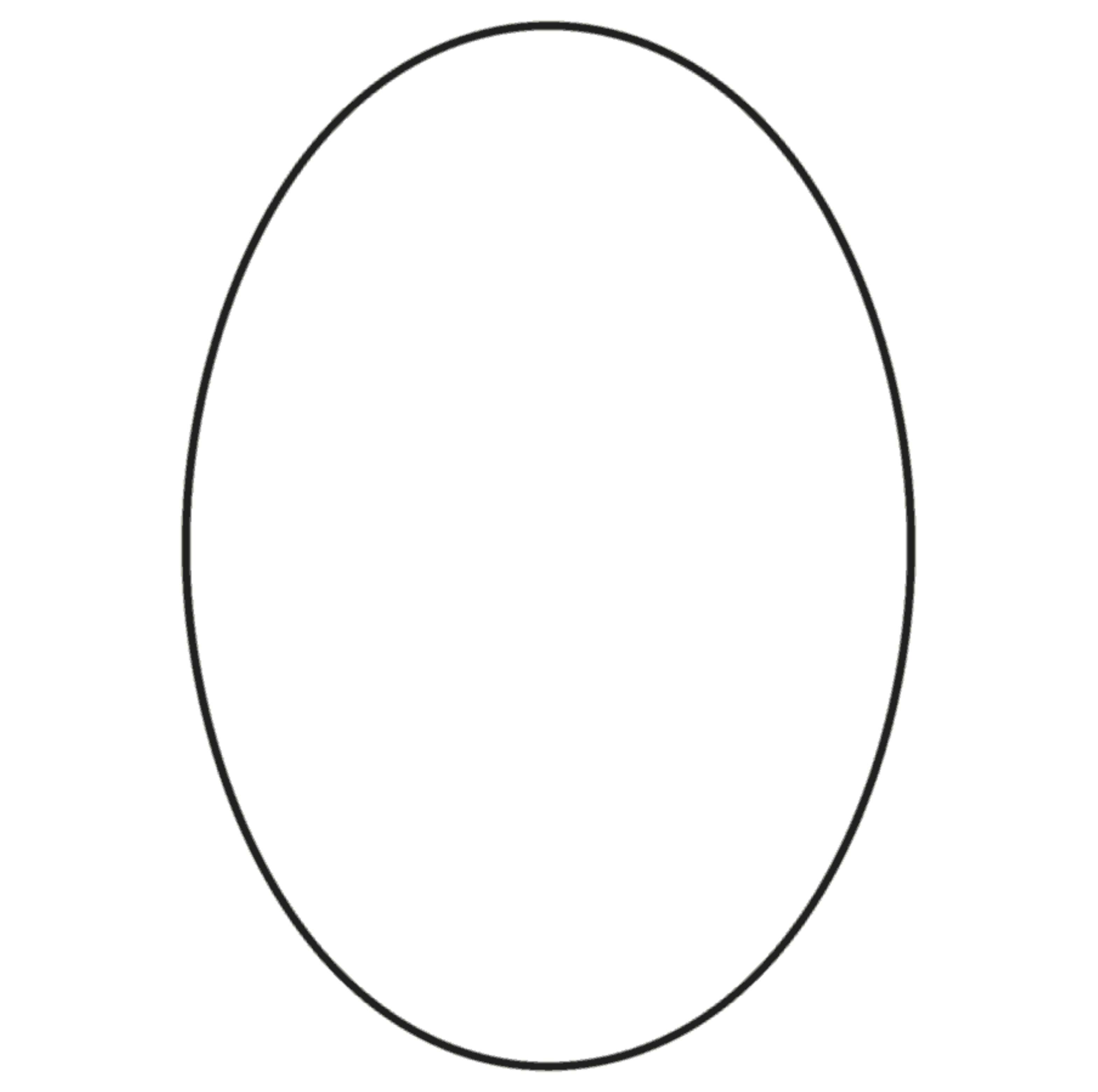}
		\subcaption{$t=2.2\cdot 10^{-4} \ \si{\second}$}
	\end{subfigure}
	\begin{subfigure}{.163\textwidth}
		\centering		
		\includegraphics[width = 0.95\textwidth]{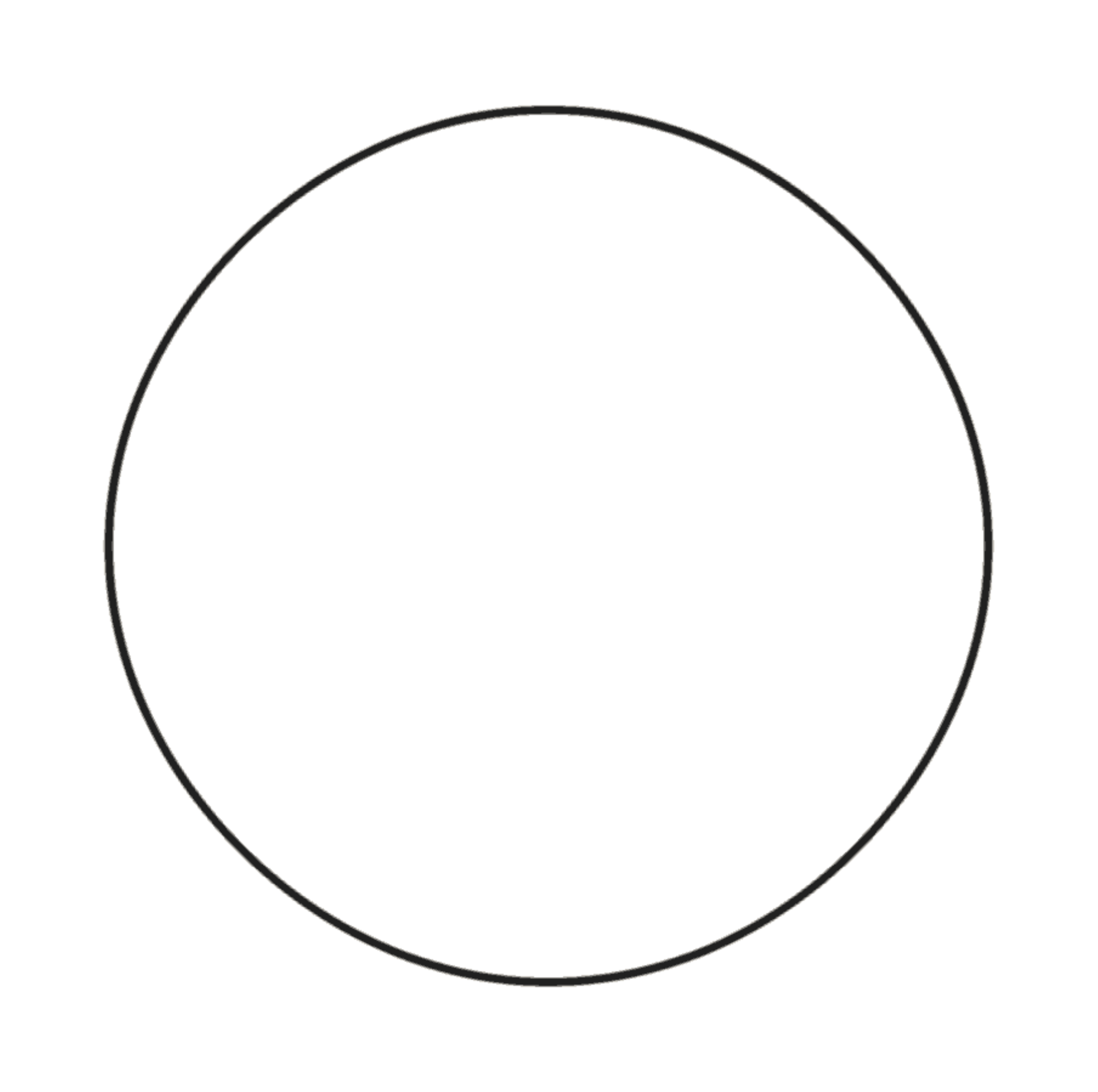}
		\subcaption{$t=3.332\cdot 10^{-4} \ \si{\second}$}
	\end{subfigure}
	\begin{subfigure}{.163\textwidth}
		\centering		
		\includegraphics[width = 0.95\textwidth]{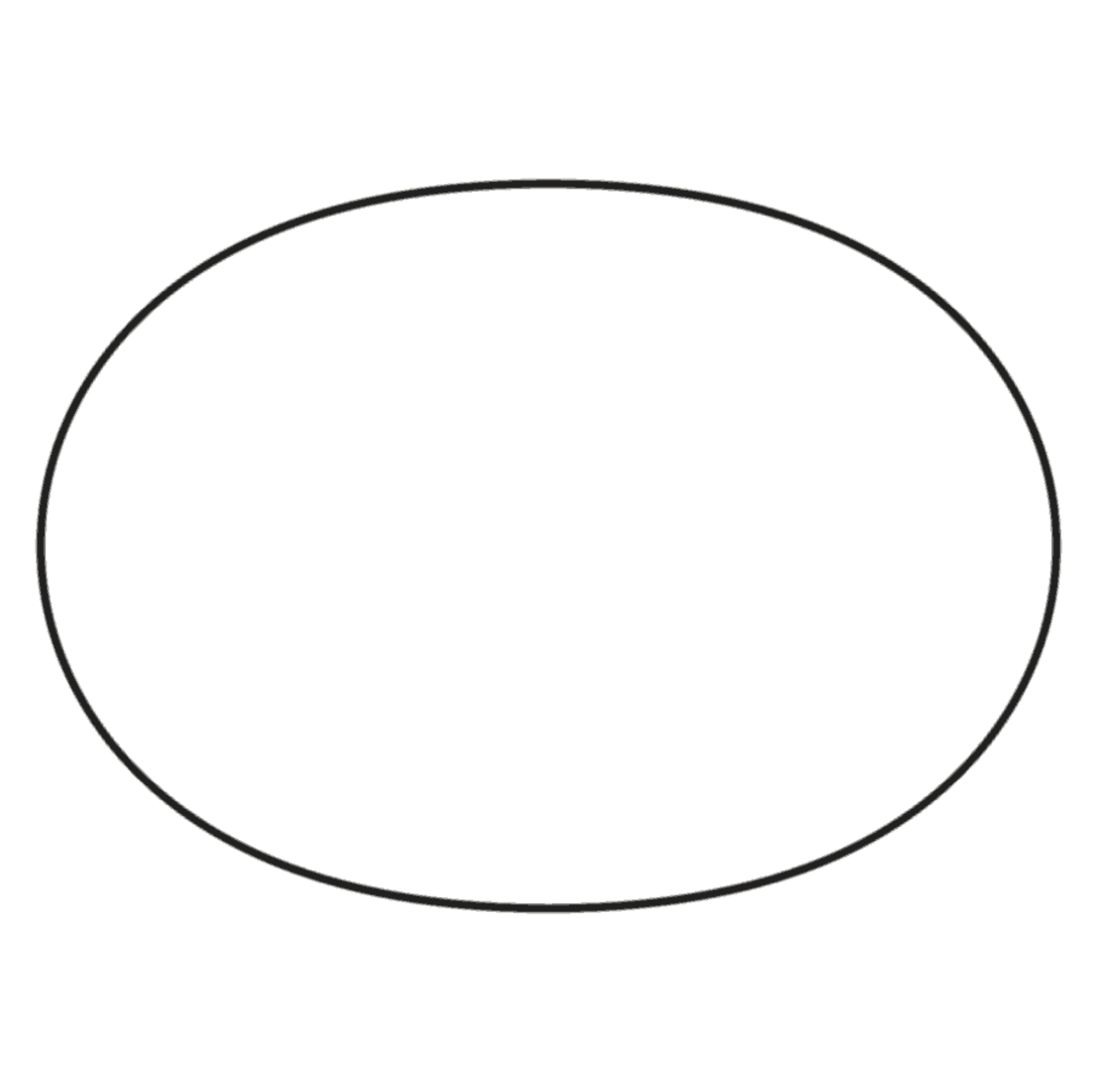}
		\subcaption{$t=4.4\cdot 10^{-4} \ \si{\second}$}
	\end{subfigure}
	\begin{subfigure}{.163\textwidth}
		\centering
		\includegraphics[width = 0.95\textwidth]{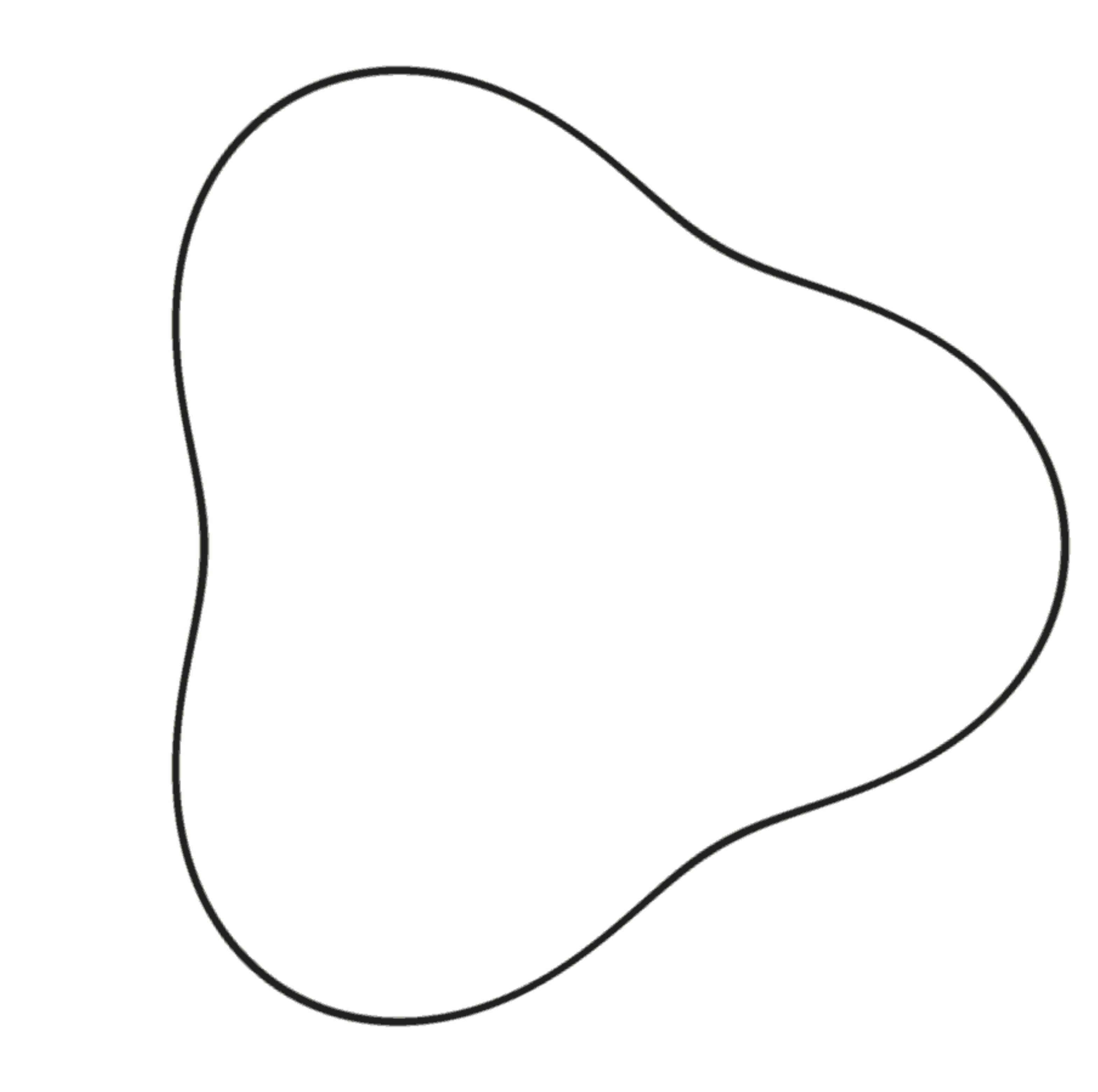}
		\subcaption{$t=0 \ \si{\second}$}
	\end{subfigure}
	\begin{subfigure}{.163\textwidth}
		\centering	
		\includegraphics[width = 0.95\textwidth]{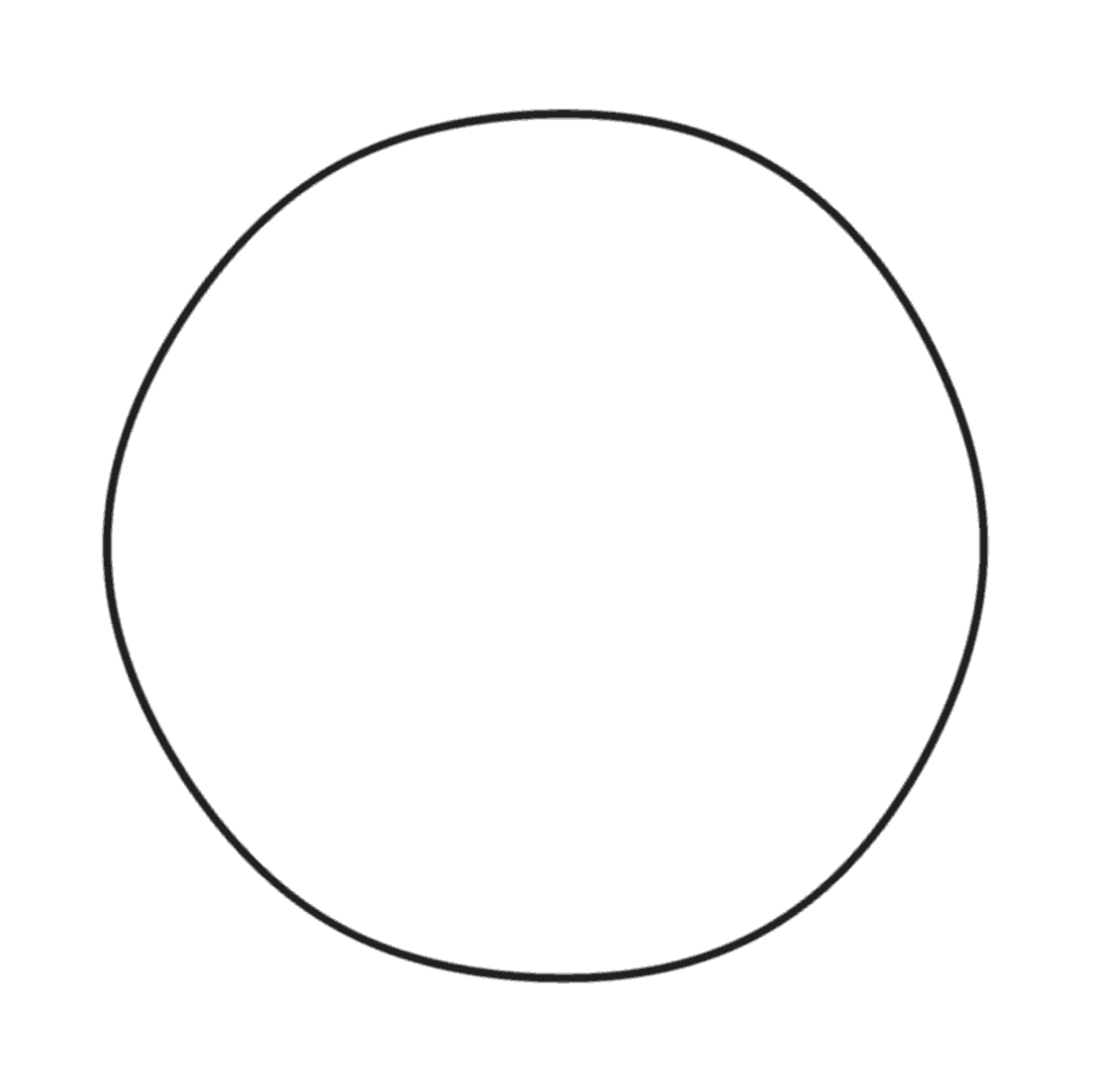}
		\subcaption{$t=6\cdot 10^{-5} \ \si{\second}$}	
	\end{subfigure}
	\begin{subfigure}{.163\textwidth}
		\centering		
		\includegraphics[width = 0.95\textwidth]{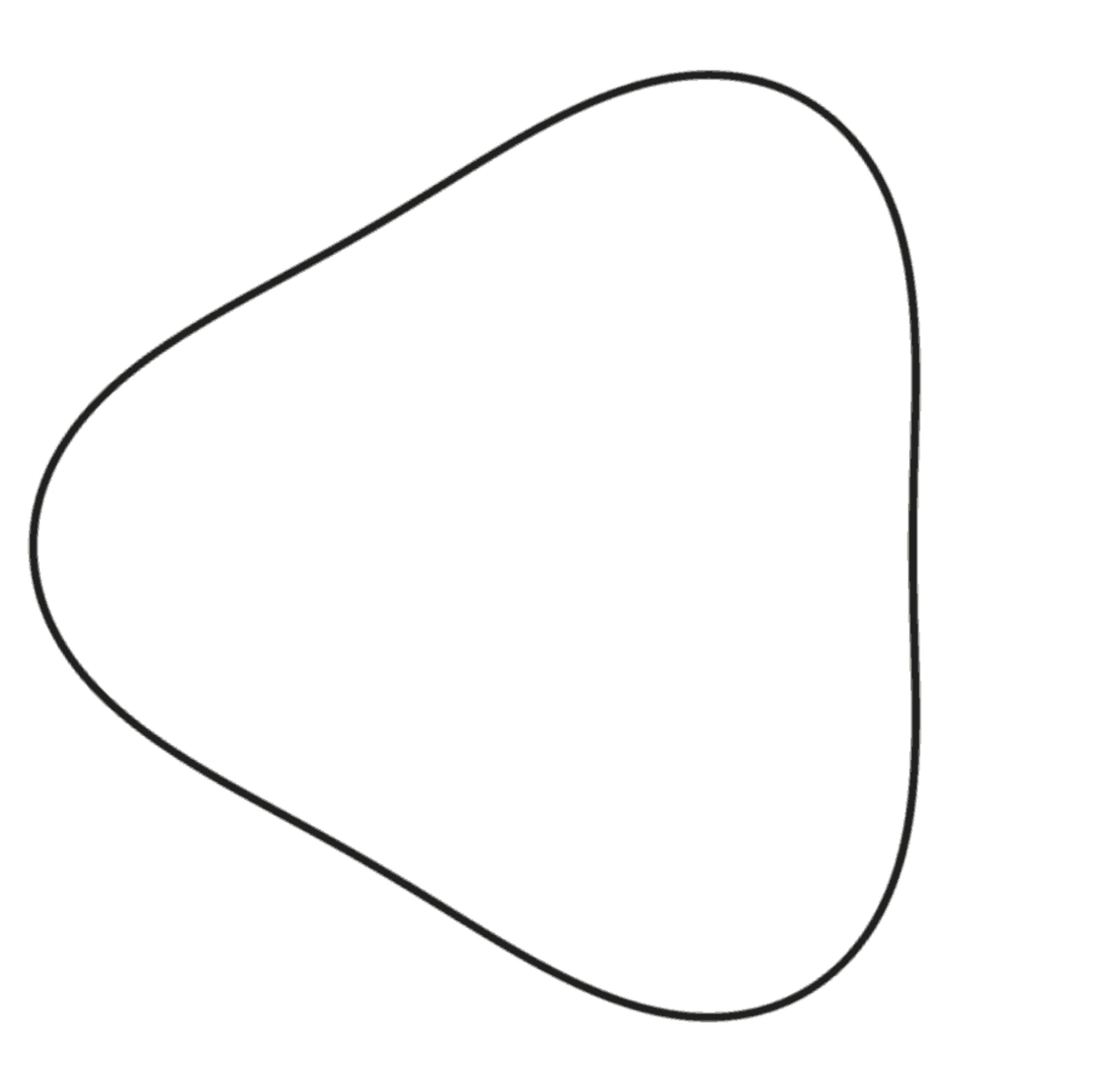}
		\subcaption{$t=1.12\cdot 10^{-4} \ \si{\second}$}	
	\end{subfigure}
	\begin{subfigure}{.163\textwidth}
		\centering		
		\includegraphics[width = 0.95\textwidth]{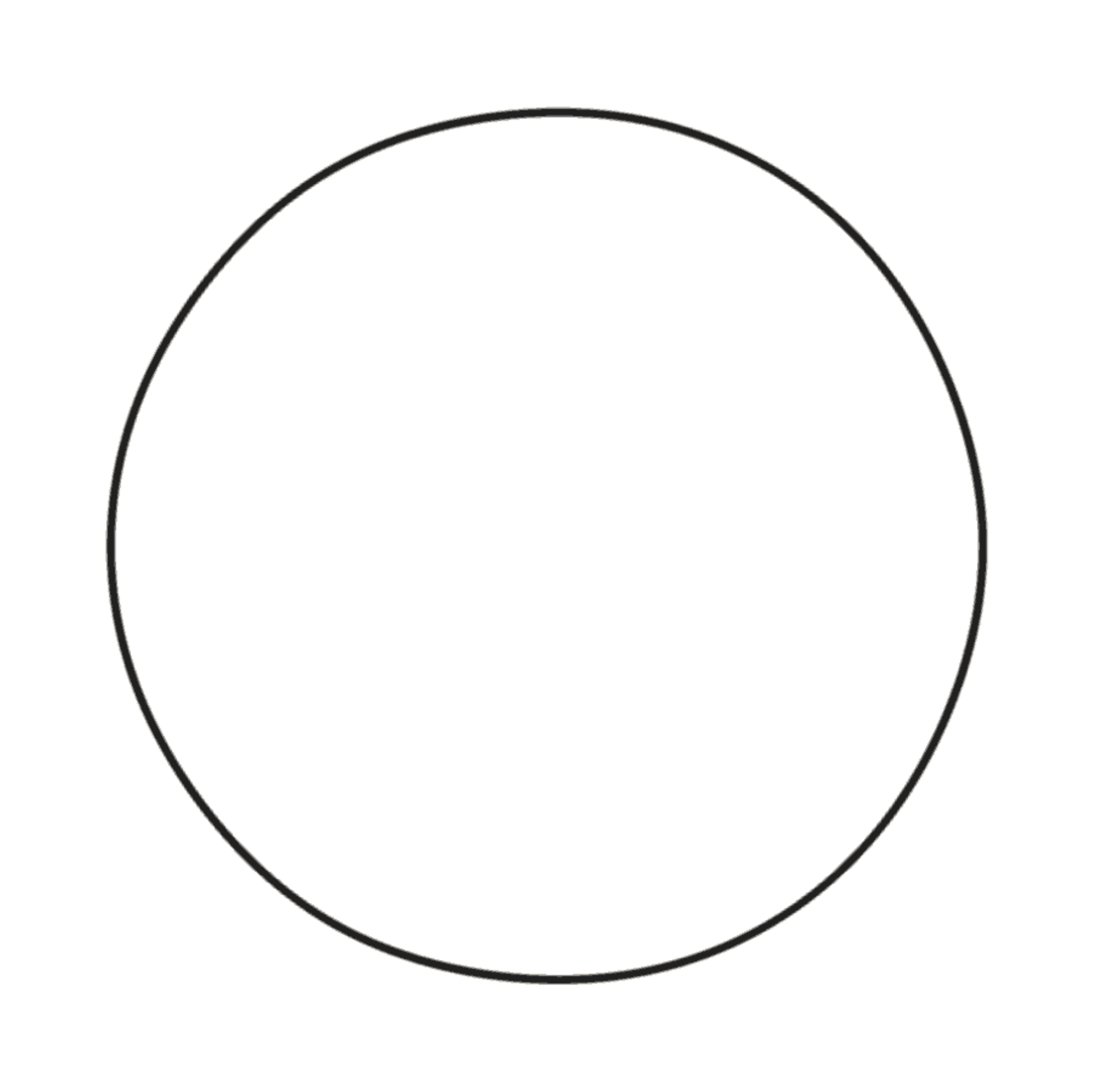}
		\subcaption{$t=1.72\cdot 10^{-4} \ \si{\second}$}	
	\end{subfigure}
	\begin{subfigure}{.163\textwidth}
		\centering		
		\includegraphics[width = 0.95\textwidth]{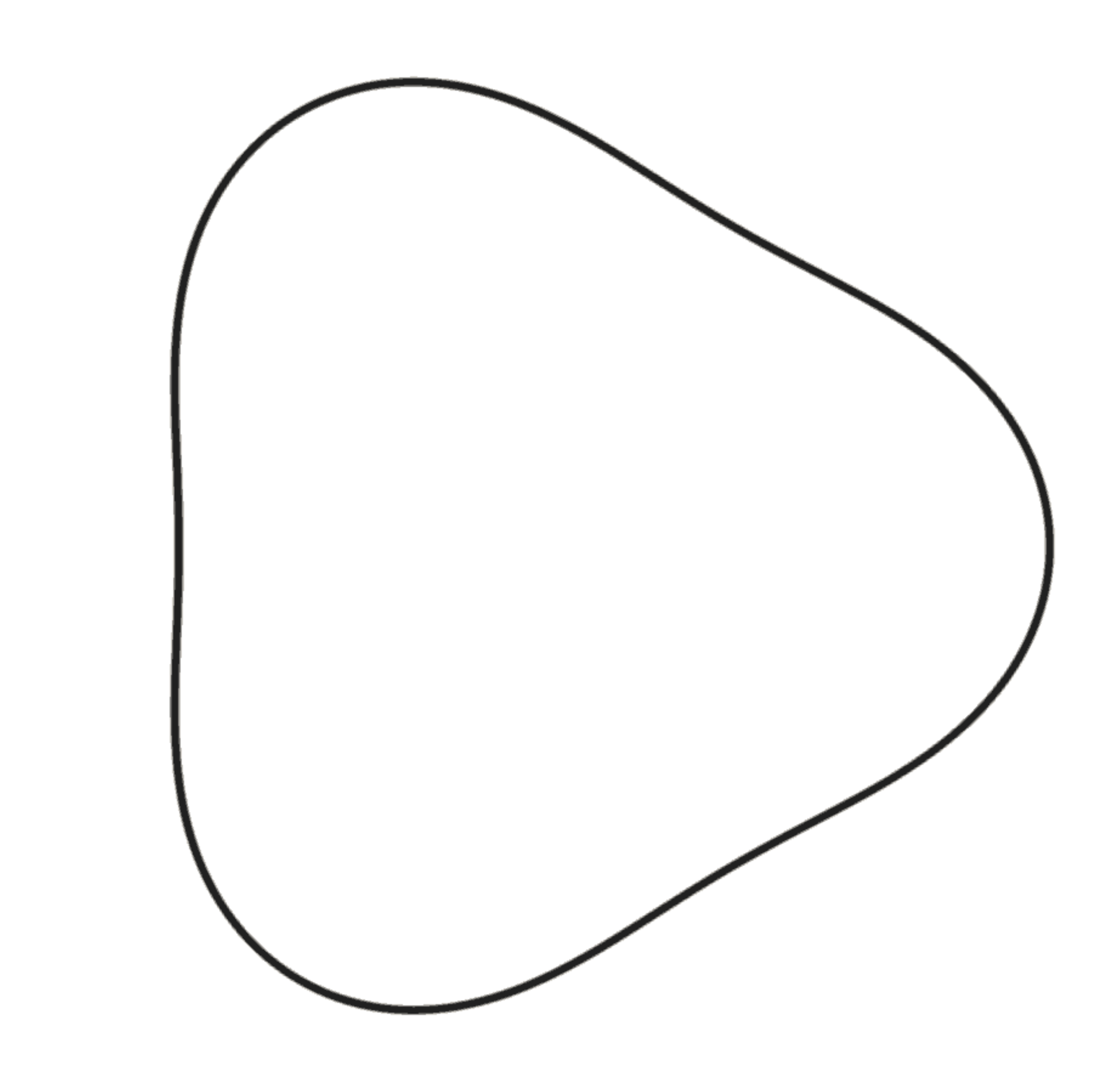}
		\subcaption{$t=2.2\cdot 10^{-4} \ \si{\second}$}		
	\end{subfigure}
	\begin{subfigure}{.163\textwidth}
		\centering
		\includegraphics[width = 0.95\textwidth]{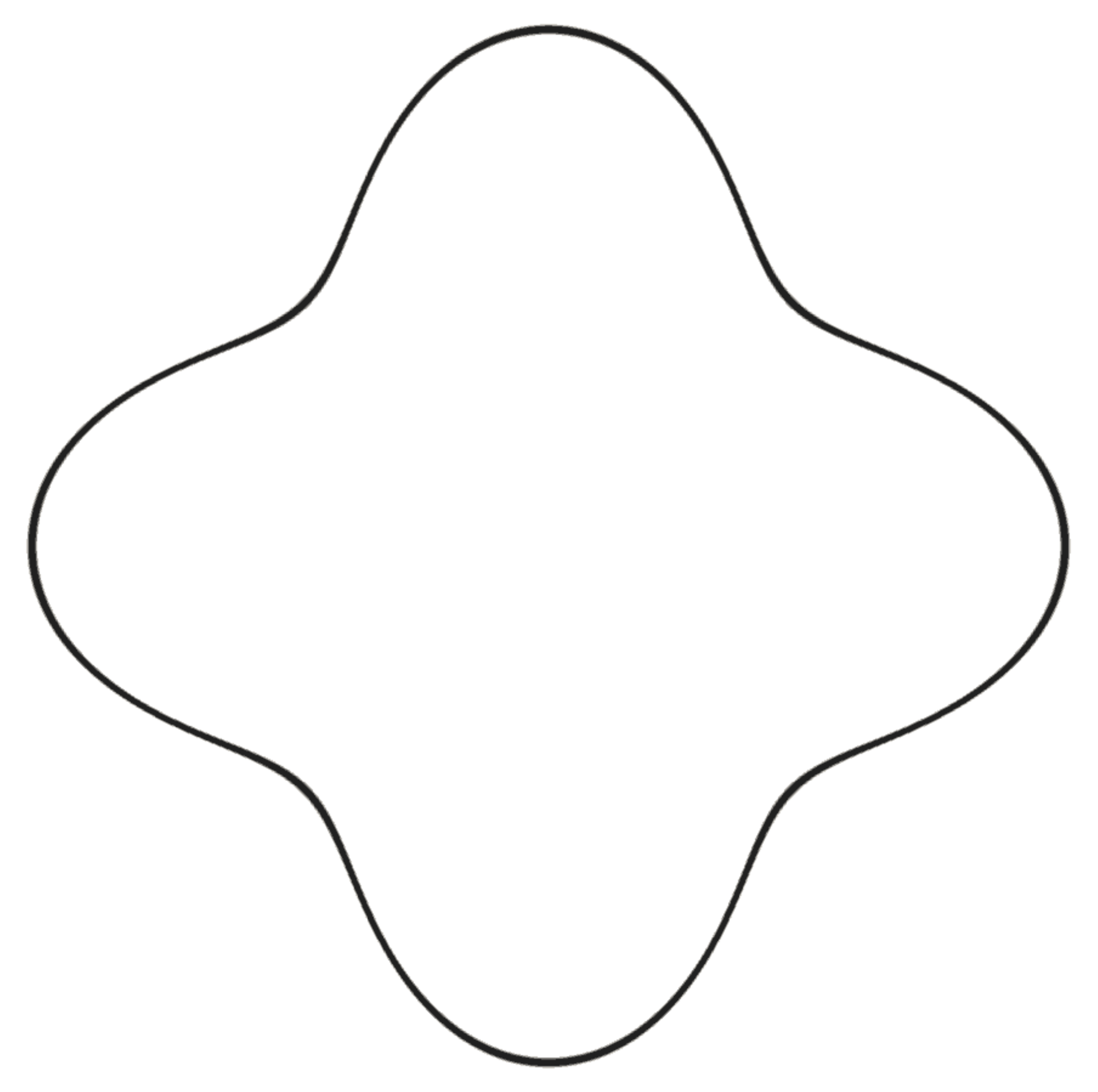}
		\subcaption{$t=0 \ \si{\second}$}
	\end{subfigure}
	\begin{subfigure}{.163\textwidth}
		\centering	
		\includegraphics[width = 0.95\textwidth]{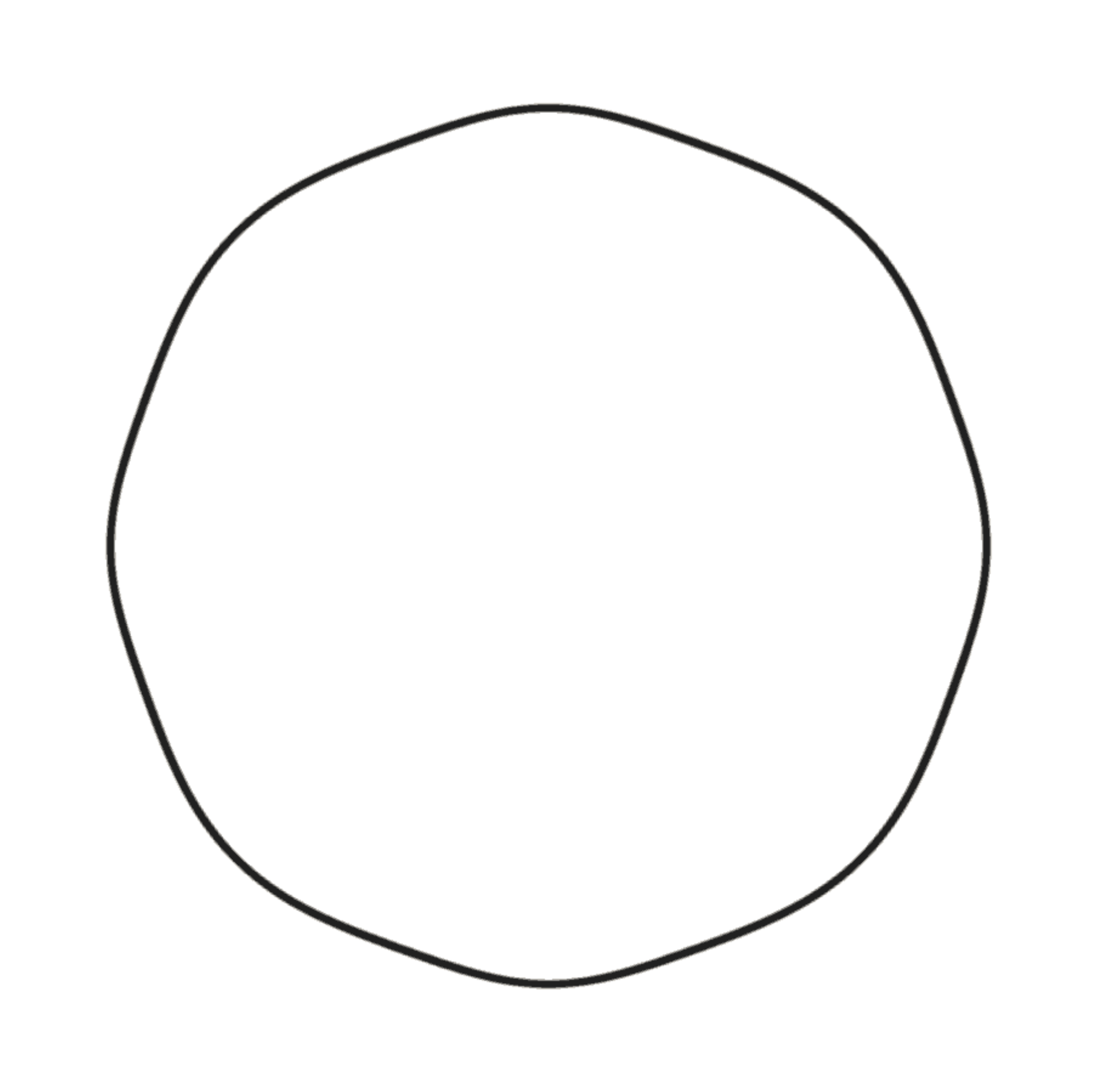}
		\subcaption{$t=4\cdot 10^{-5} \ \si{\second}$}			
	\end{subfigure}
	\begin{subfigure}{.163\textwidth}
		\centering		
		\includegraphics[width = 0.95\textwidth]{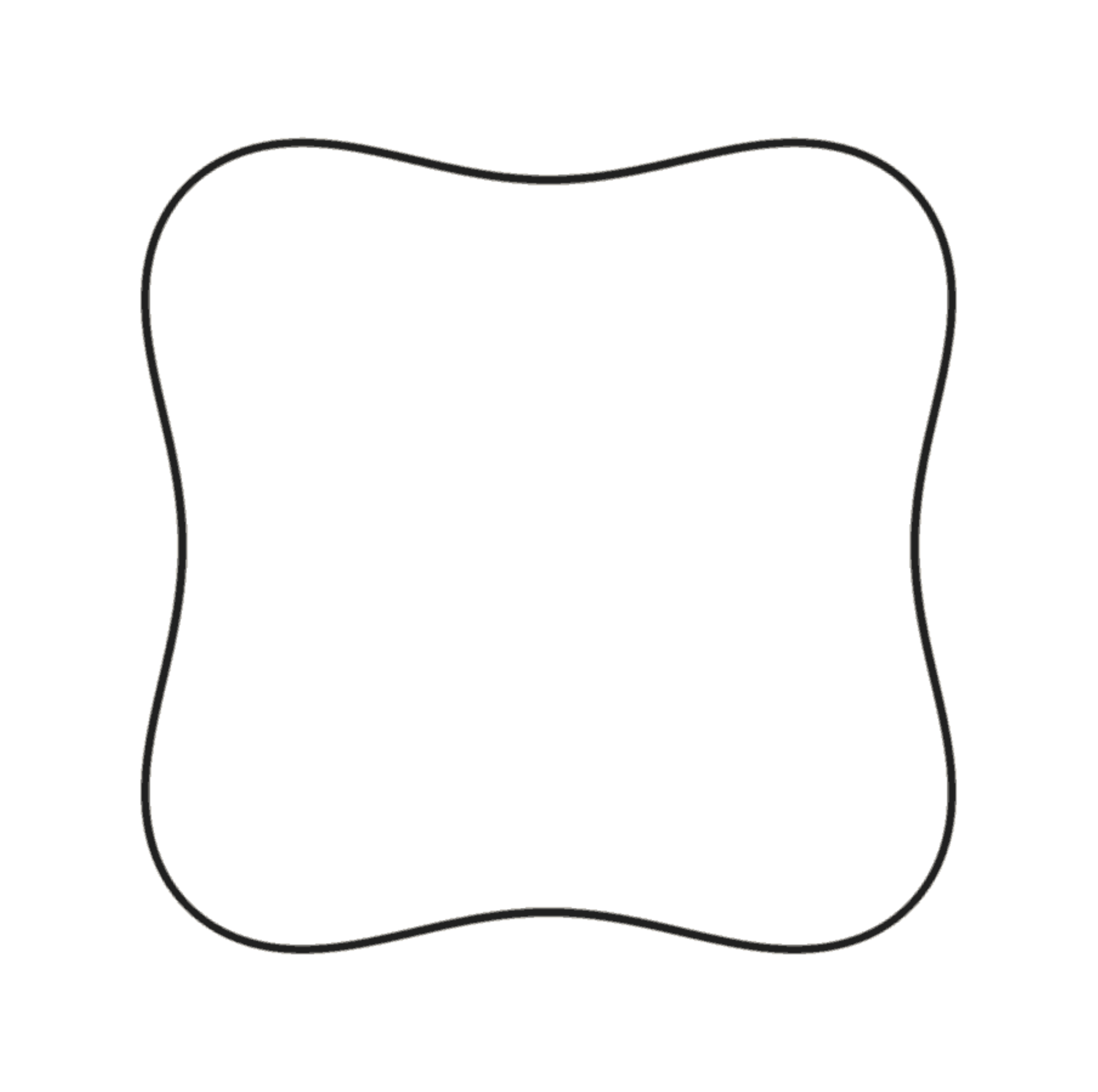}
		\subcaption{$t=7.6\cdot 10^{-5} \ \si{\second}$}			
	\end{subfigure}
	\begin{subfigure}{.163\textwidth}
		\centering		
		\includegraphics[width = 0.95\textwidth]{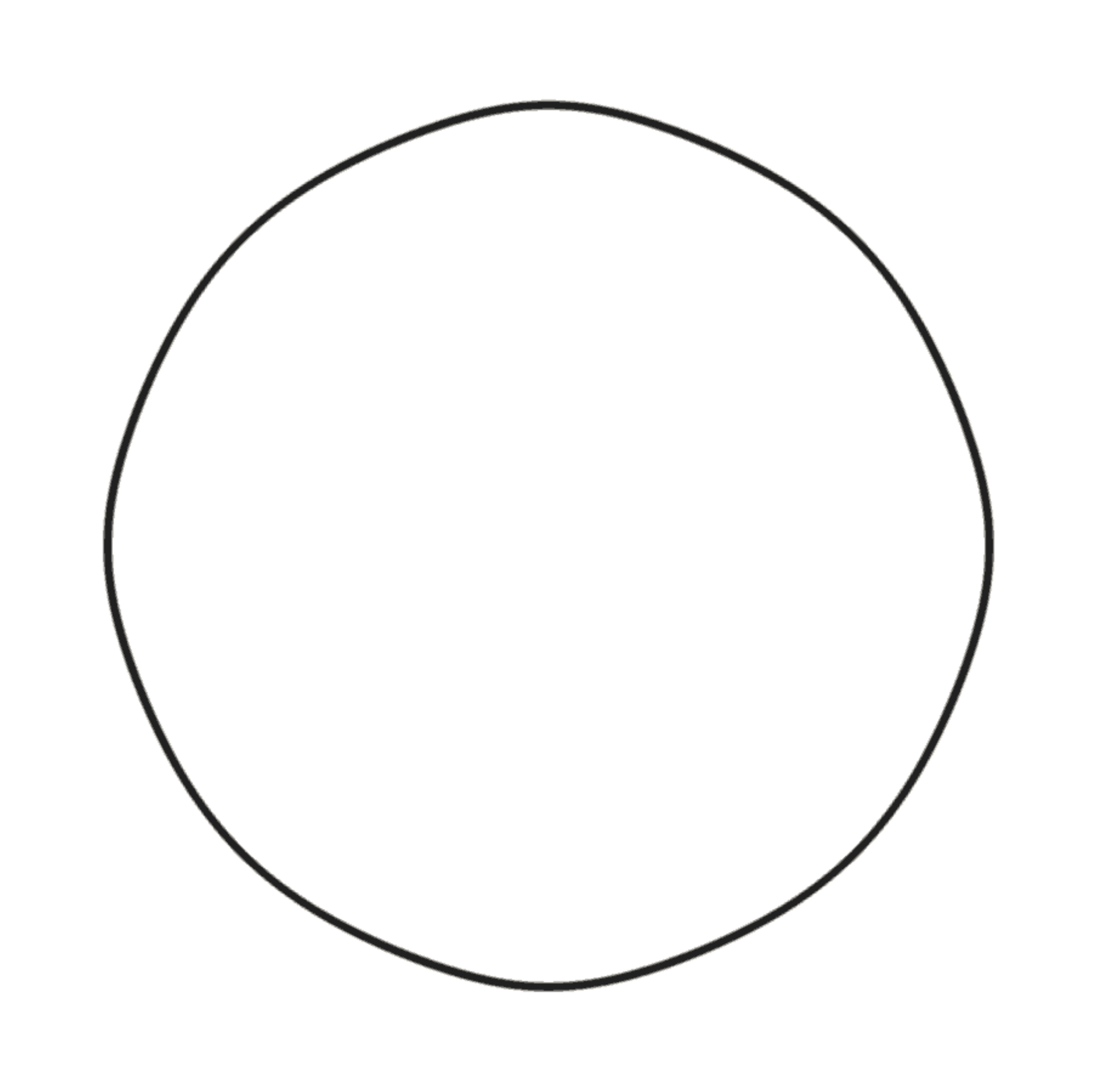}
		\subcaption{$t=1.16\cdot 10^{-4} \ \si{\second}$}			
	\end{subfigure}
	\begin{subfigure}{.163\textwidth}
		\centering		
		\includegraphics[width = 0.95\textwidth]{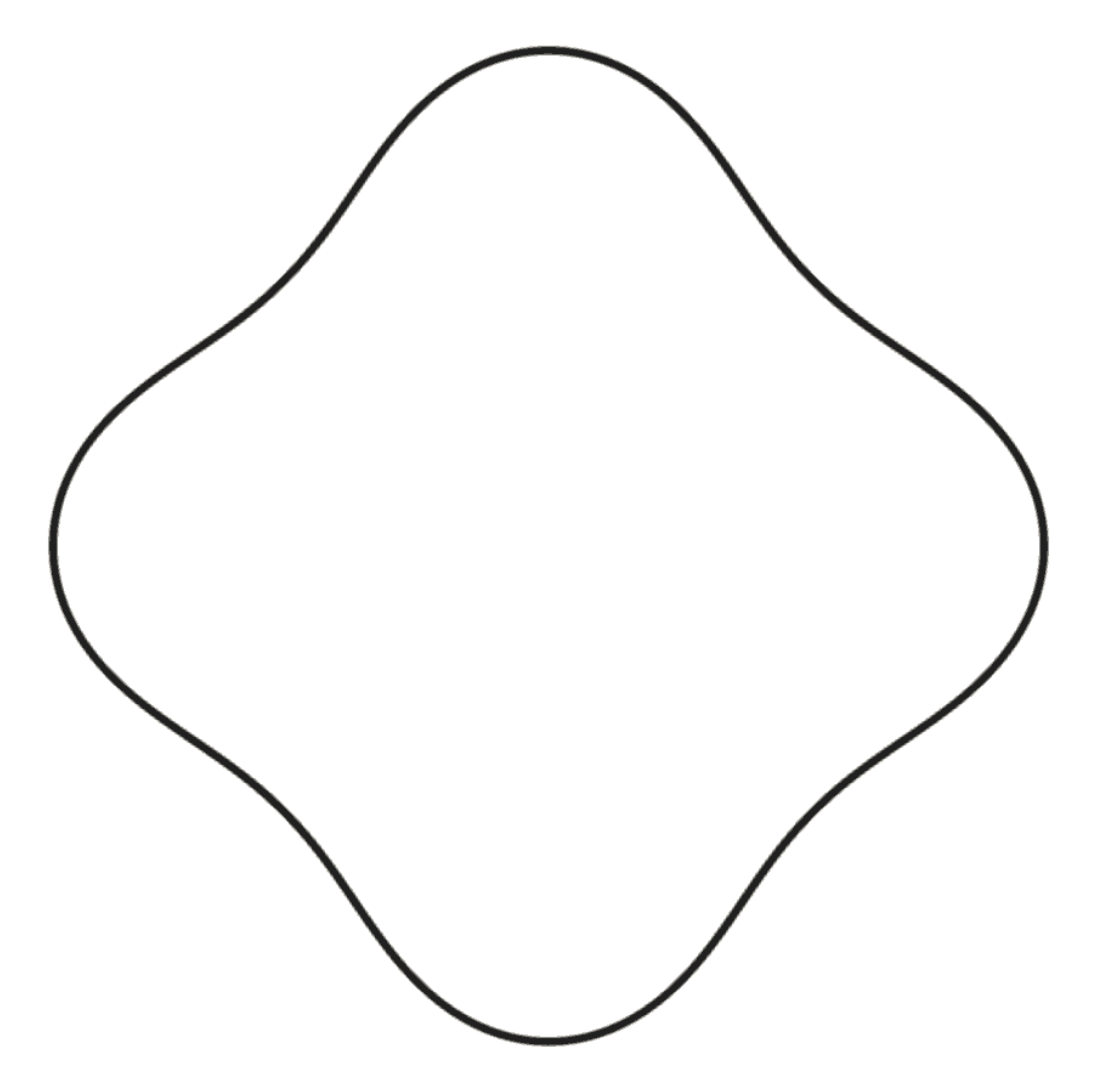}
		\subcaption{$t=1.52\cdot 10^{-4} \ \si{\second}$}			
	\end{subfigure}		
	\caption{Large amplitude oscillation of a two-dimensional drop: Interface isolines for $n=2,3,4$.}
	\label{fig:large_oscillations_phi}	
\end{figure}

\begin{figure}[tb!]
	\captionsetup[subfigure]{labelformat=empty}
	\centering
	\begin{subfigure}{.2\textwidth}
		\centering
		\includegraphics[width = \textwidth]{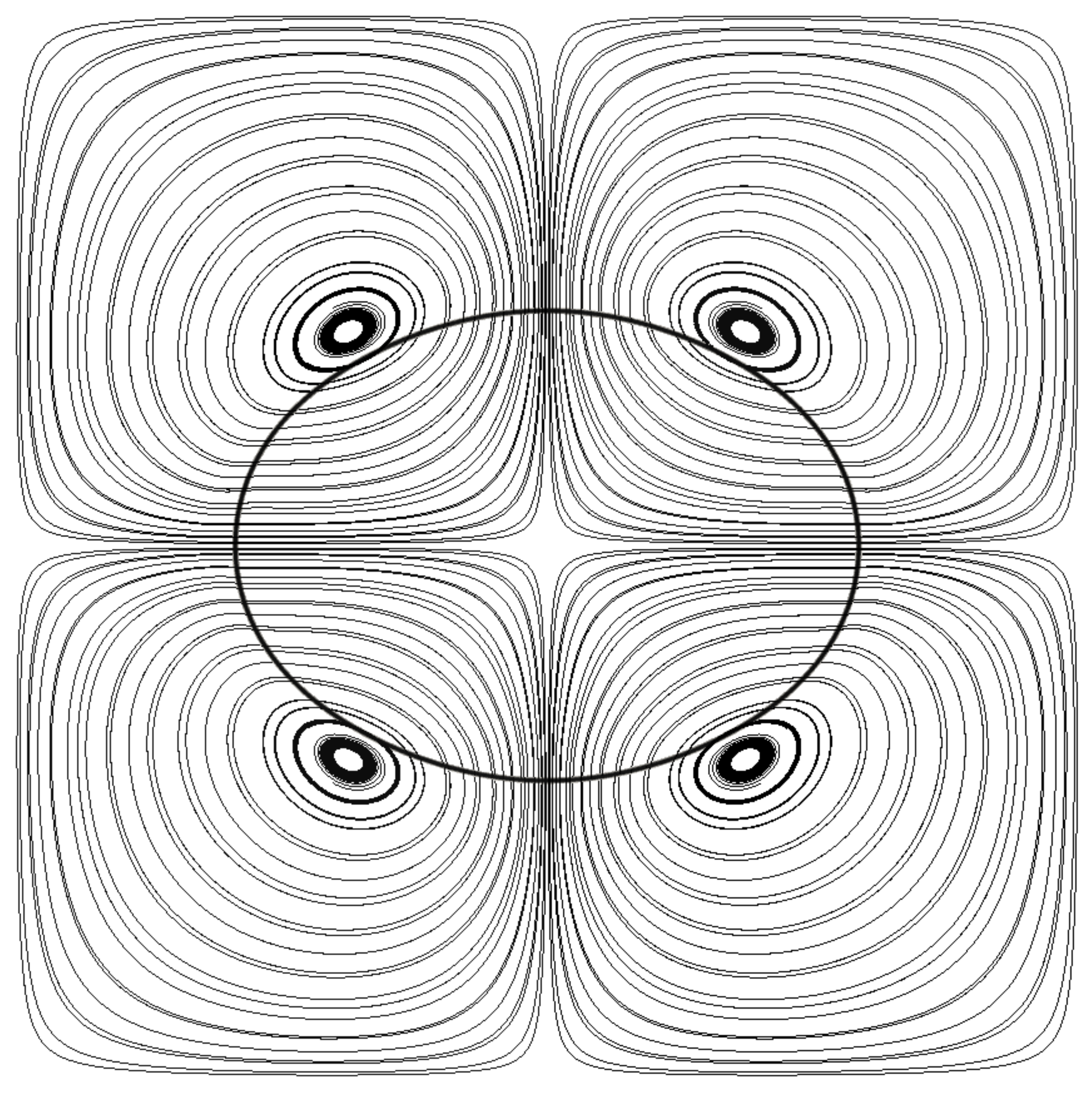}
		\subcaption{$t=7.2\cdot 10^{-5} \ \si{\second}$}
	\end{subfigure}
	\begin{subfigure}{.2\textwidth}
		\centering
		\includegraphics[width = \textwidth]{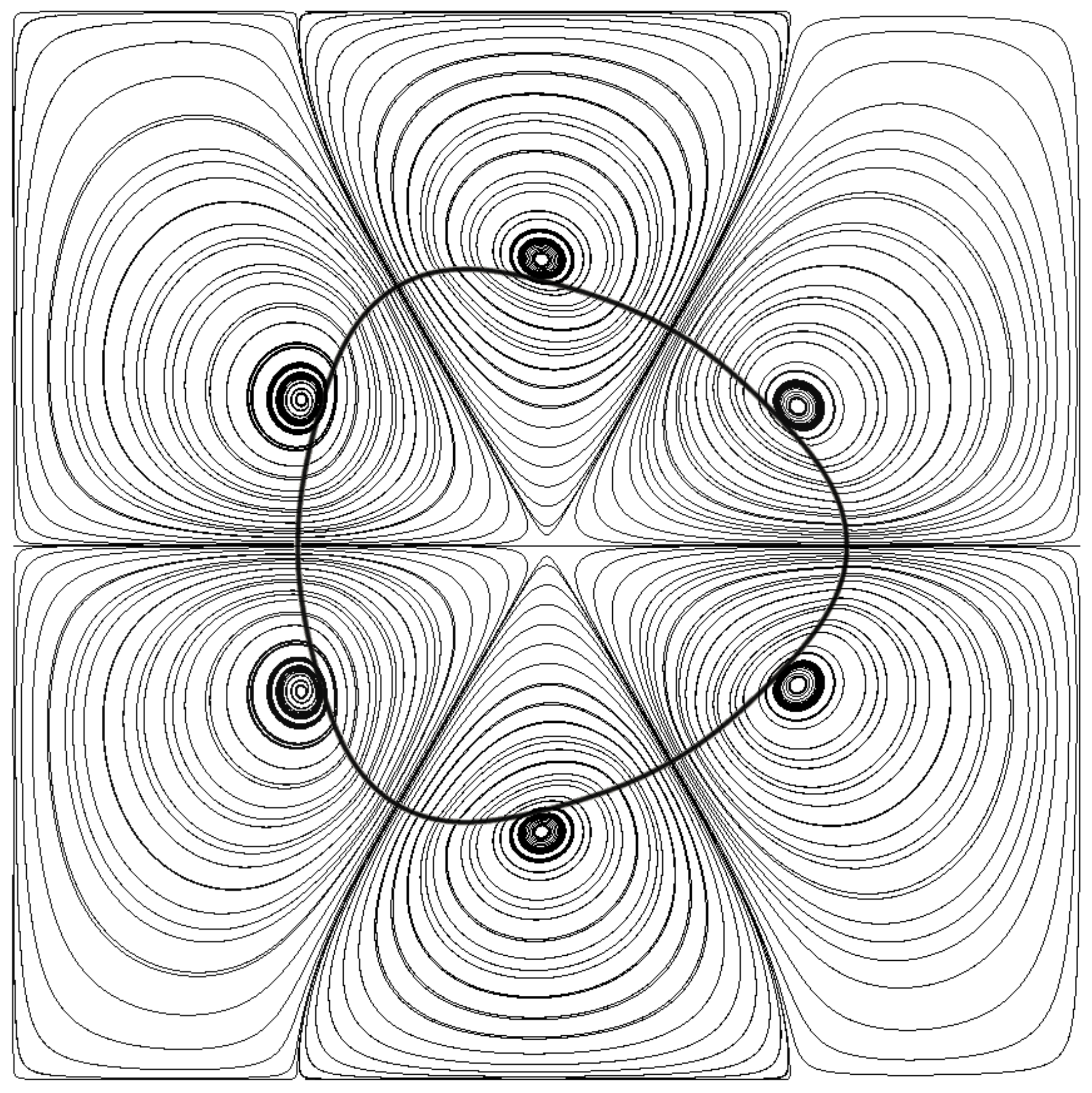}
		\subcaption{$t=4\cdot 10^{-5} \ \si{\second}$}
	\end{subfigure}	
	\begin{subfigure}{.2\textwidth}
		\centering
		\includegraphics[width = \textwidth]{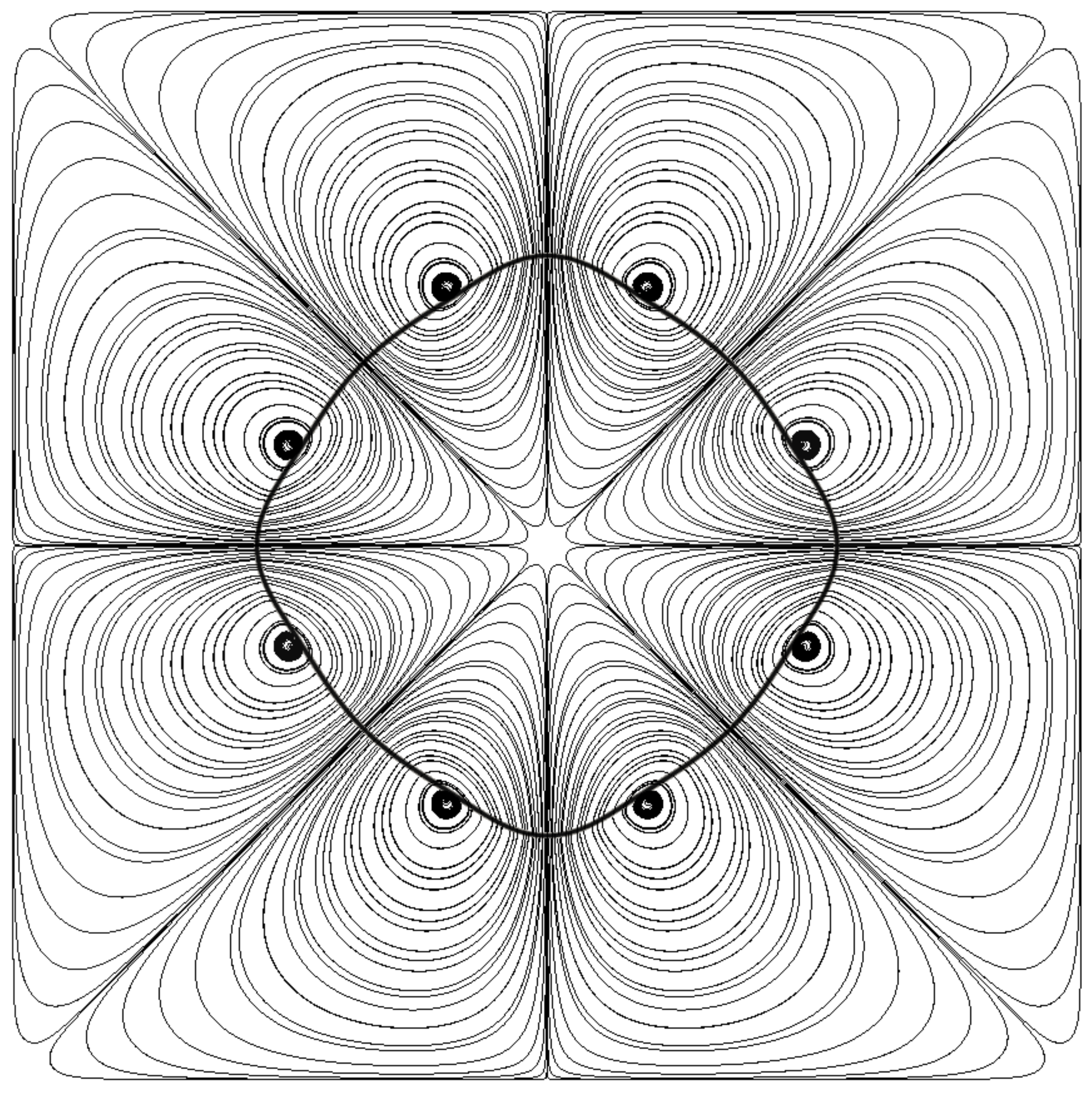}
		\subcaption{$t=3.2\cdot 10^{-5} \ \si{\second}$}
	\end{subfigure}
	\caption{Large amplitude oscillations of a two-dimensional drop: Velocity streamlines at different time instances for $n=2,3,4$}
	\label{fig:large_oscillations_velocity}	
\end{figure}

\begin{figure}[tb!]
	\captionsetup[subfigure]{labelformat=empty}
	\centering
	\begin{subfigure}{.25\textwidth}
		\centering
		$n=2$\\[5.5pt]
		\includegraphics[width = 0.79\textwidth]{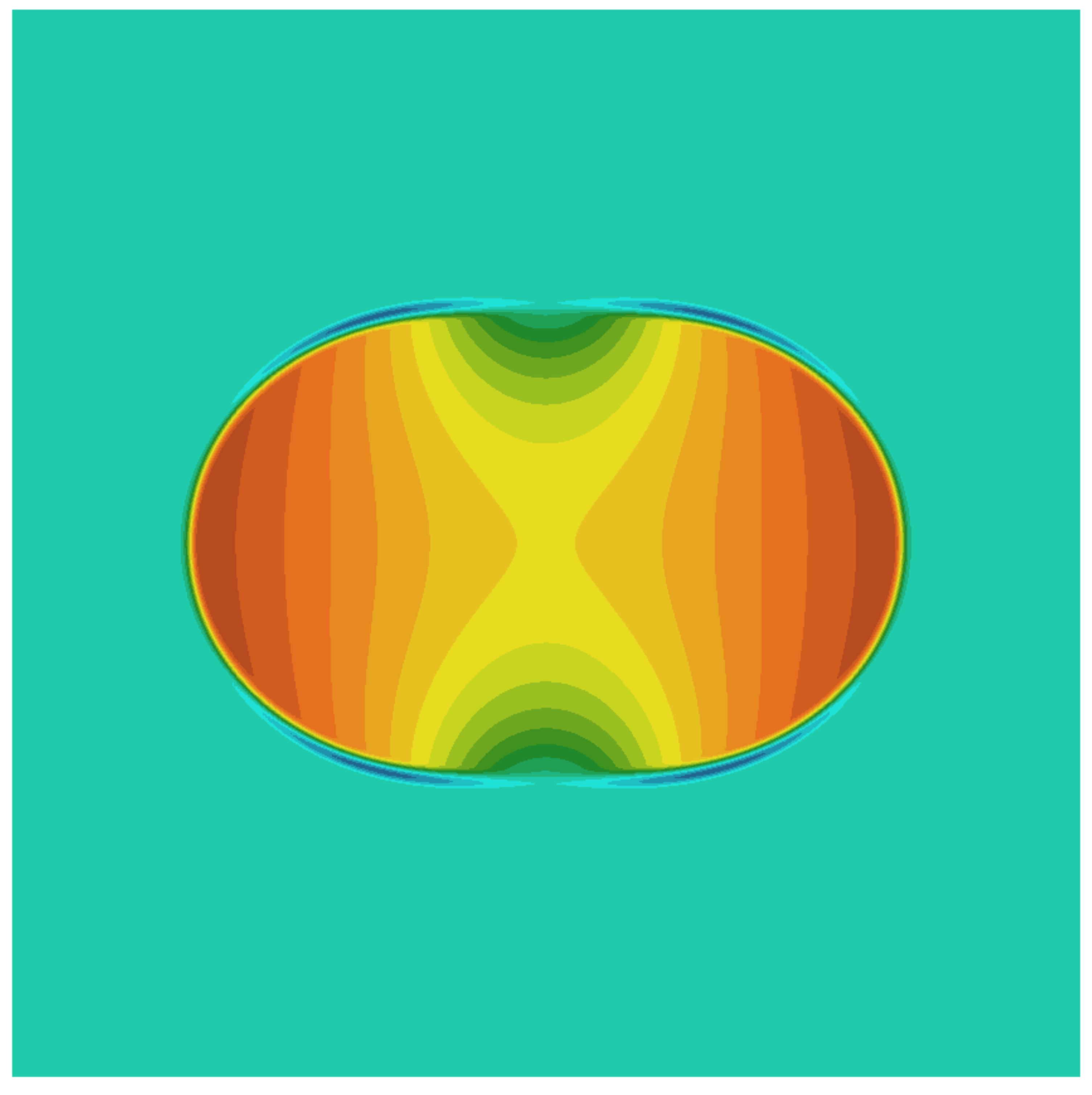}
		\subcaption{$t=4\cdot 10^{-6} \ \si{\second}$}
	\end{subfigure}
	\begin{subfigure}{.25\textwidth}
		\centering
		$n=3$\\[5.5pt]	
		\includegraphics[width = 0.79\textwidth]{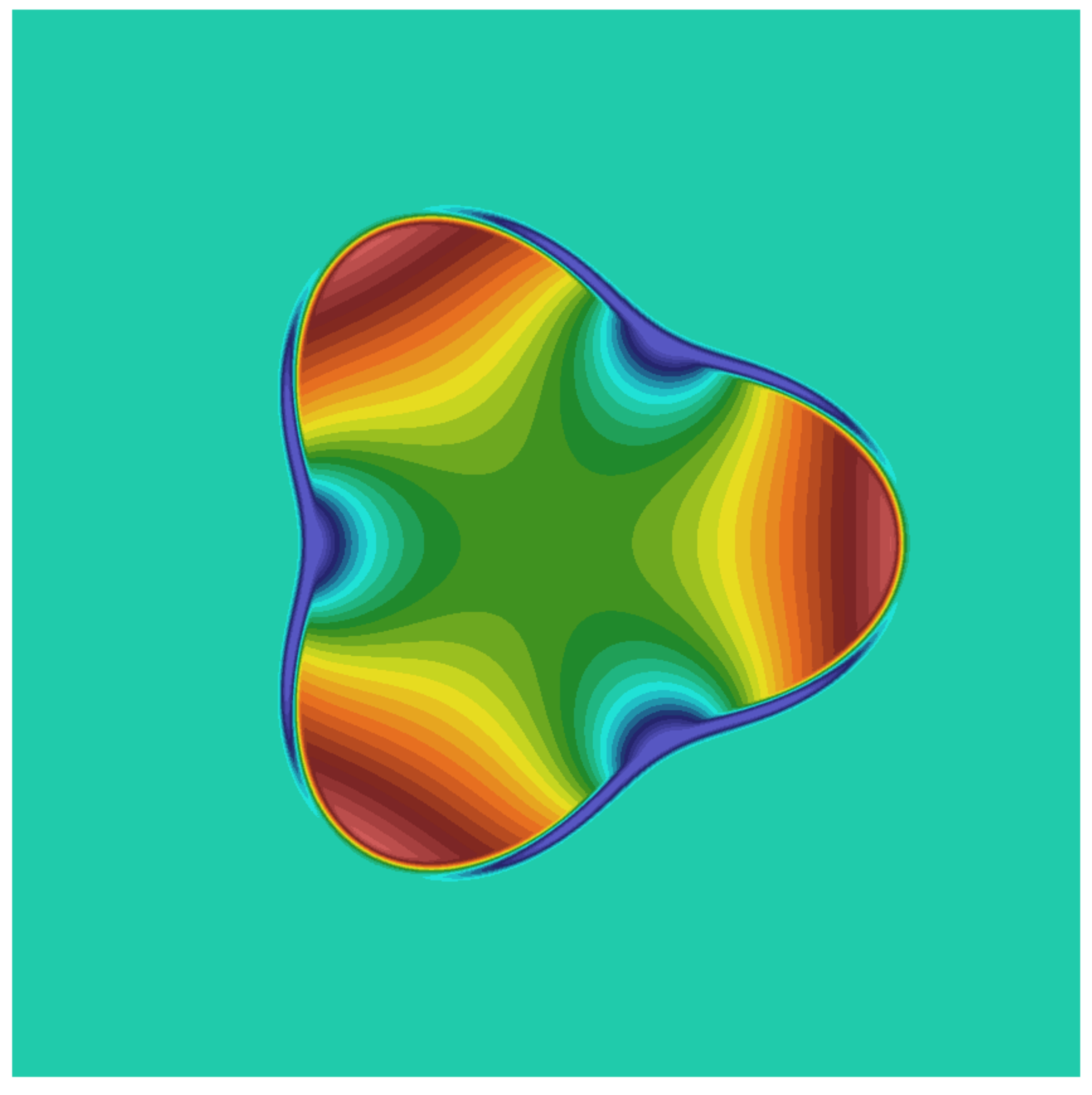}
		\subcaption{$t=4\cdot 10^{-6} \ \si{\second}$}	
	\end{subfigure}
	\begin{subfigure}{.25\textwidth}
		\centering		
		$n=4$\\[5.5pt]	
		\includegraphics[width = 0.79\textwidth]{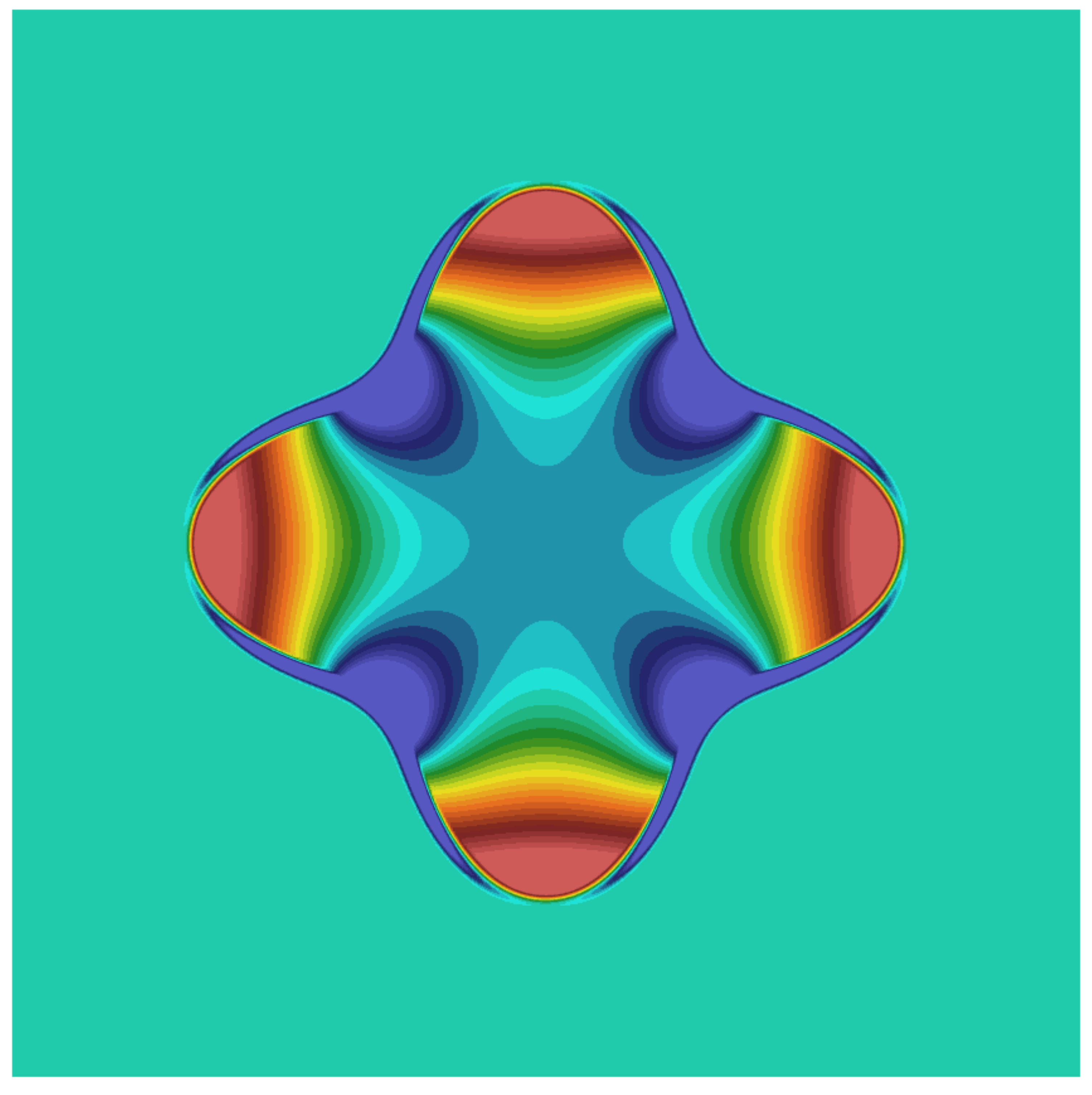}
		\subcaption{$t=4\cdot 10^{-6} \ \si{\second}$}
	\end{subfigure}
	\begin{subfigure}{.25\textwidth}
		\centering
		\includegraphics[width = 0.79\textwidth]{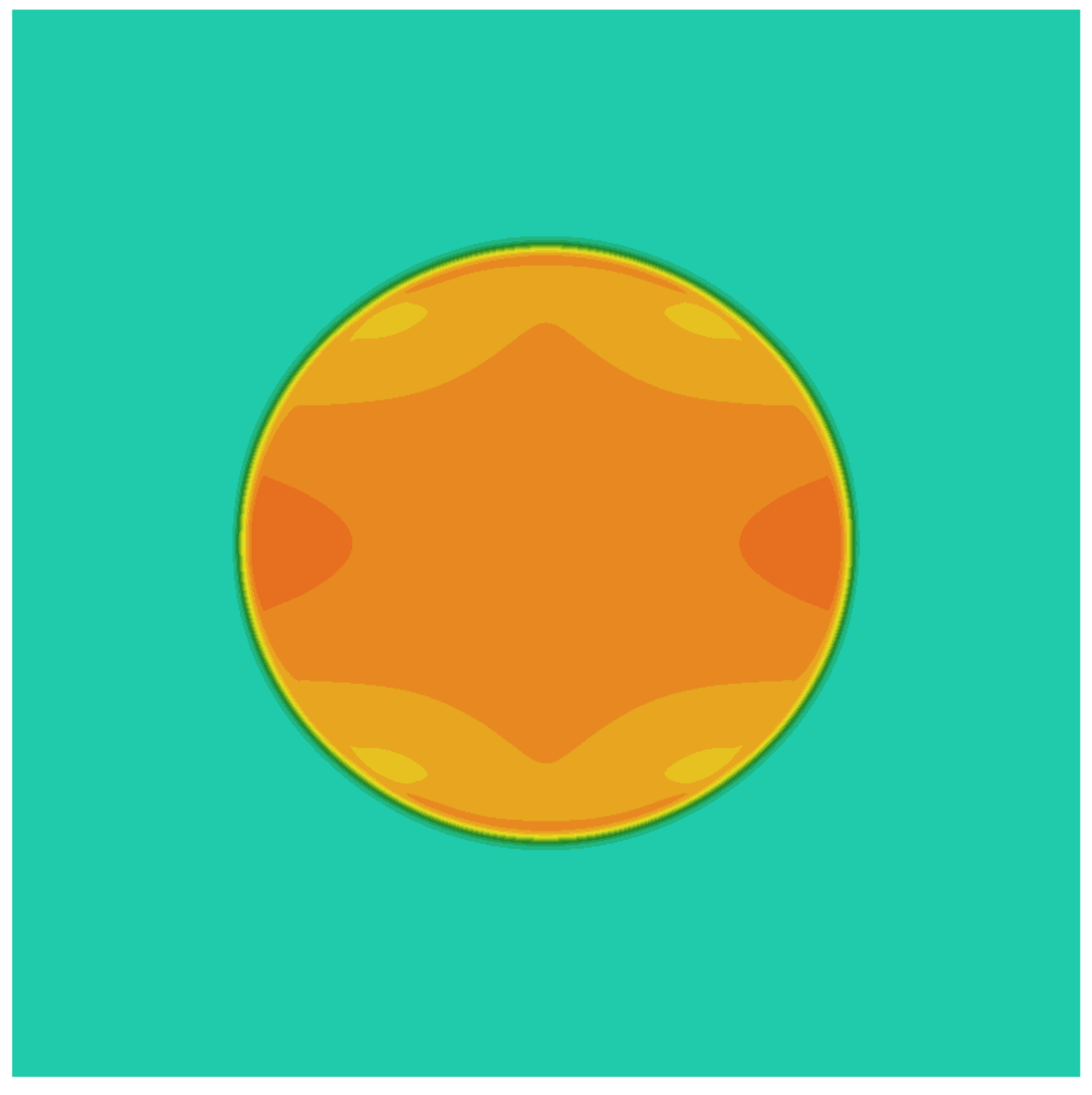}
		\subcaption{$t=1.08\cdot 10^{-4} \ \si{\second}$}
	\end{subfigure}
	\begin{subfigure}{.25\textwidth}
		\centering	
		\includegraphics[width = 0.79\textwidth]{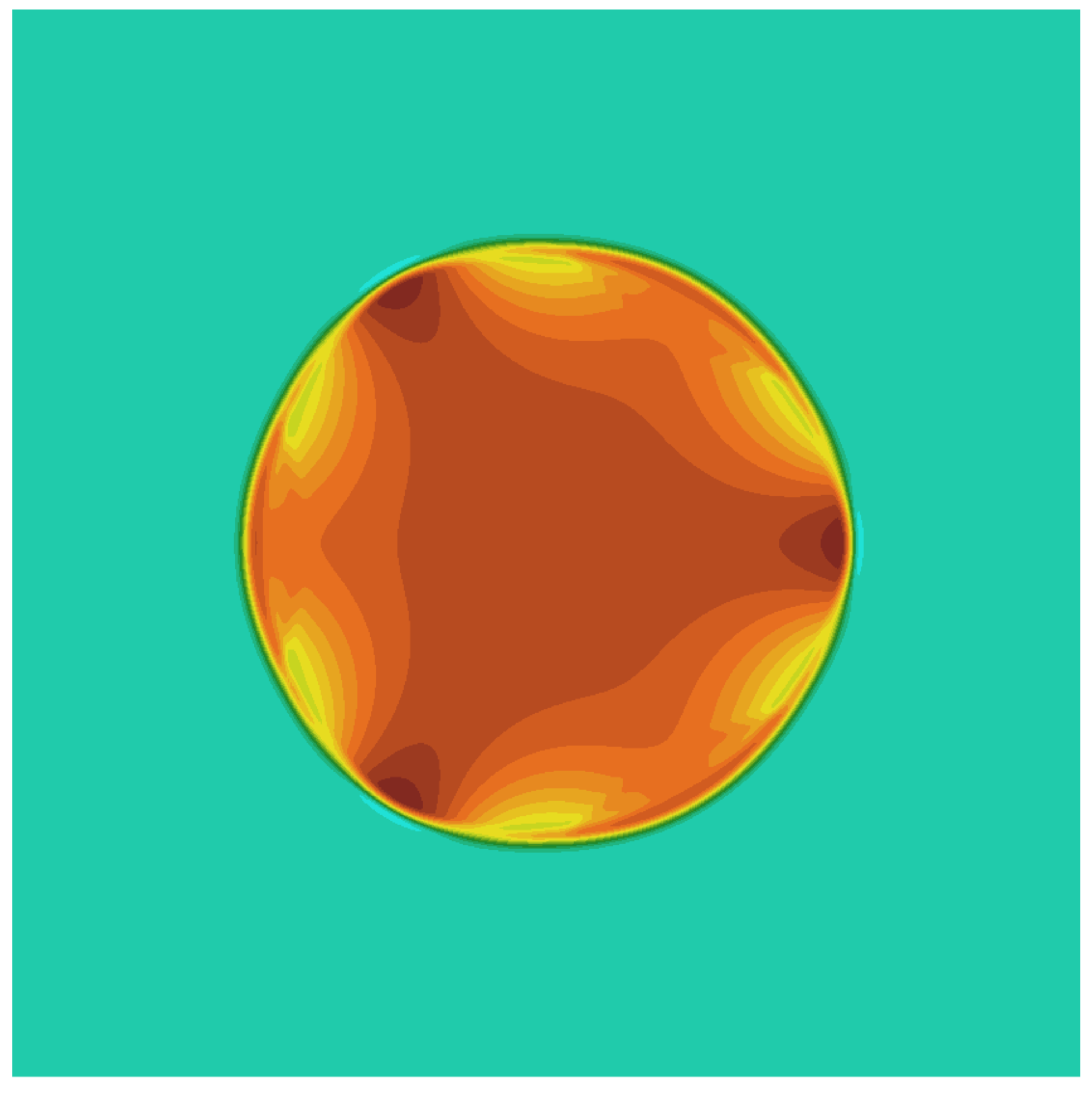}
		\subcaption{$t=6\cdot 10^{-5} \ \si{\second}$}	
	\end{subfigure}
	\begin{subfigure}{.25\textwidth}
		\centering		
		\includegraphics[width = 0.79\textwidth]{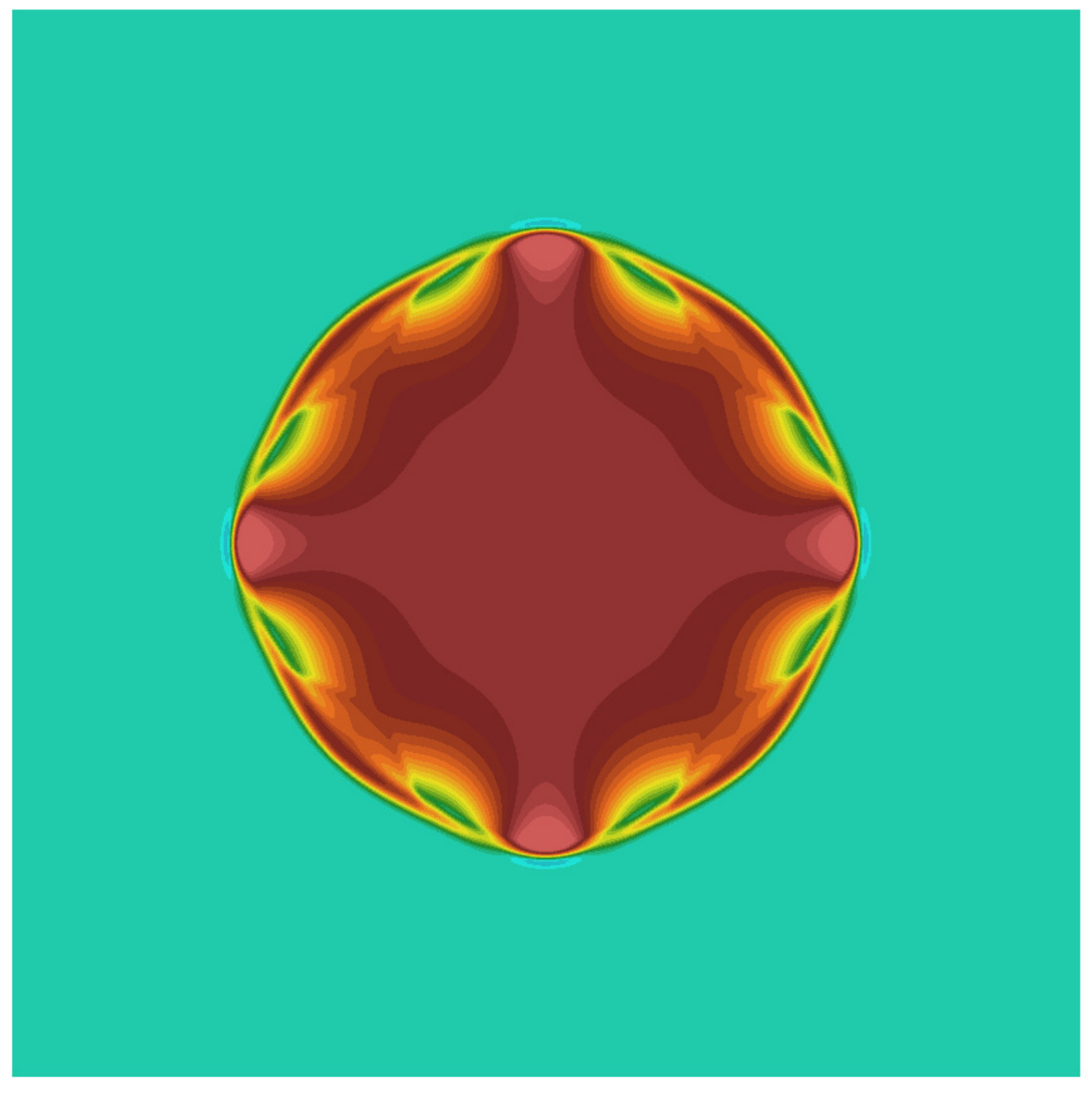}
		\subcaption{$t=4\cdot 10^{-5} \ \si{\second}$}	
	\end{subfigure}
	\begin{subfigure}{.25\textwidth}
		\centering
		\includegraphics[width = 0.79\textwidth]{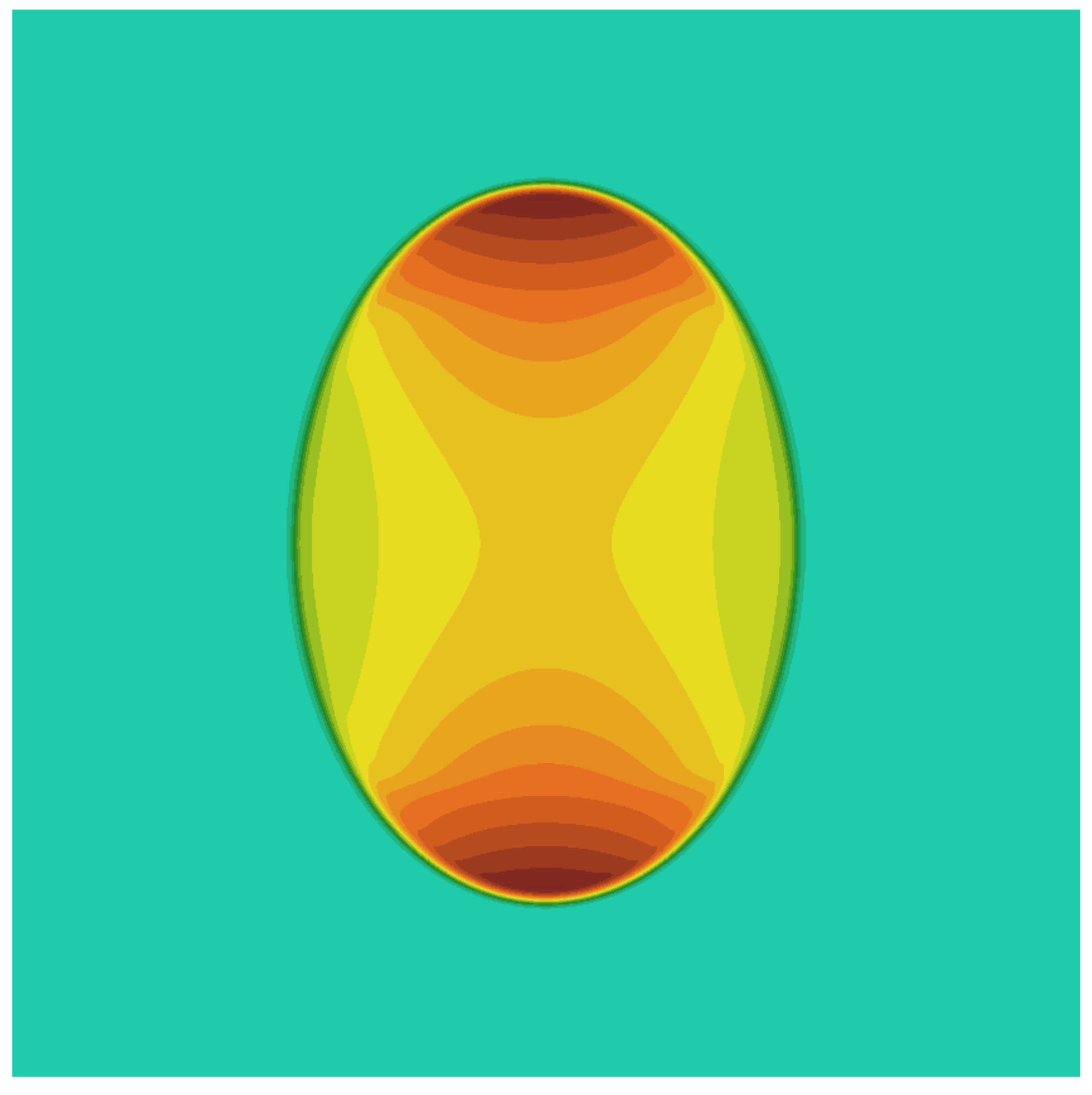}
		\subcaption{$t=2.2\cdot 10^{-4} \ \si{\second}$}
	\end{subfigure}
	\begin{subfigure}{.25\textwidth}
		\centering	
		\includegraphics[width = 0.79\textwidth]{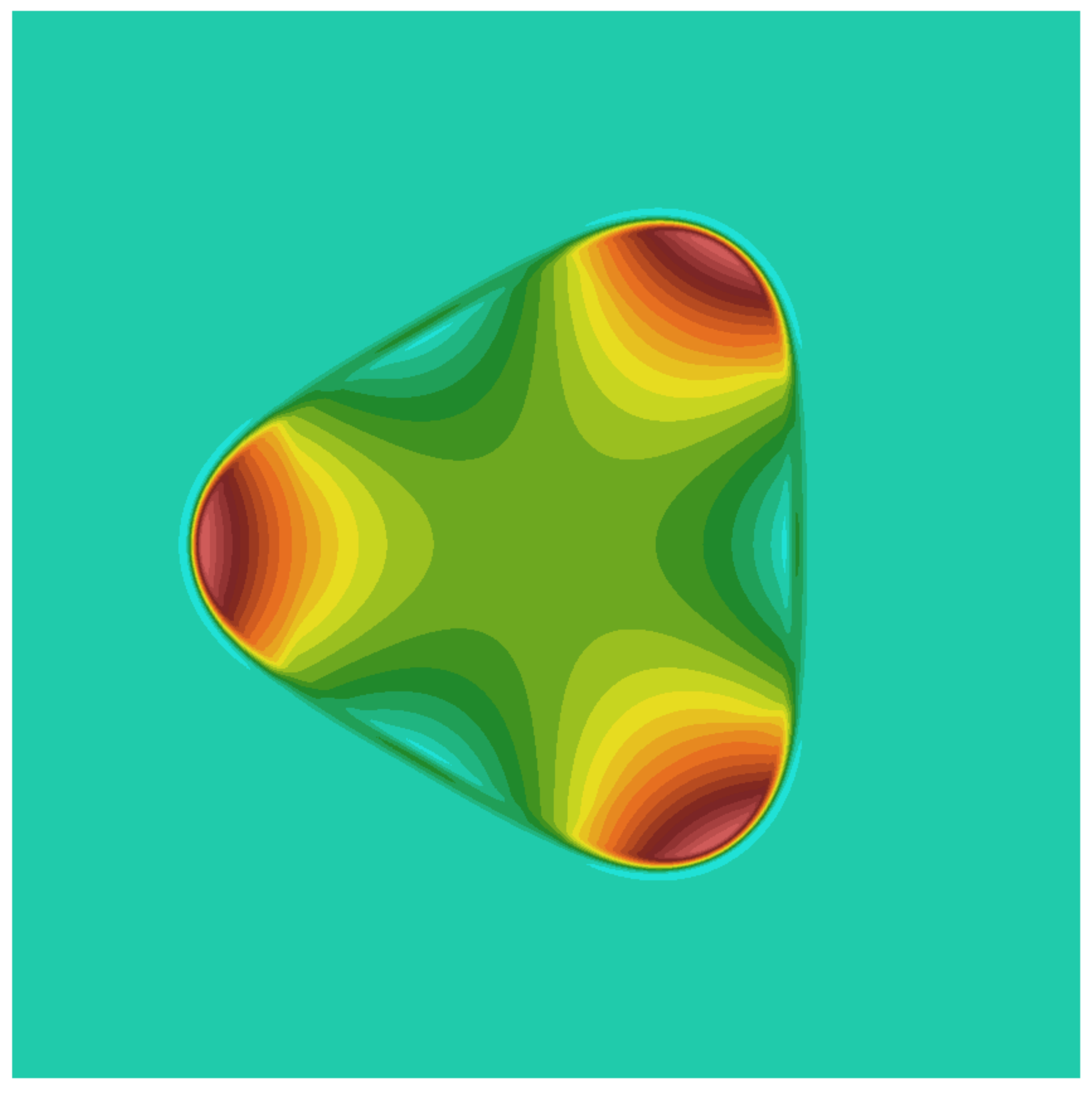}
		\subcaption{$t=1.12\cdot 10^{-4} \ \si{\second}$}			
	\end{subfigure}
	\begin{subfigure}{.25\textwidth}
		\centering		
		\includegraphics[width = 0.79\textwidth]{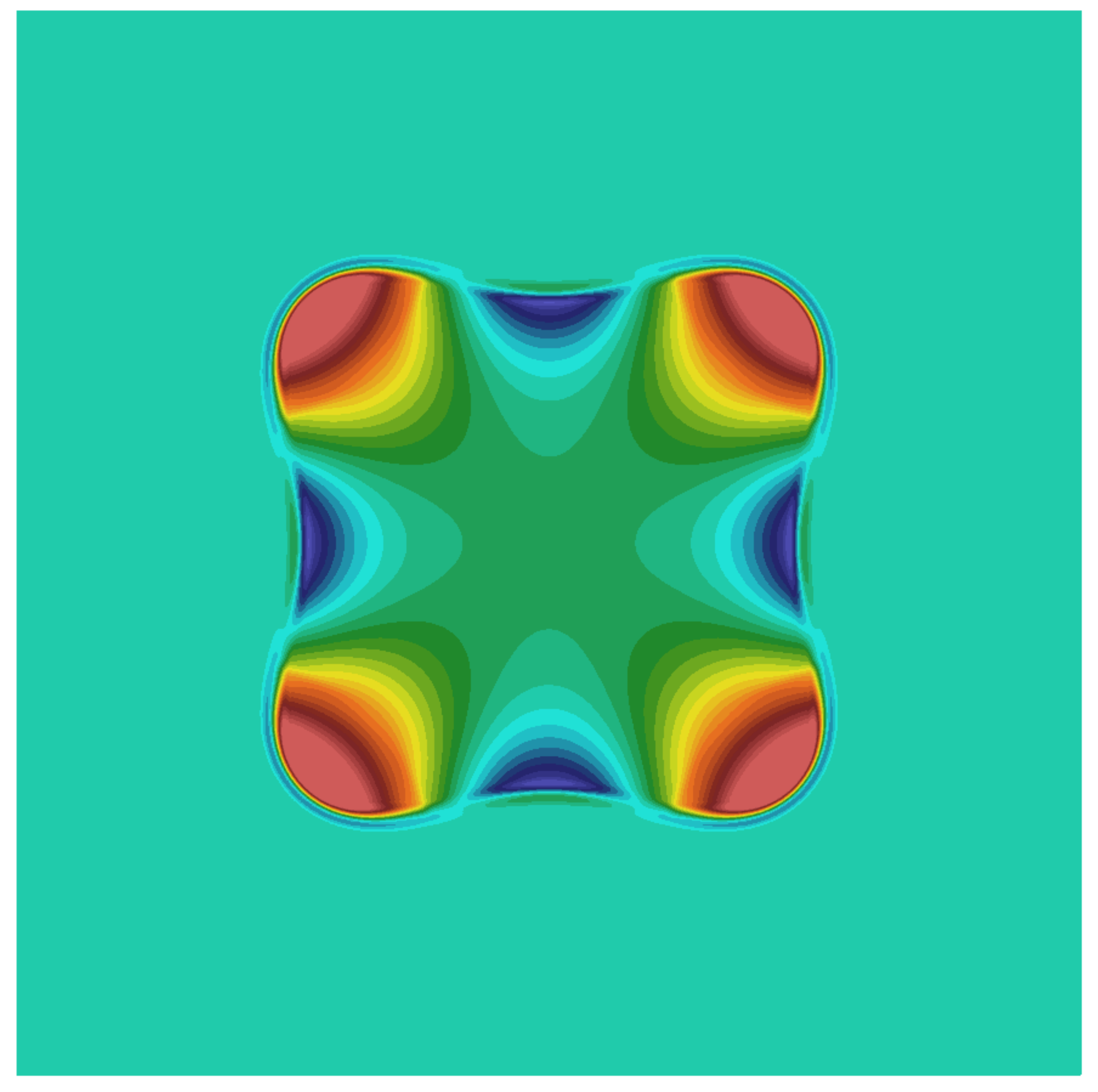}
		\subcaption{$t=7.6\cdot 10^{-5} \ \si{\second}$}			
	\end{subfigure}
	\begin{subfigure}{.25\textwidth}
		\centering
		\includegraphics[width = 0.79\textwidth]{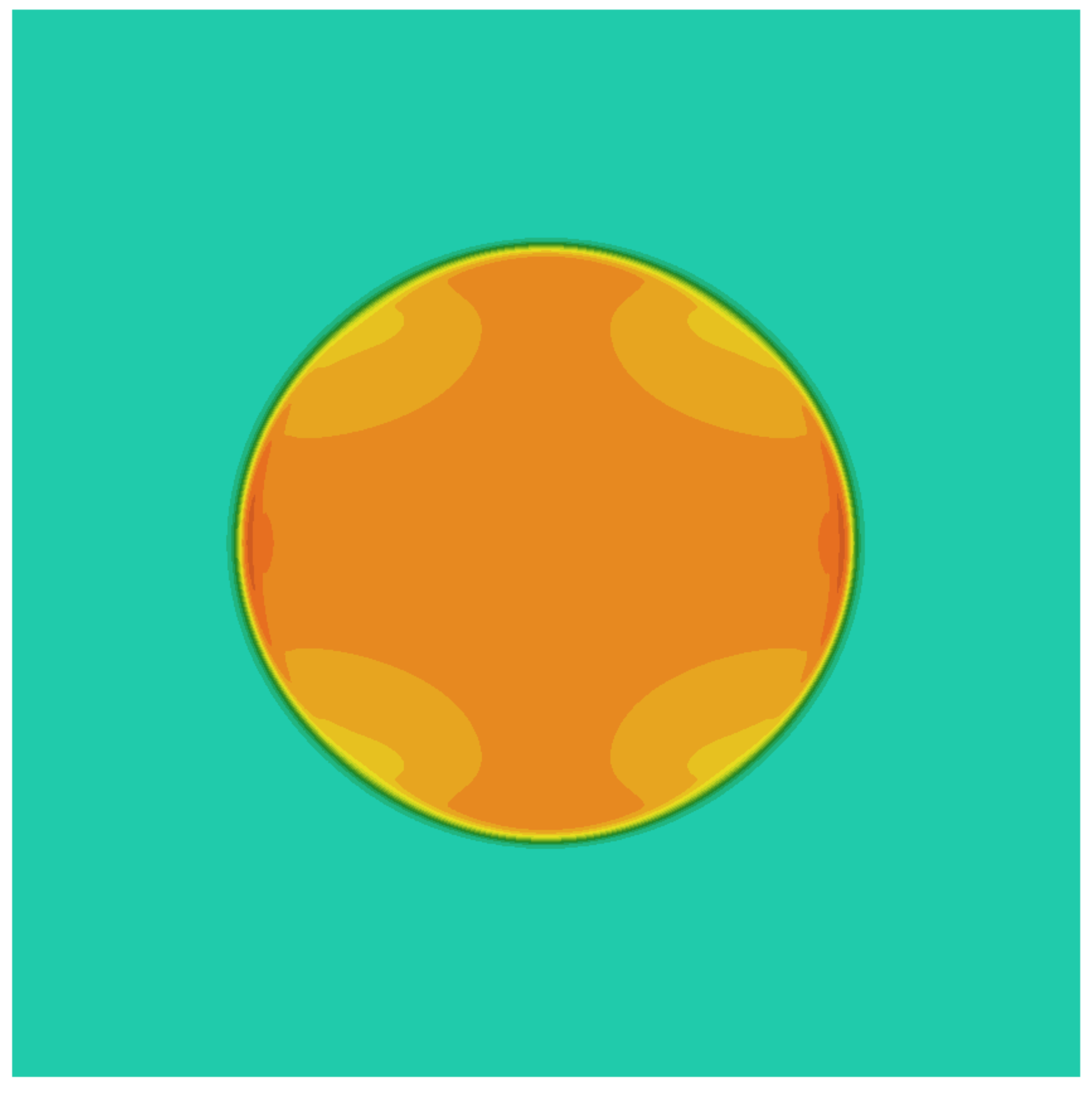}
		\subcaption{$t=3.332\cdot 10^{-4} \ \si{\second}$}
	\end{subfigure}
	\begin{subfigure}{.25\textwidth}
		\centering	
		\includegraphics[width = 0.79\textwidth]{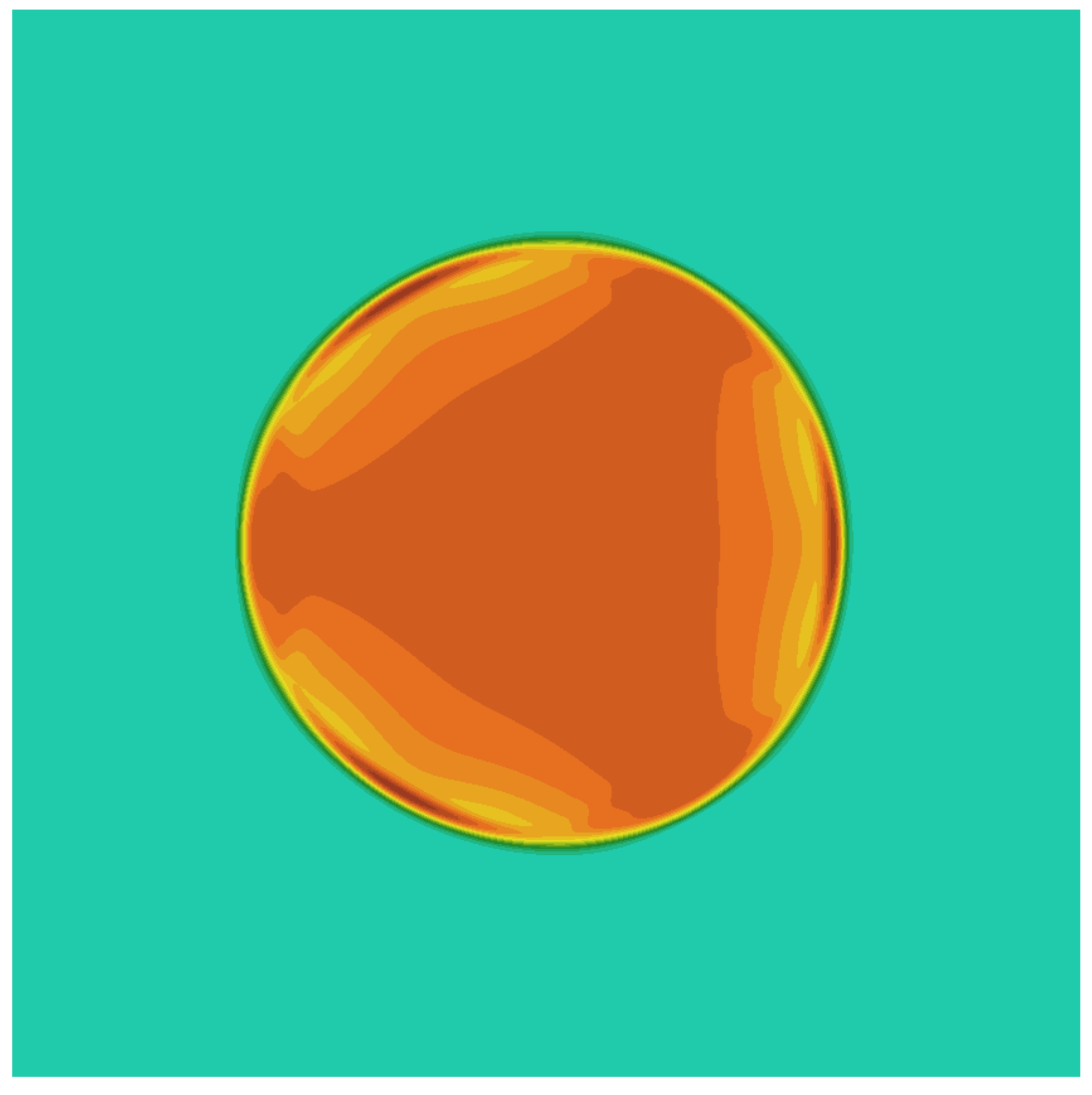}
		\subcaption{$t=1.72\cdot 10^{-4} \ \si{\second}$}			
	\end{subfigure}
	\begin{subfigure}{.25\textwidth}
		\centering		
		\includegraphics[width = 0.79\textwidth]{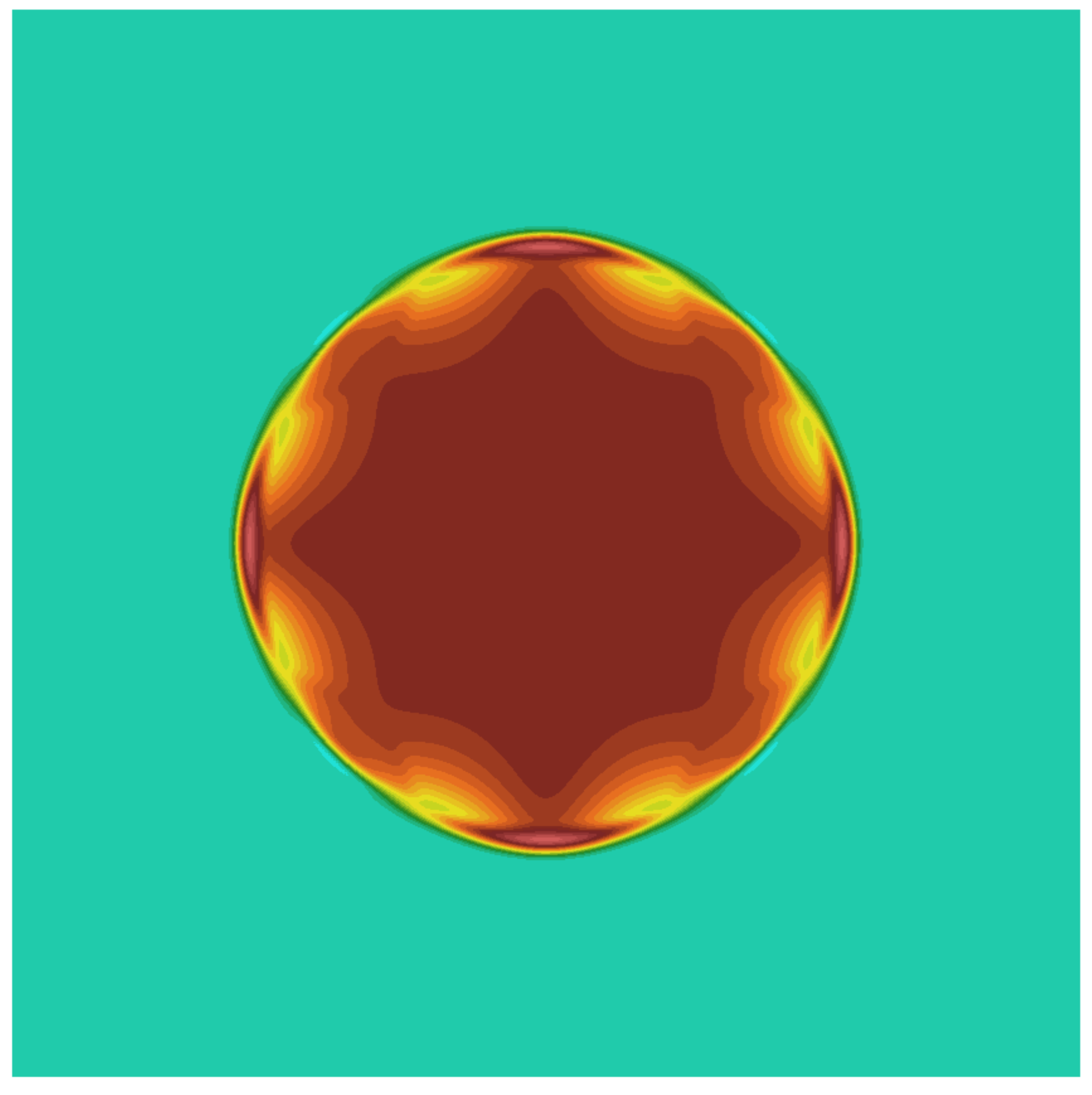}
		\subcaption{$t=1.16\cdot 10^{-4} \ \si{\second}$}			
	\end{subfigure}
	\begin{subfigure}{.25\textwidth}
		\centering
		\includegraphics[width = 0.79\textwidth]{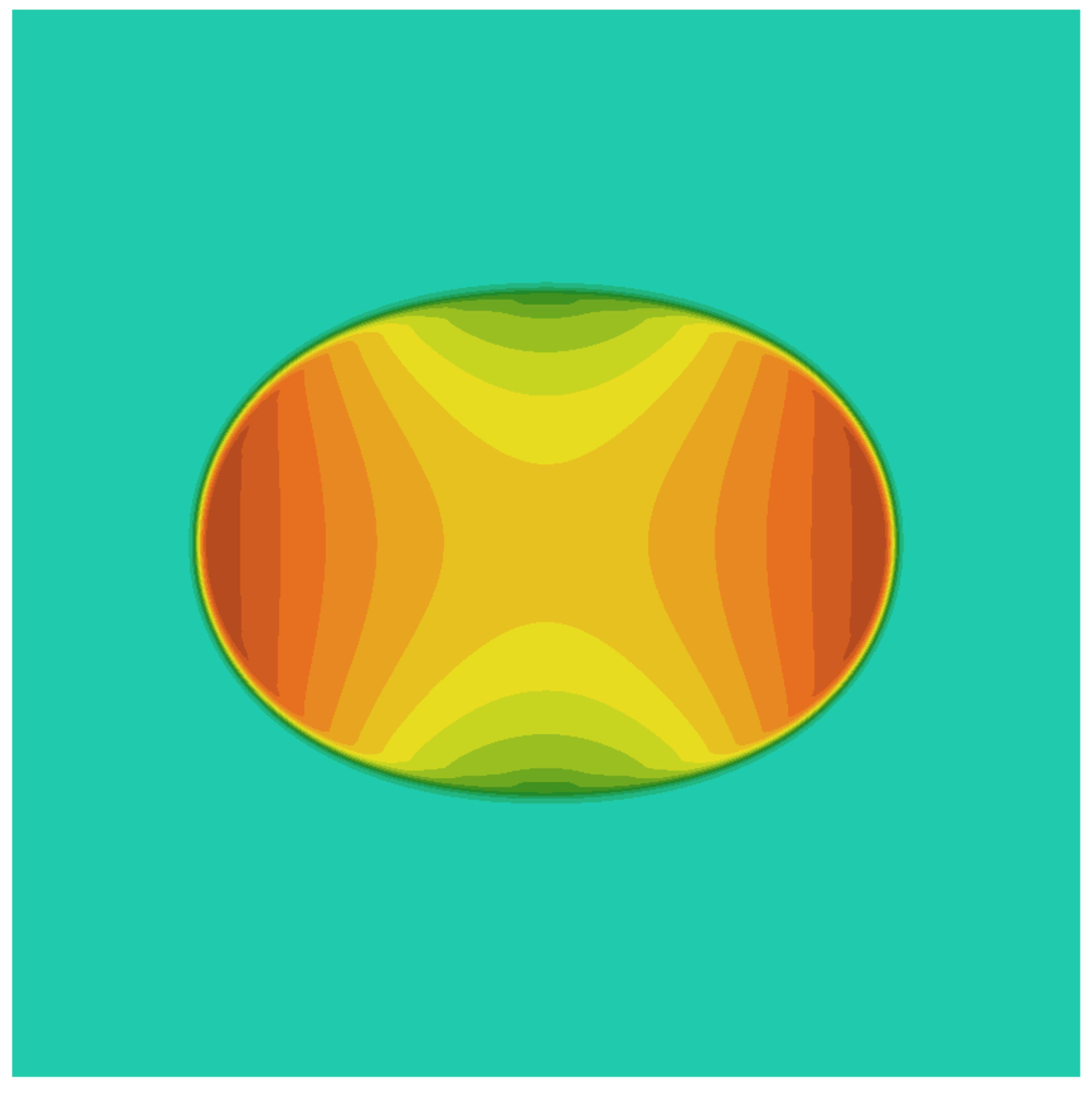}
		\subcaption{$t=4.4\cdot 10^{-4} \ \si{\second}$}
	\end{subfigure}
	\begin{subfigure}{.25\textwidth}
		\centering	
		\includegraphics[width = 0.79\textwidth]{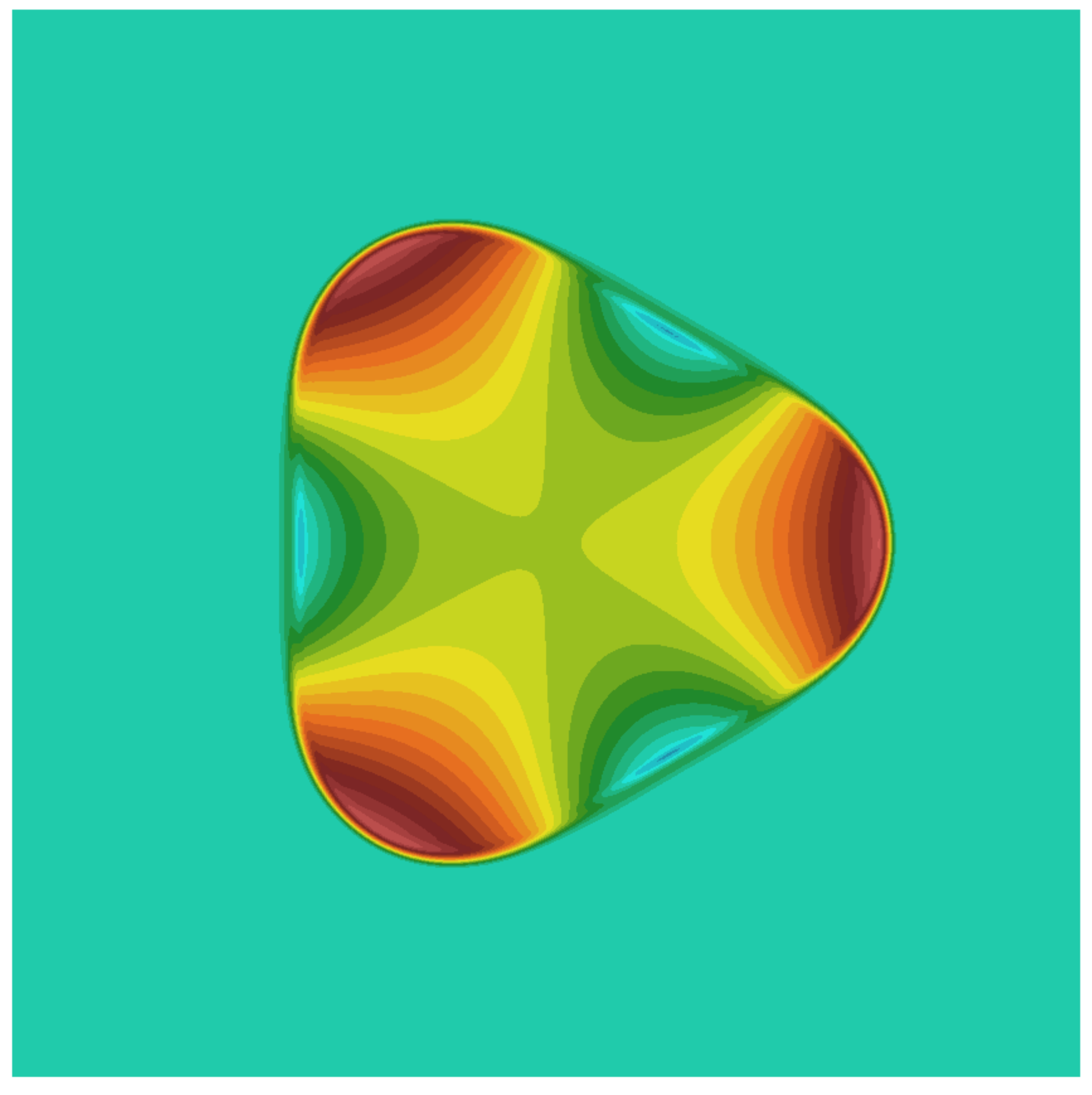}
		\subcaption{$t=2.2\cdot 10^{-4} \ \si{\second}$}			
	\end{subfigure}
	\begin{subfigure}{.25\textwidth}
		\centering		
		\includegraphics[width = 0.79\textwidth]{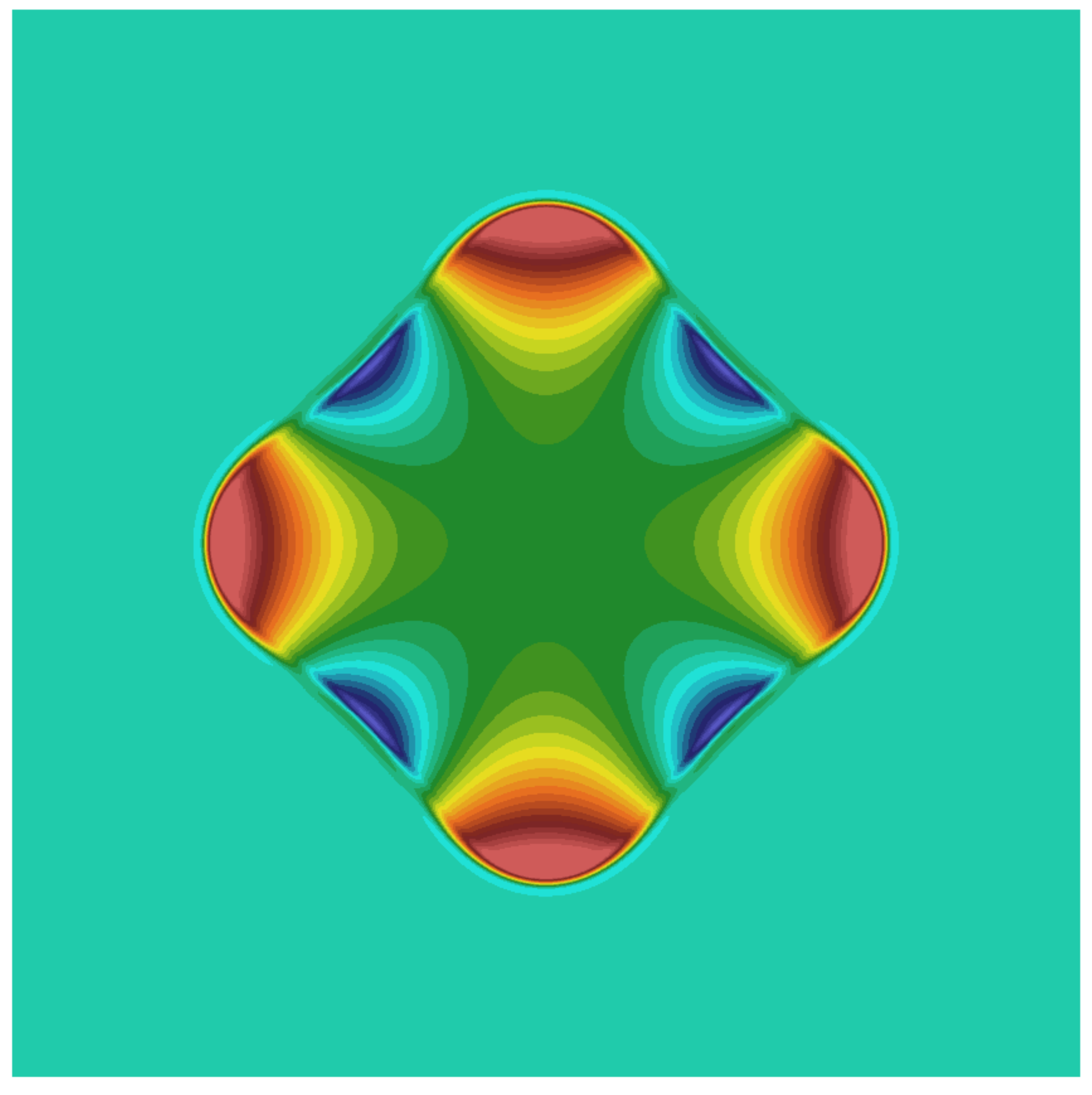}
		\subcaption{$t=1.52\cdot 10^{-4} \ \si{\second}$}			
	\end{subfigure}
	\begin{subfigure}{.5\textwidth}
		\centering		
		\includegraphics[width = \textwidth]{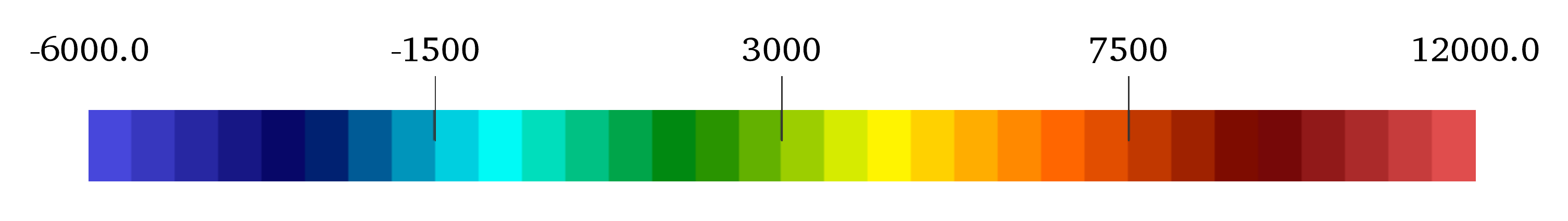}		
	\end{subfigure}		
	\caption{Large amplitude oscillation of a two-dimensional drop: Pressure contour plots.}
	\label{fig:large_oscillations_pres}	
\end{figure}

\begin{figure}[bt!]
	\captionsetup[subfigure]{labelformat=empty}
	\centering
	\begin{subfigure}{.40\textwidth}
		\centering	
		\includegraphics[width = \textwidth]{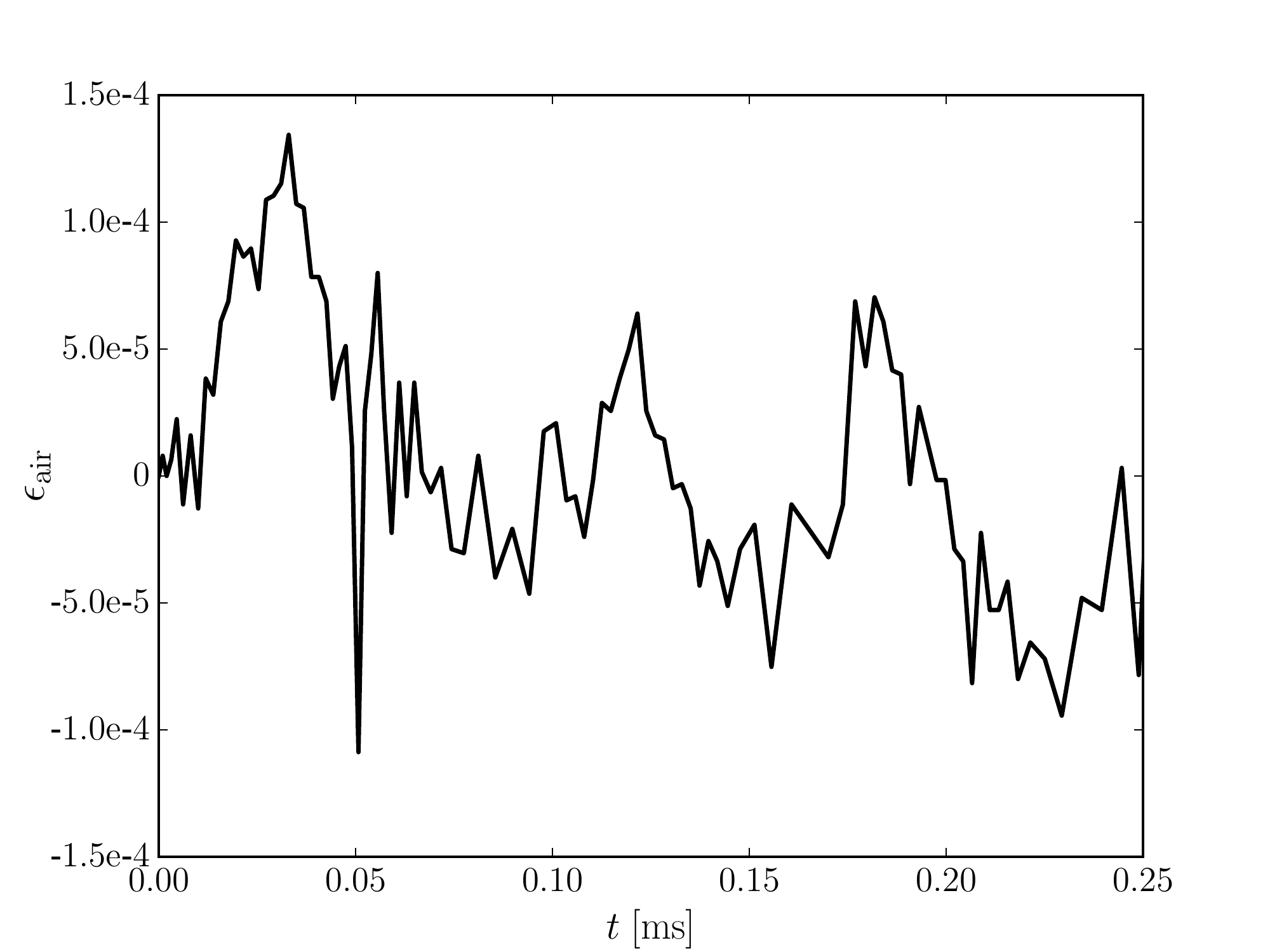}	
	\end{subfigure}
	\begin{subfigure}{.40\textwidth}
		\centering	
		\includegraphics[width = \textwidth]{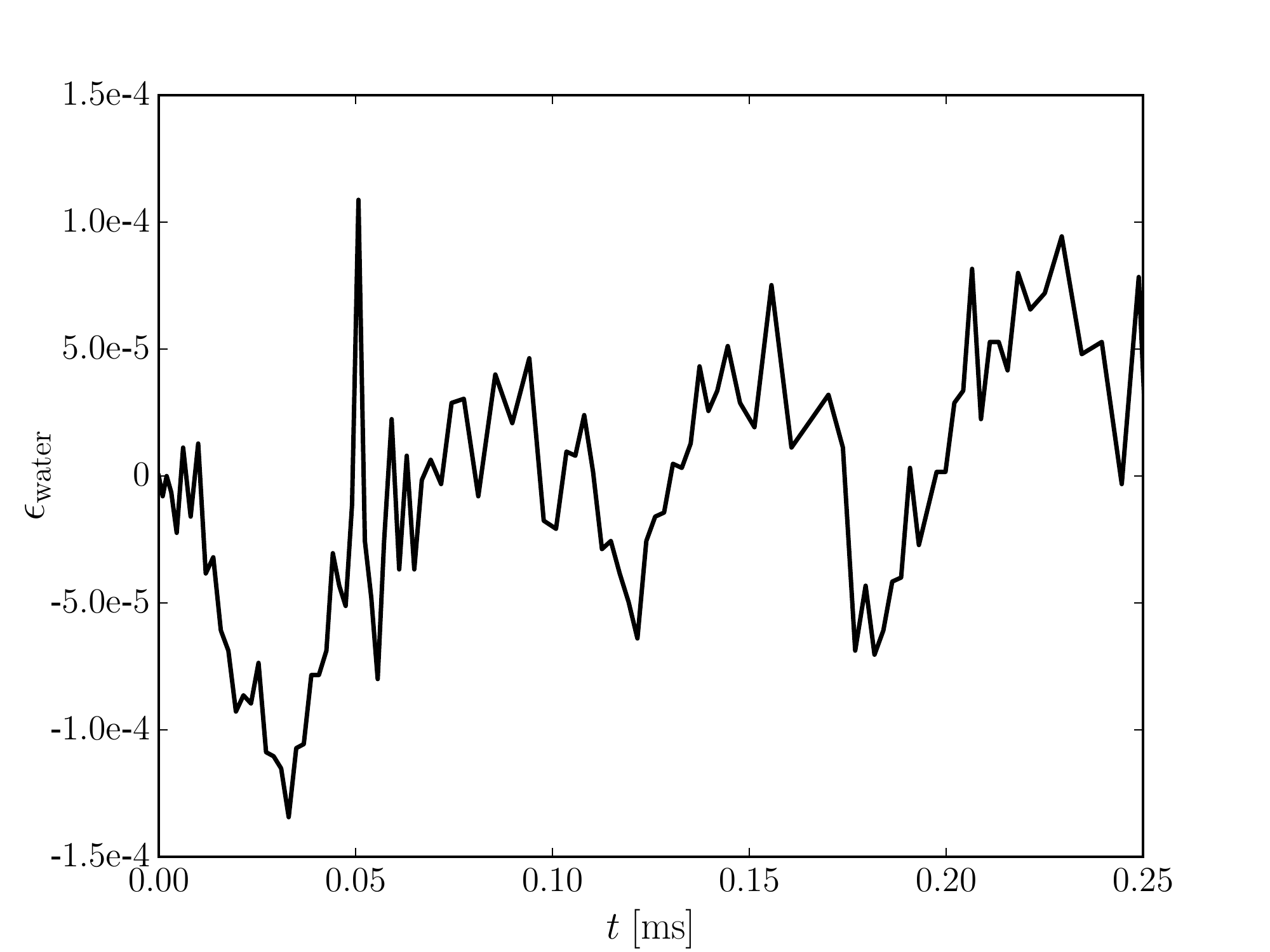}
	\end{subfigure}
	\caption{Large amplitude oscillation of a two-dimensional drop: Relative volume changes for air and water for $n=4$.}
	\label{fig:large_oscillations_vol}	
\end{figure}

 Figure \ref{fig:large_oscillations_amplitude} shows the evolution of the amplitude for the cases $n=2,3,4$. It is clear that the periods observed in the present study slightly lag behind the analytical period. This is due to the larger amplitudes (compared to Section \ref{sec: numerical examples: small oscillations}), which exceed the linear range and cause strongly nonlinear behaviour.
Figure \ref{fig:large_oscillations_phi} shows the interface in terms of the isolines of $\varphi$ at $\varphi=0$.
Figure \ref{fig:large_oscillations_velocity} shows the velocity streamlines at different time instances. Notably the axes of symmetry for all three modes are clearly defined.
Figure \ref{fig:large_oscillations_pres} shows the pressure distributions at different times within a single period of oscillation.
Clearly the pressure variation in the air is minimal, and higher pressure concentrations are present in regions with more curvature, as expected from the Young-Laplace equation. 
The accurate conservation of the volumes of water and air are demonstrated in Figure \ref{fig:large_oscillations_vol} for $n=4$, where the relative volume error is defined as $\epsilon(t_n) = (V_n - V_0)/V_0$. It is observed that for both air and water, the relative error does not exceed $0.015\%$. 
%
\subsection{Capillary rise} \label{sec: numerical examples: capillary rise}
\begin{figure}[bt!]
	\captionsetup[subfigure]{labelformat=empty}
	\centering
	\begin{subfigure}{.4\textwidth}
		\centering
		\includegraphics[width=.85\textwidth]{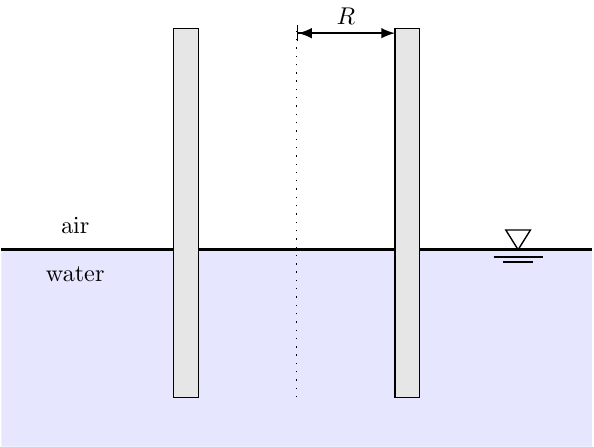}
	\end{subfigure}
	\begin{subfigure}{.4\textwidth}
		\centering
		\includegraphics[width=.85\textwidth]{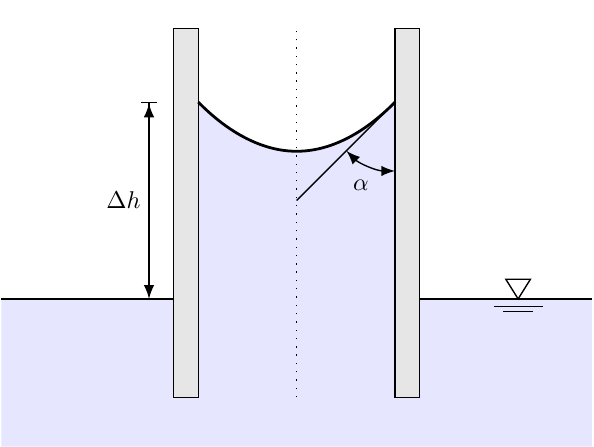}
	\end{subfigure}	
	\caption{Capillary rise: Initial (left) and final (right) configurations.}
	\label{fig:capillary_rise_geom}
\end{figure}

\begin{figure}[bt!]
	\captionsetup[subfigure]{labelformat=empty}
	\centering
	\begin{subfigure}{.425\textwidth}
		\centering
		\includegraphics[width=0.9\textwidth]{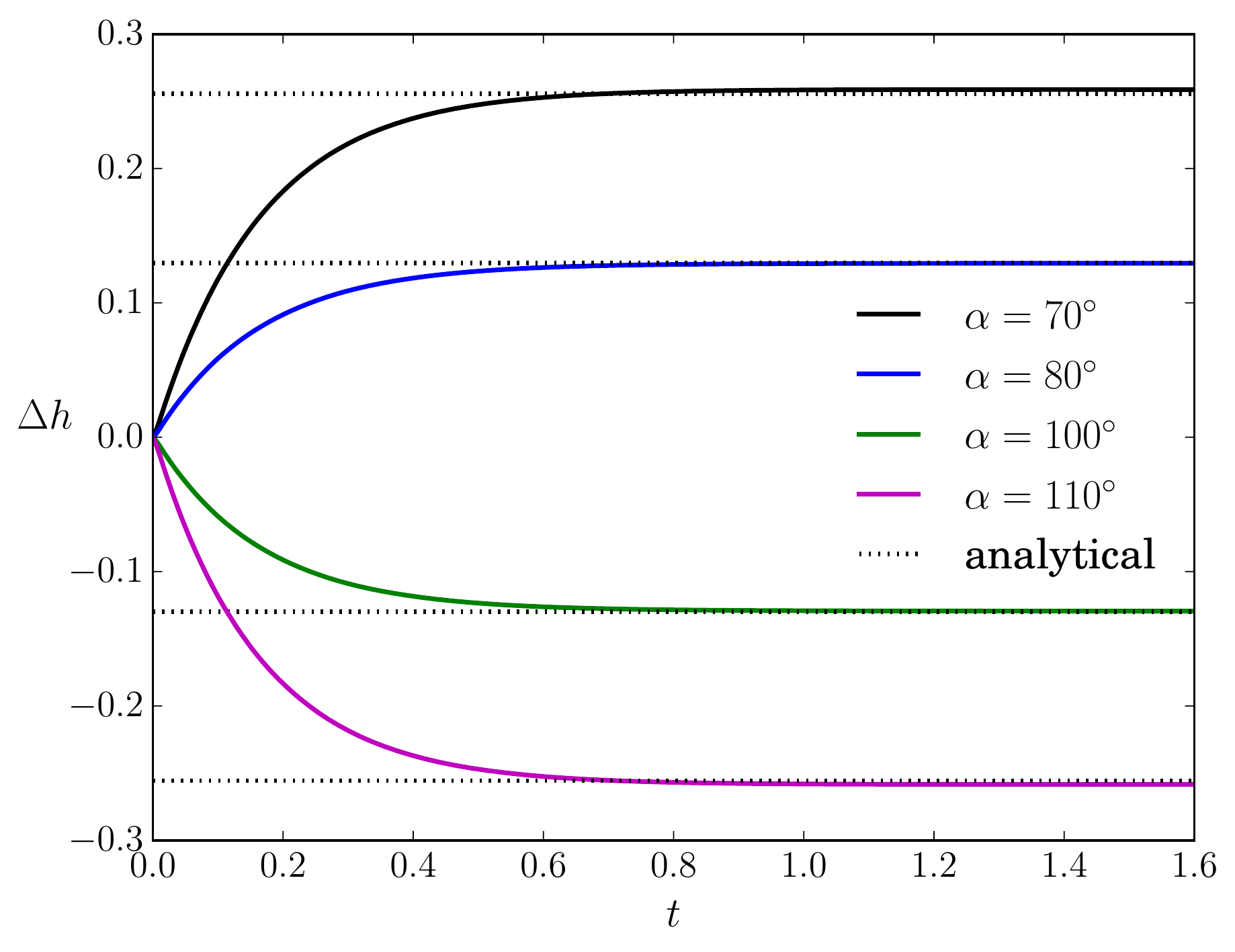}
	\end{subfigure}
	\begin{subfigure}{.425\textwidth}
		\flushleft
		\includegraphics[width=0.9\textwidth]{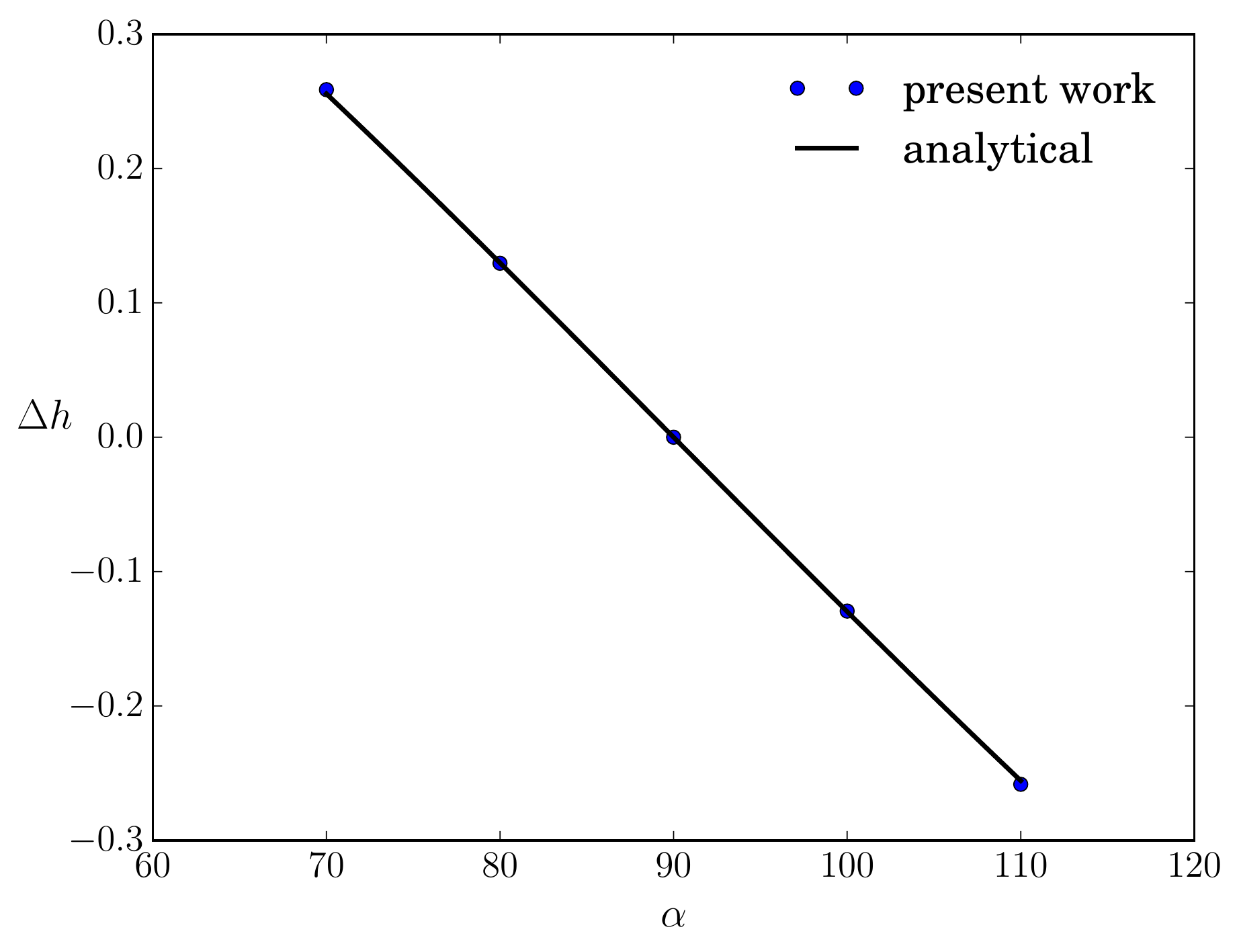}
	\end{subfigure}	
	\caption{Capillary rise between two parallel plates: Fluid level evolution with respect to time (left) and contact angle (right) compared to Jurin's law, with $256\times96$ linear elements.}
	\label{fig:capillary_rise_evo}
\end{figure}

\begin{figure}[tb!]
	\captionsetup[subfigure]{labelformat=empty}
	\centering
	\begin{subfigure}{.20\textwidth}
		\centering
		\includegraphics[width = 0.9\textwidth]{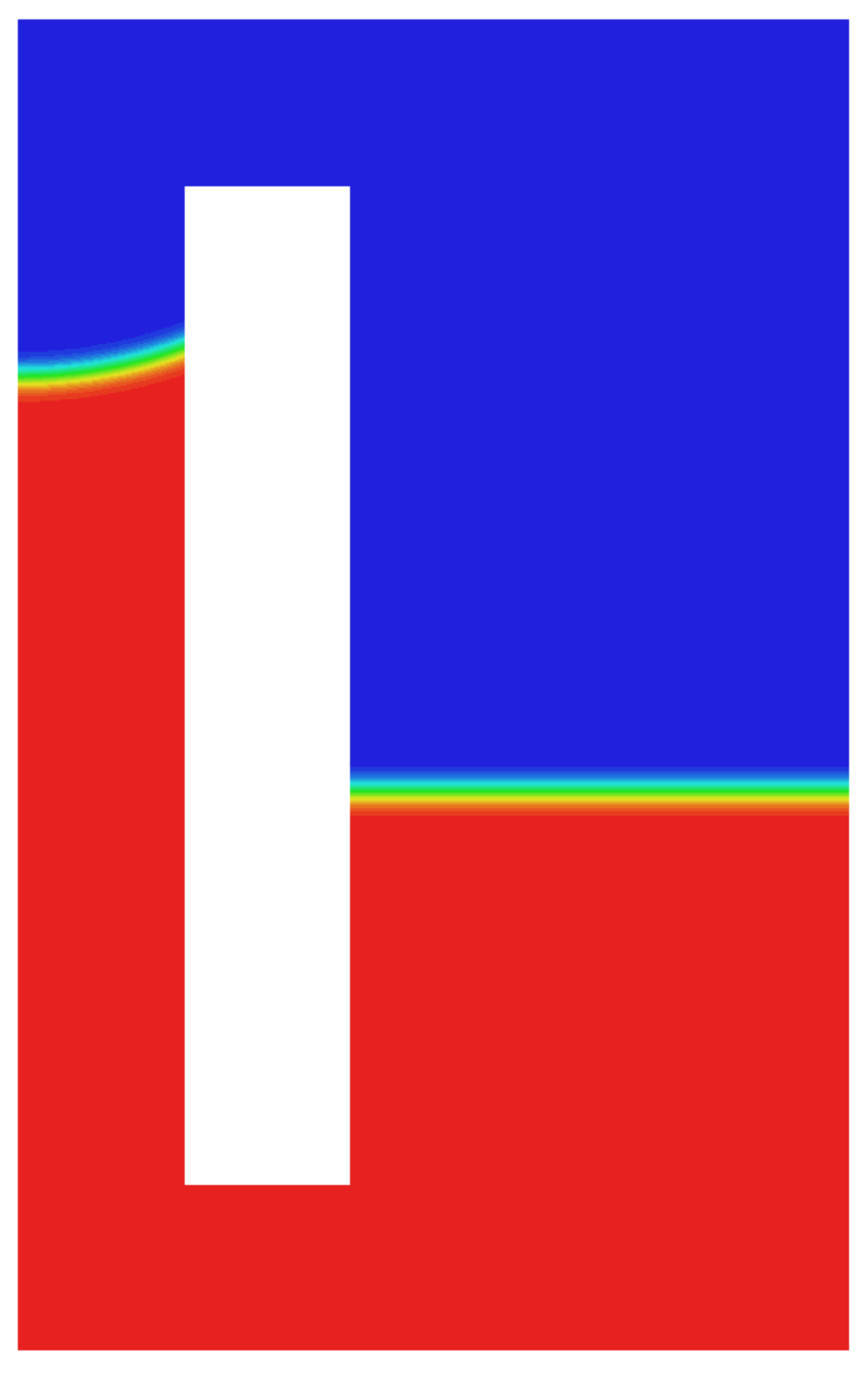}
		\subcaption{$\alpha=70^{\circ}$}
	\end{subfigure}
	\begin{subfigure}{.20\textwidth}
		\centering	
		\includegraphics[width = 0.9\textwidth]{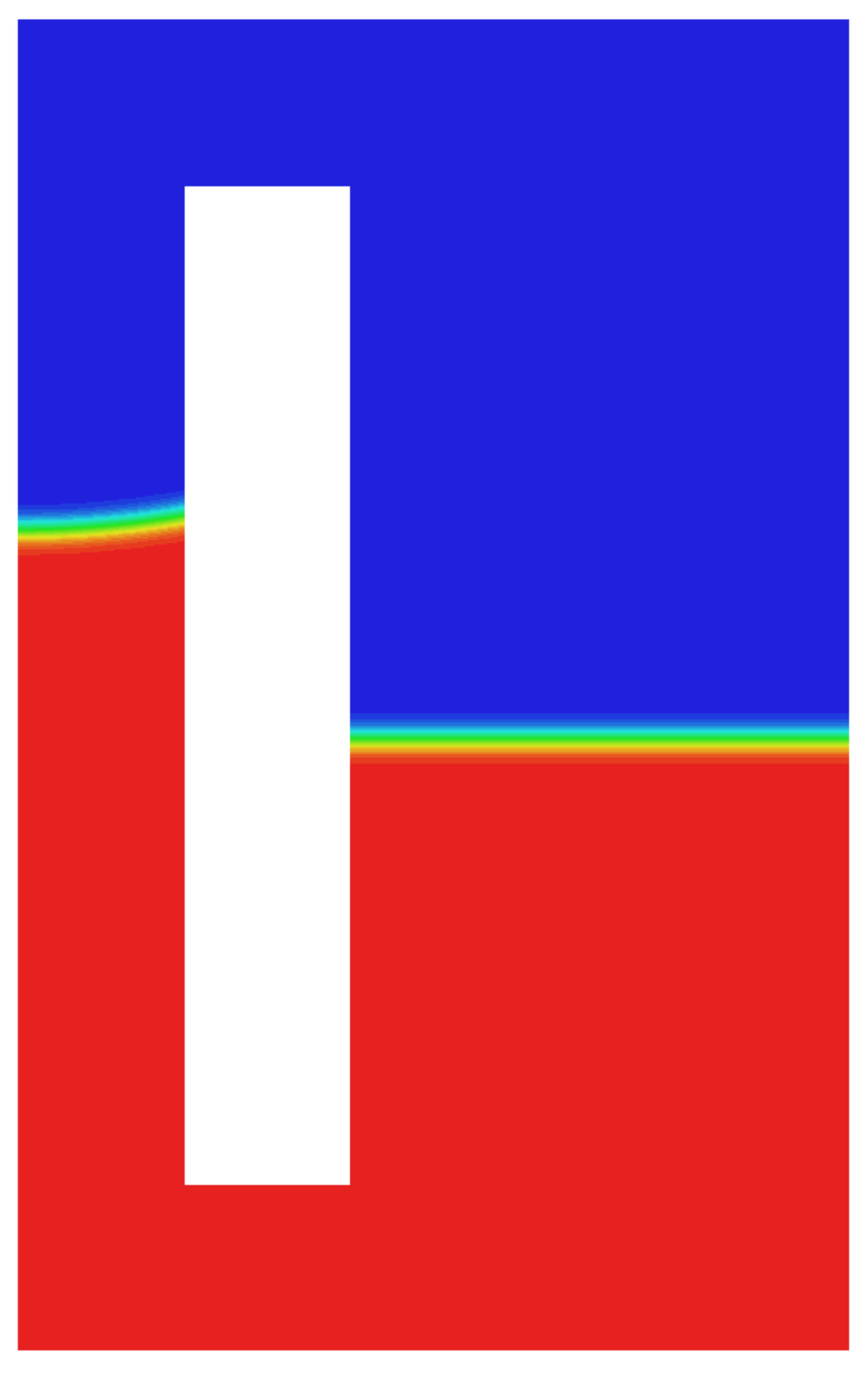}
		\subcaption{$\alpha=80^{\circ}$}			
	\end{subfigure}
	\begin{subfigure}{.20\textwidth}
		\centering		
		\includegraphics[width = 0.9\textwidth]{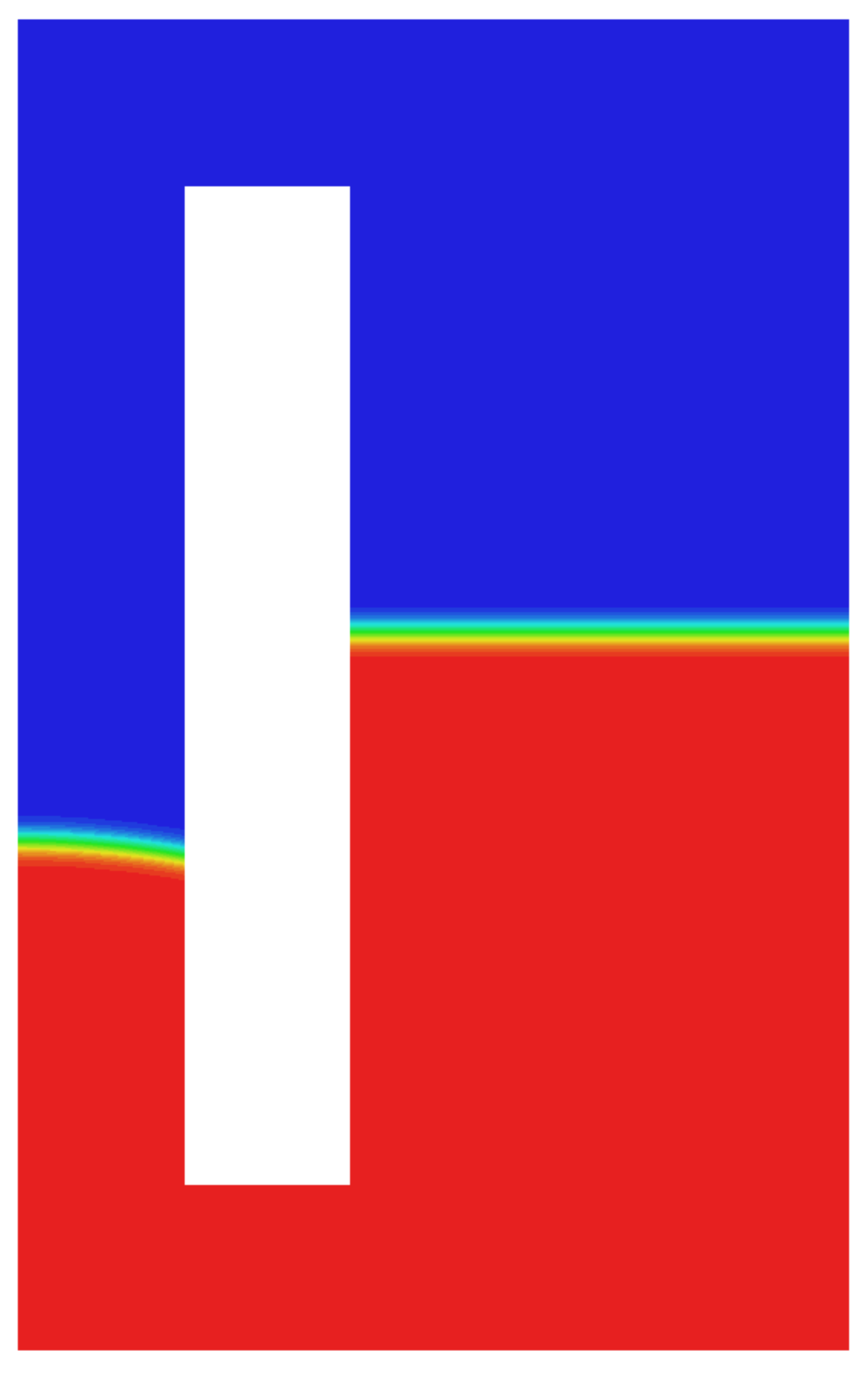}
		\subcaption{$\alpha=100^{\circ}$}			
	\end{subfigure}
	\begin{subfigure}{.20\textwidth}
		\centering		
		\includegraphics[width = 0.9\textwidth]{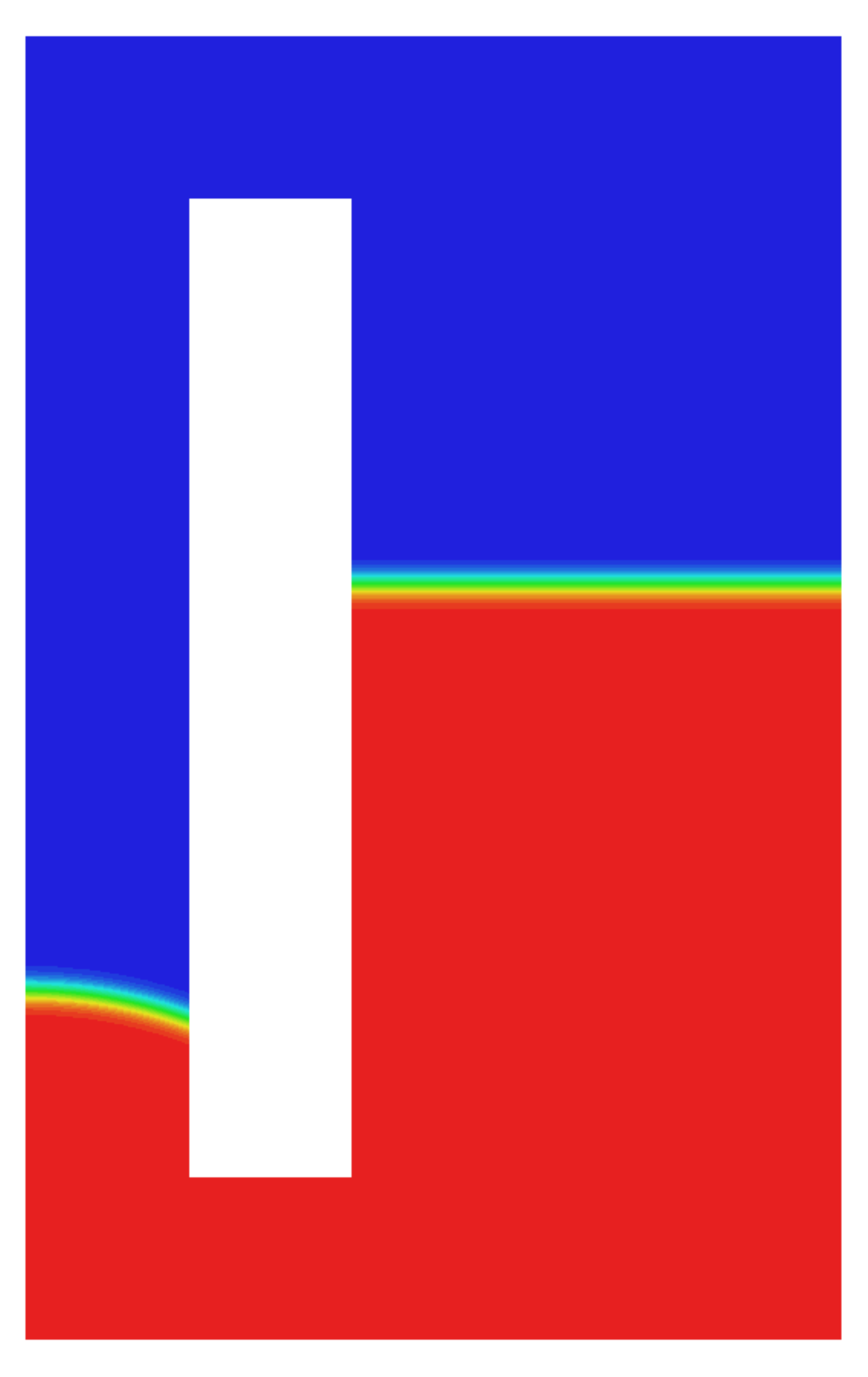}
		\subcaption{$\alpha=110^{\circ}$}			
	\end{subfigure}	
	\caption{Capillary rise between two parallel plates: One half of the equilibrium configurations of $\varphi$ for various contact angles, with $256\times96$ linear elements.}
	\label{fig:capillary_rise}	
\end{figure}

%
The capillary rise of fluid between two parallel plates is simulated. Figure \ref{fig:capillary_rise_geom} illustrates typical initial and final configurations of this problem. 
The column height can accurately be approximated using Jurin's law \cite{Liu2018}:
\begin{align}
	\Delta h &= \frac{\gamma \cos{\alpha}}{\rho g R} \quad \textrm{(2D)}
	\label{eq:JurinsLawa}
	\\\Delta h &= \frac{2\gamma \cos{\alpha}}{\rho g R} \quad \textrm{(axisymmetric)},
	\label{eq:JurinsLawb}
\end{align}
where $R$ is the pipe radius and $\alpha$ is the contact angle.
The initial configuration of the two-dimensional problem is as follows: The initial water level in a $[0,10]\times[0,8] \ \si{\milli\metre^2}$ container filled with air is set to $4 \ \si{\milli\metre}$. Two parallel plates at a distance $R=1 \ \si{\milli\metre}$, are placed in the centre of the container. Two meshes are considered: a coarse mesh with $64\times24$ linear quadrilateral elements and a fine mesh with $256\times96$ linear quadrilateral elements. Slip boundary conditions are applied on the side and top boundaries, as well as on the tube walls. The contact angle $\alpha$ is applied on the inner tube walls. The external air pressure is set to zero. Adaptive time stepping is used, with the initial time step size being $\Delta t = 0.001$.
The fluid levels $\Delta h$ are shown with respect to various hydrophilic and hydrophobic contact angles in Figure \ref{fig:capillary_rise_evo}. The numerical fluid levels are obtained as
\begin{align}
	\Delta h = \frac{1}{3}\left[ \Delta h(0) + \Delta h(R/2) + \Delta h(R) \right].
\end{align}
The fluid level heights are tabulated in Table \ref{tab:capillary_rise}.
%
%
%
\begin{table}[bt!]
	\centering
	\caption{Capillary rise between two parallel plates: Fluid level heights.}
	
	\begin{tabular}{ cccc }
		\toprule
		$\alpha$ & $\Delta h$ [coarse] & $\Delta h$ [fine] & $\Delta h$ [Jurin's law \eqref{eq:JurinsLawa}] \\
		\midrule
		$70 ^{\circ}$ & $0.2587$  & $0.2567$  & $0.2553$  \\
		$80 ^{\circ}$ & $0.1295$  & $0.1292$ & $0.1296$   \\
		$100^{\circ}$ & $-0.1294$ & $-0.1292$ & $-0.1296$ \\
		$110^{\circ}$ & $-0.2583$ & $-0.2567$ & $-0.2553$ \\	
		\bottomrule				
	\end{tabular}
	\label{tab:capillary_rise}
\end{table}
%
The equilibrium configurations are shown in Figure \ref{fig:capillary_rise}.
\par
\begin{figure}[bt!]
	\captionsetup[subfigure]{labelformat=empty}
	\centering
	\begin{subfigure}{.35\textwidth}
		\centering
		\includegraphics[width=0.85\textwidth]{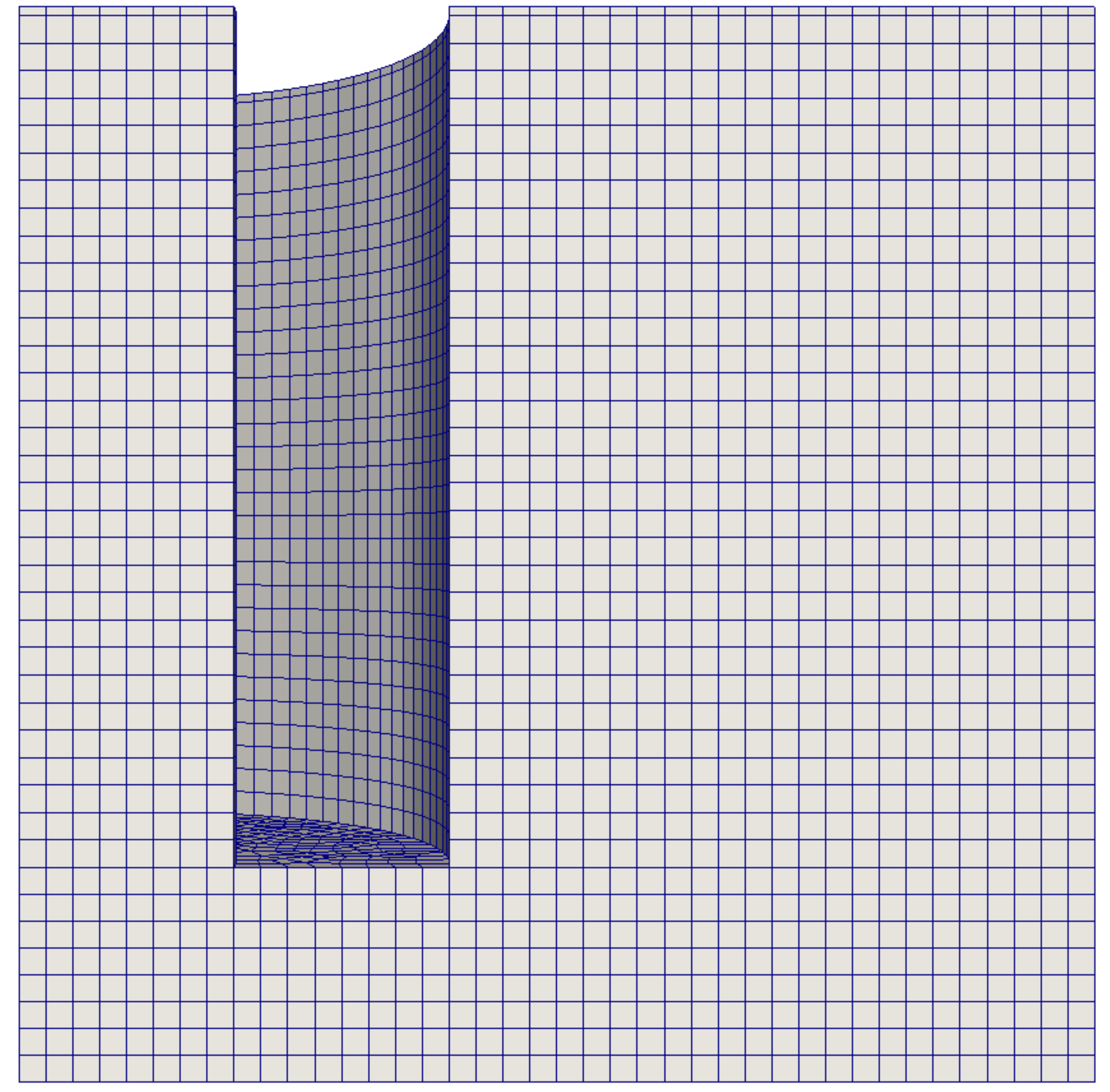}
	\end{subfigure}
	\begin{subfigure}{.35\textwidth}
		\centering
		\includegraphics[width=0.85\textwidth]{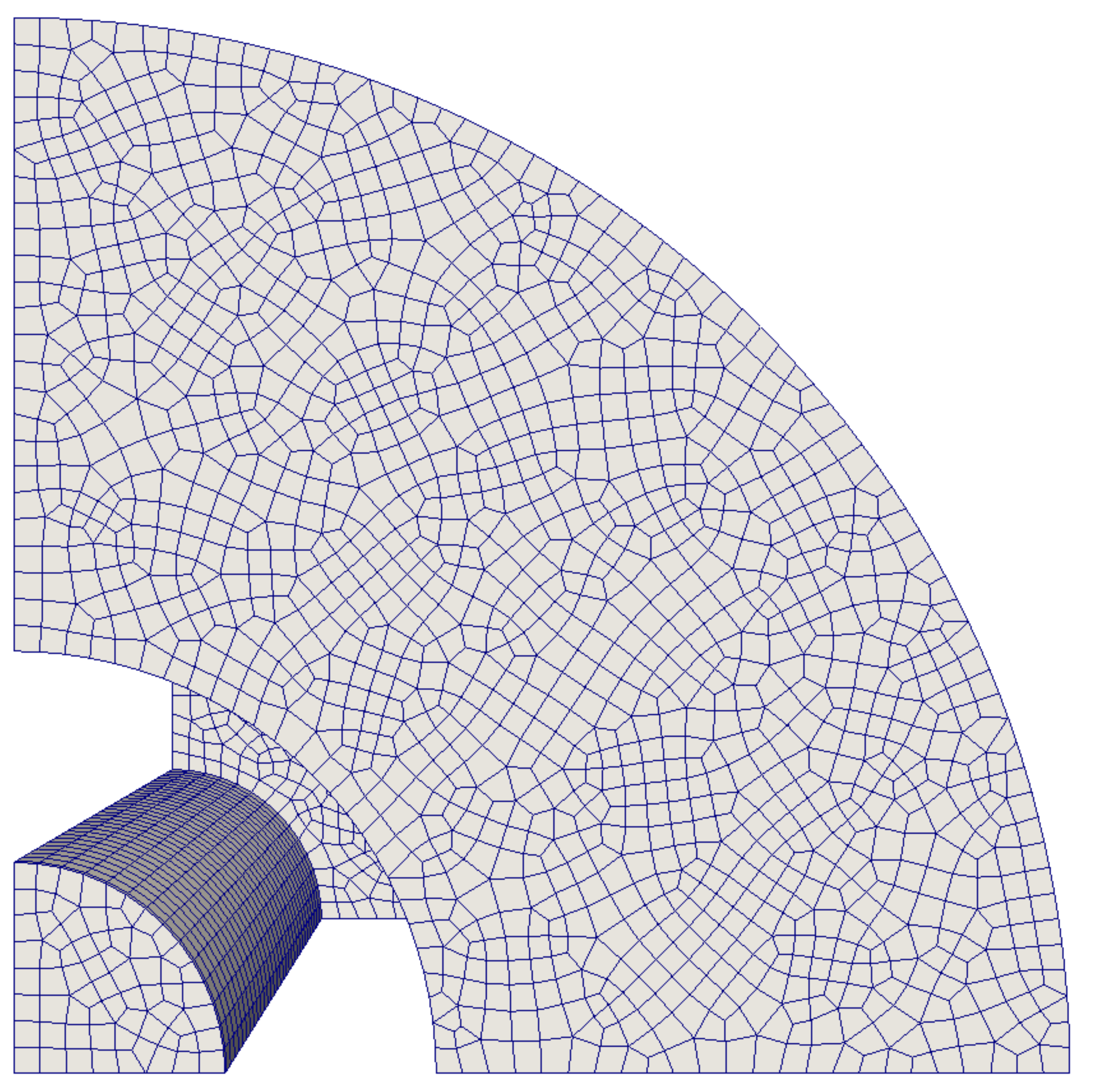}
	\end{subfigure}	
	\caption{Capillary rise in a circular pipe: Slice through horizontal axis of finite element mesh with 1,559,048 linear hexahedron elements.}
	\label{fig:capillary_rise_3D_geom}
\end{figure}

\begin{figure}[bt!]
	\captionsetup[subfigure]{labelformat=empty}
	\centering
	\begin{subfigure}[t]{.42\textwidth}
		\centering
		\includegraphics[width=0.9\textwidth]{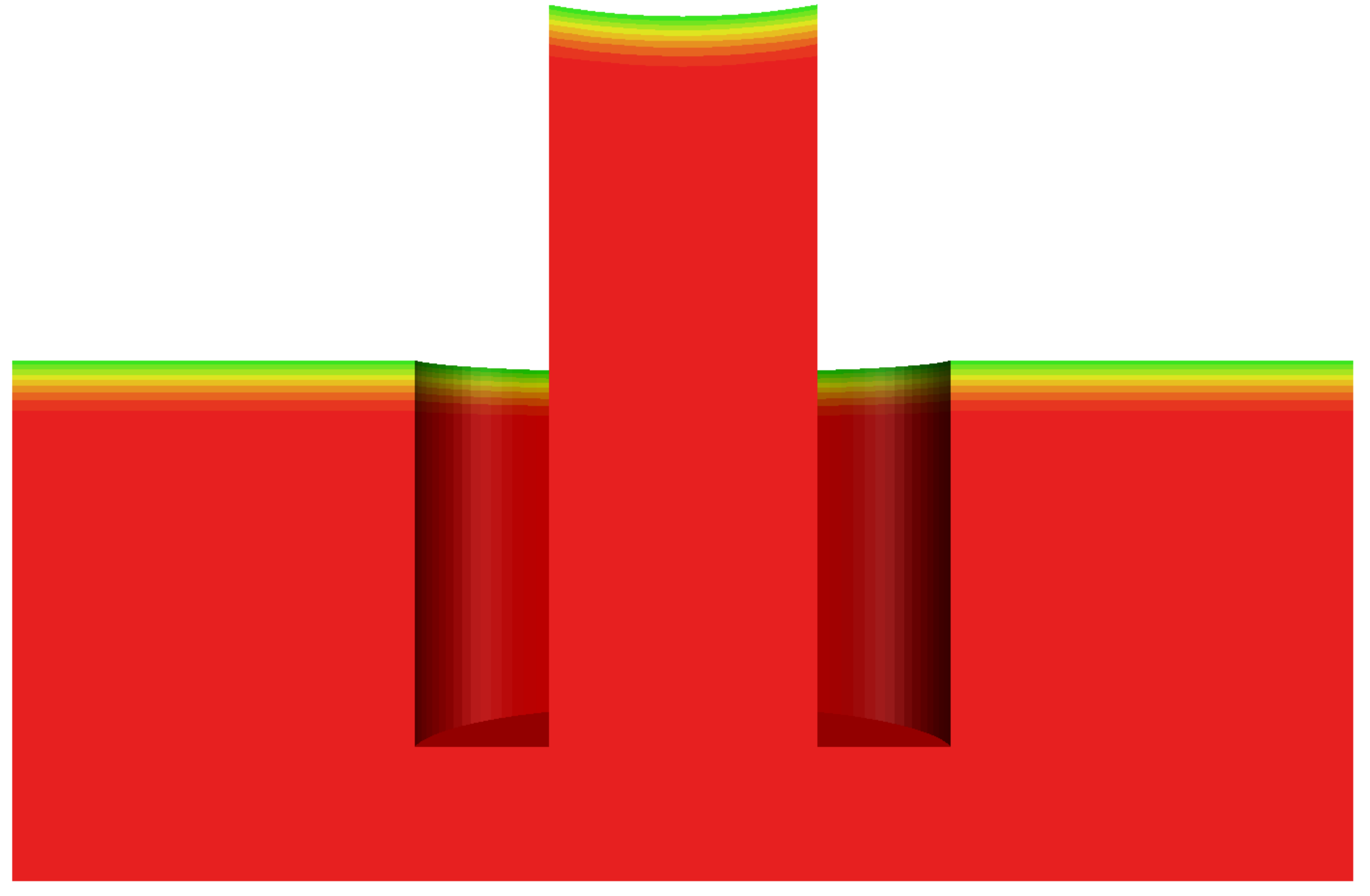}
		\subcaption{$\alpha=80^{\circ}$}	
	\end{subfigure}
	\begin{subfigure}[t]{.42\textwidth}
		\centering
		\includegraphics[width=0.9\textwidth]{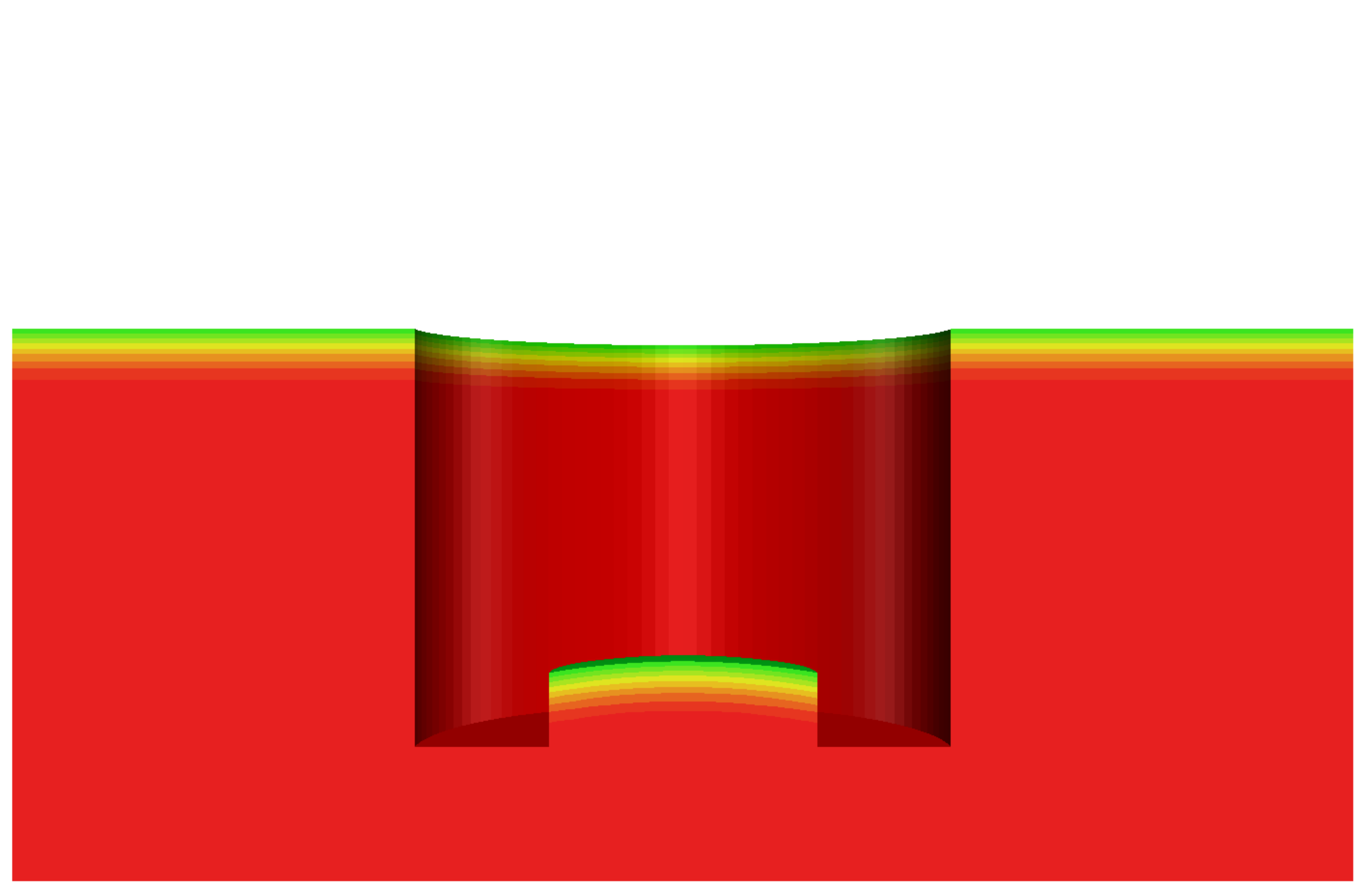}
		\subcaption{$\alpha=100^{\circ}$}	
	\end{subfigure}	
	\caption{Capillary rise in a circular pipe: Equilibrium configurations of $\varphi$, the part of the domain filled with air is not shown.}
	\label{fig:capillary_rise_3D_sol}
\end{figure}

 Capillary rise in a circular pipe is also considered. Following Equation \eqref{eq:JurinsLawb} it is expected that twice the fluid level height is reached. The geometry and mesh are shown in Figure \ref{fig:capillary_rise_3D_geom}.
The steady state solutions to $\varphi$ are shown in Figure \ref{fig:capillary_rise_3D_sol} and
Table \ref{tab:capillary_rise_3D} shows corresponding fluid levels for the present numerical solution and the analytical solution.
\begin{table}[tb!]
	\centering
	\caption{Capillary rise in a circular pipe: Fluid level heights.}
	
	\begin{tabular}{ ccc }
		\toprule
		$\alpha$ & $\Delta h$ & $\Delta h$ [Jurin's law \eqref{eq:JurinsLawb}] \\
		\midrule
		$80 ^{\circ}$ & $0.2574$  & $0.2595$   \\
		$100^{\circ}$ & $-0.2557 $ & $-0.2595$ \\
		\bottomrule				
	\end{tabular}
	\label{tab:capillary_rise_3D}
\end{table}
Clearly, excellent correspondence between numerical and analytical results is observed in all comparisons.
%
\subsection{Sessile drop} \label{sec: numerical examples: sessile drop}
\begin{figure}[bt!]
	\captionsetup[subfigure]{labelformat=empty}
	\centering
	\includegraphics[width=0.4\textwidth]{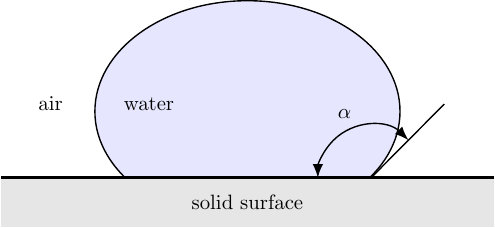}
	\caption{Sessile drops: Geometry.}
	\label{fig:sessile_drop_geom}
\end{figure}

%
A small water bubble in the presence of gravity is placed on a solid surface at the bottom of a rectangular domain filled with air, as shown in Figure \ref{fig:sessile_drop_geom}. 
The rectangular domain is of dimensions $[0,1.5]\times[0,0.5] \ \si{cm}^2$ and the drop radius is $R=0.25 \ \si{\centi\metre}$. Both two and three-dimensional sessile drops are considered with the same radius. In order to reduce dynamic effects and accelerate the computation of the equilibrium configuration, the viscosities are multiplied by a factor of $100$. Slip boundary conditions are applied at the lower boundary.
A mesh with $96\times64$ linear quadrilateral elements is considered. Adaptive time stepping is used, with an initial time step size of $\Delta t = 0.001$.
\par
\begin{figure}[bt!]
	\captionsetup[subfigure]{labelformat=empty}
	\centering
	\begin{subfigure}{.32\textwidth}
		\centering
		\includegraphics[width = 0.95\textwidth]{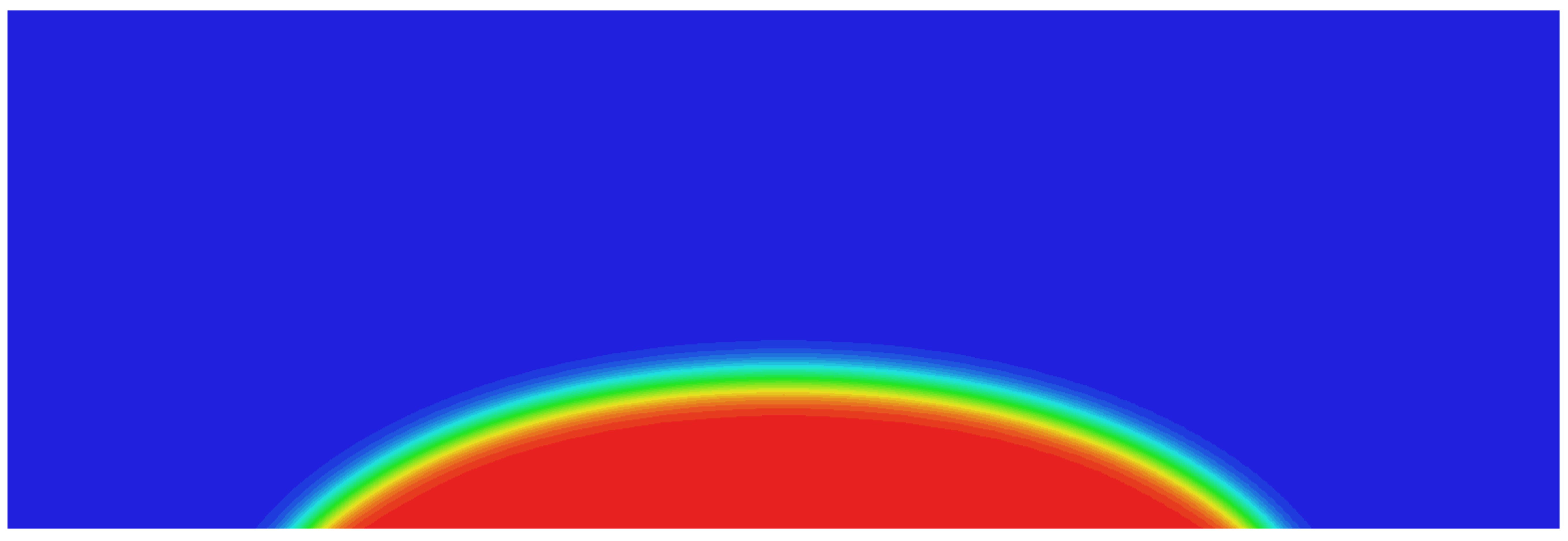}
		\subcaption{$\alpha=45^{\circ}$}
	\end{subfigure}
	\begin{subfigure}{.32\textwidth}
		\centering	
		\includegraphics[width = 0.95\textwidth]{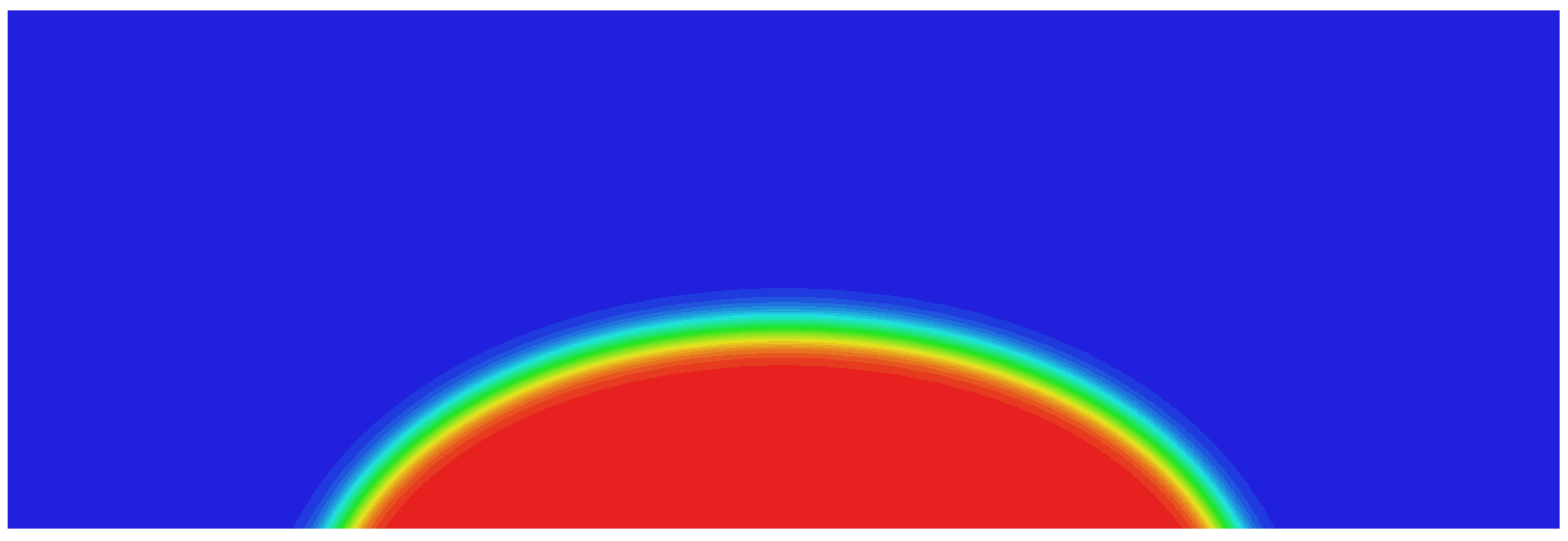}
		\subcaption{$\alpha=60^{\circ}$}			
	\end{subfigure}
	\begin{subfigure}{.32\textwidth}
		\centering		
		\includegraphics[width = 0.95\textwidth]{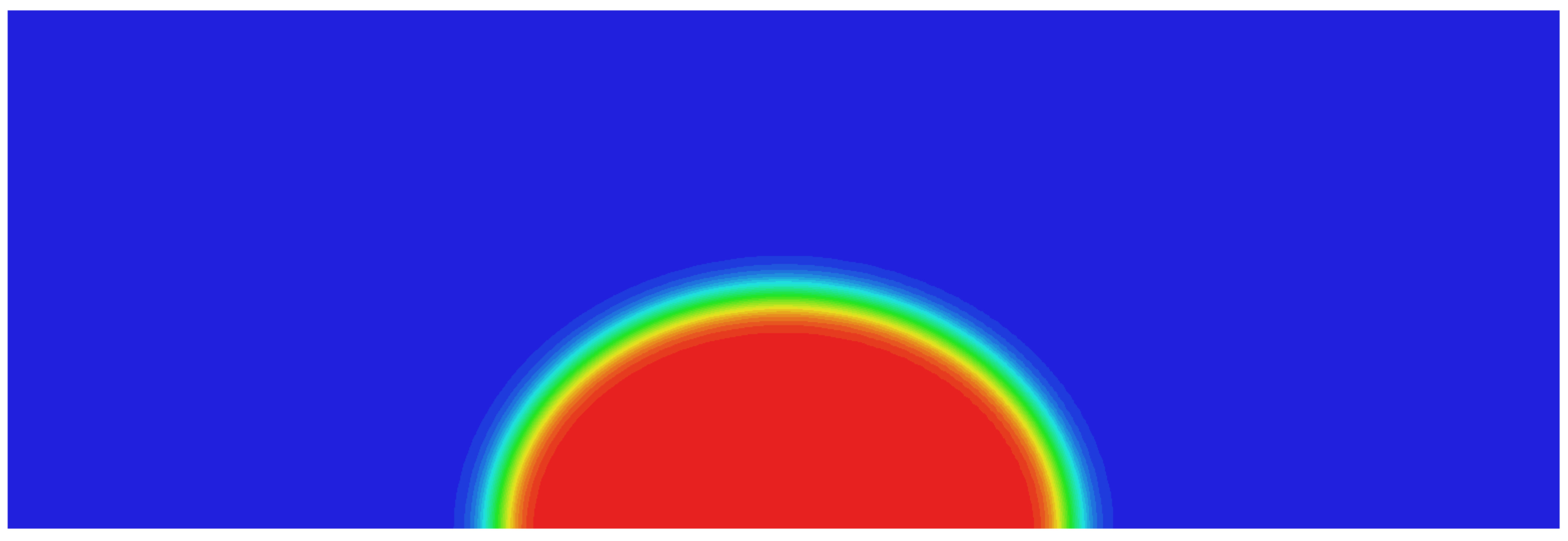}
		\subcaption{$\alpha=90^{\circ}$}			
	\end{subfigure}
	\begin{subfigure}{.32\textwidth}
		\centering		
		\includegraphics[width = 0.95\textwidth]{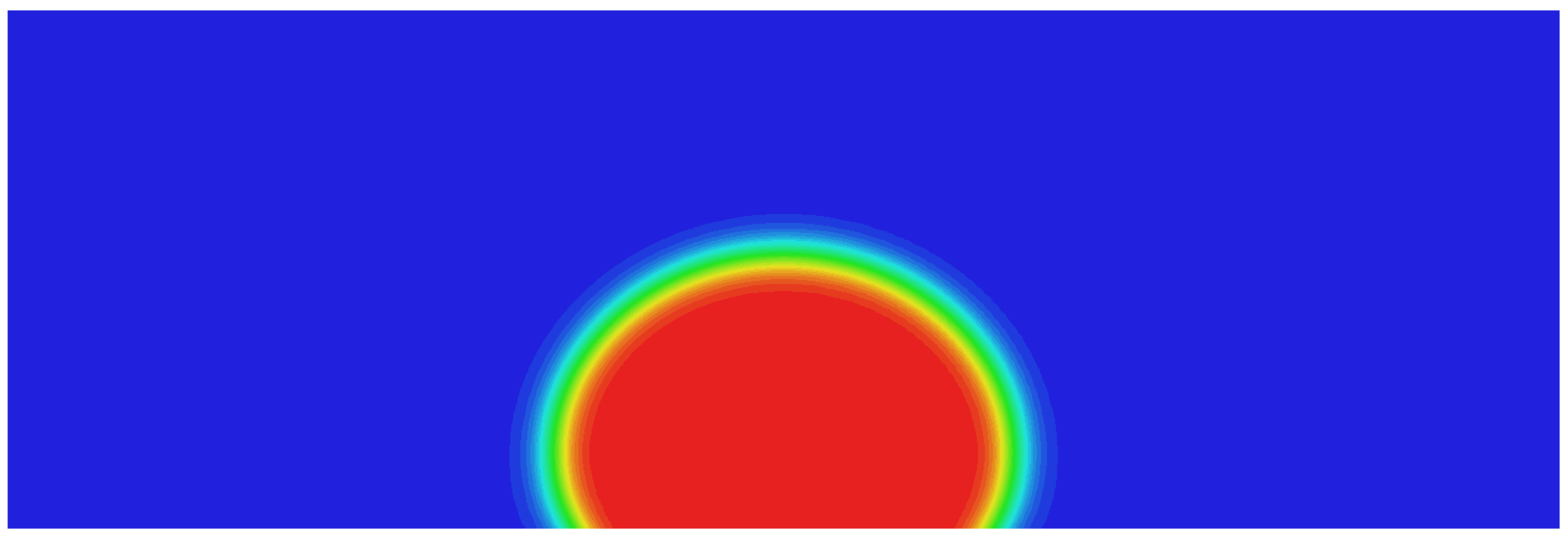}
		\subcaption{$\alpha=120^{\circ}$}			
	\end{subfigure}	
	\begin{subfigure}{.32\textwidth}
		\centering		
		\includegraphics[width = 0.95\textwidth]{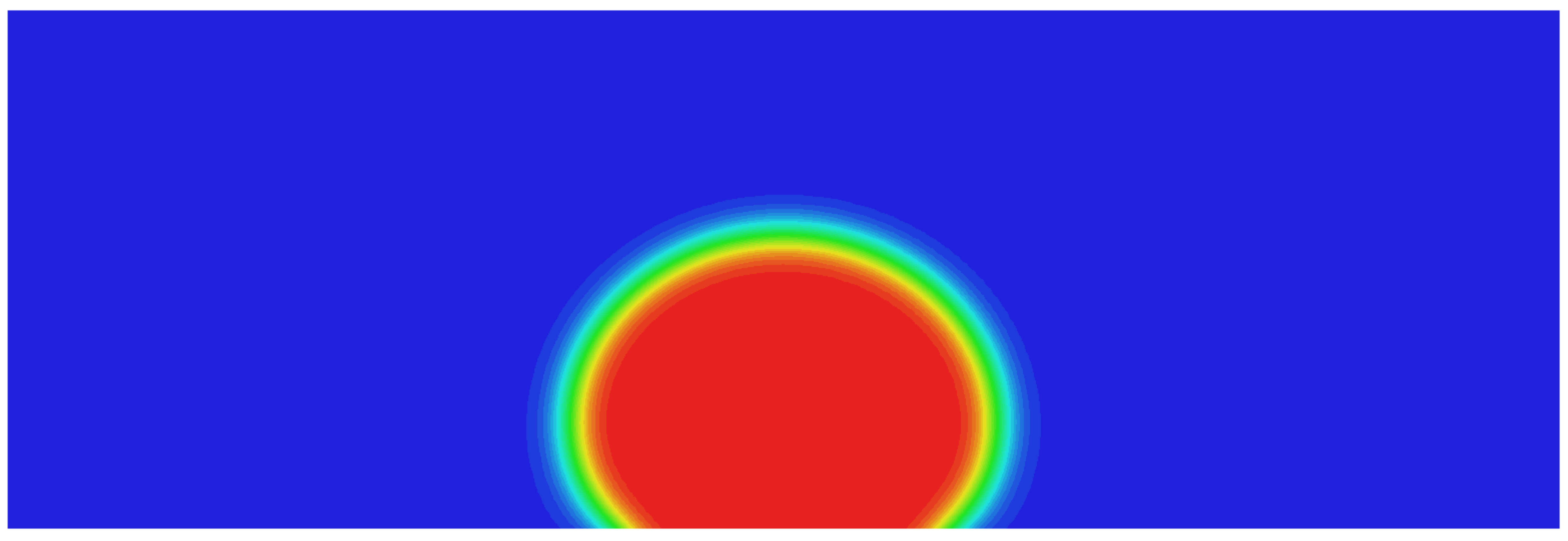}
		\subcaption{$\alpha=135^{\circ}$}			
	\end{subfigure}		
	\caption{Sessile drop in two dimensions: Equilibrium configuration of $\varphi$ using $96\times64$ linear elements.}
	\label{fig:sessile_drop_evolution}	
\end{figure}

\begin{figure}[bt!]
	\captionsetup[subfigure]{labelformat=empty}
	\centering
	\includegraphics[width=0.75\textwidth]{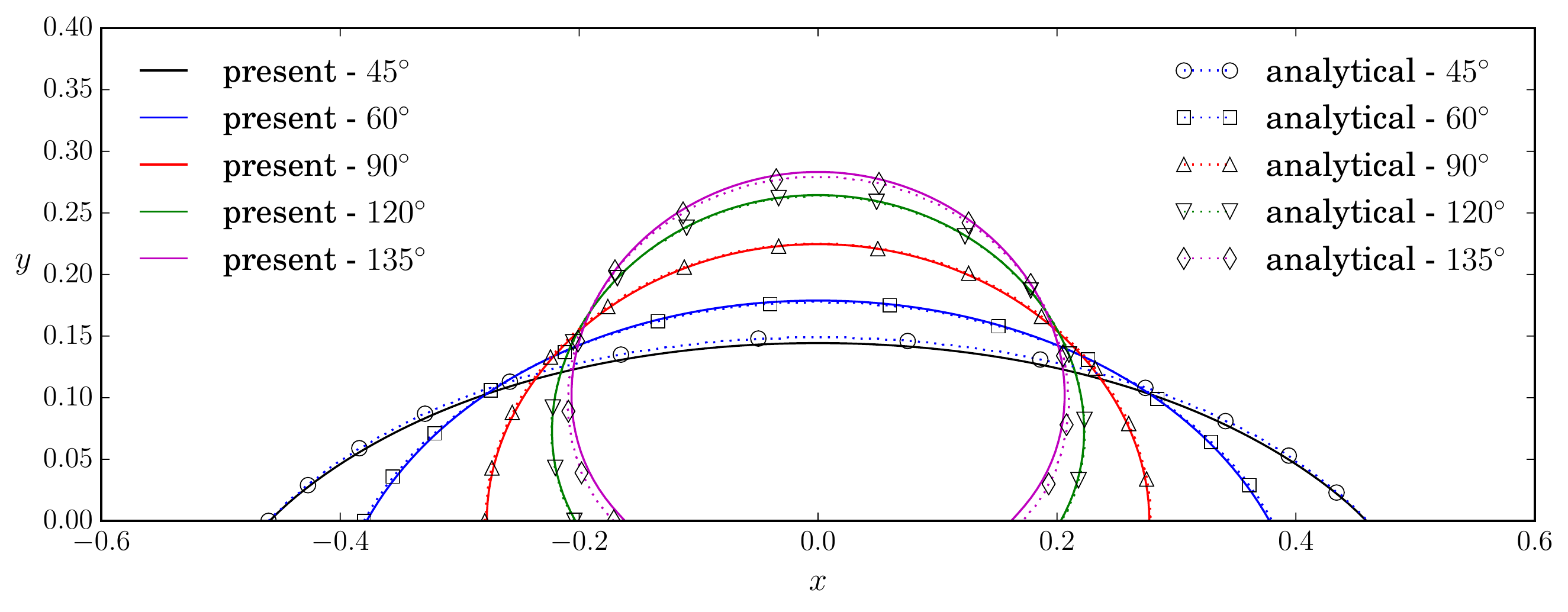}
	\caption{Sessile drop in two dimensions: Comparison of numerical results with the analytical solution given by Pozrikidis \cite{Pozrikidis1997}, using $96\times64$ linear elements.}
	\label{fig:sessile_drop_comp}
\end{figure}

\begin{figure}[tb!]
	\captionsetup[subfigure]{labelformat=empty}
	\centering
	\begin{subfigure}{.32\textwidth}
		\centering		
		\includegraphics[width = 1\textwidth,trim={0cm 1cm 0cm 0cm}]{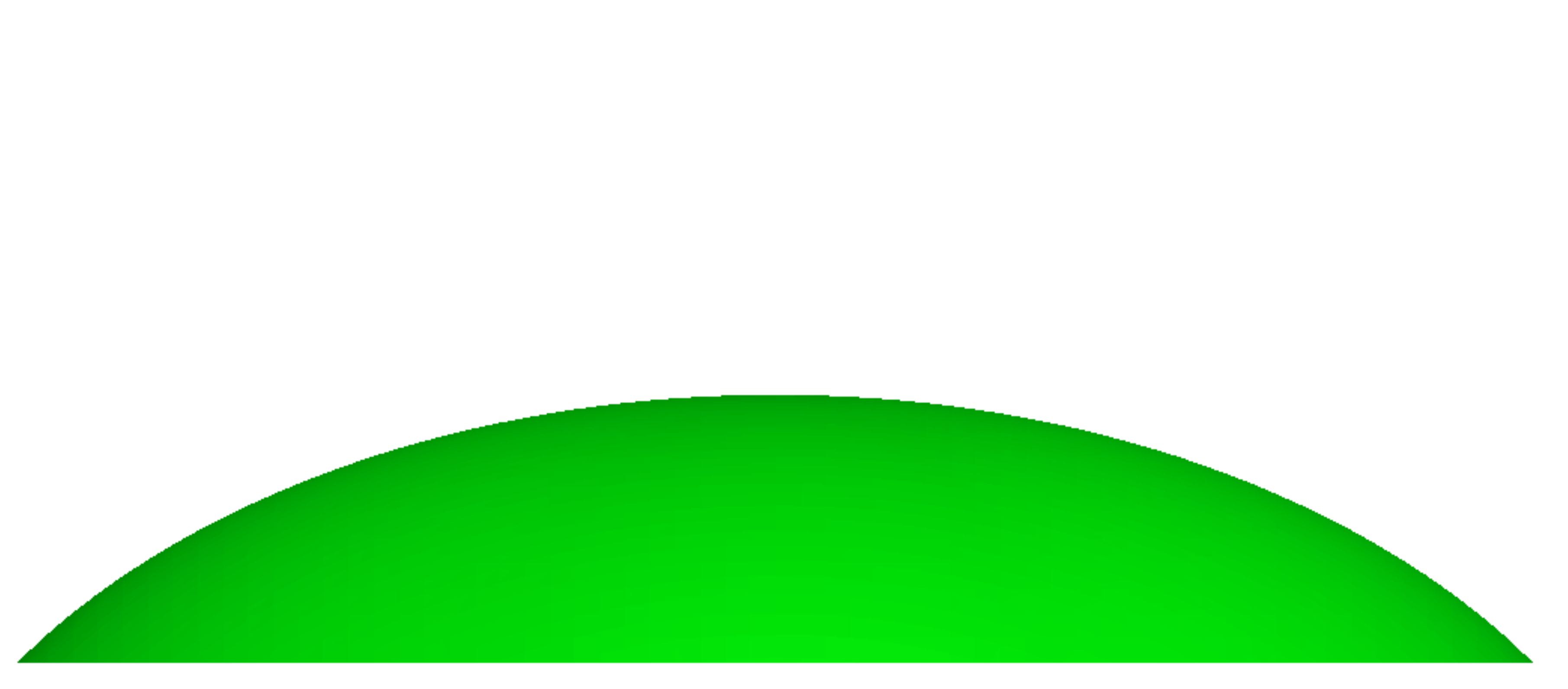}
		\subcaption{$\alpha=45^{\circ}$}			
	\end{subfigure}		
	\begin{subfigure}{.32\textwidth}
		\centering
		\includegraphics[width = 1\textwidth,trim={0cm 1cm 0cm 0cm}]{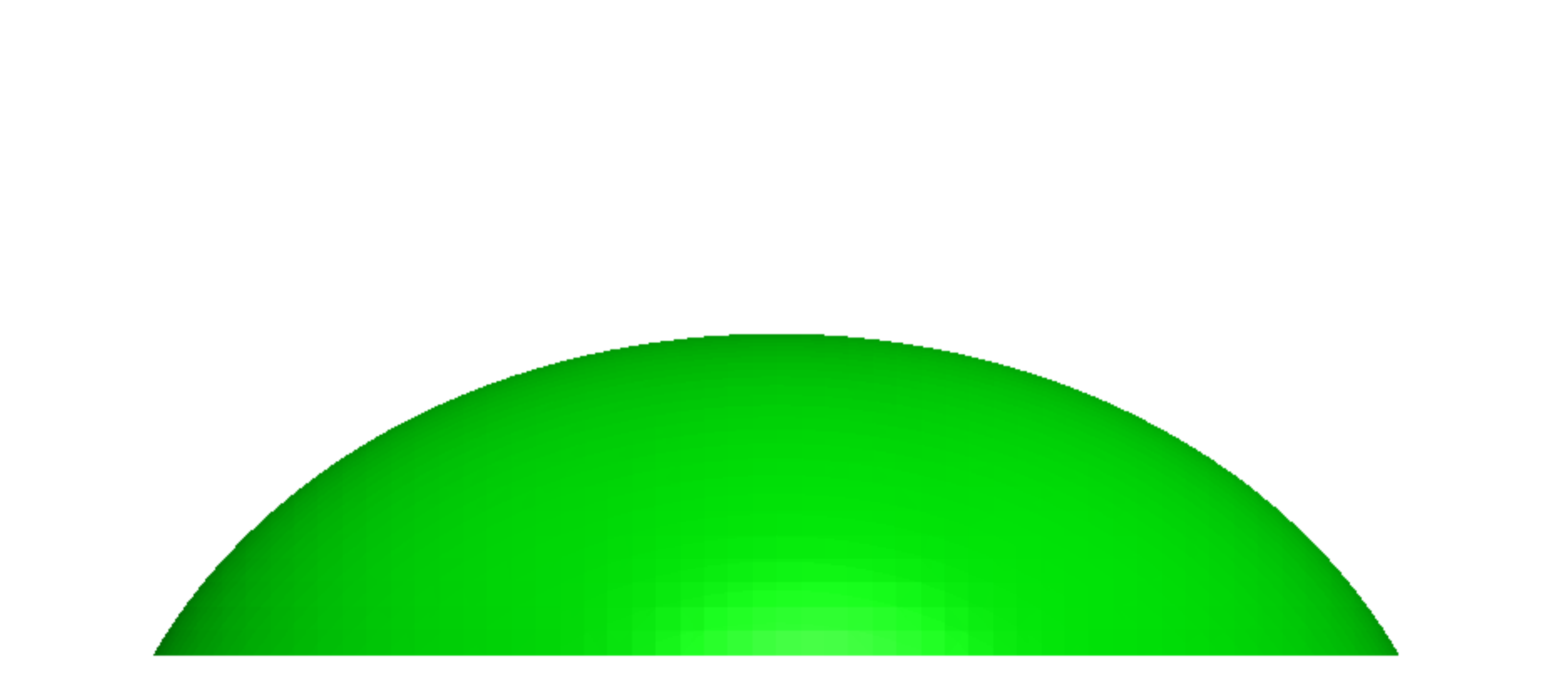}
		\subcaption{$\alpha=60^{\circ}$}
	\end{subfigure}
	\begin{subfigure}{.32\textwidth}
		\centering	
		\includegraphics[width = 1\textwidth,trim={0cm 1cm 0cm 0cm}]{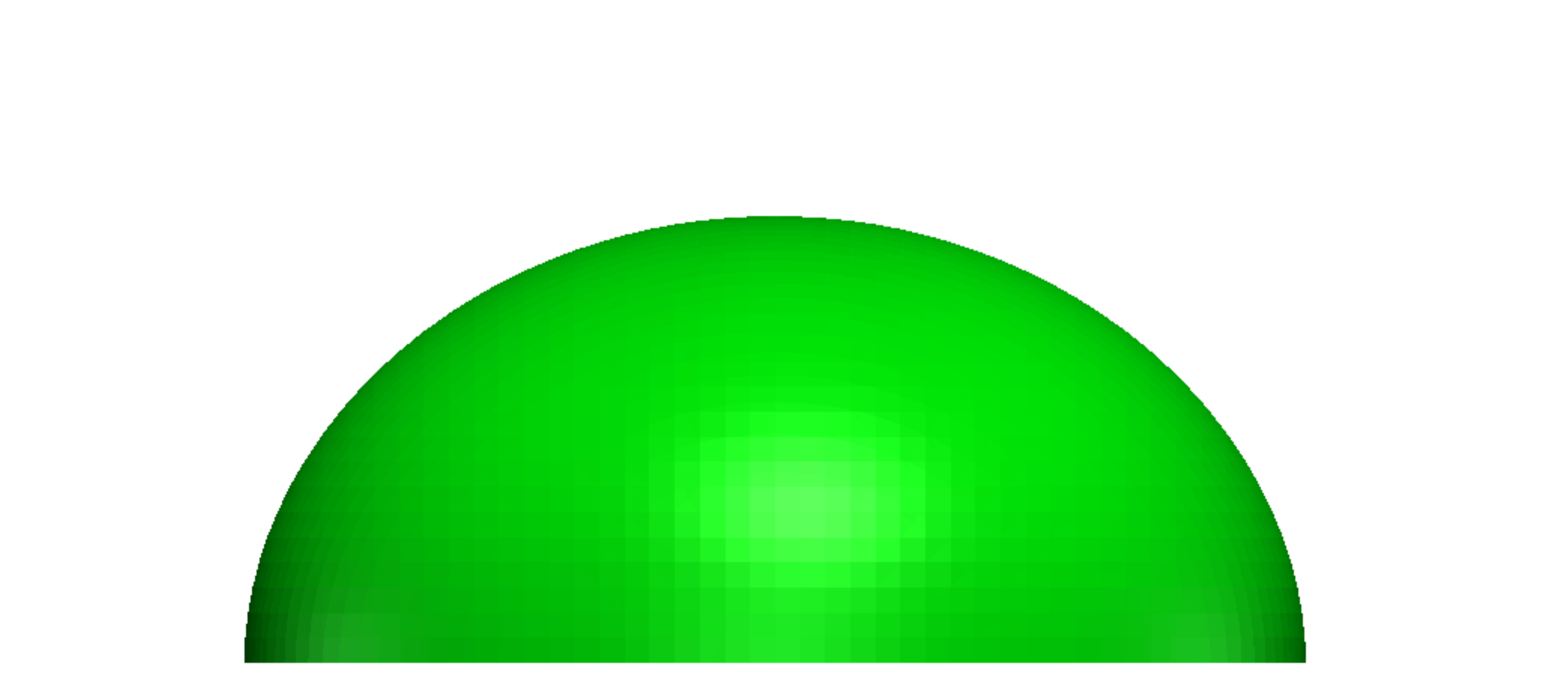}
		\subcaption{$\alpha=90^{\circ}$}			
	\end{subfigure}
	\begin{subfigure}{.32\textwidth}
		\centering		
		\includegraphics[width = 1\textwidth,trim={0cm 1cm 0cm 0cm}]{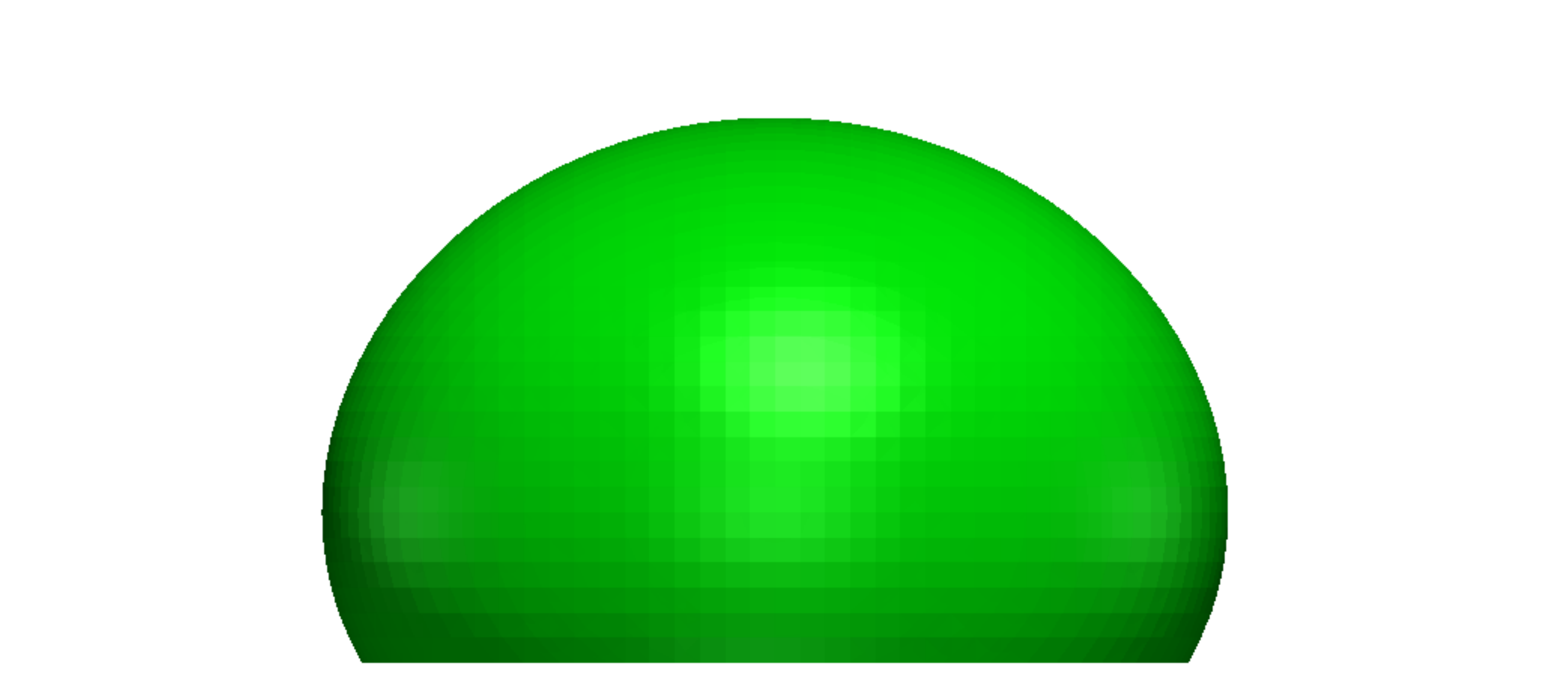}
		\subcaption{$\alpha=120^{\circ}$}			
	\end{subfigure}
	\begin{subfigure}{.32\textwidth}
		\centering		
		\includegraphics[width = 1\textwidth,trim={0cm 1cm 0cm 0cm}]{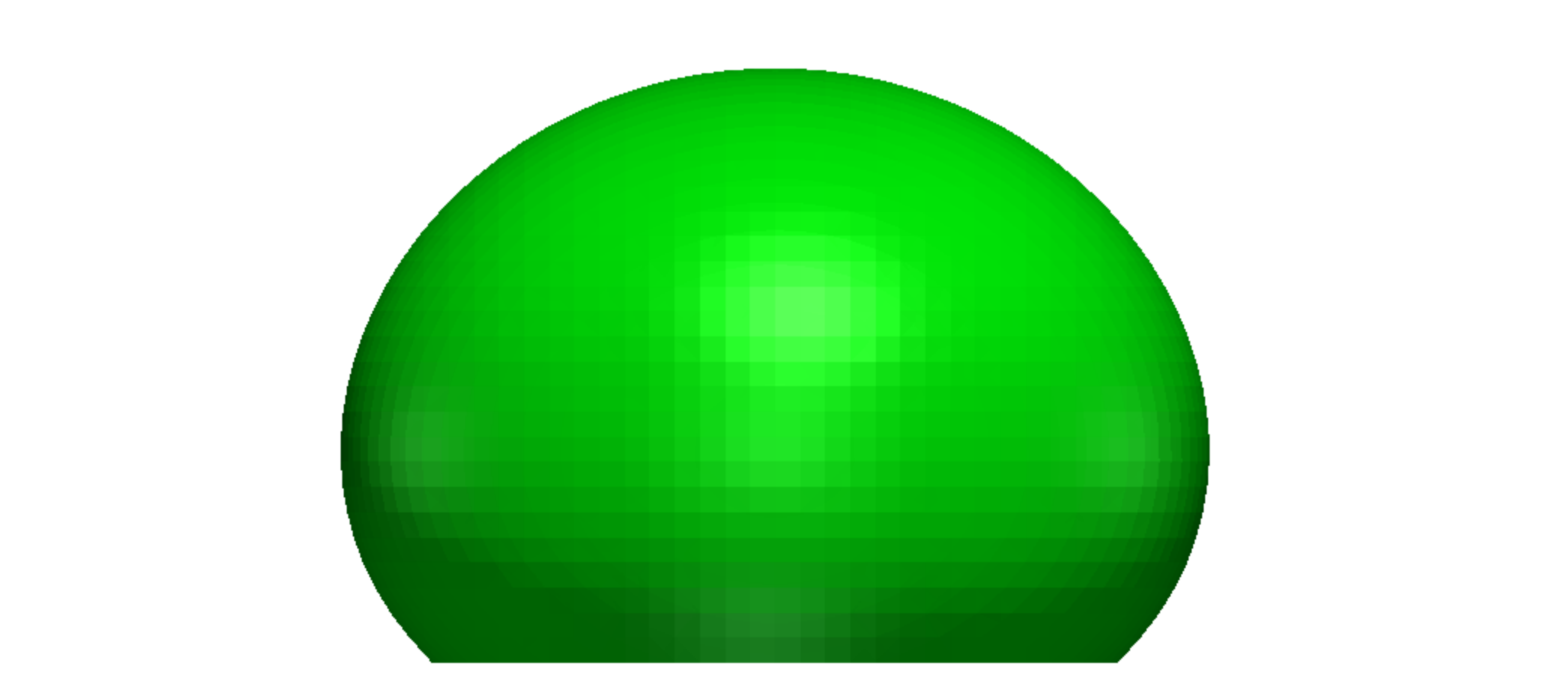}
		\subcaption{$\alpha=135^{\circ}$}			
	\end{subfigure}	
	\caption{Sessile drop: Steady state $\varphi$ solution for a small three-dimensional sessile drops with various contact angles, using $48\times32\times48$ linear elements.}
	\label{fig:sessile_drop_evolution_3D}	
\end{figure}

\begin{figure}[tb!]
	\captionsetup[subfigure]{labelformat=empty}
	\centering
	\includegraphics[width=0.7\textwidth]{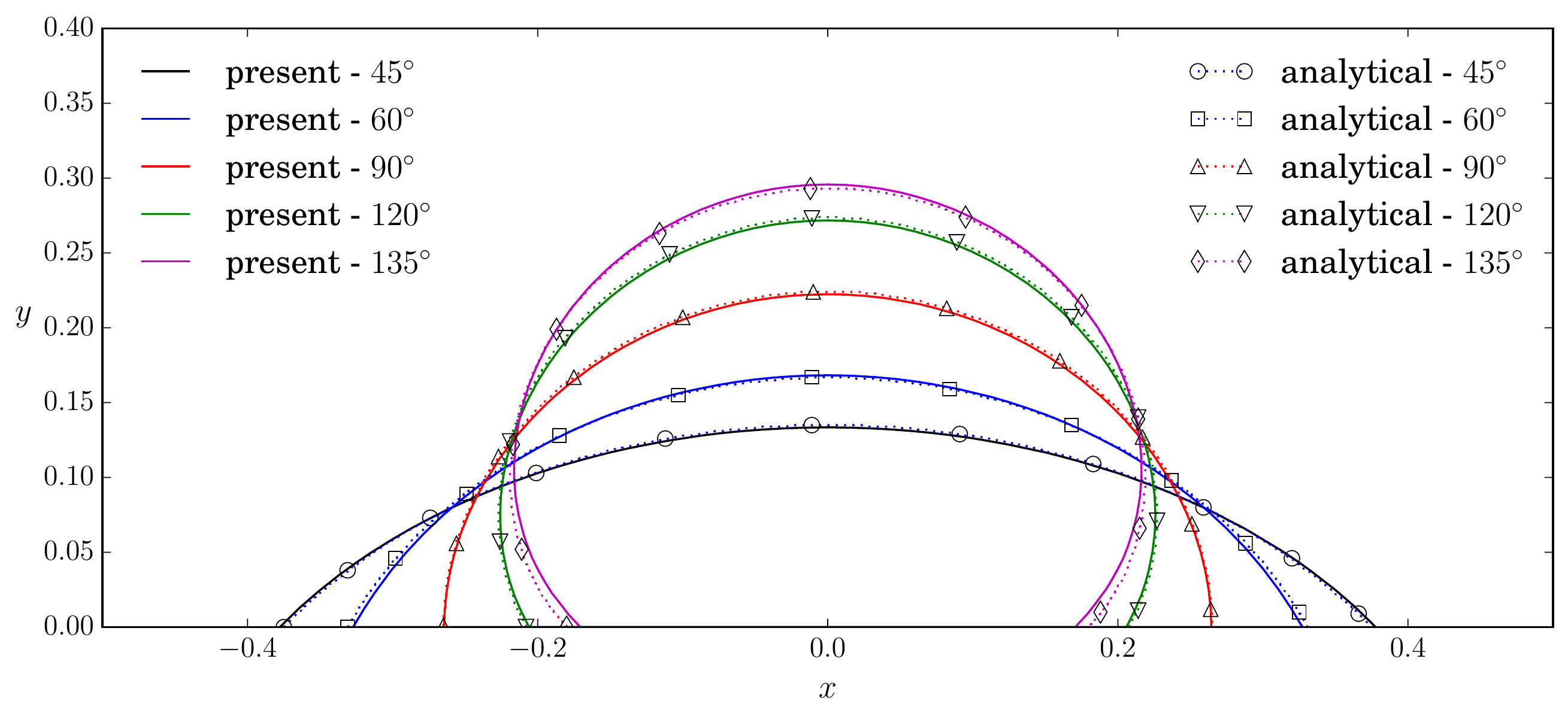}
	\caption{Sessile drop: Comparison of a small three-dimensional sessile drops with various contact angles to the analytical solution given by Pozrikidis \cite{Pozrikidis1997}, using $48\times32\times48$ linear elements.}
	\label{fig:sessile_drop_comp_3D_small}
\end{figure}

 The final configurations of the two-dimensional drops subject to various contact angles are shown in Figure \ref{fig:sessile_drop_evolution}.
In Figure \ref{fig:sessile_drop_comp}, the results are compared to the exact equilibrium configuration obtained by numerically solving a set of ordinary differential equations \cite{Pozrikidis1997}.
Figures \ref{fig:sessile_drop_evolution_3D} and \ref{fig:sessile_drop_comp_3D_small} show the results obtained for the three-dimensional case. A very good match is observable for both two and three-dimensional cases.
%
%
\subsection{Rising bubble} \label{sec: numerical examples: rising bubble}
%
In this example a bubble of radius $R=0.25$ is placed near the bottom of a rectangular domain of size $[0,1]\times[0,2]$ filled with a heavier fluid. The parameters are set as given in \cite{Hosseini2017}, with the heavier fluid properties $\rho = 1000$, $\mu = 10$, and the lighter bubble properties $\rho=1$, $\mu=1$. The surface tension parameter and the acceleration vector are set, respectively, to $\gamma = 1.96$ and $b = [0,-0.98]^T$. The mesh consists of $128\times256$ linear elements, and a fixed time step size of $\Delta t = 0.01$ is used. The boundaries are set up such that the fluids are allowed to slip along the left and right boundaries, but not along the top and bottom boundaries.
\par
\begin{figure}[tb!]
	\captionsetup[subfigure]{labelformat=empty}
	\centering
	\begin{subfigure}{.19\textwidth}
		\centering
		\includegraphics[width = 0.95\textwidth]{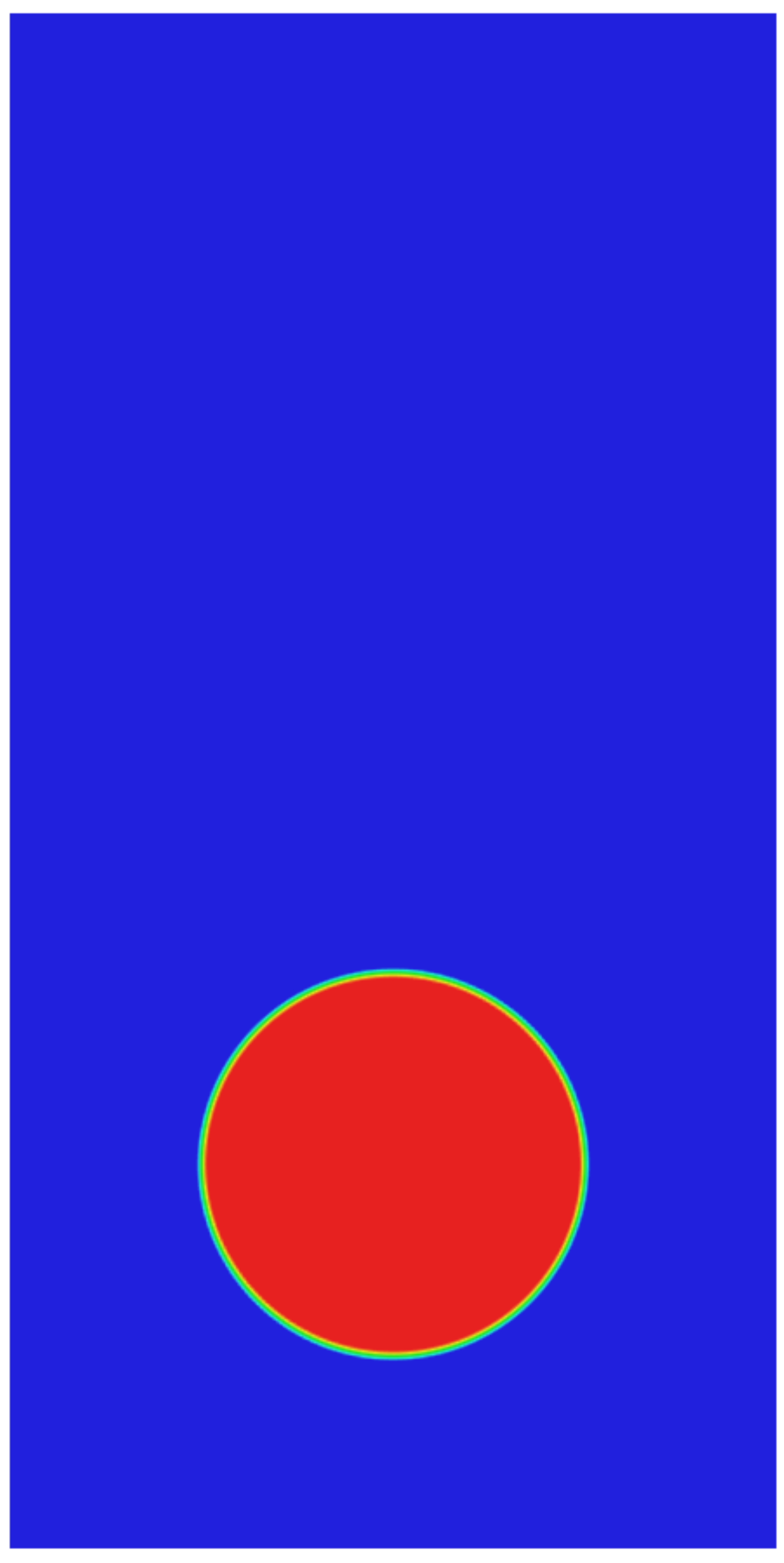}
		\subcaption{$t=0$}
	\end{subfigure}
	\begin{subfigure}{.19\textwidth}
		\centering	
		\includegraphics[width = 0.95\textwidth]{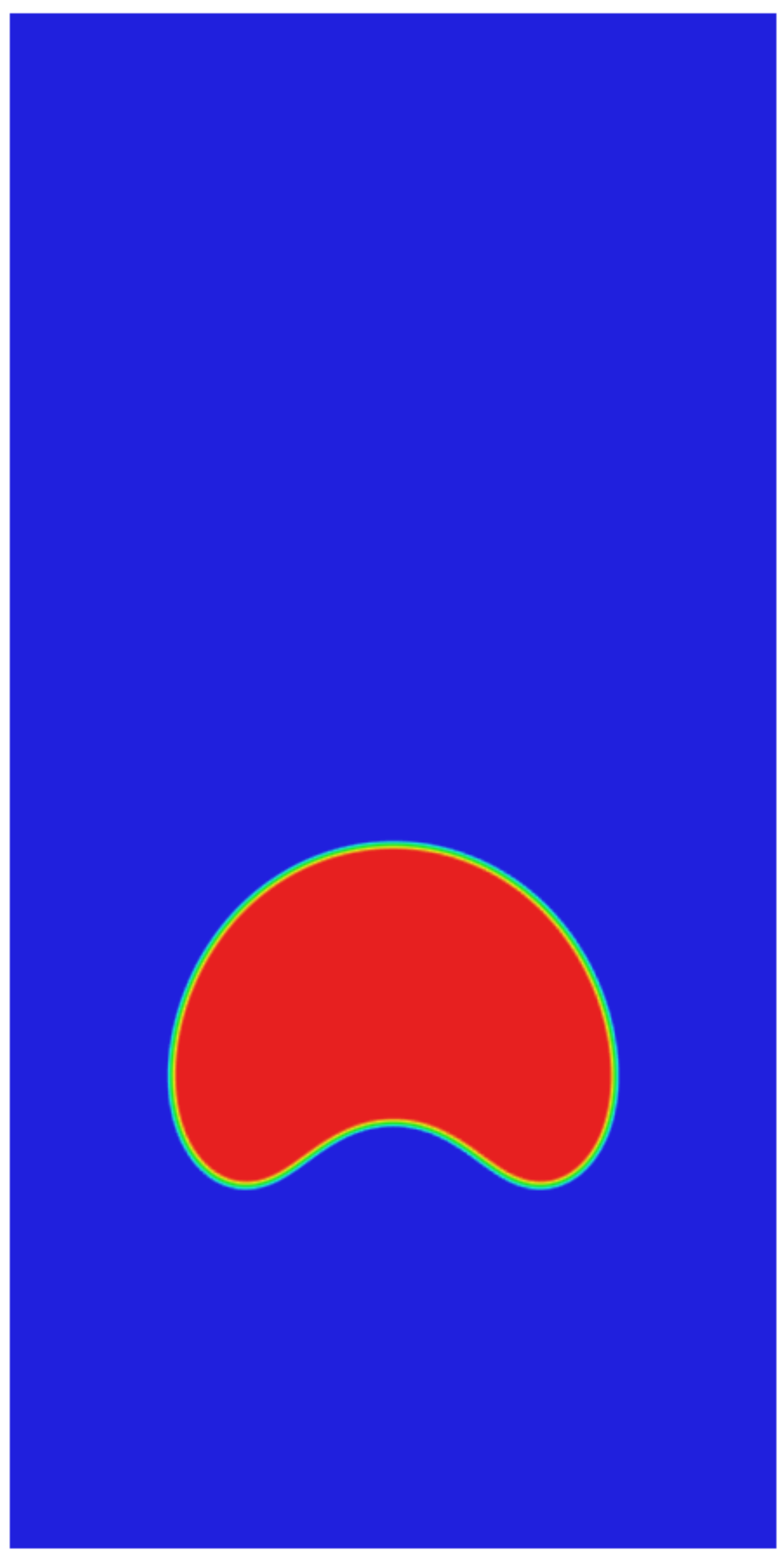}
		\subcaption{$t=1$}			
	\end{subfigure}
	\begin{subfigure}{.19\textwidth}
		\centering		
		\includegraphics[width = 0.95\textwidth]{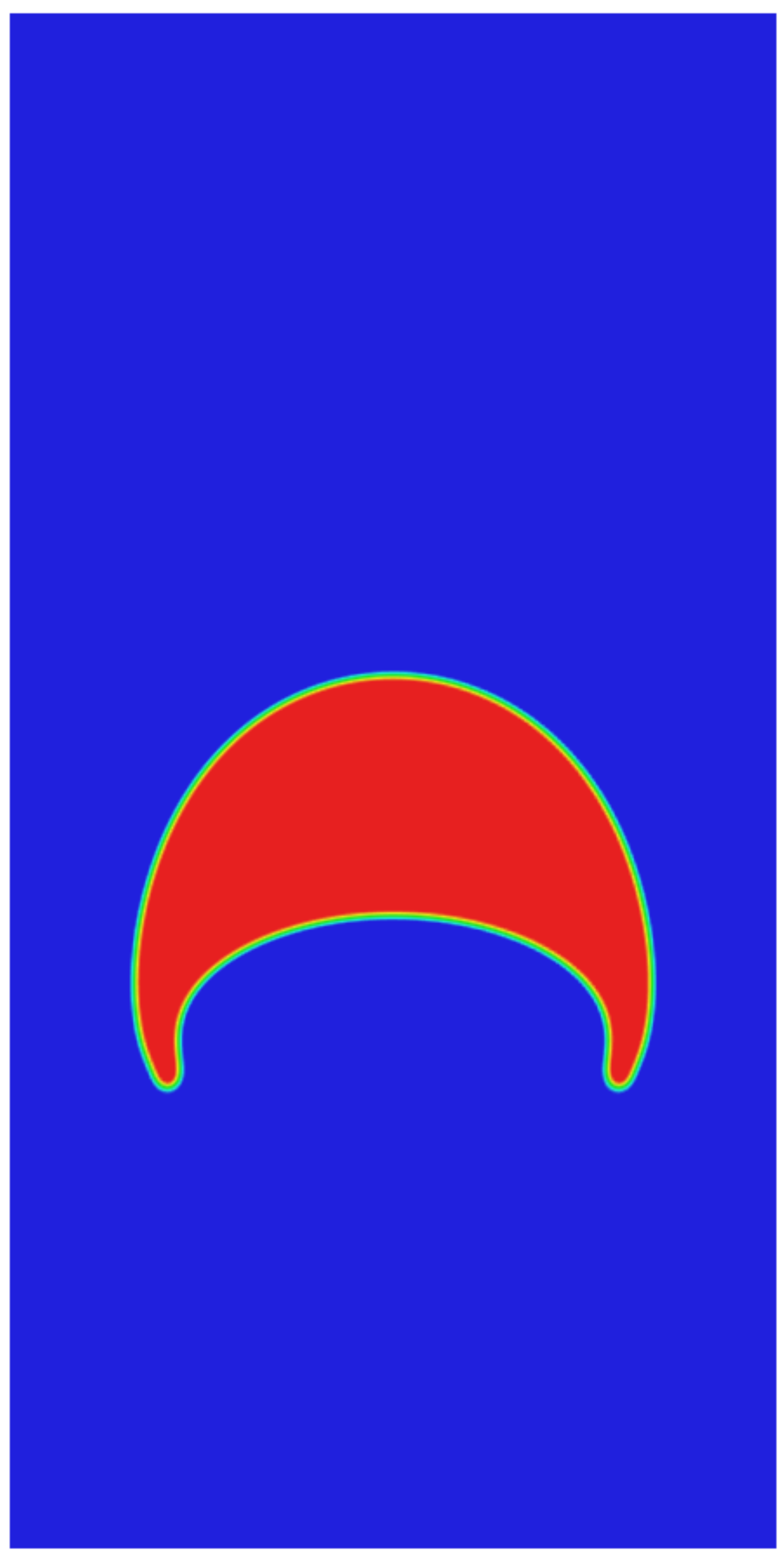}
		\subcaption{$t=2$}			
	\end{subfigure}
	\begin{subfigure}{.19\textwidth}
		\centering		
		\includegraphics[width = 0.95\textwidth]{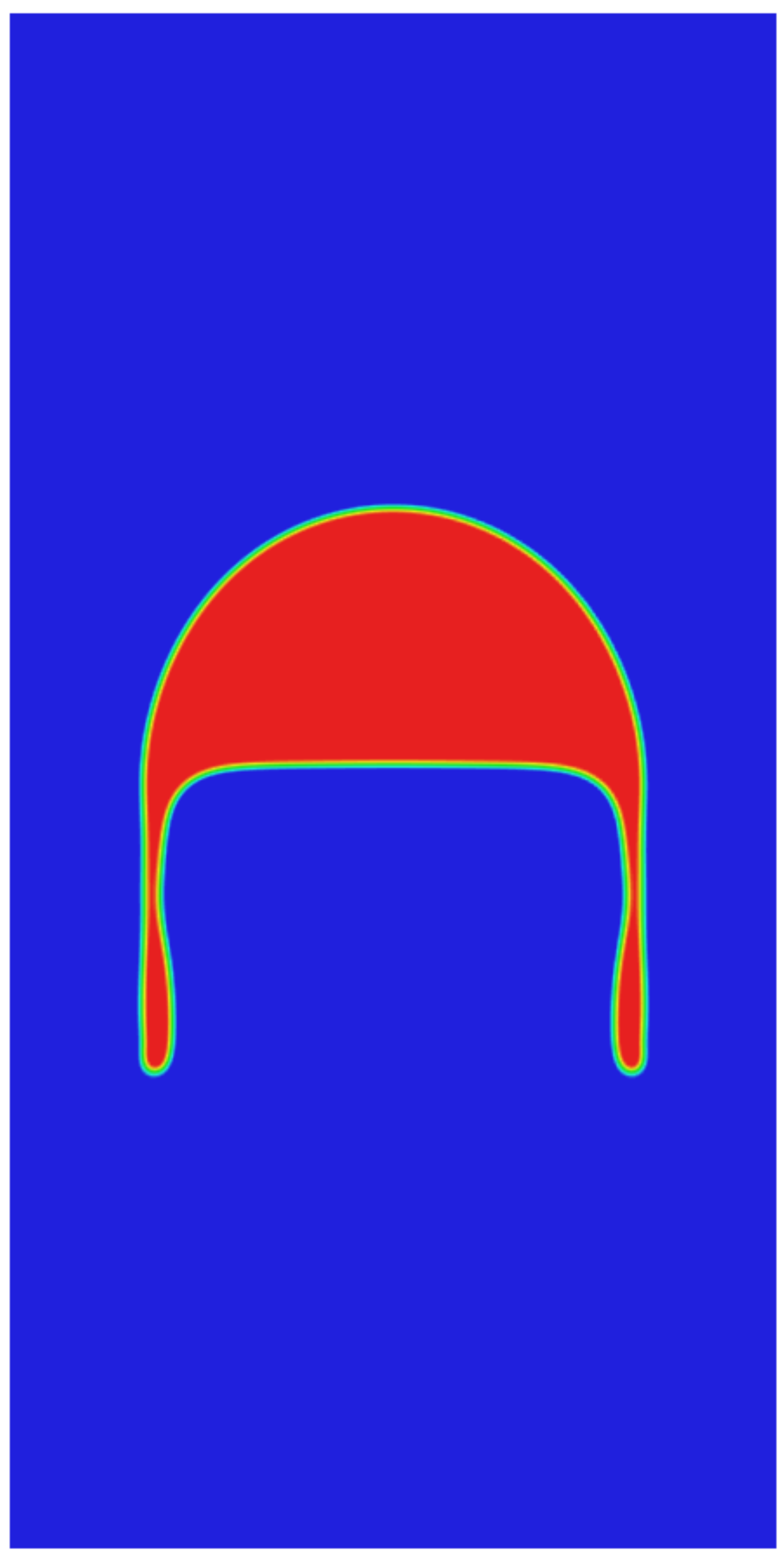}
		\subcaption{$t=3$}			
	\end{subfigure}
	\begin{subfigure}{.19\textwidth}
		\centering		
		\includegraphics[width = 0.95\textwidth]{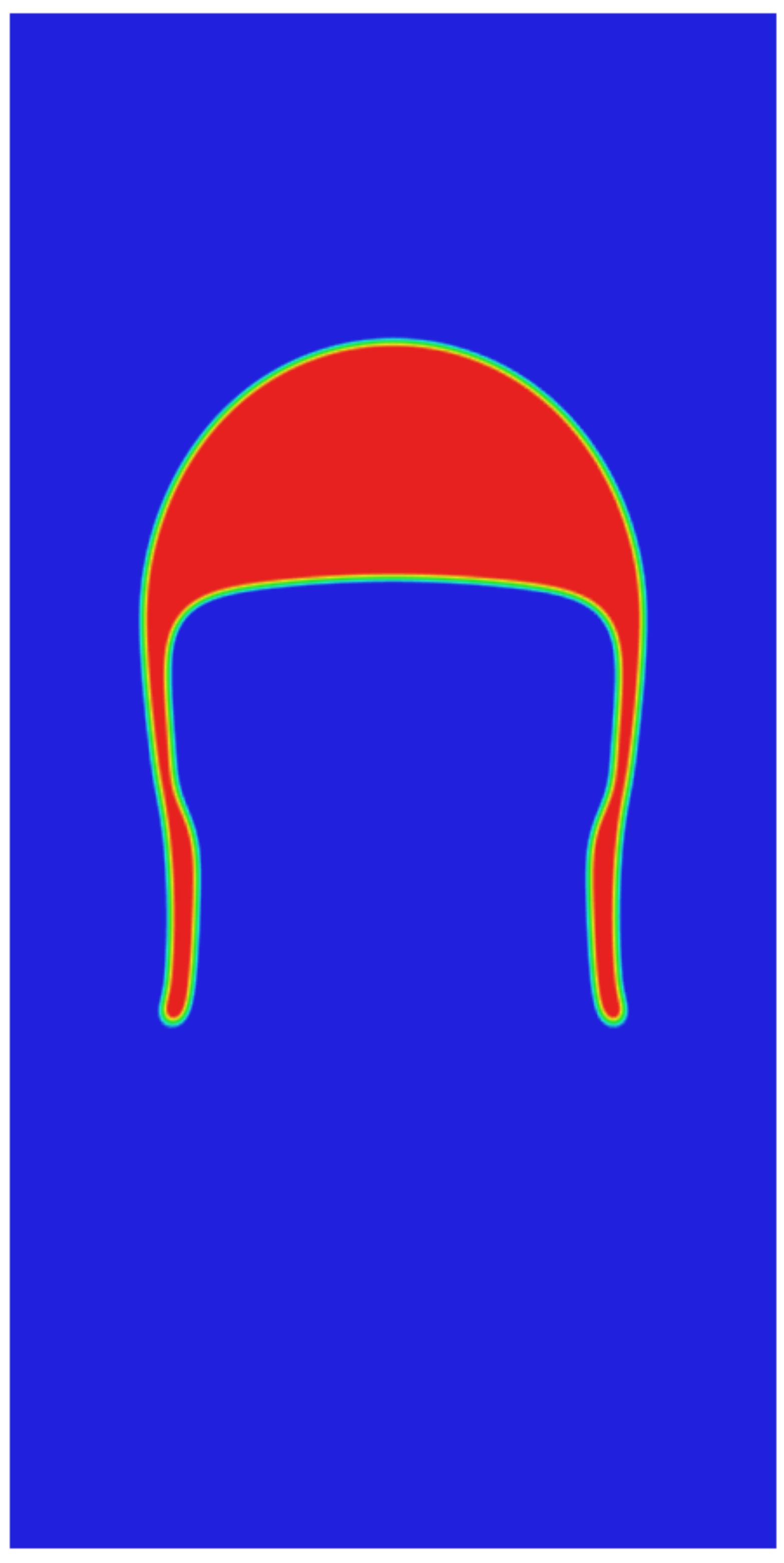}
		\subcaption{$t=4$}			
	\end{subfigure}	
	\caption{Rising bubble: $\varphi$ evolution using $128\times256$ linear elements.}
	\label{fig:rising bubble}	
\end{figure}

 It is expected that due to the large density difference, there would be an overpowering body force contribution acting on either side of the interface in comparison to the rather negligible surface tension forces. Thus the bubble would experience large deformations characterised by long and narrow filaments at the edges. This behaviour is observable in Figure \ref{fig:rising bubble}, where the solution to $\varphi$ is shown at different time instances. The observations match well with those made in \cite{Hosseini2017}.
%
%
\subsection{Faucet leak in two dimensions} \label{sec: numerical examples: filling drop_2D}
\begin{figure}[b!]
	\captionsetup[subfigure]{labelformat=empty}
	\centering
	\includegraphics[width=0.3\textwidth]{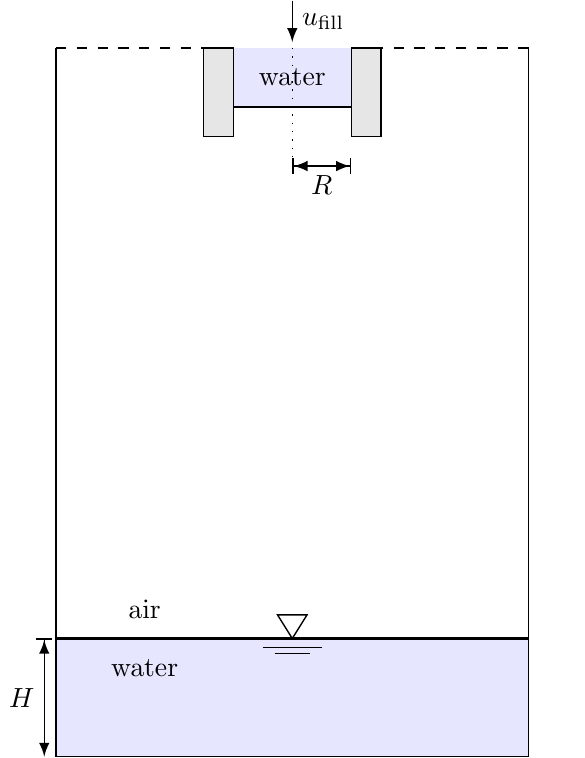}
	\caption{Filling drop in two dimensions: Geometry.}
	\label{fig:filling_drop_geom}
\end{figure}

%
In this example we consider a \textit{leaky faucet} problem, which consists of water dripping from a faucet into a pool of water at the bottom of a rectangular box filled with air, as shown in Figure \ref{fig:filling_drop_geom}. The dimensions of the box are $[0,2] \times [0,4.5] \ \si{\centi\metre^2}$, the height of the pool is $H=0.4 \ \si{\centi\metre}$, and the radius of the faucet is $R=0.26 \ \si{\centi\metre}$. The simulations are run with a filling velocity of $\vu_{\mathrm{fill}} = 10 \ \si{\centi\metre\per\second}$, which is set uniformly at the inlet. 
At the beginning of the simulation the faucet is completely closed, after which the inlet velocity is ramped smoothly up to the the filling velocity with the function $\frac{1}{2} \left( 1 - \cos(\pi t/T) \right)$, where $T=0.01 \ \si{\second}$. A mesh comprising of 35,712 linear elements is considered in all simulations. Adaptive time stepping is used, with an initial time step size of $\Delta t = 0.0001$. The mobility is set as $M=0.5 \ \si{\centi\metre\per\second\squared}$. Slip boundary conditions are applied on all surfaces.
\par
\begin{figure}[tb!]
	\captionsetup[subfigure]{labelformat=empty}
	\centering
	\begin{subfigure}{.160\textwidth}
		\centering
		\includegraphics[width = 0.95\textwidth]{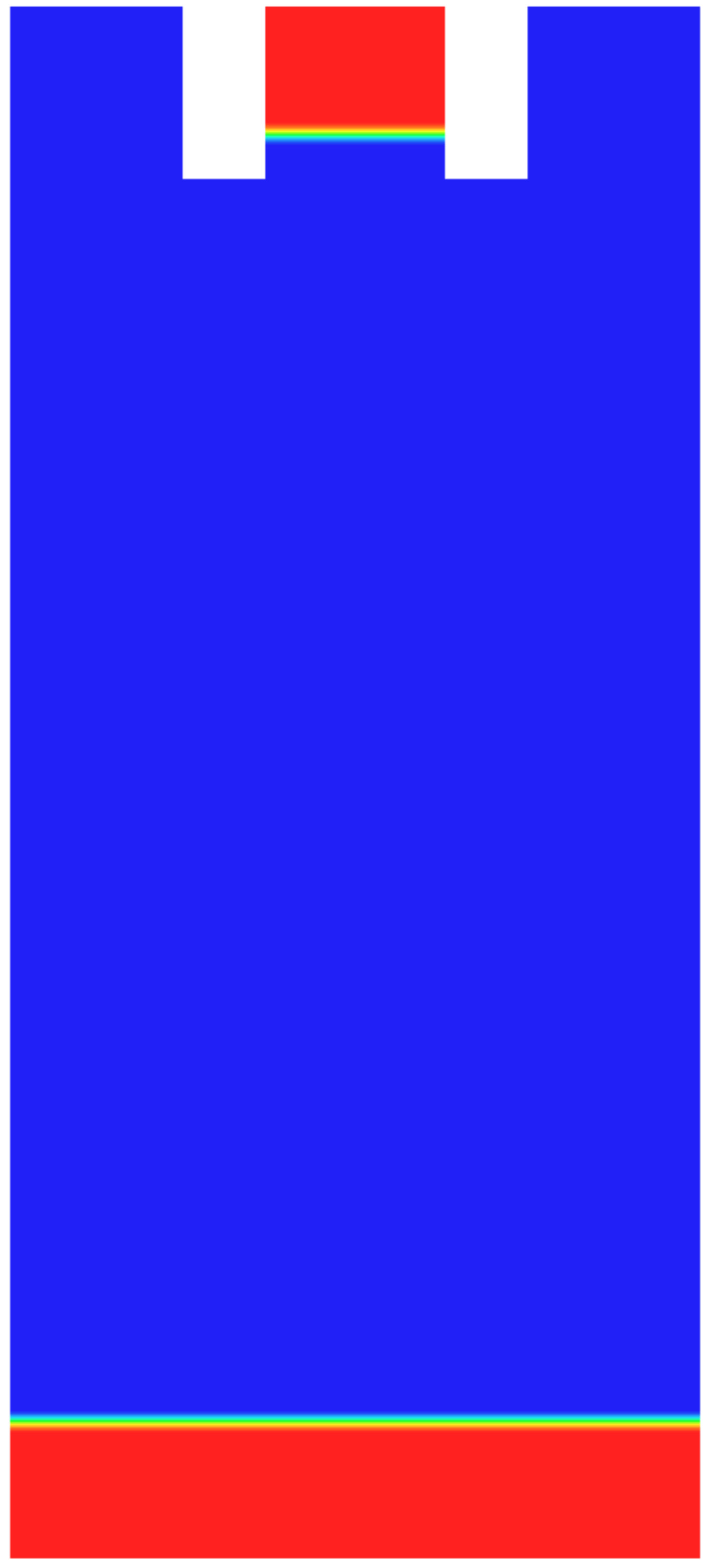}
		\subcaption{{$t=0 \ \si{\second}$}}
	\end{subfigure}	
	\begin{subfigure}{.160\textwidth}
		\centering
		\includegraphics[width = 0.95\textwidth]{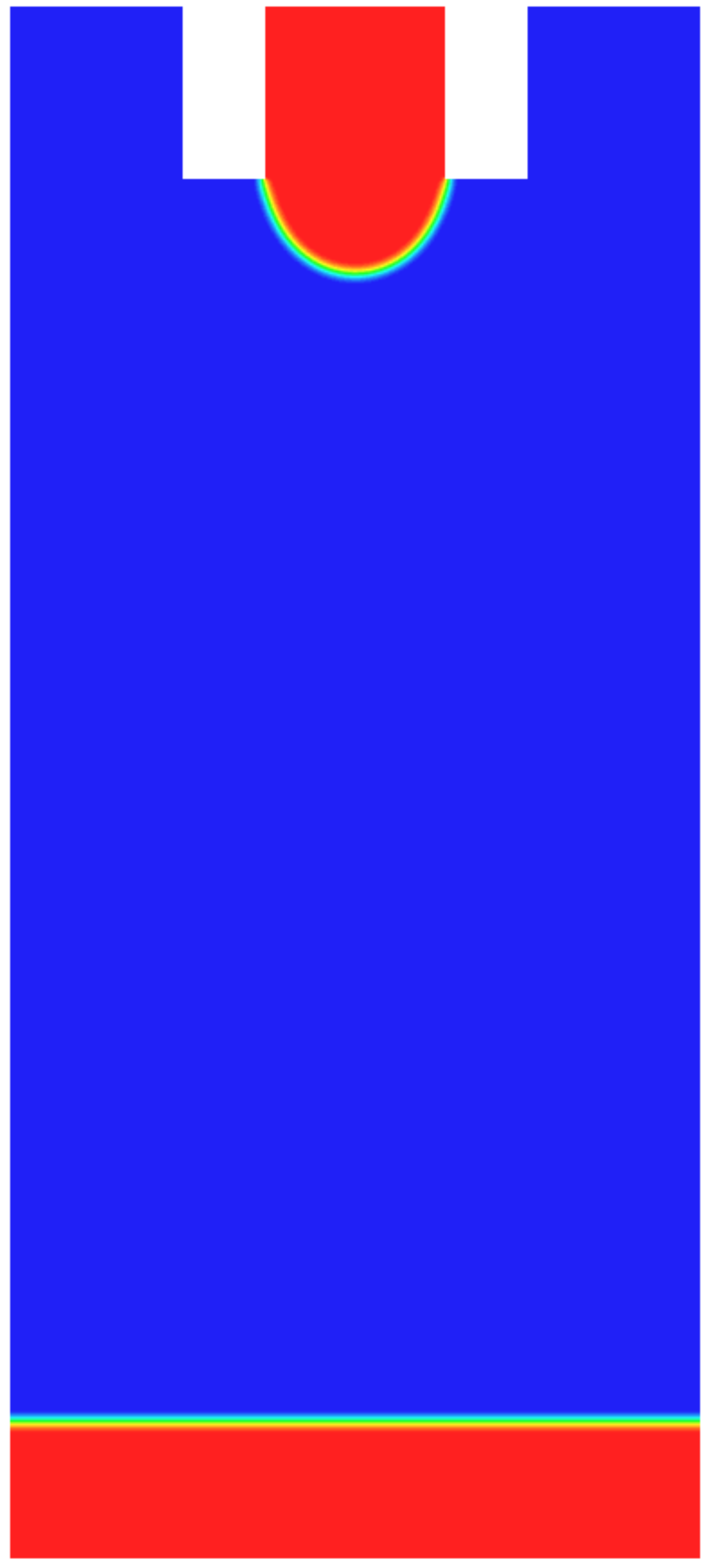}
		\subcaption{{$t=0.0354 \ \si{\second}$}}
	\end{subfigure}
	\begin{subfigure}{.160\textwidth}
		\centering	
		\includegraphics[width = 0.95\textwidth]{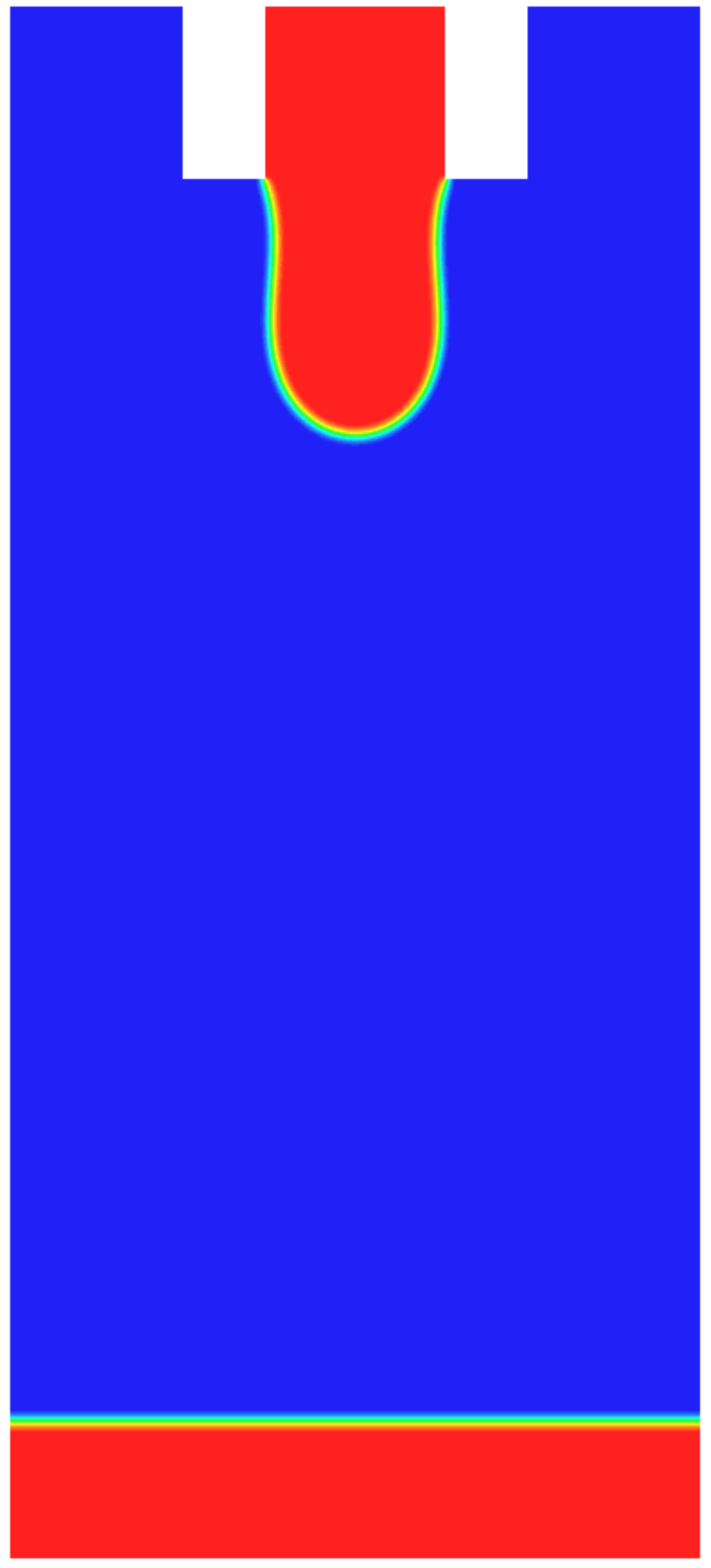}
		\subcaption{$t=0.0816 \ \si{\second}$}			
	\end{subfigure}
	\begin{subfigure}{.160\textwidth}
		\centering		
		\includegraphics[width = 0.95\textwidth]{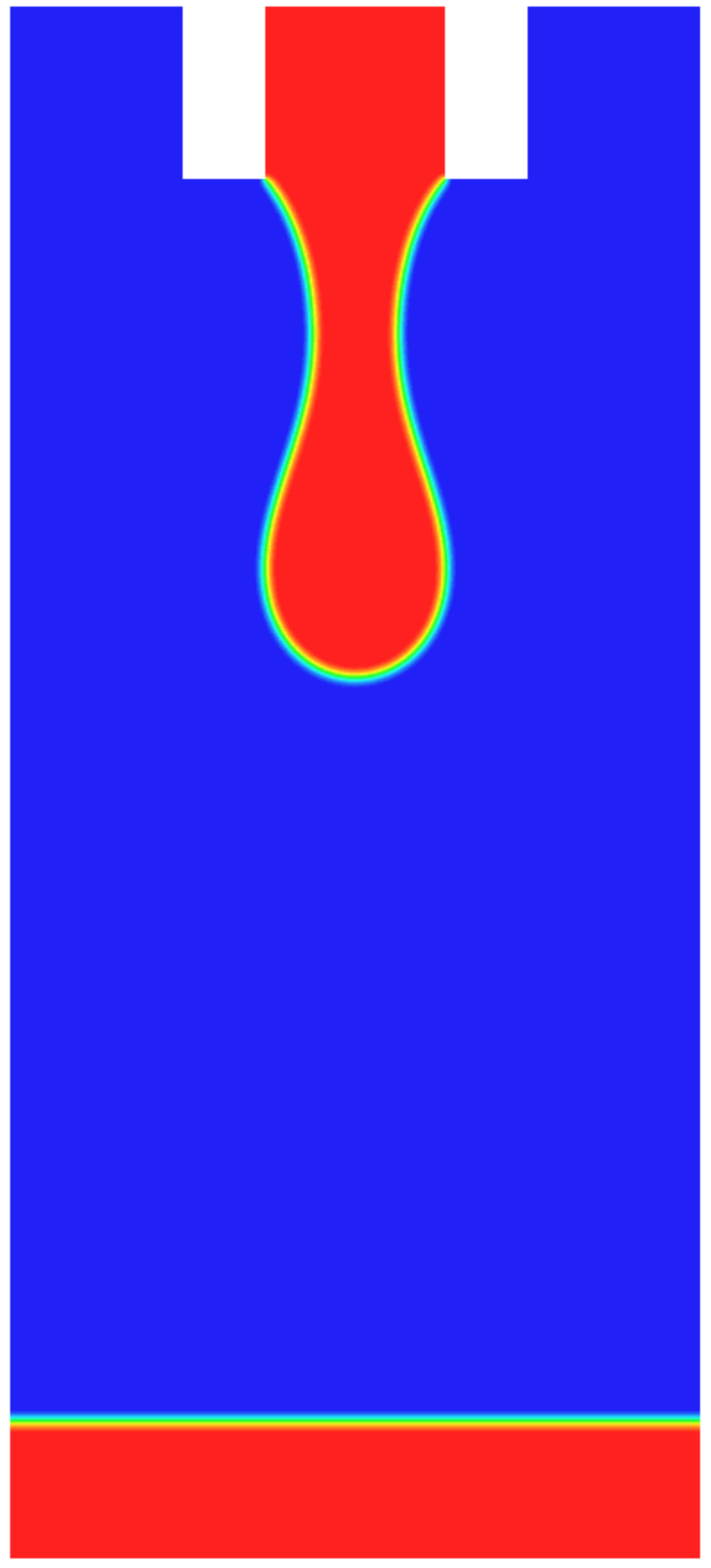}
		\subcaption{$t=0.1198 \ \si{\second}$}			
	\end{subfigure}
	\begin{subfigure}{.160\textwidth}
		\centering		
		\includegraphics[width = 0.95\textwidth]{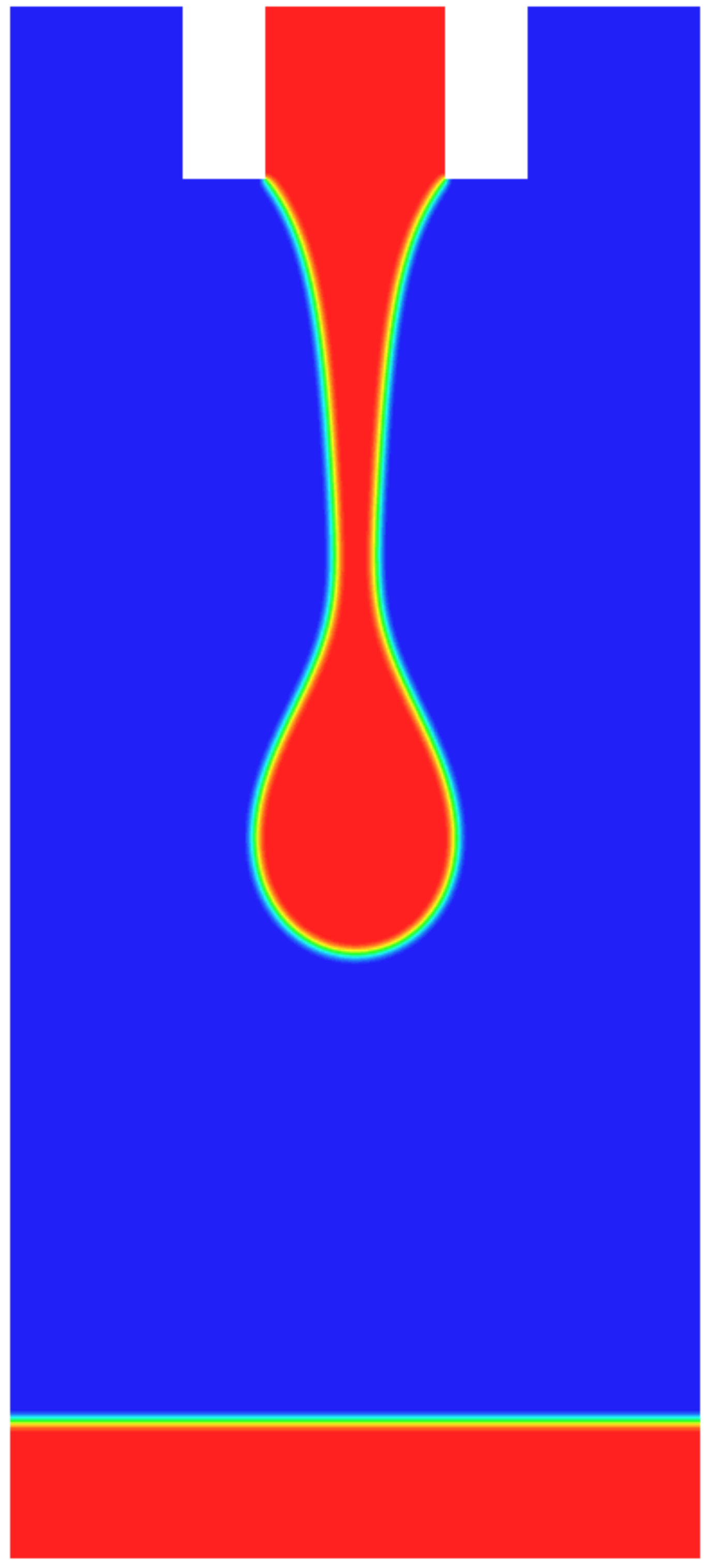}
		\subcaption{$t=0.1444 \ \si{\second}$}			
	\end{subfigure}
	\begin{subfigure}{.160\textwidth}
		\centering		
		\includegraphics[width = 0.95\textwidth]{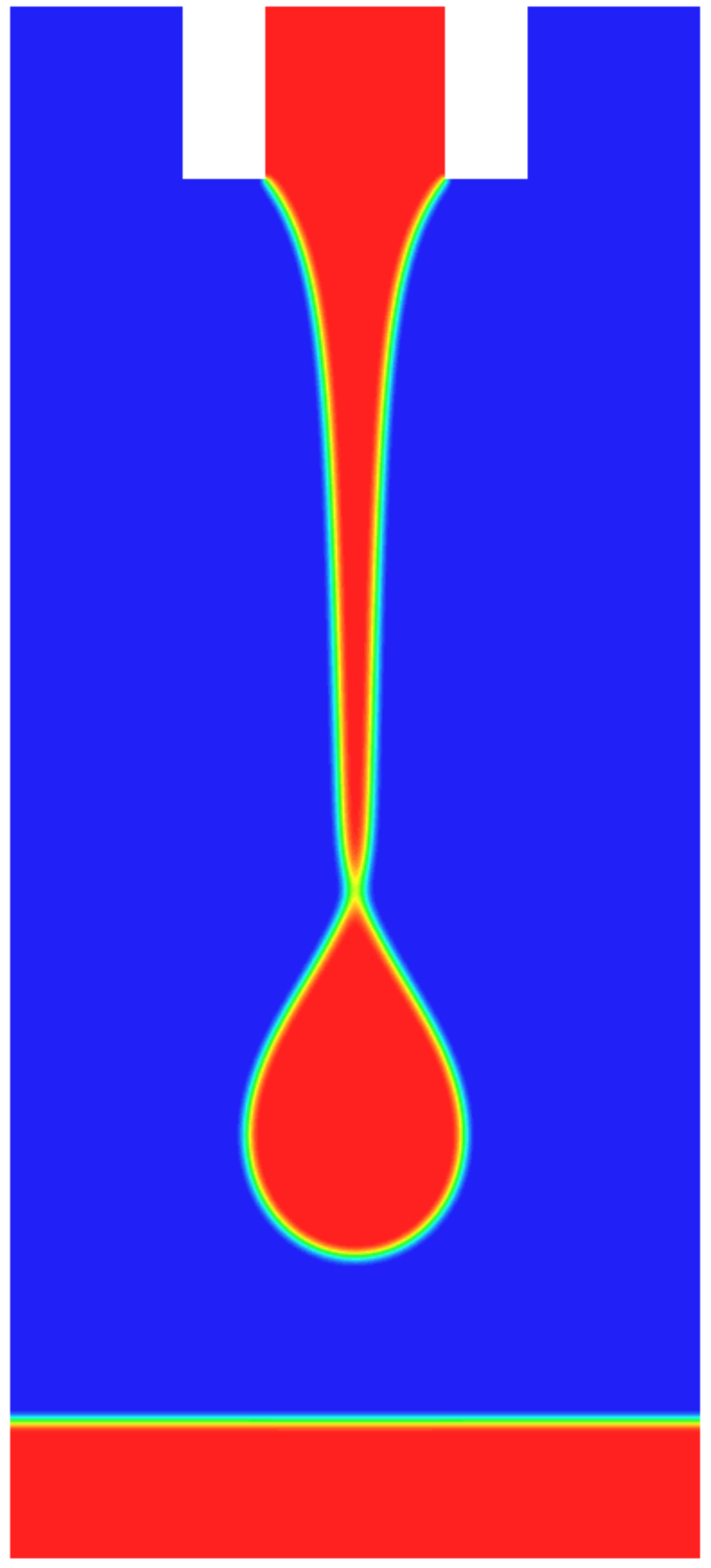}
		\subcaption{$t=0.1641 \ \si{\second}$}			
	\end{subfigure}
	\begin{subfigure}{.160\textwidth}
		\centering
		\includegraphics[width = 0.95\textwidth]{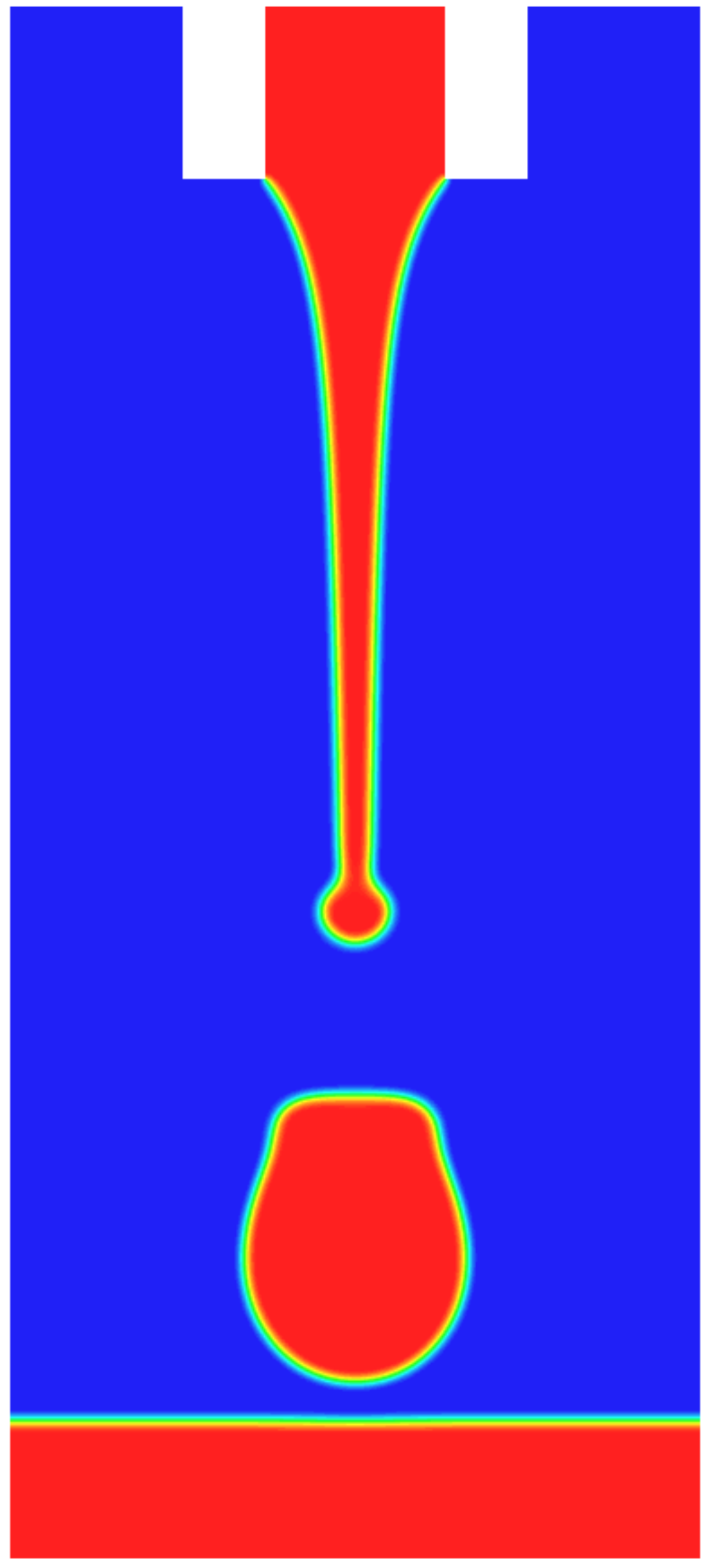}
		\subcaption{$t=0.1708 \ \si{\second}$}
	\end{subfigure}
	\begin{subfigure}{.160\textwidth}
		\centering		
		\includegraphics[width = 0.95\textwidth]{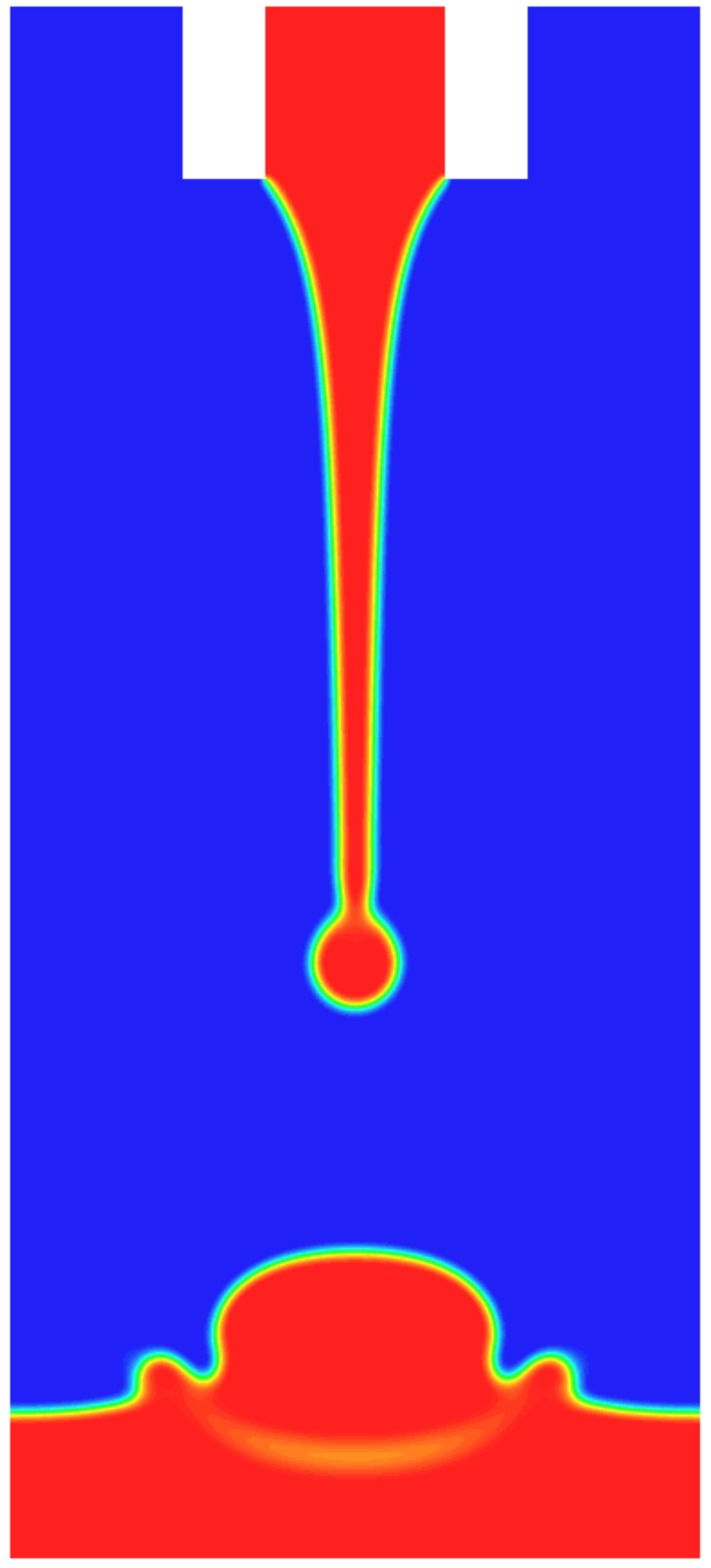}
		\subcaption{$t=0.1771 \ \si{\second}$}			
	\end{subfigure}
	\begin{subfigure}{.160\textwidth}
		\centering		
		\includegraphics[width = 0.95\textwidth]{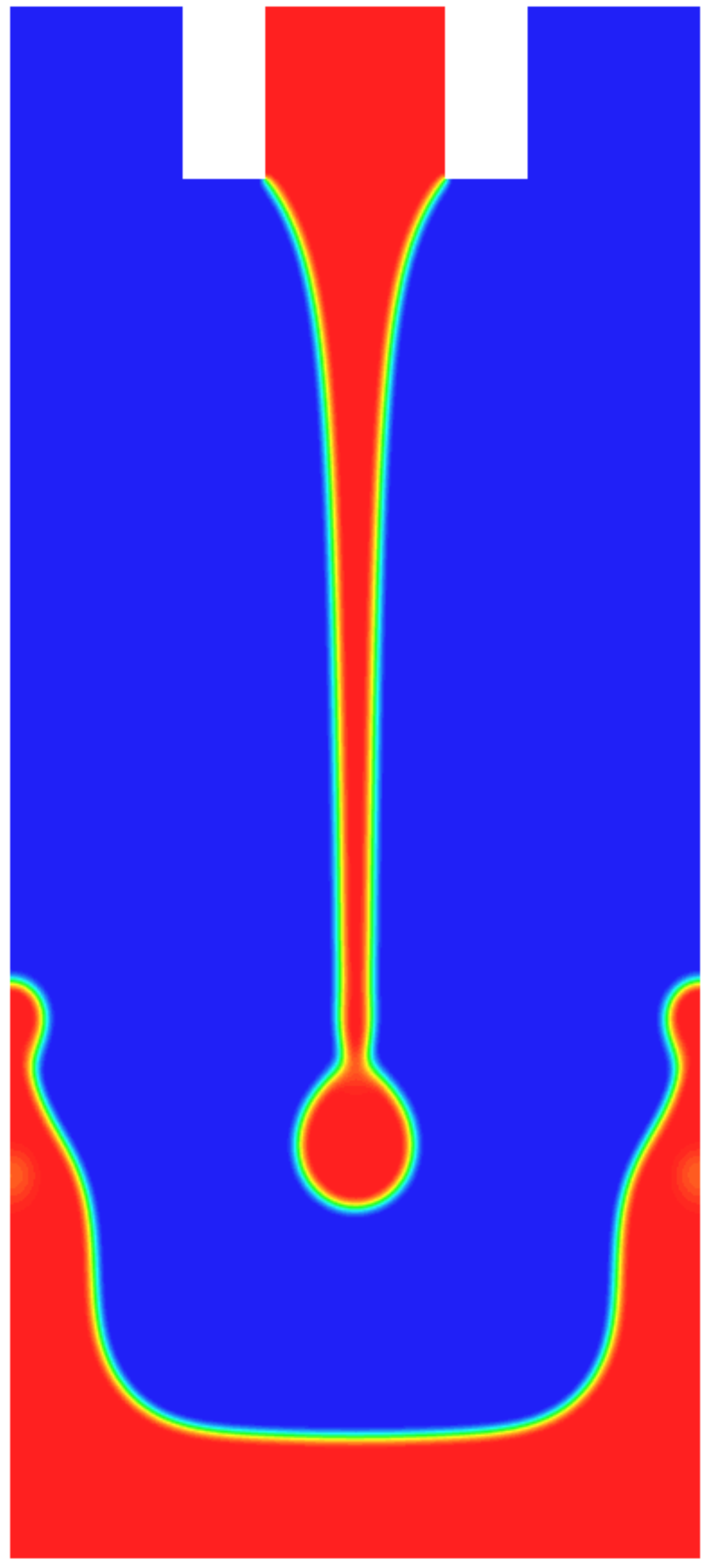}
		\subcaption{$t=0.1940 \ \si{\second}$}			
	\end{subfigure}
	\begin{subfigure}{.160\textwidth}
		\centering		
		\includegraphics[width = 0.95\textwidth]{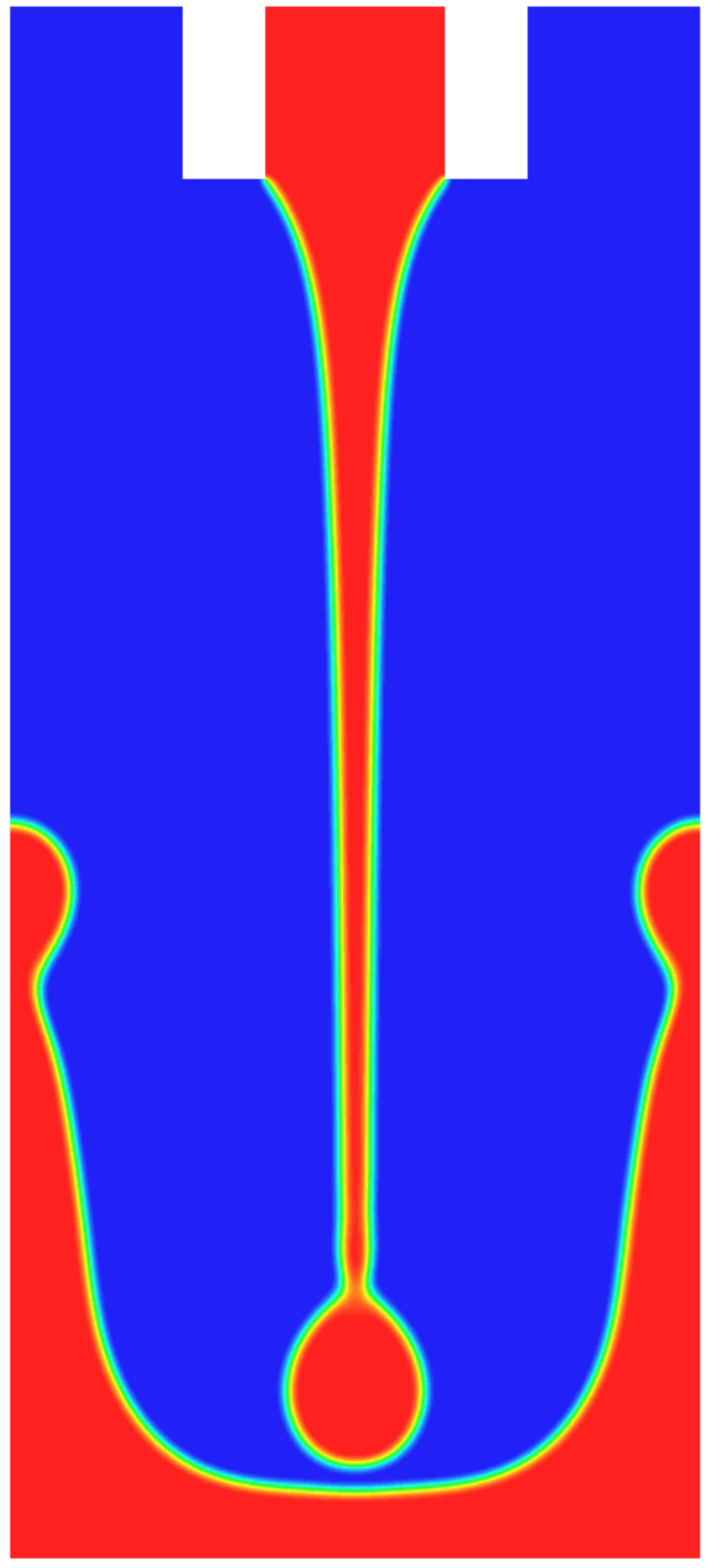}
		\subcaption{$t=0.2118 \ \si{\second}$}			
	\end{subfigure}
	\begin{subfigure}{.160\textwidth}
		\centering
		\includegraphics[width = 0.95\textwidth]{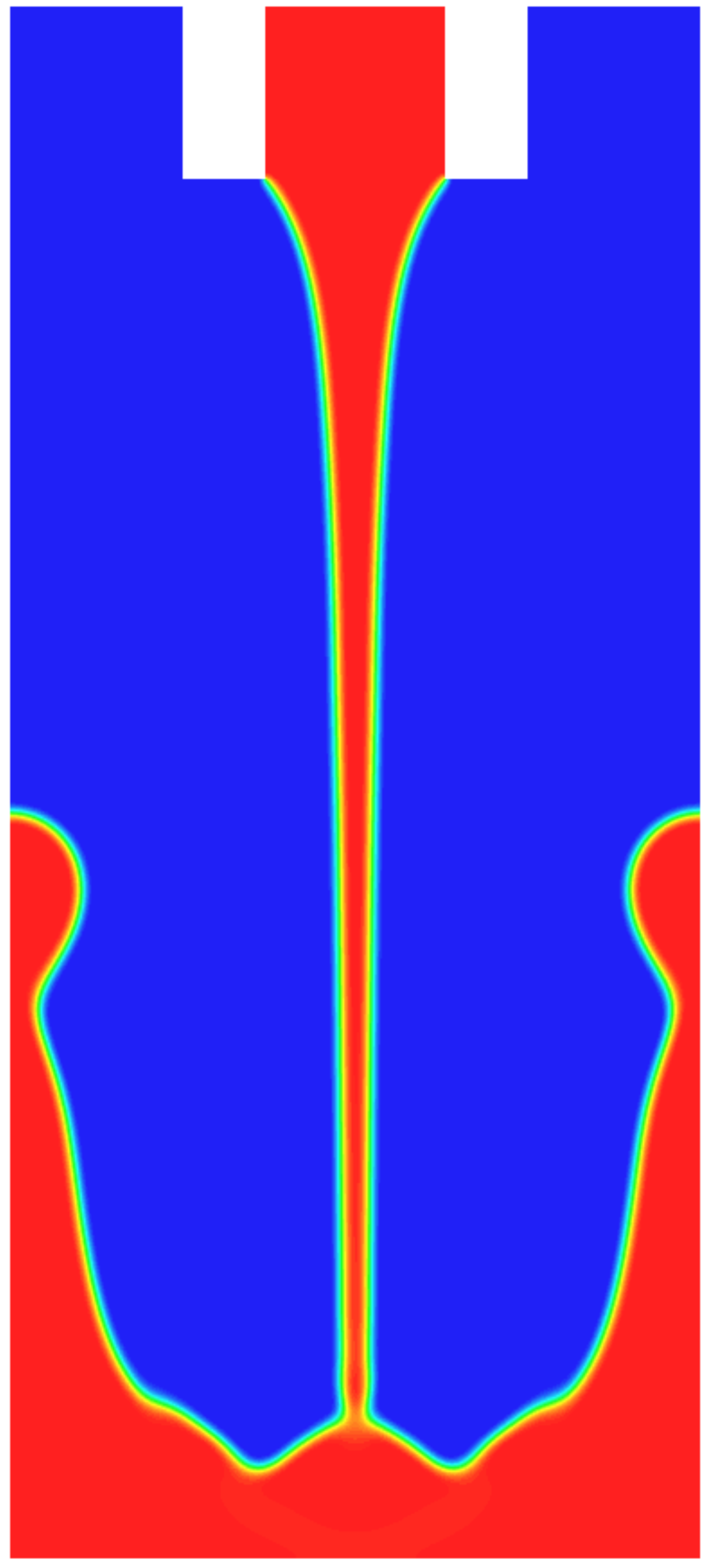}
		\subcaption{$t=0.2204 \ \si{\second}$}
	\end{subfigure}
	\begin{subfigure}{.160\textwidth}
		\centering	
		\includegraphics[width = 0.95\textwidth]{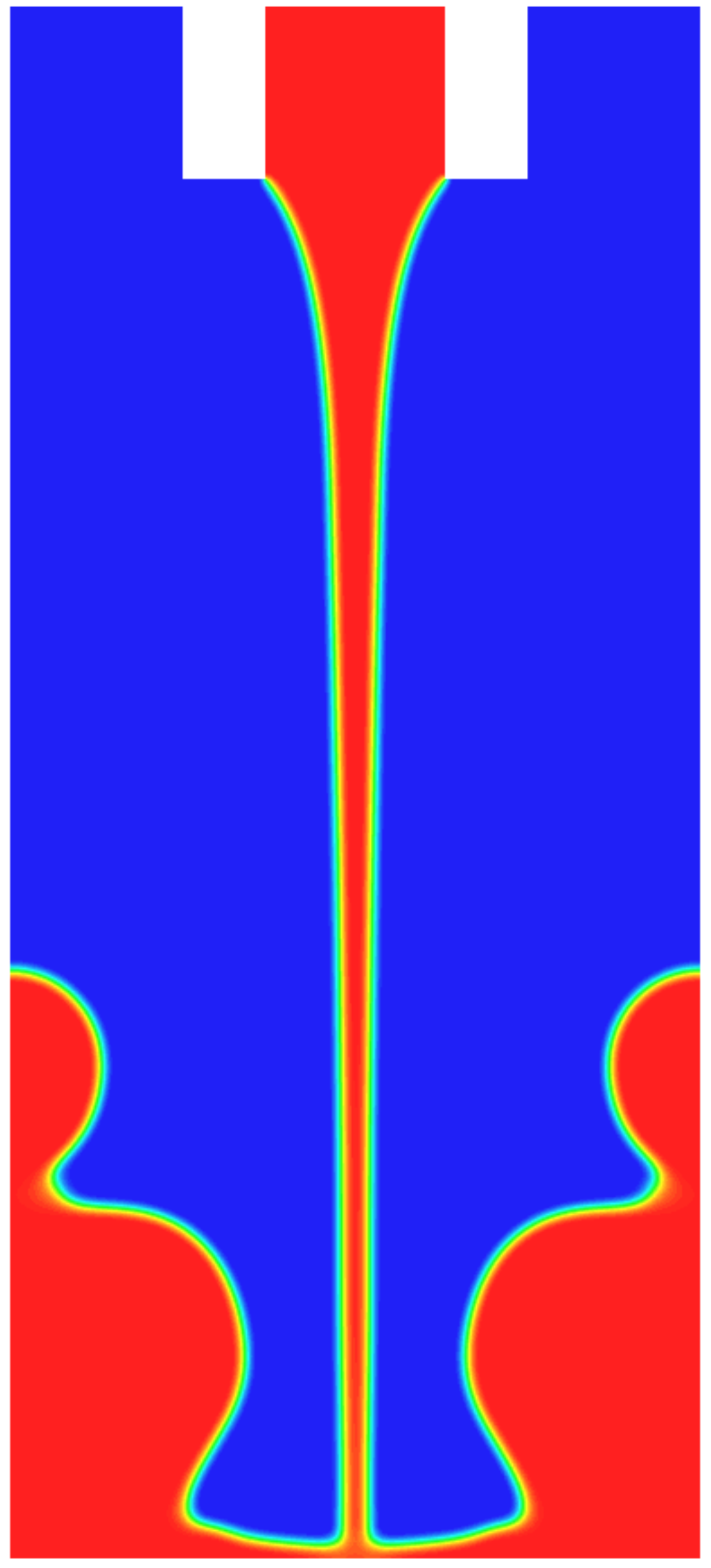}
		\subcaption{$t=0.2442 \ \si{\second}$}			
	\end{subfigure}
	\begin{subfigure}{.160\textwidth}
		\centering		
		\includegraphics[width = 0.95\textwidth]{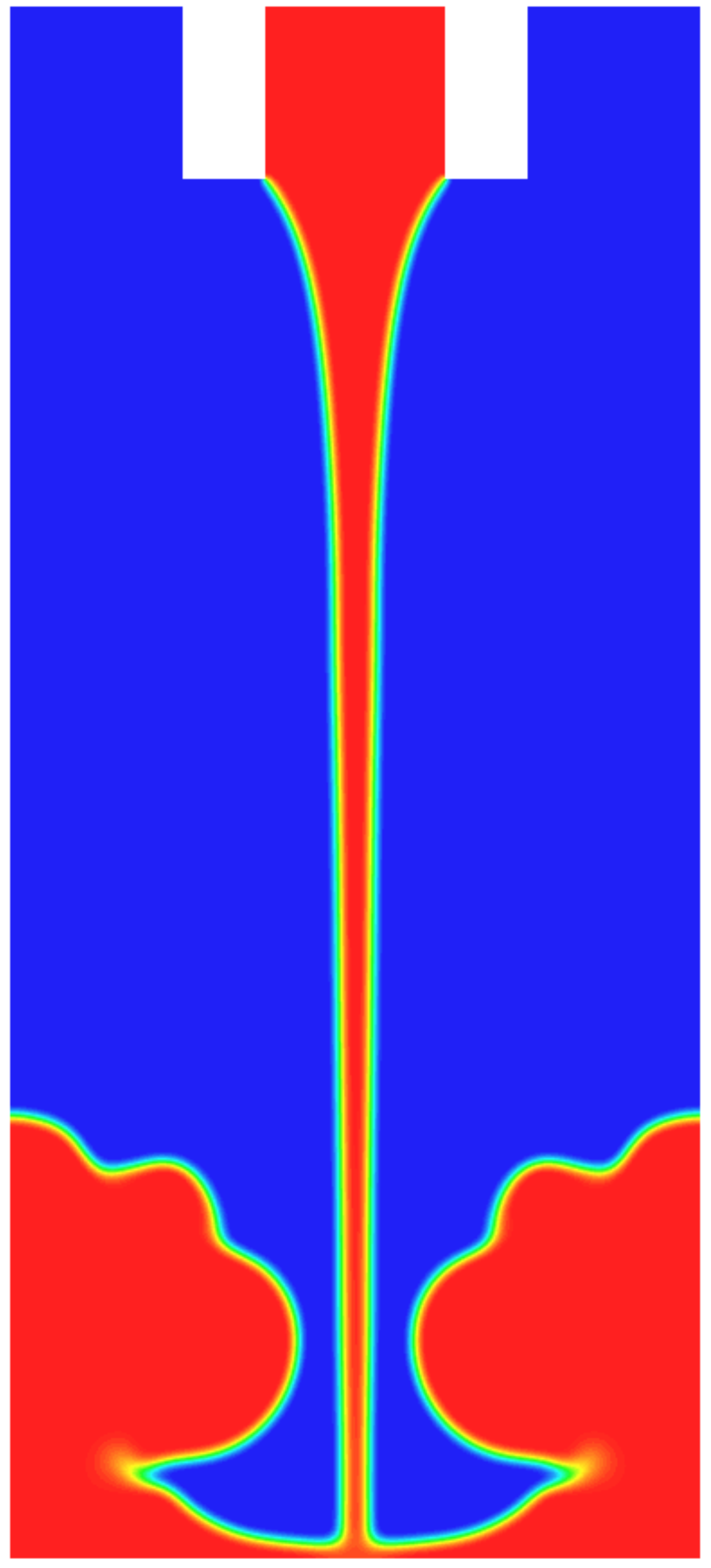}
		\subcaption{$t=0.2549 \ \si{\second}$}			
	\end{subfigure}
	\begin{subfigure}{.160\textwidth}
		\centering		
		\includegraphics[width = 0.95\textwidth]{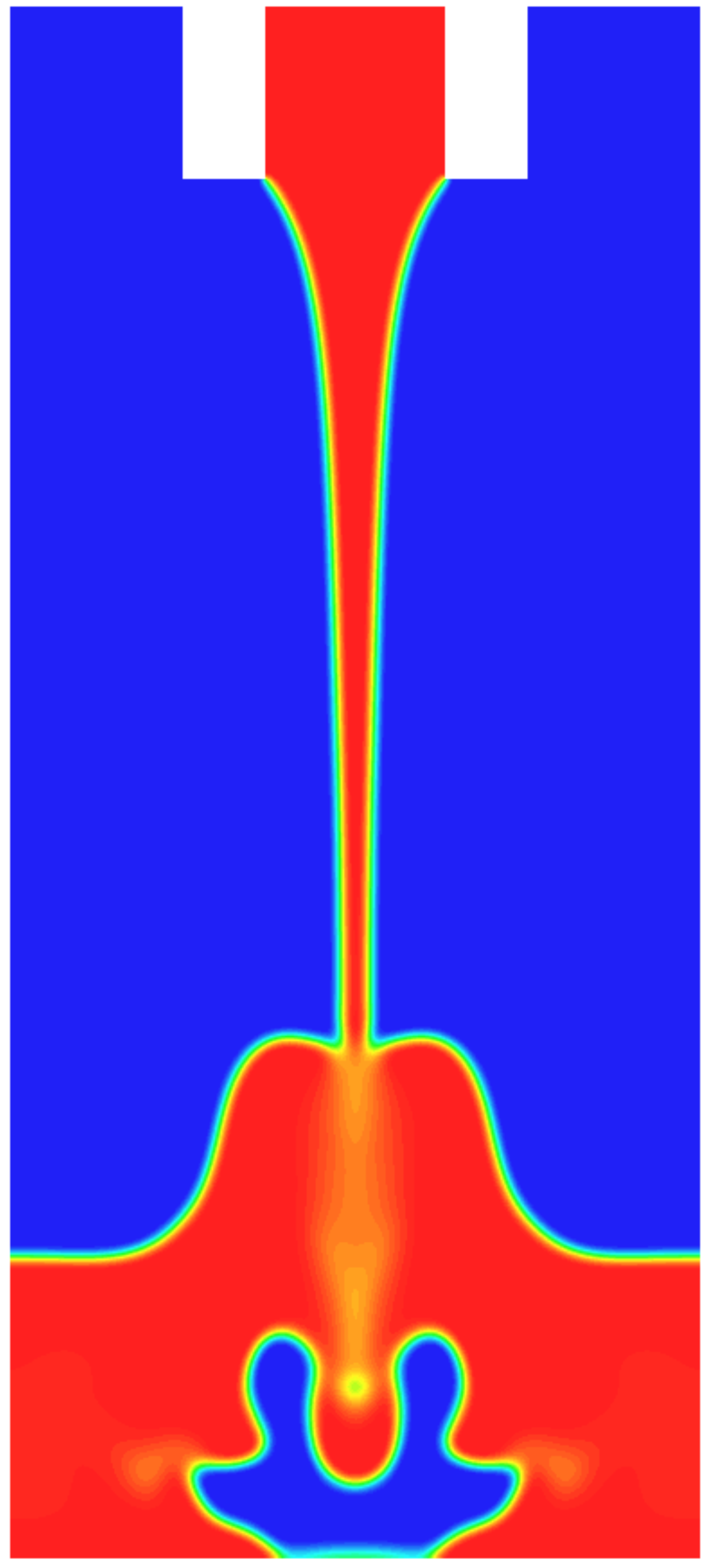}
		\subcaption{$t=0.2837 \ \si{\second}$}			
	\end{subfigure}	
	\begin{subfigure}{.160\textwidth}
		\centering		
		\includegraphics[width = 0.95\textwidth]{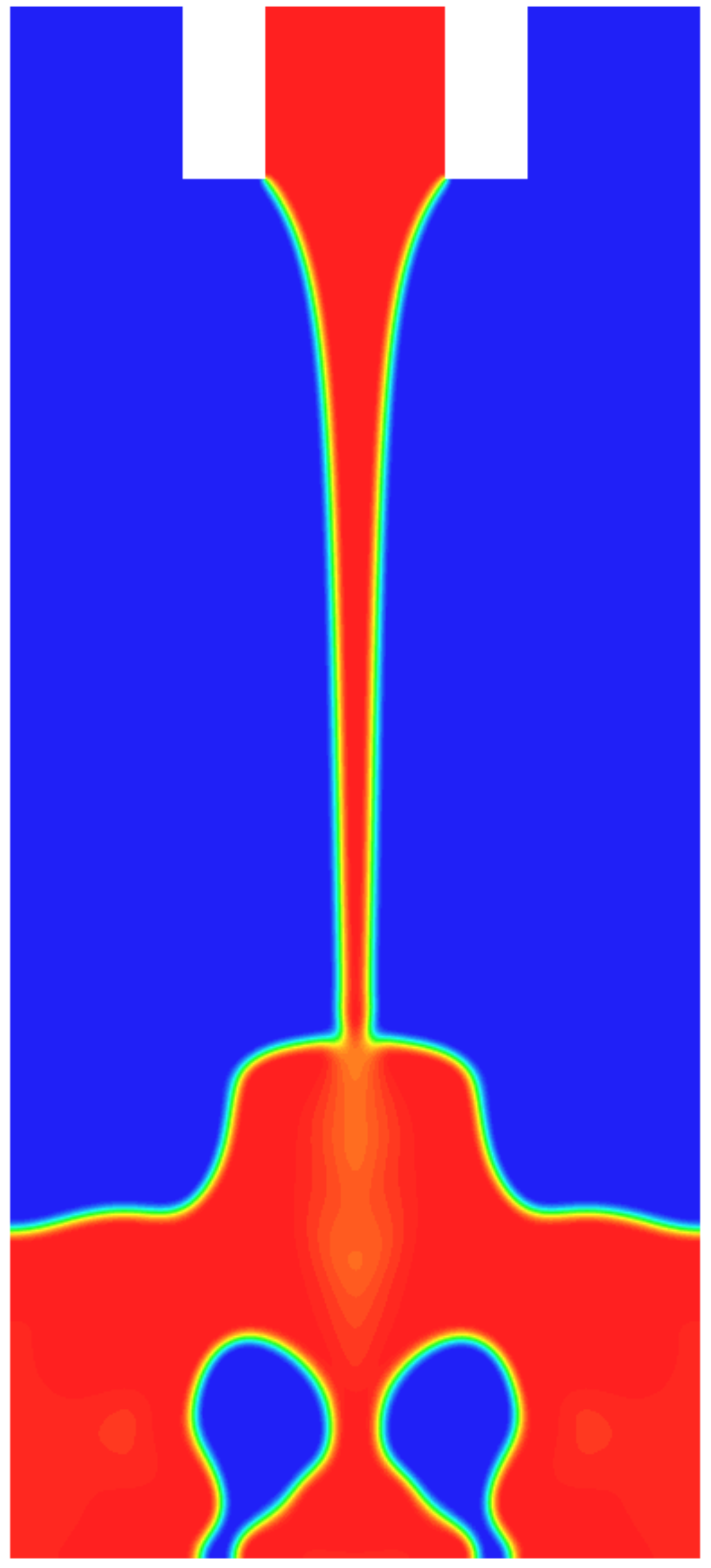}
		\subcaption{$t=0.2974 \ \si{\second}$}			
	\end{subfigure}
	\begin{subfigure}{.160\textwidth}
		\centering		
		\includegraphics[width = 0.95\textwidth]{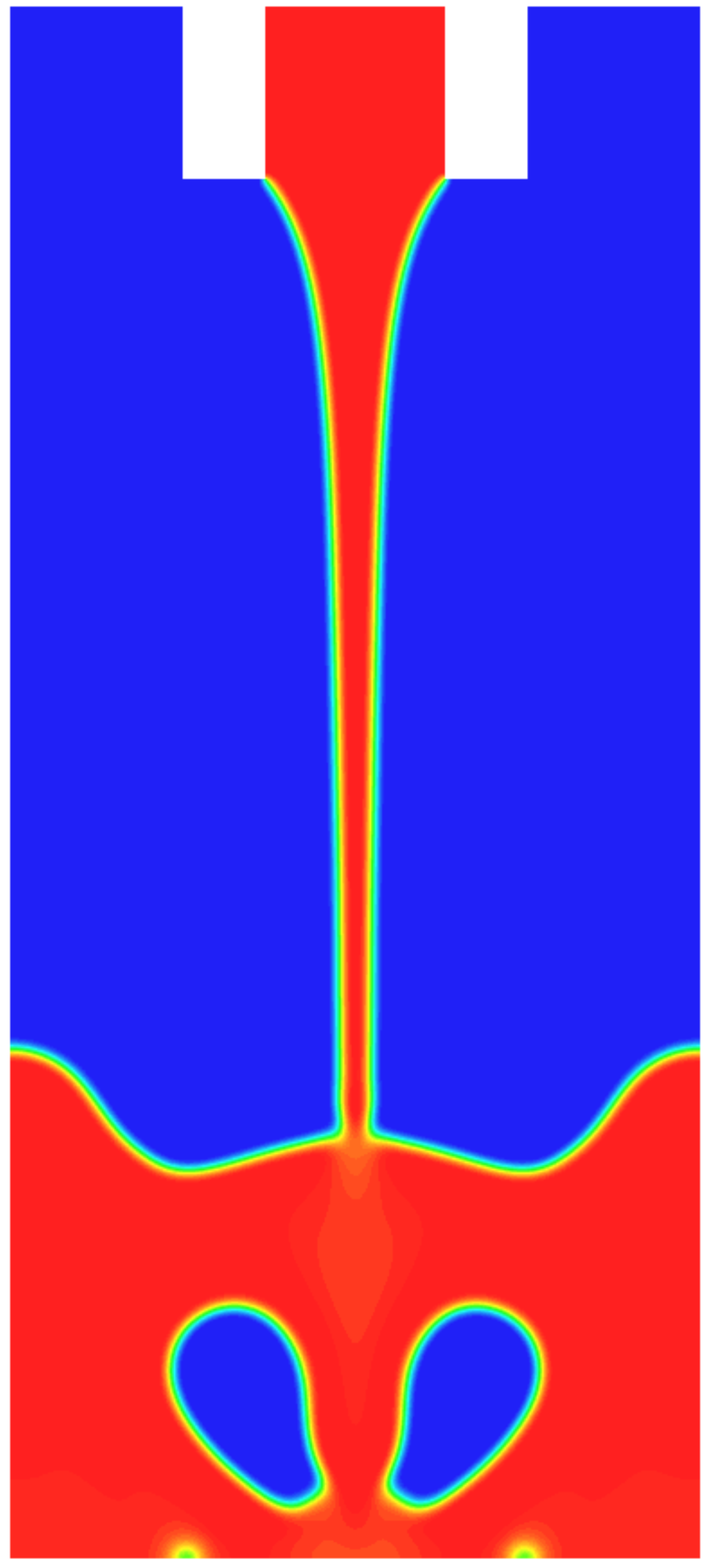}
		\subcaption{$t=0.3242 \ \si{\second}$}			
	\end{subfigure}
	\begin{subfigure}{.160\textwidth}
		\centering		
		\includegraphics[width = 0.95\textwidth]{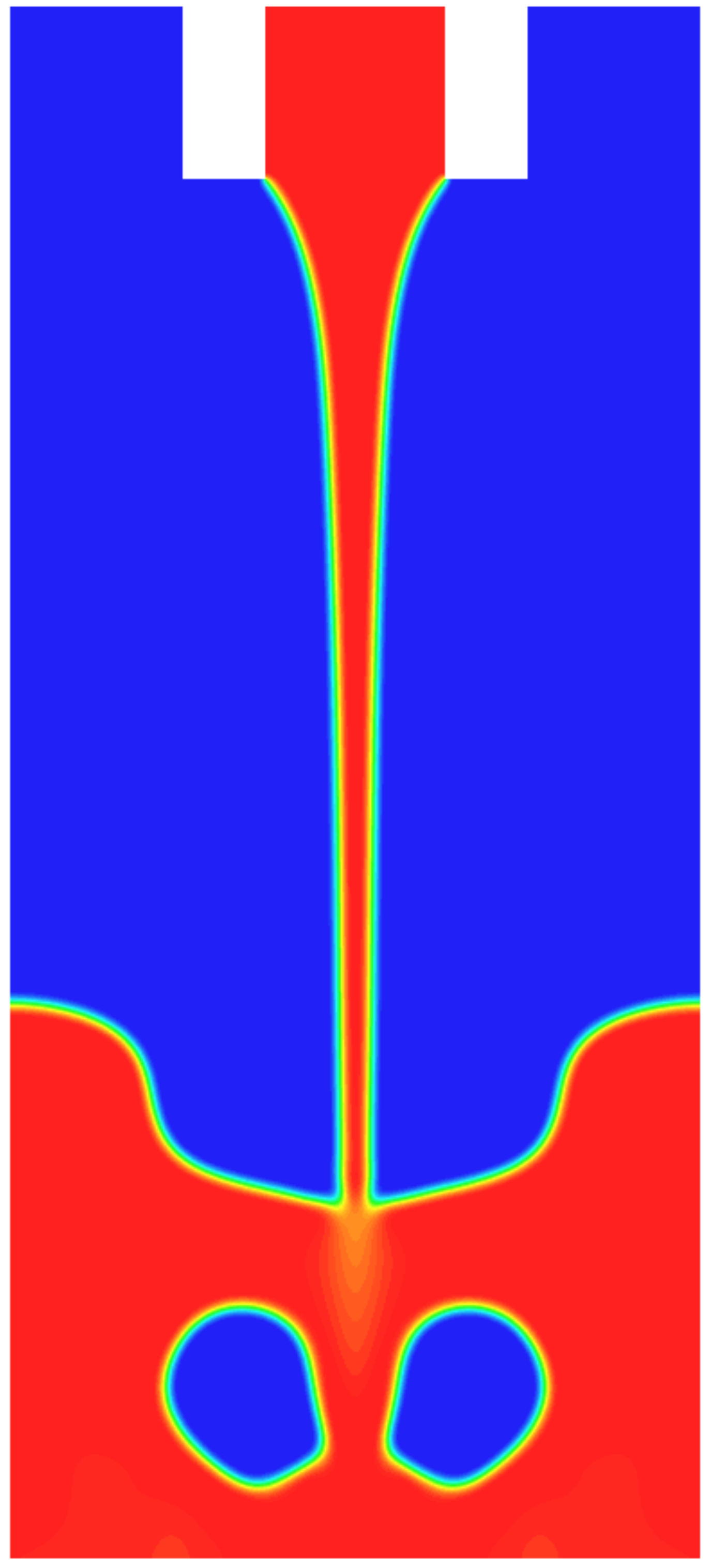}
		\subcaption{$t=0.3400 \ \si{\second}$}			
	\end{subfigure}	
	\begin{subfigure}{.160\textwidth}
		\centering		
		\includegraphics[width = 0.95\textwidth]{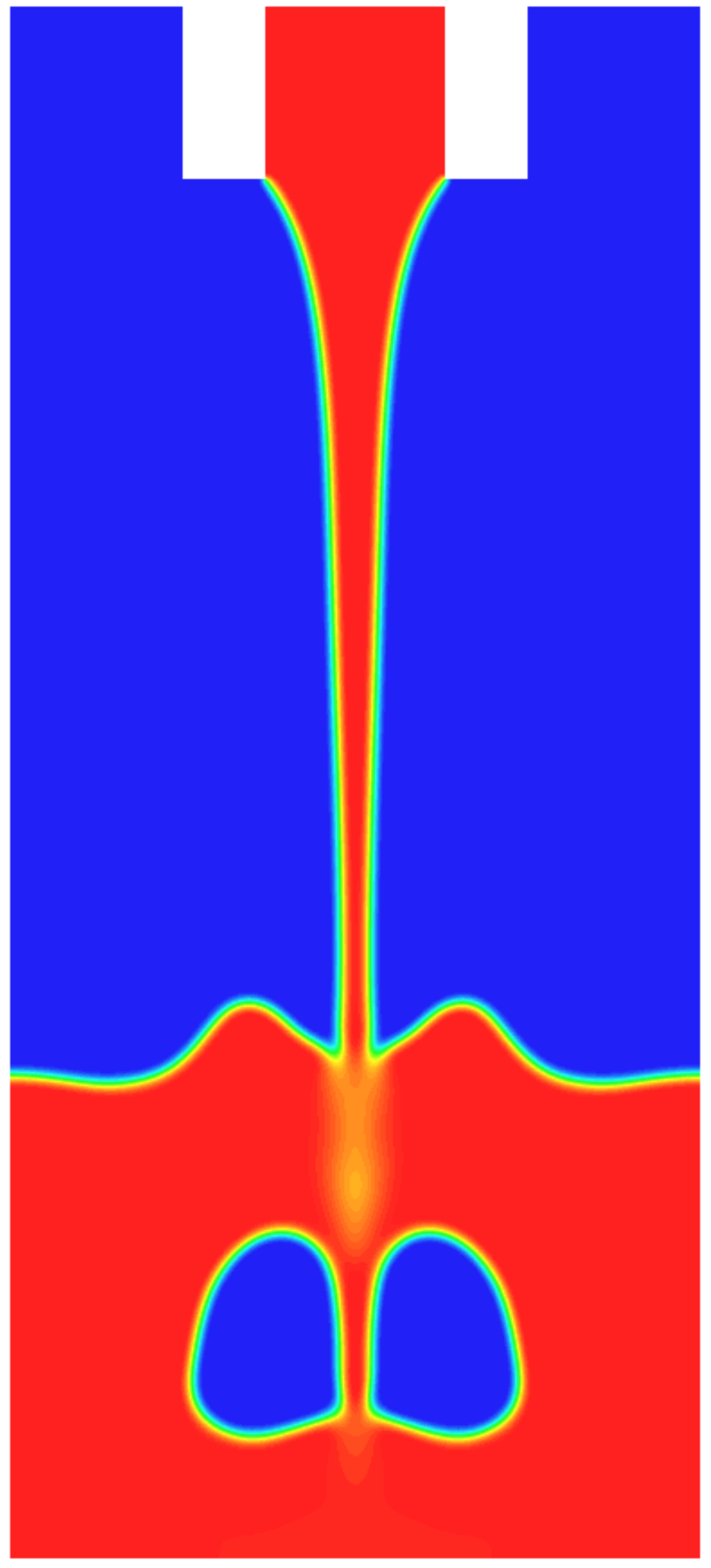}
		\subcaption{$t=0.3850 \ \si{\second}$}			
	\end{subfigure}		
	
	\caption{Faucet leak in two dimensions: Evolution of $\varphi$, using $35,712$ linear elements.}
	\label{fig:filling_drop_v10_2D}	
\end{figure}

 The evolution of $\varphi$ is shown in Figure \ref{fig:filling_drop_v10_2D}, illustrating the ability of the described formulation to deal with highly complex physical problems resulting in large geometry modifications including significant topology changes.
It is noted that small filling velocities lead to the development of non-physical flows in the air domain that originate from the tip of the drop. This phenomenon is likely to be triggered by the combination of the massively different values of density and viscosity of water and air with an imbalance of the selected mobility, interface thickness, mesh density and relatively large time domain of interest. It is currently further investigated by the authors.
%
\subsection{Faucet leak in three dimensions} \label{sec: numerical examples: filling drop_3D}
\begin{figure}[bt!]
	\captionsetup[subfigure]{labelformat=empty}
	\centering
	\includegraphics[width=0.25\textwidth]{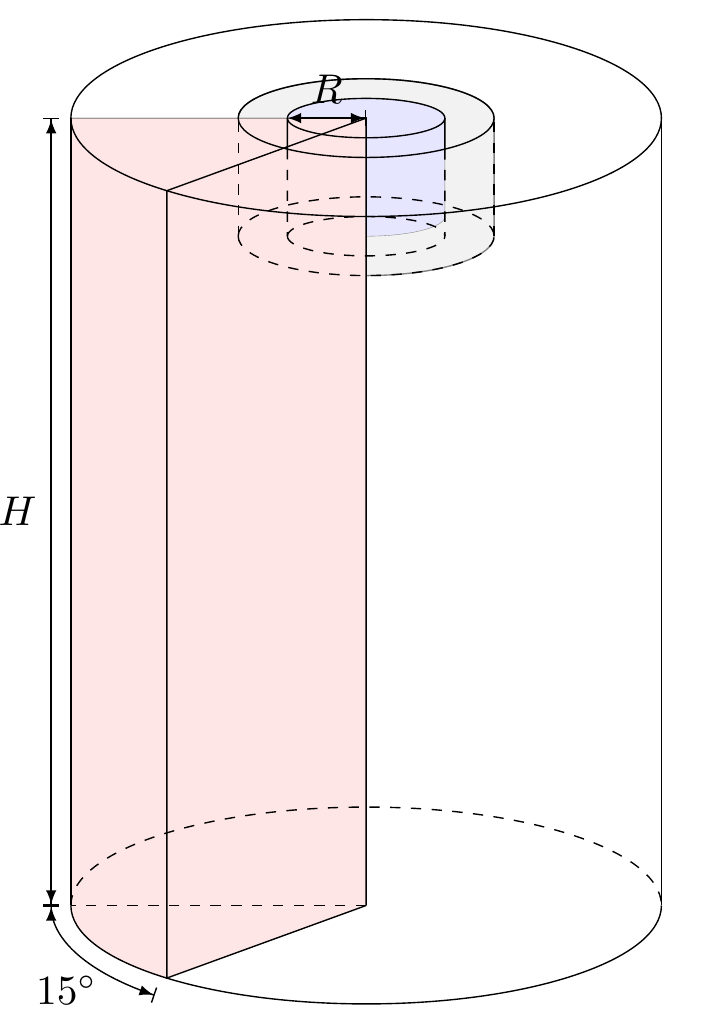}
	
	\caption{Faucet leak in three dimensions: Geometry with $15^{\circ}$ wedge.}
	\label{fig:filling_drop_3D_geom}
\end{figure}

%
A three-dimensional faucet leak is now considered with a similar setup to Section \ref{sec: numerical examples: filling drop_2D}. In this analysis the drop break-off is particularly of interest, thus the domain size is reduced to $[0,2 \ \si{\centi\metre}] \times [0,3 \ \si{\centi\metre}]$, and the pool at the bottom has been removed. A $15^\circ$ wedge is considered due to the symmetry of the problem, as shown in Figure \ref{fig:filling_drop_3D_geom}, which consists of 1,004,665 linear tetrahedron elements. Adaptive time stepping is used, with an initial time step size of $\Delta t = 0.0001$. The filling velocity is set as $\vu_{\mathrm{fill}} = 10 \ \si{\centi\metre\per\second}$.
\par
\begin{figure}[tb!]
	\captionsetup[subfigure]{labelformat=empty}
	\centering
	\begin{subfigure}{.08\textwidth}
		\centering
		\subcaption{{$0 \ \si{\second}$}}
		\includegraphics[width = 0.95\textwidth]{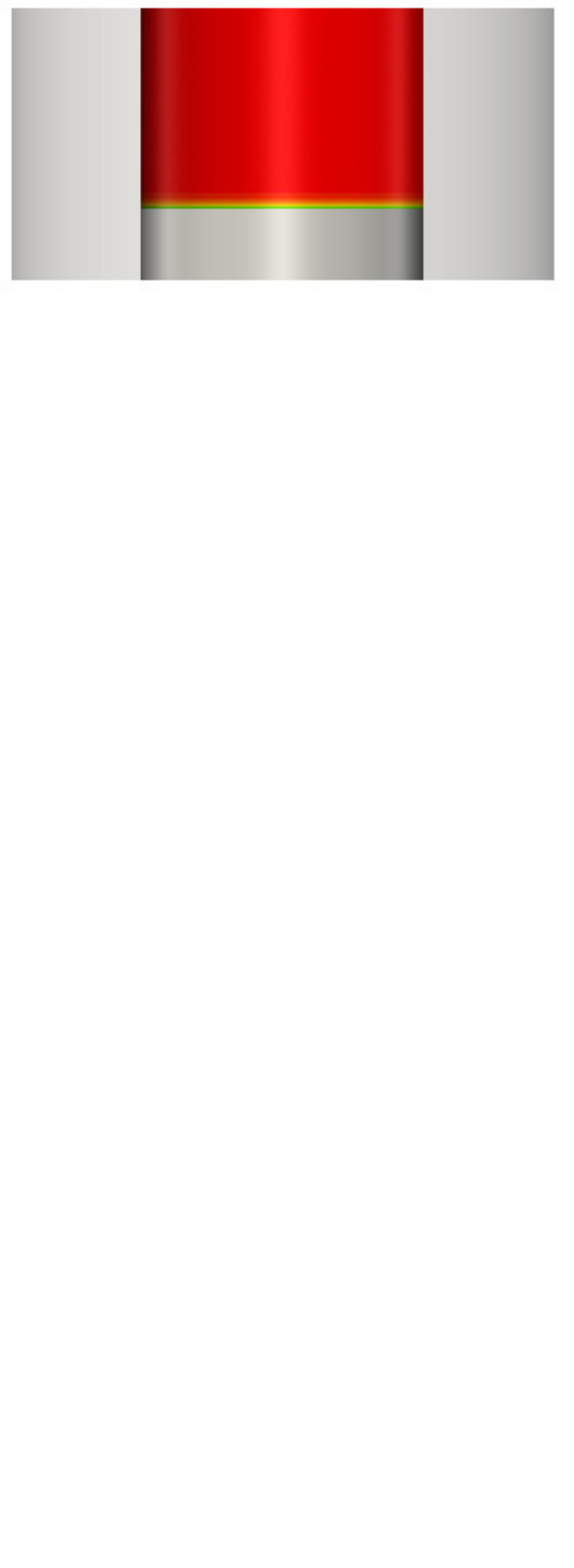}
	\end{subfigure}
	\begin{subfigure}{.08\textwidth}
		\centering	
		\subcaption{$0.0363 \ \si{\second}$}	
		\includegraphics[width = 0.95\textwidth]{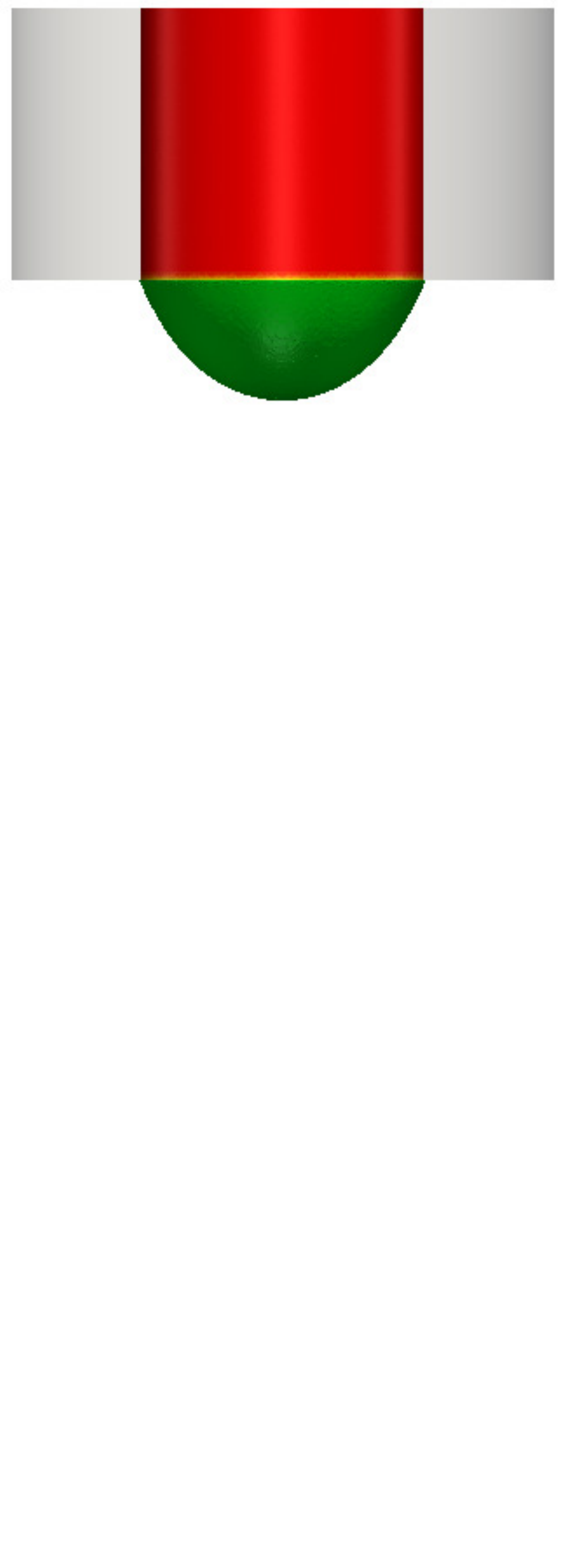}		
	\end{subfigure}
	\begin{subfigure}{.08\textwidth}
		\centering	
		\subcaption{$0.0550 \ \si{\second}$}	
		\includegraphics[width = 0.95\textwidth]{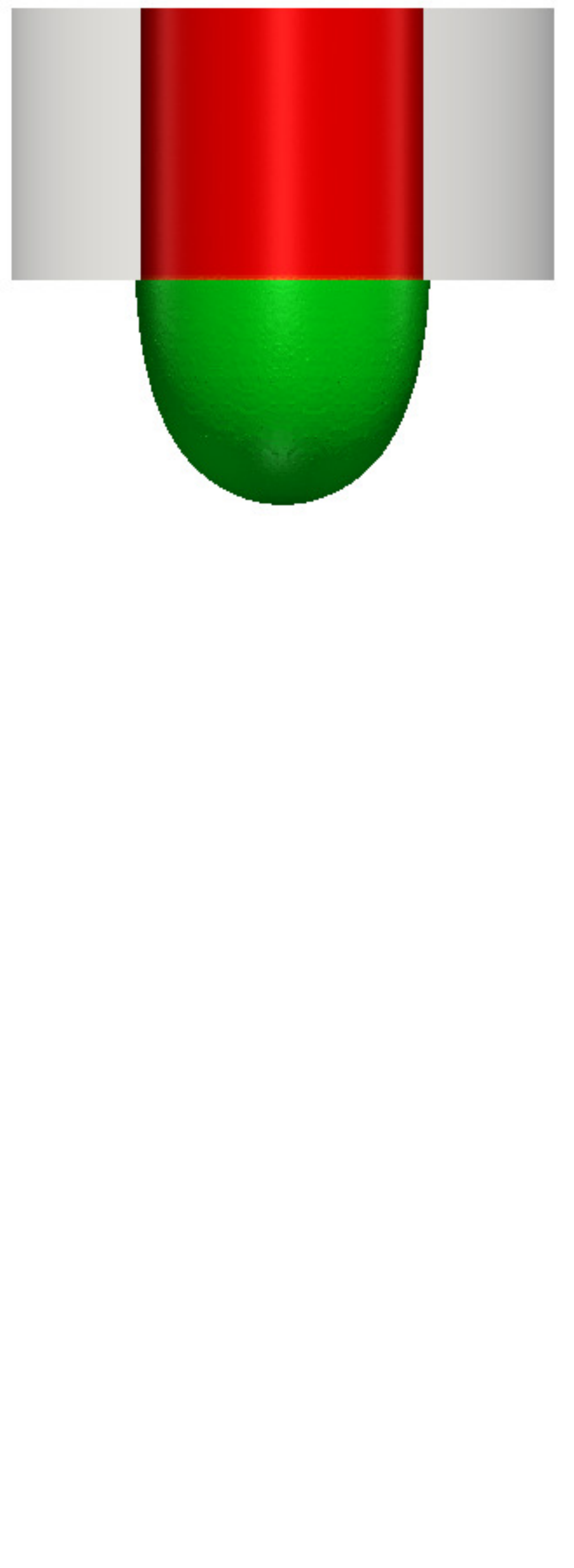}			
	\end{subfigure}
	\begin{subfigure}{.08\textwidth}
		\centering	
		\subcaption{$0.0869 \ \si{\second}$}	
		\includegraphics[width = 0.95\textwidth]{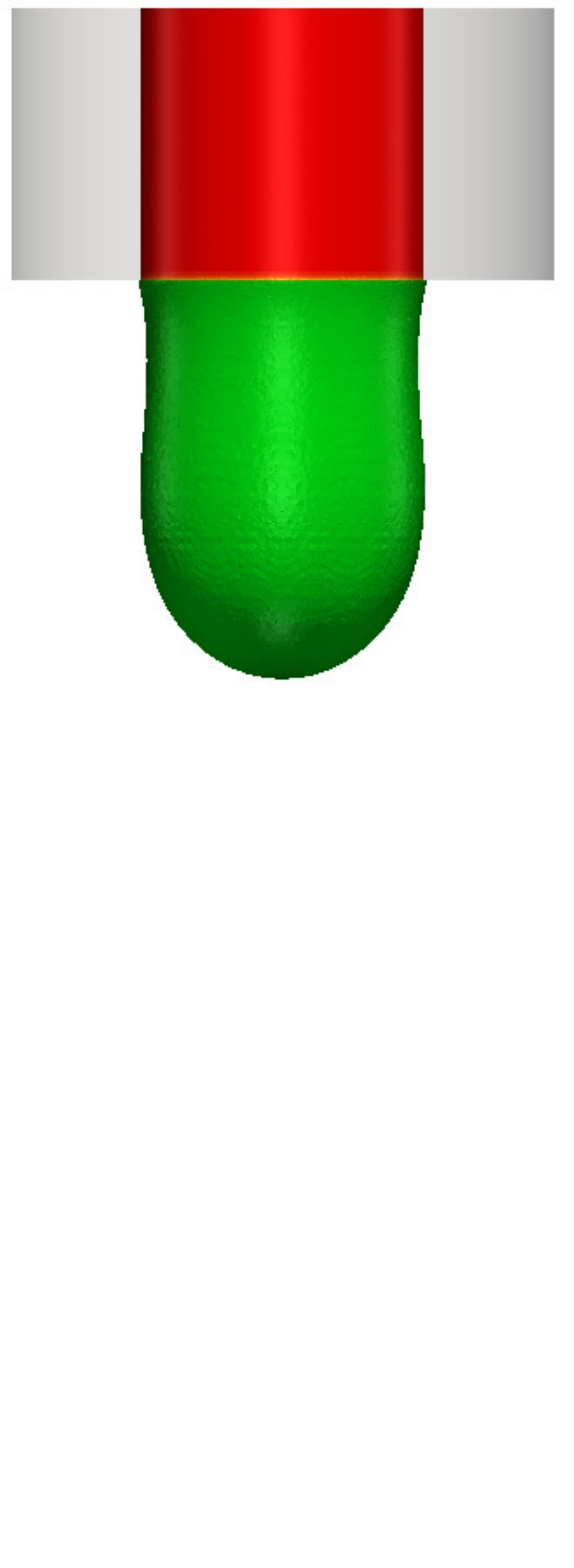}			
	\end{subfigure}
	\begin{subfigure}{.08\textwidth}
		\centering	
		\subcaption{$0.1098 \ \si{\second}$}		
		\includegraphics[width = 0.95\textwidth]{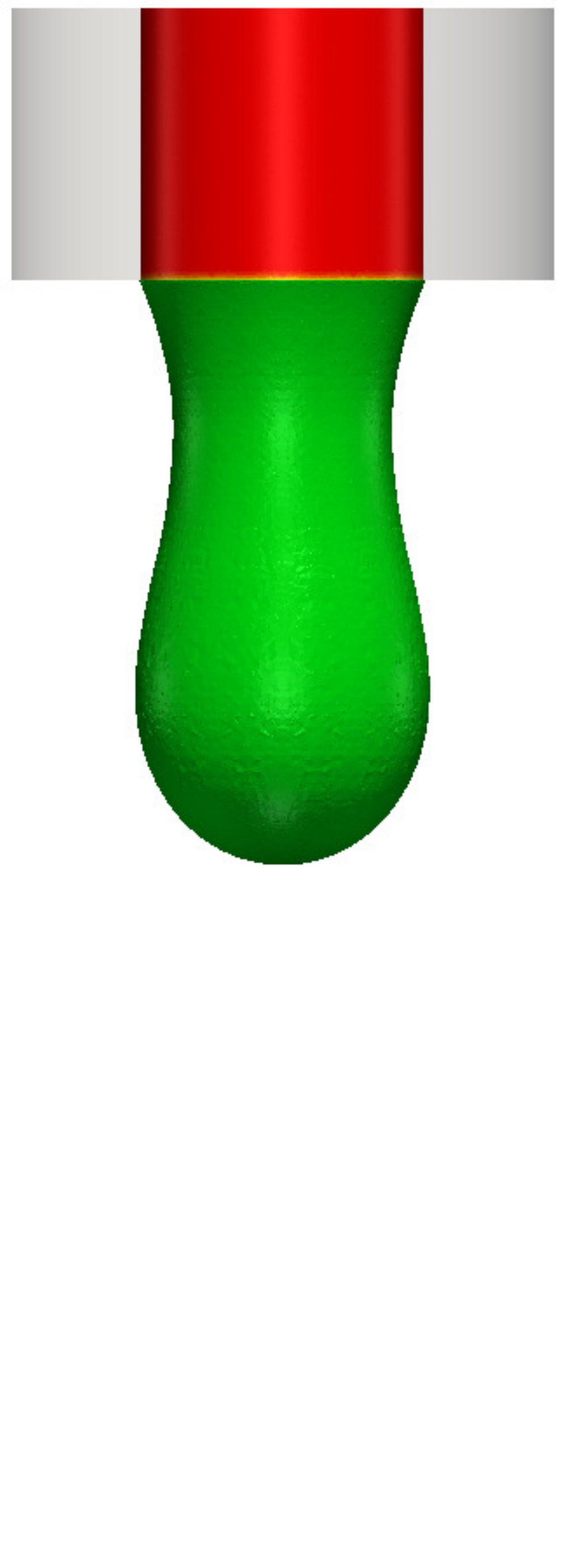}		
	\end{subfigure}
	\begin{subfigure}{.08\textwidth}
		\centering
		\subcaption{$0.1241 \ \si{\second}$}
		\includegraphics[width = 0.95\textwidth]{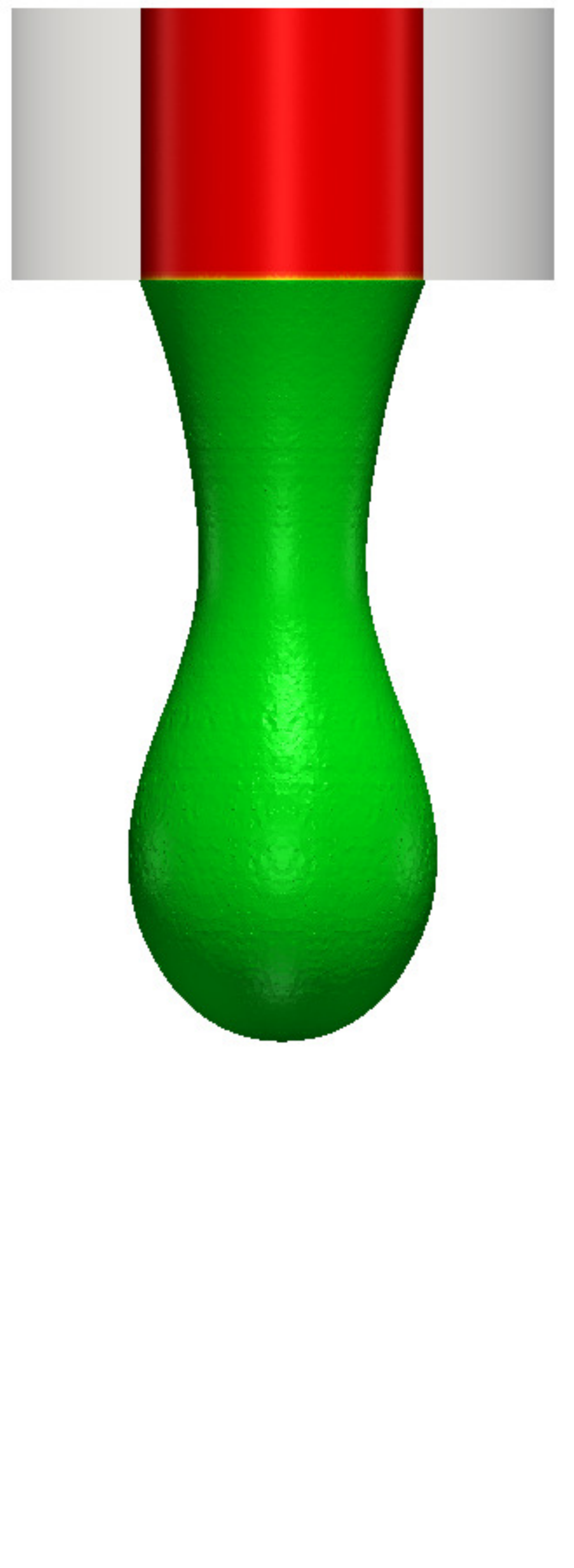}
	\end{subfigure}
	\begin{subfigure}{.08\textwidth}
		\centering	
		\subcaption{$0.1359 \ \si{\second}$}	
		\includegraphics[width = 0.95\textwidth]{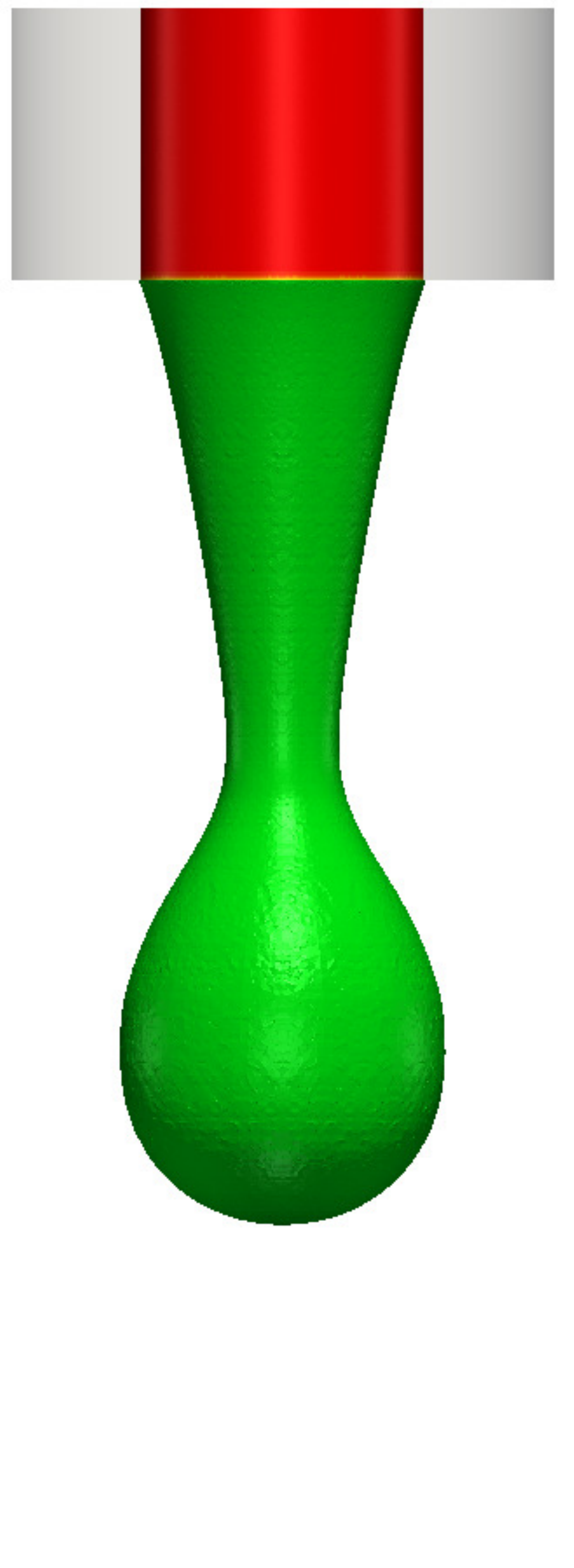}		
	\end{subfigure}
	\begin{subfigure}{.08\textwidth}
		\centering	
		\subcaption{$0.1453 \ \si{\second}$}	
		\includegraphics[width = 0.95\textwidth]{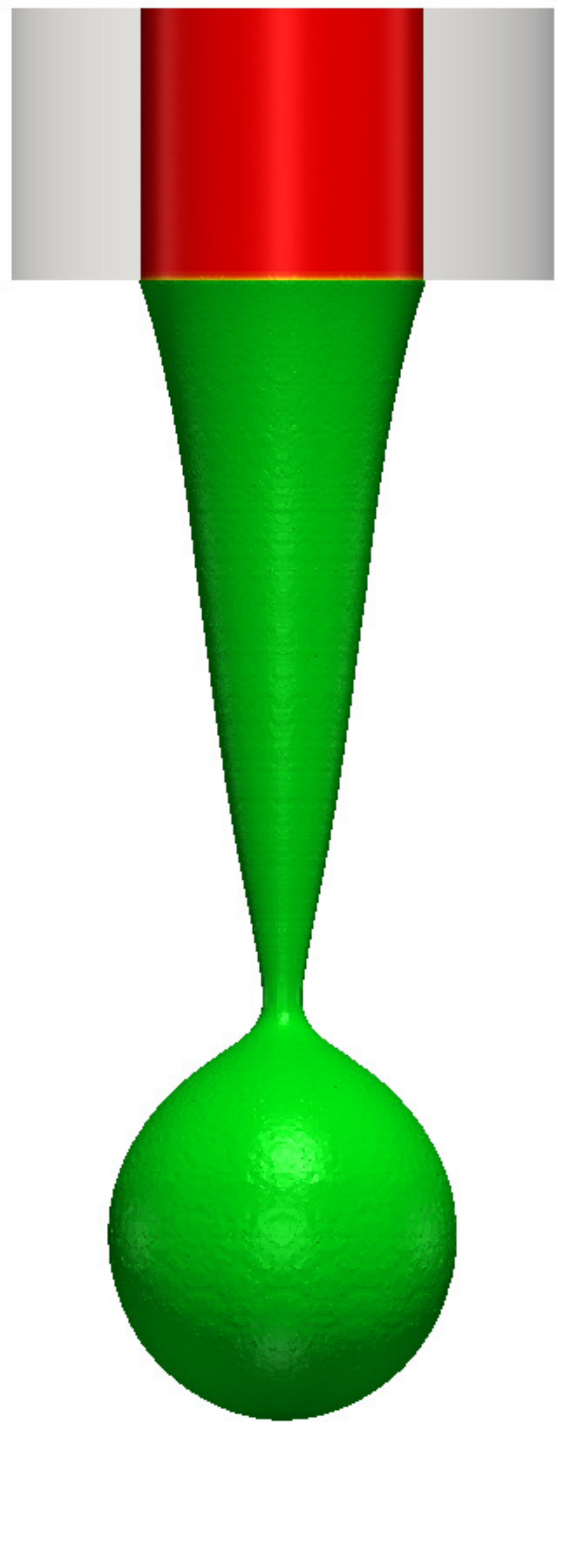}			
	\end{subfigure}
	\begin{subfigure}{.08\textwidth}
		\centering	
		\subcaption{$0.1467 \ \si{\second}$}	
		\includegraphics[width = 0.95\textwidth]{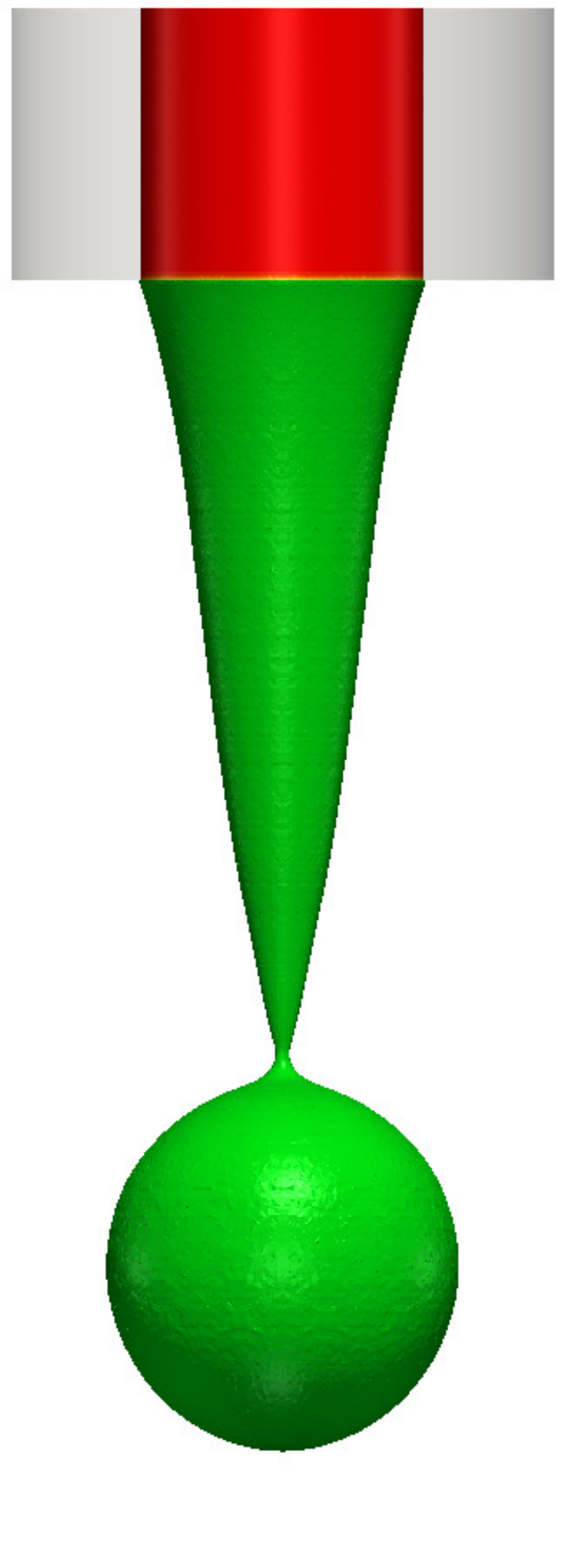}			
	\end{subfigure}
	\begin{subfigure}{.08\textwidth}
		\centering
		\subcaption{$0.1478 \ \si{\second}$}			
		\includegraphics[width = 0.95\textwidth]{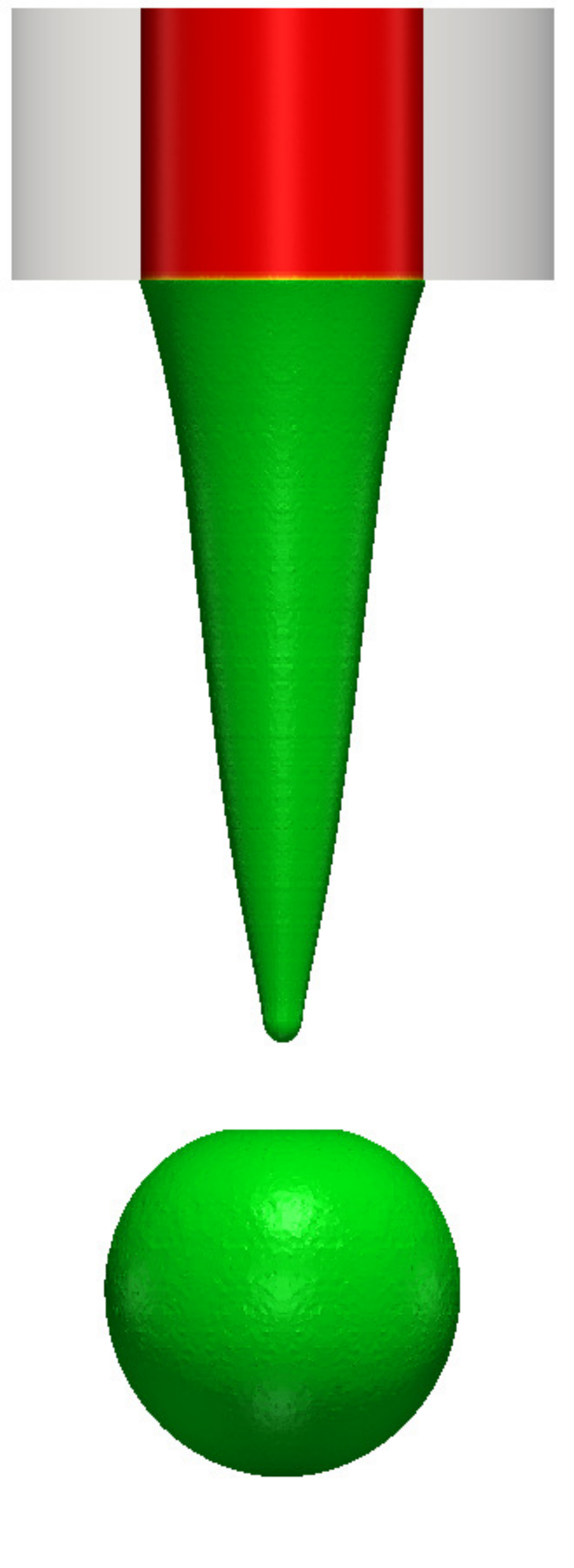}		
	\end{subfigure}
	\begin{subfigure}{.08\textwidth}
		\centering	
		\subcaption{$0.1499 \ \si{\second}$}	
		\includegraphics[width = 0.95\textwidth]{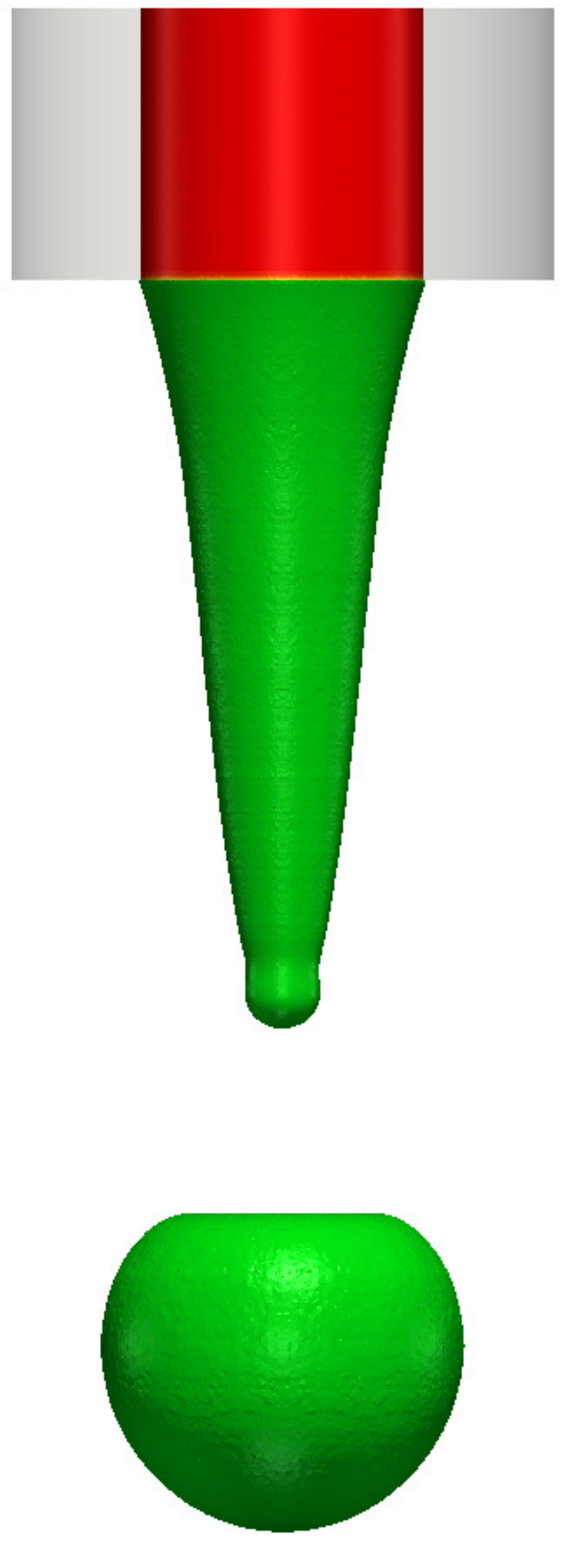}			
	\end{subfigure}
	\caption{Faucet leak in three dimensions: Evolution of $\varphi$, using $1,004,665$ linear elements.}
	\label{fig:filling_drop_v10_3D}	
\end{figure}

\begin{figure}[!t]
	\centering
	\captionsetup[subfigure]{labelformat=empty}
	{\adjustbox{minipage=0em,valign=t}{\subcaption{}}%
	\begin{subfigure}[t]{\dimexpr.110\linewidth-0.55em\relax}
		\centering
		\includegraphics[width = 1.00\textwidth,valign=t]{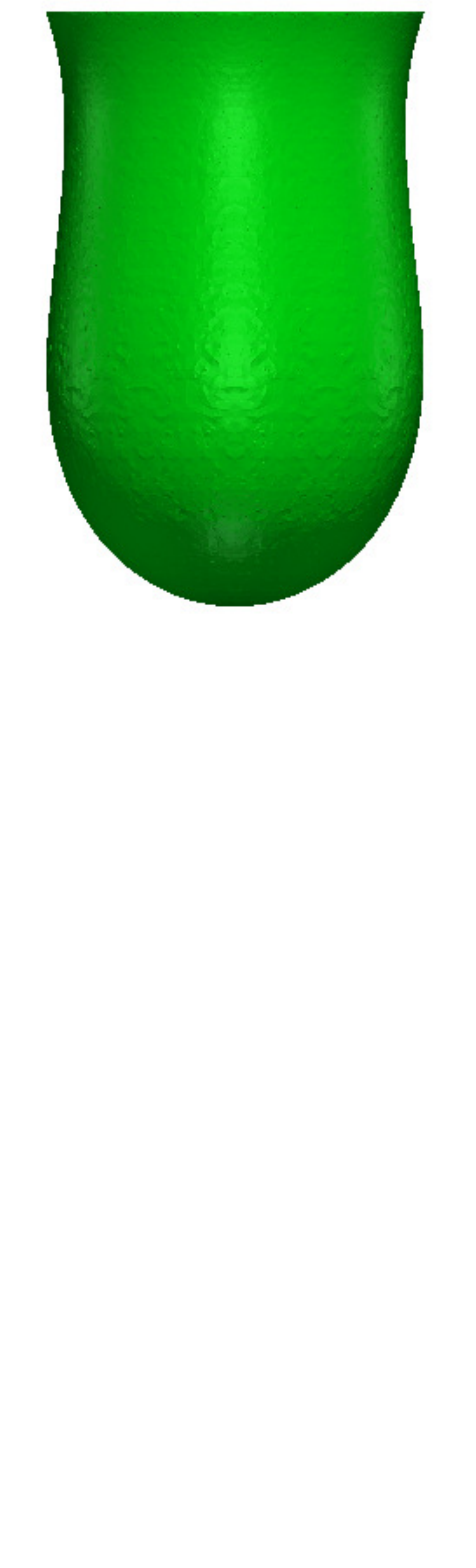}
	\end{subfigure}}
	{\adjustbox{minipage=0em,valign=t}{\subcaption{}}%
	\begin{subfigure}[t]{\dimexpr.110\linewidth-0.55em\relax}
		\centering	
		\includegraphics[width = 0.88\textwidth,valign=t]{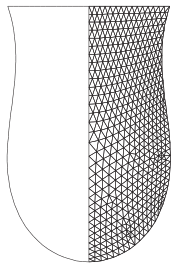}
	\end{subfigure}}	
	{\adjustbox{minipage=0em,valign=t}{\subcaption{}}%
	\begin{subfigure}[t]{\dimexpr.110\linewidth-0.55em\relax}
		\centering
		\includegraphics[width = 0.93\textwidth,valign=t]{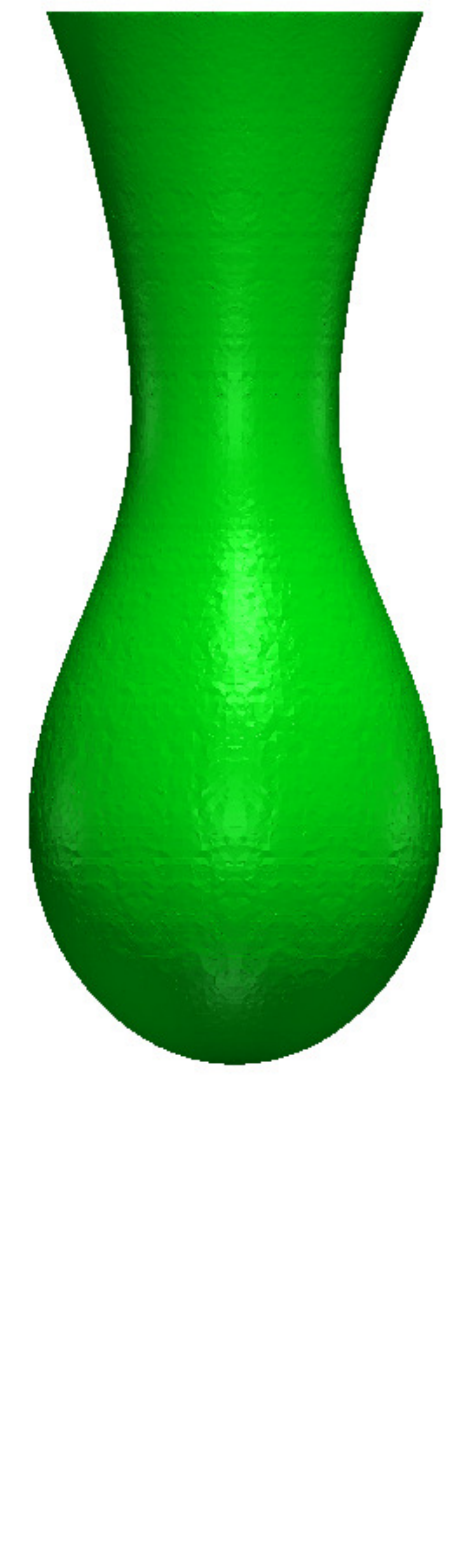}
	\end{subfigure}}
	{\adjustbox{minipage=0em,valign=t}{\subcaption{}}%
	\begin{subfigure}[t]{\dimexpr.110\linewidth-0.55em\relax}
		\centering	
		\includegraphics[width = 0.88\textwidth,valign=t]{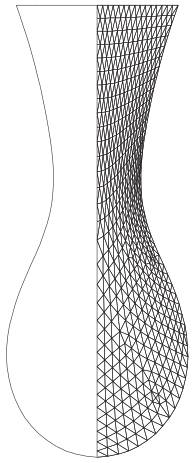}
	\end{subfigure}}
	{\adjustbox{minipage=0em,valign=t}{\subcaption{}}%
	\begin{subfigure}[t]{\dimexpr.110\linewidth-0.55em\relax}
		\centering
		\includegraphics[width = 0.86\textwidth,valign=t]{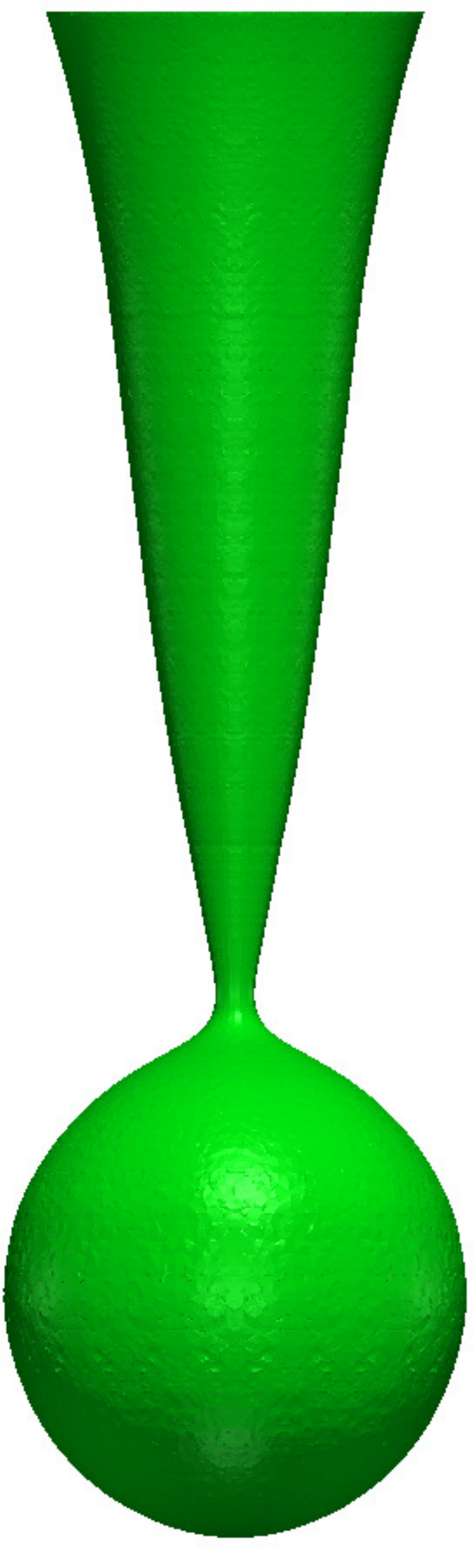}
	\end{subfigure}}
	{\adjustbox{minipage=0em,valign=t}{\subcaption{}}%
	\begin{subfigure}[t]{\dimexpr.110\linewidth-0.55em\relax}
		\centering	
		\includegraphics[width = 0.9\textwidth,valign=t]{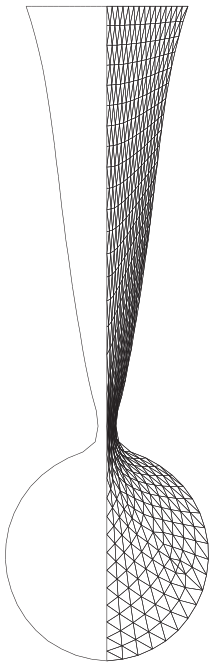}
	\end{subfigure}}	
	\caption{Faucet leak in three dimensions: Comparison of interface isolines with \cite{Dettmer2006}.}
	\label{fig:filling_drop_v10_comp_3D}	
\end{figure}

\begin{figure}[bt!]
	\captionsetup[subfigure]{labelformat=empty}
	\centering
	\begin{subfigure}{.1125\textwidth}
		\centering
		\subcaption{{$0.1438 \ \si{\second}$}}
		\includegraphics[width = 0.95\textwidth]{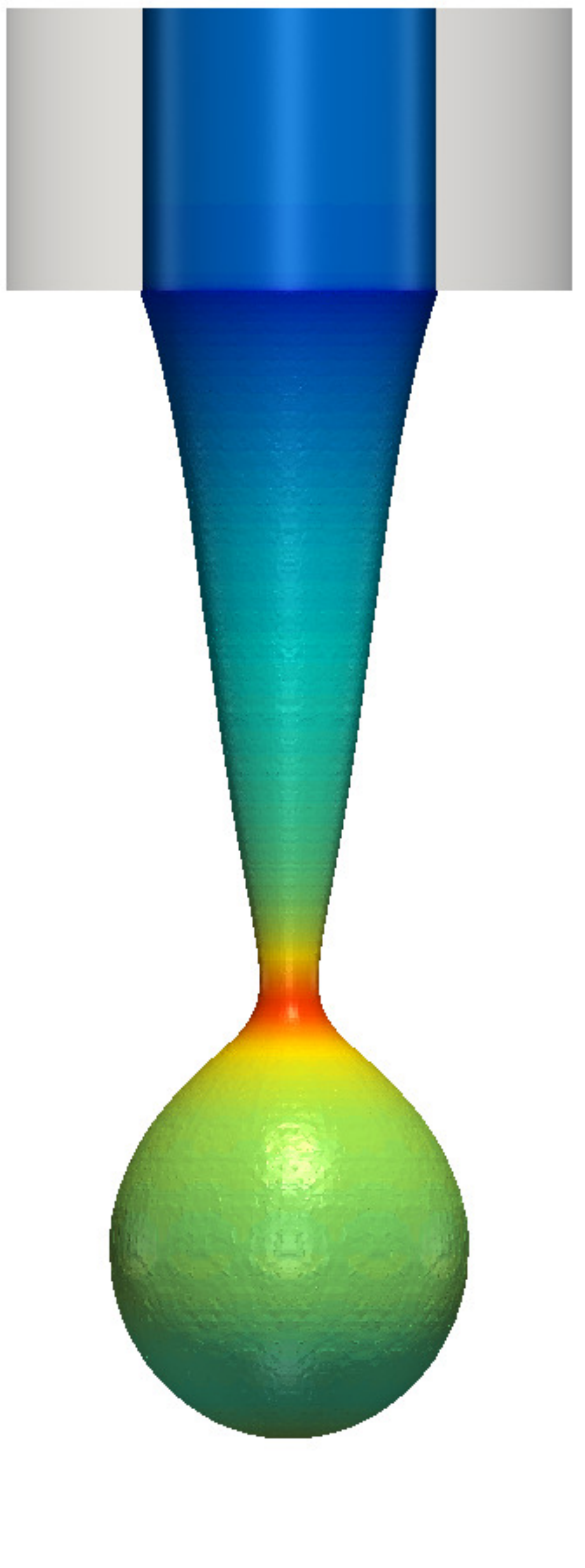}
	\end{subfigure}
	\begin{subfigure}{.1125\textwidth}
		\centering	
		\subcaption{$0.1463 \ \si{\second}$}	
		\includegraphics[width = 0.95\textwidth]{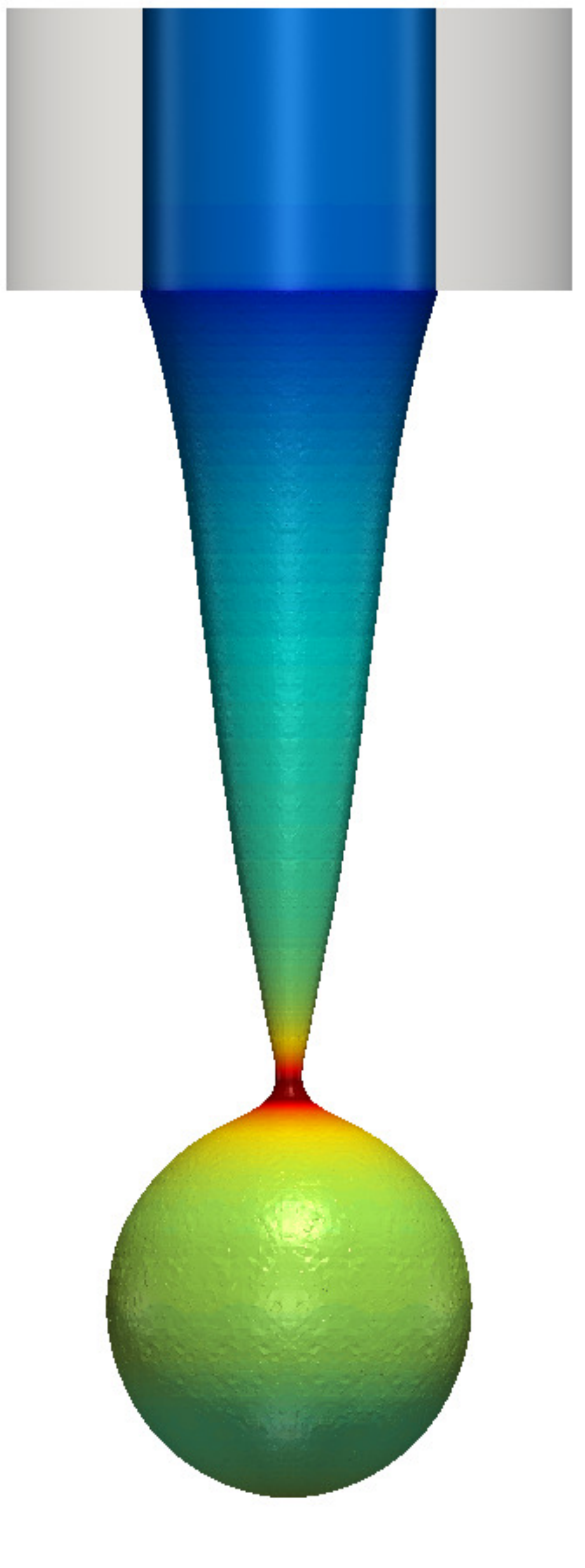}		
	\end{subfigure}
	\begin{subfigure}{.1125\textwidth}
		\centering	
		\subcaption{$0.1468 \ \si{\second}$}	
		\includegraphics[width = 0.95\textwidth]{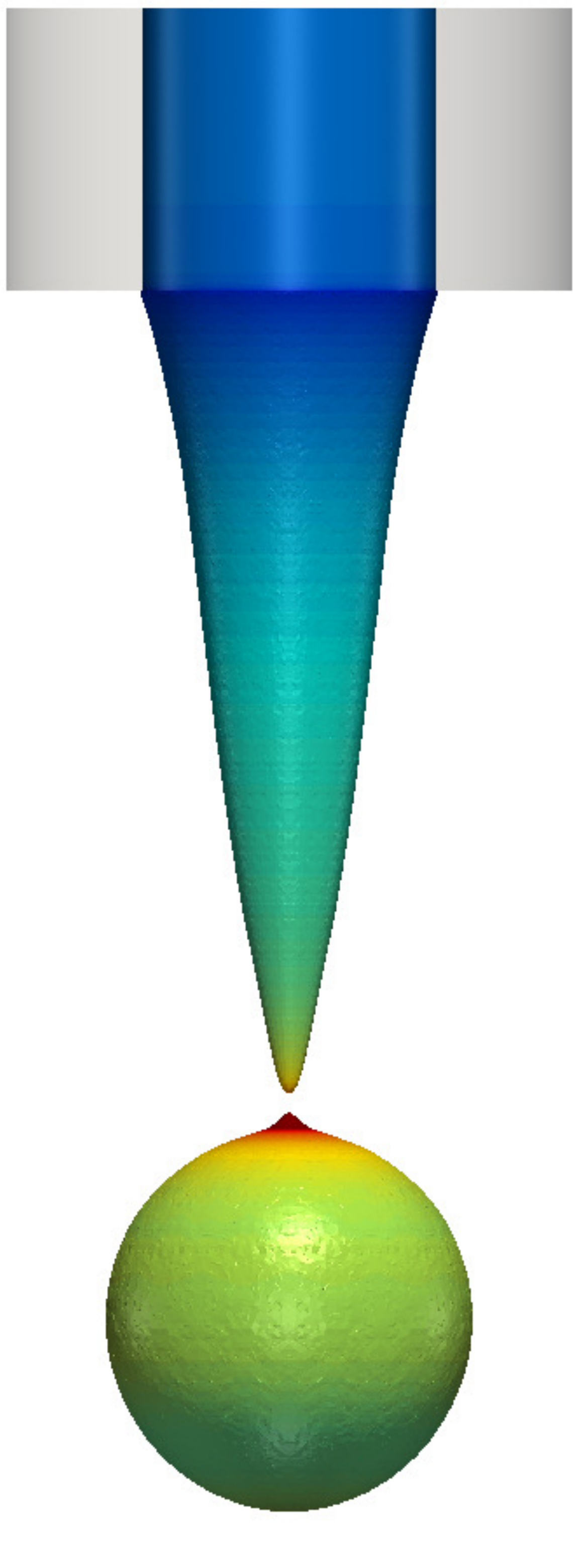}			
	\end{subfigure}
	\begin{subfigure}{.1125\textwidth}
		\centering	
		\subcaption{$0.1486 \ \si{\second}$}	
		\includegraphics[width = 0.95\textwidth]{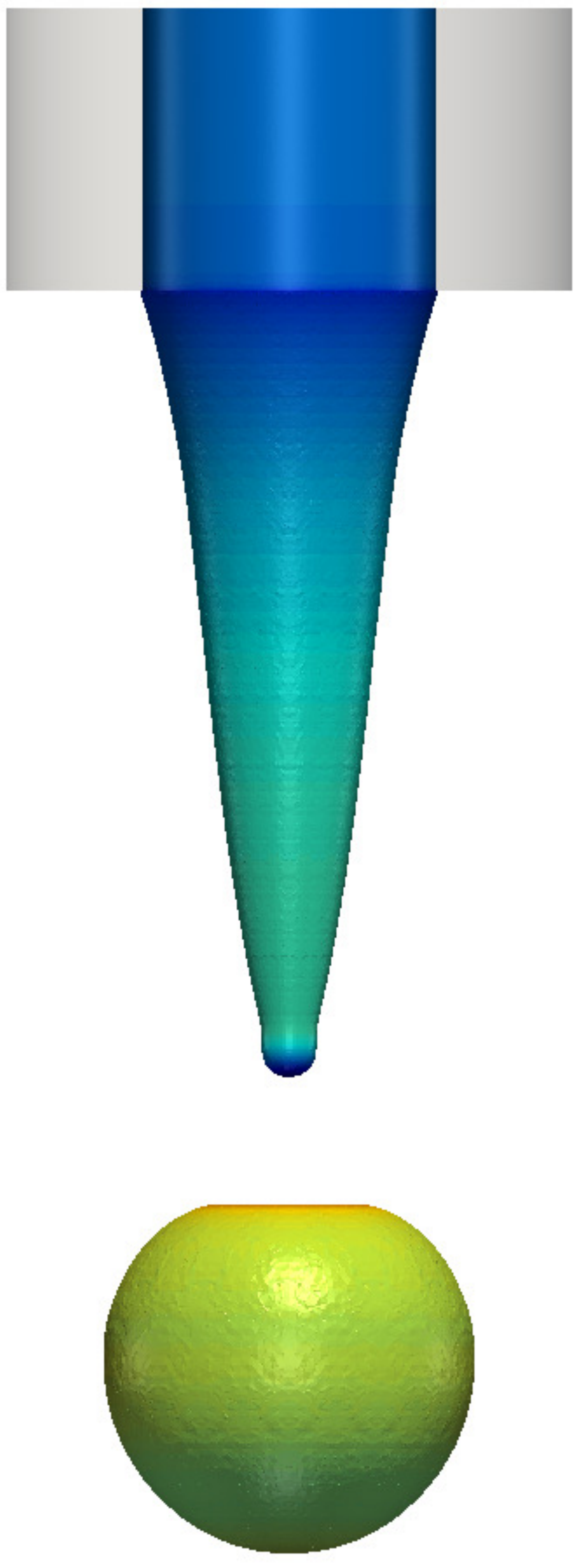}			
	\end{subfigure}
	\begin{subfigure}{.55\textwidth}
		\centering	
		\includegraphics[width = 0.95\textwidth]{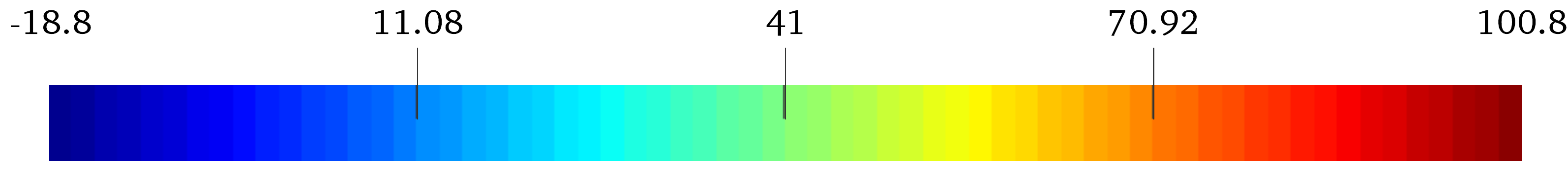}			
	\end{subfigure}	
	\caption{Faucet leak in three dimensions: Evolution of vertical velocity, using $1,004,665$ linear elements.}
	\label{fig:filling_drop_v10_vel_3D}	
\end{figure}

\begin{figure}[tb!]
	\centering
	\captionsetup[subfigure]{labelformat=empty}
	\begin{subfigure}{.40\textwidth}
		\centering
		\includegraphics[width = 0.9\textwidth]{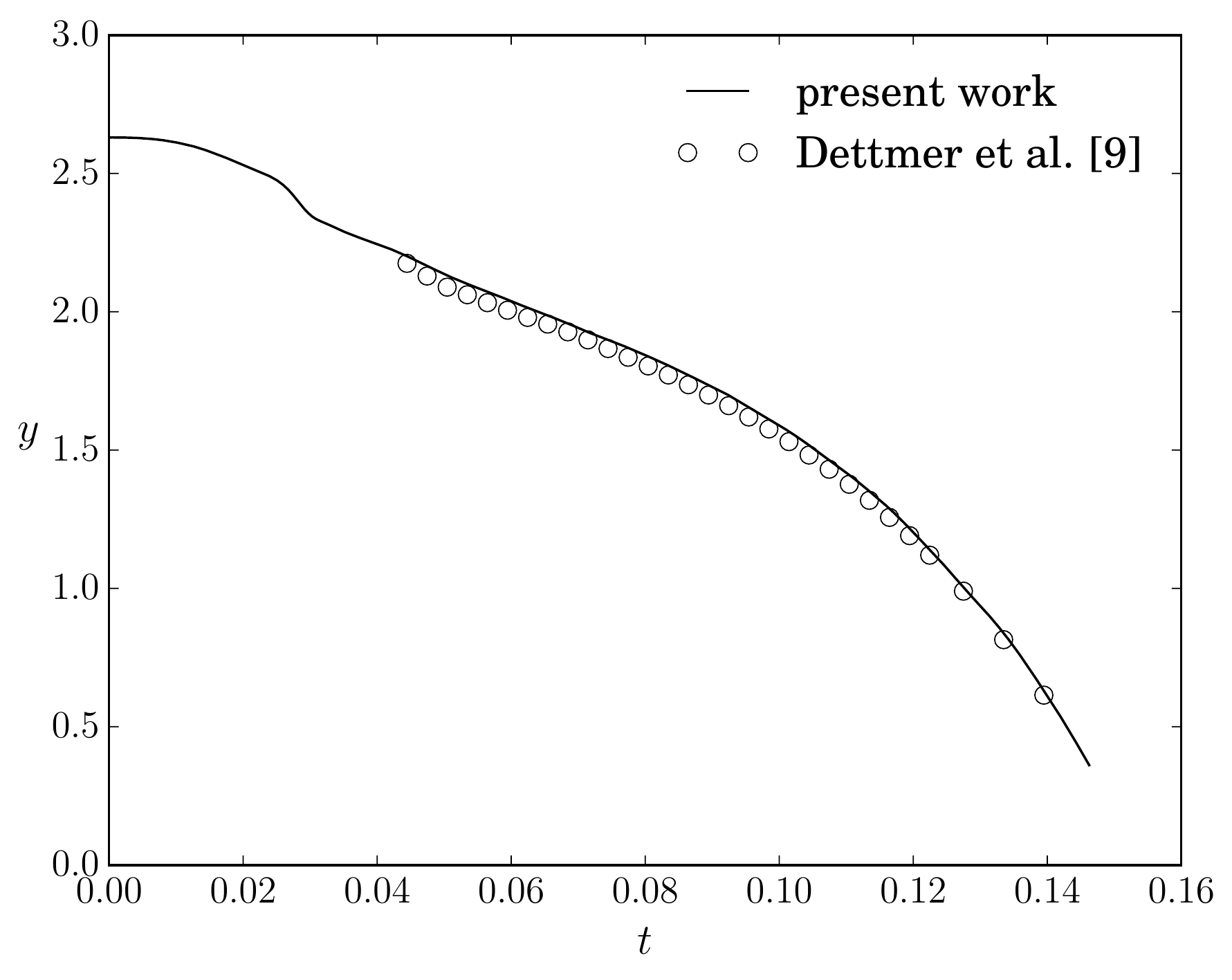}
	\end{subfigure}
	\begin{subfigure}{.40\textwidth}
		\centering
		\includegraphics[width = 0.9\textwidth]{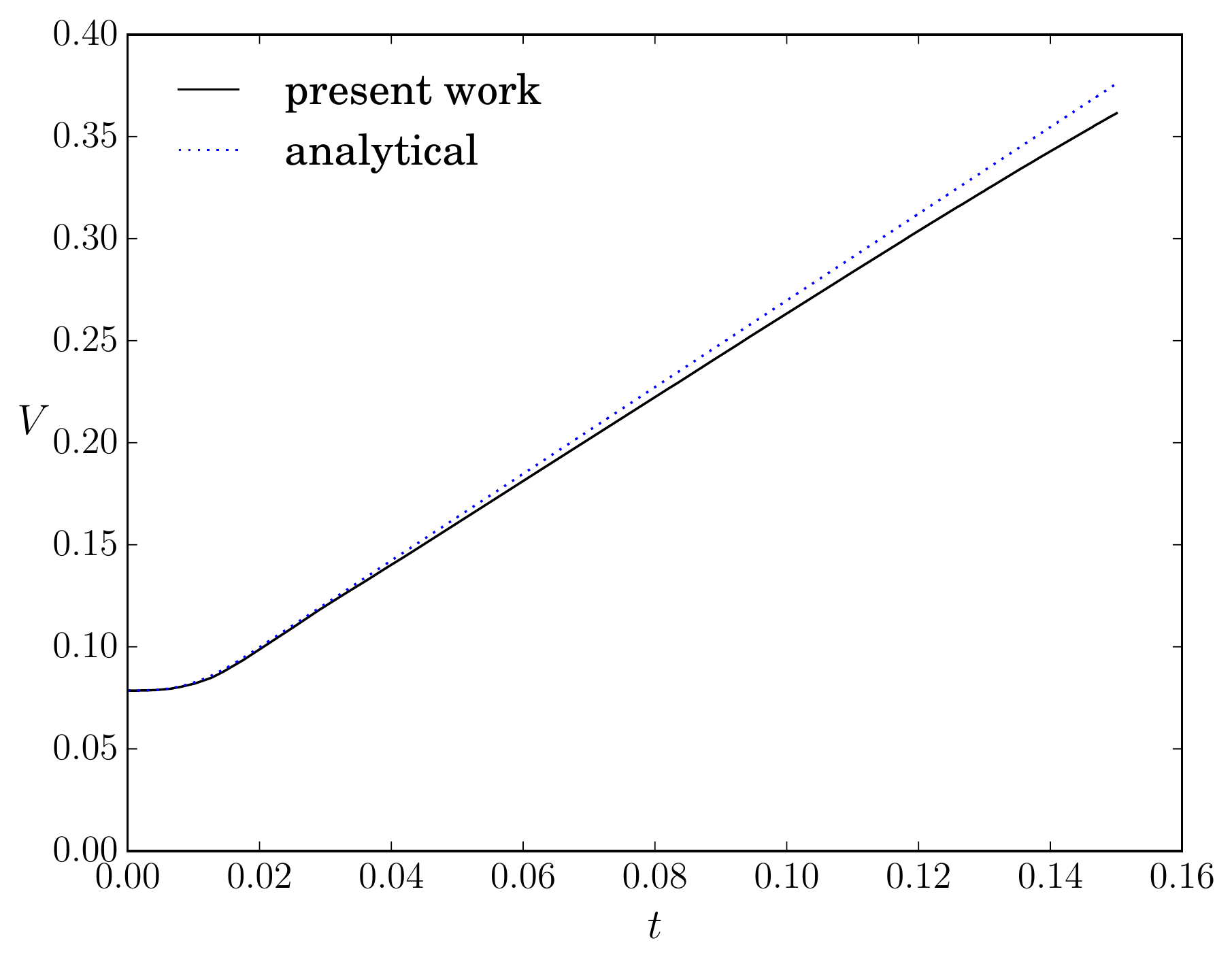}
	\end{subfigure}
	\caption{Faucet leak in three dimensions: Tip position comparison with \cite{Dettmer2006} (left), and water volume evolution comparison with analytical solution (right).}
	\label{fig:filling_drop_wulf_v10_comp_3D}
\end{figure}

 The evolution of $\varphi$ is shown in Figure \ref{fig:filling_drop_v10_3D}.
A visual comparison of the drop configuration at different time instances is made with Dettmer and Peri\'c \cite{Dettmer2006} in Figure \ref{fig:filling_drop_v10_comp_3D}. 
The vertical velocity component during break-off is visible in Figure \ref{fig:filling_drop_v10_vel_3D}. 
In Figure \ref{fig:filling_drop_wulf_v10_comp_3D}, the tip position evolution compared with \cite{Dettmer2006} is shown along with the water the volume evolution. Unlike in the present study, the initial configuration in \cite{Dettmer2006} is set as an equilibrium pendant drop, thus these results are shifted in time to the appropriate position. It is observable that the tip position evolution matches well with \cite{Dettmer2006}. 
%
%
\subsection{Broken dam} \label{sec: numerical examples: broken dam}
%
 A fluid column ($\rho = 1$, $\mu=0.01$) of width $b = 3.5$ and height $h = 7$ is placed in a rectangular domain ($\rho = 0.001$, $\mu=0.0001$) of dimensions $[0,15] \times [0,10]$. The gravitational acceleration is set by using $\bm{b} = [0,-1]^T$, and the surface tension coefficient is set to zero, \textit{i.e.} $\gamma=0$. The fluid is allowed to slip on the horizontal and vertical surfaces, and is allowed to adopt any contact angle by replacing the boundary terms in Equations \eqref{eq:NSCHweakD} and \eqref{eq:NSCHstabWeakD}, such that the terms become
 \begin{align}
	\intom s^h \left( \eta^h - f \right) - \nabla s^h \cdot \epsilon^2 \nabla \varphi^h \dom 
	- \intgam s^h \ \epsilon^2 \nabla \varphi^h \cdot \vn \dgam = 0.
 	\label{eq:free boundary}
 \end{align}
 \par
\begin{figure}[bt!]
	\centering
	\begin{subfigure}{.40\textwidth}
		\centering
		\vspace{0.25cm}
		\includegraphics[width=0.9\textwidth]{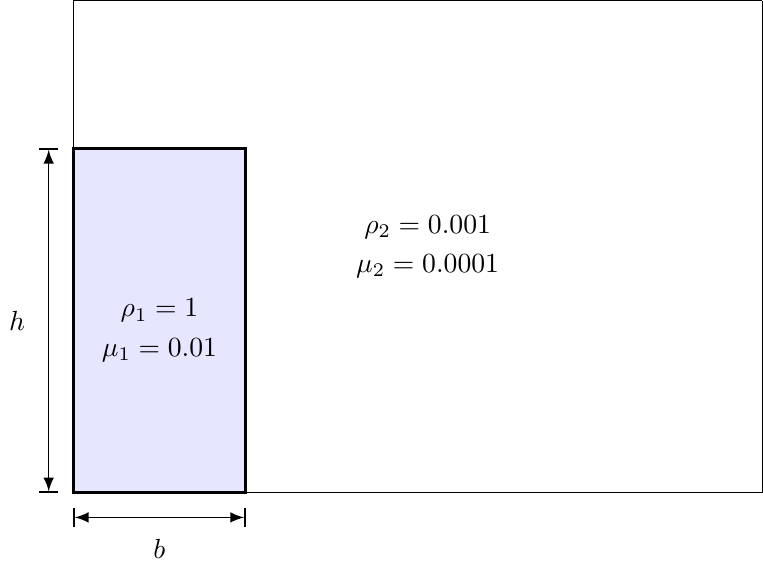}
	\end{subfigure}
	\begin{subfigure}{.40\textwidth}
		\centering
		\includegraphics[width=0.85\textwidth]{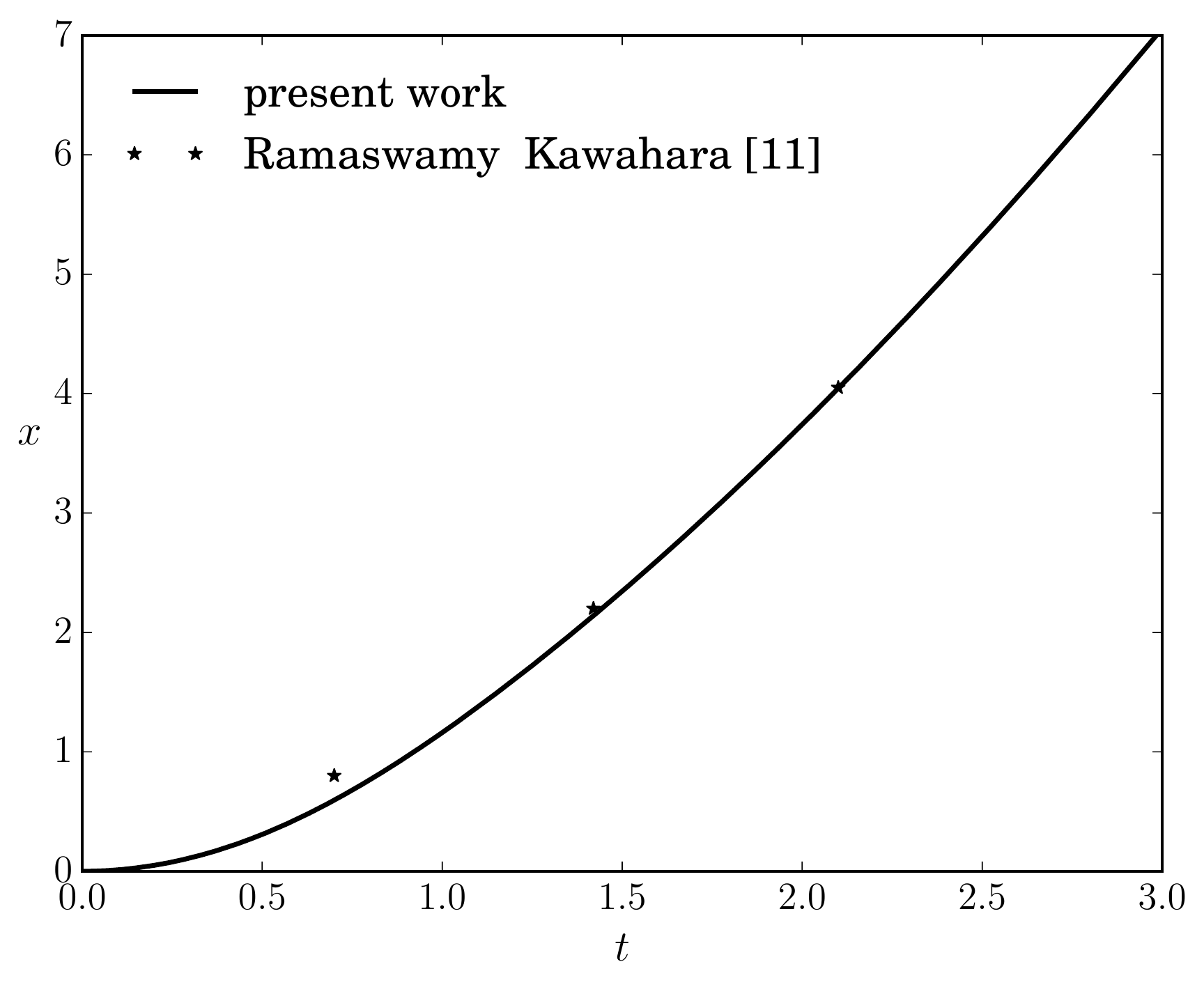}
	\end{subfigure}	
	\caption{Broken dam: Geometry (left) and evolution of tip displacement (right).}
	\label{fig:broken_dam_geom}	
\end{figure}

\begin{figure}[bt!]
	\captionsetup[subfigure]{labelformat=empty}
	\centering
	\begin{subfigure}{.32\textwidth}
		\centering
		\includegraphics[width = 0.95\textwidth]{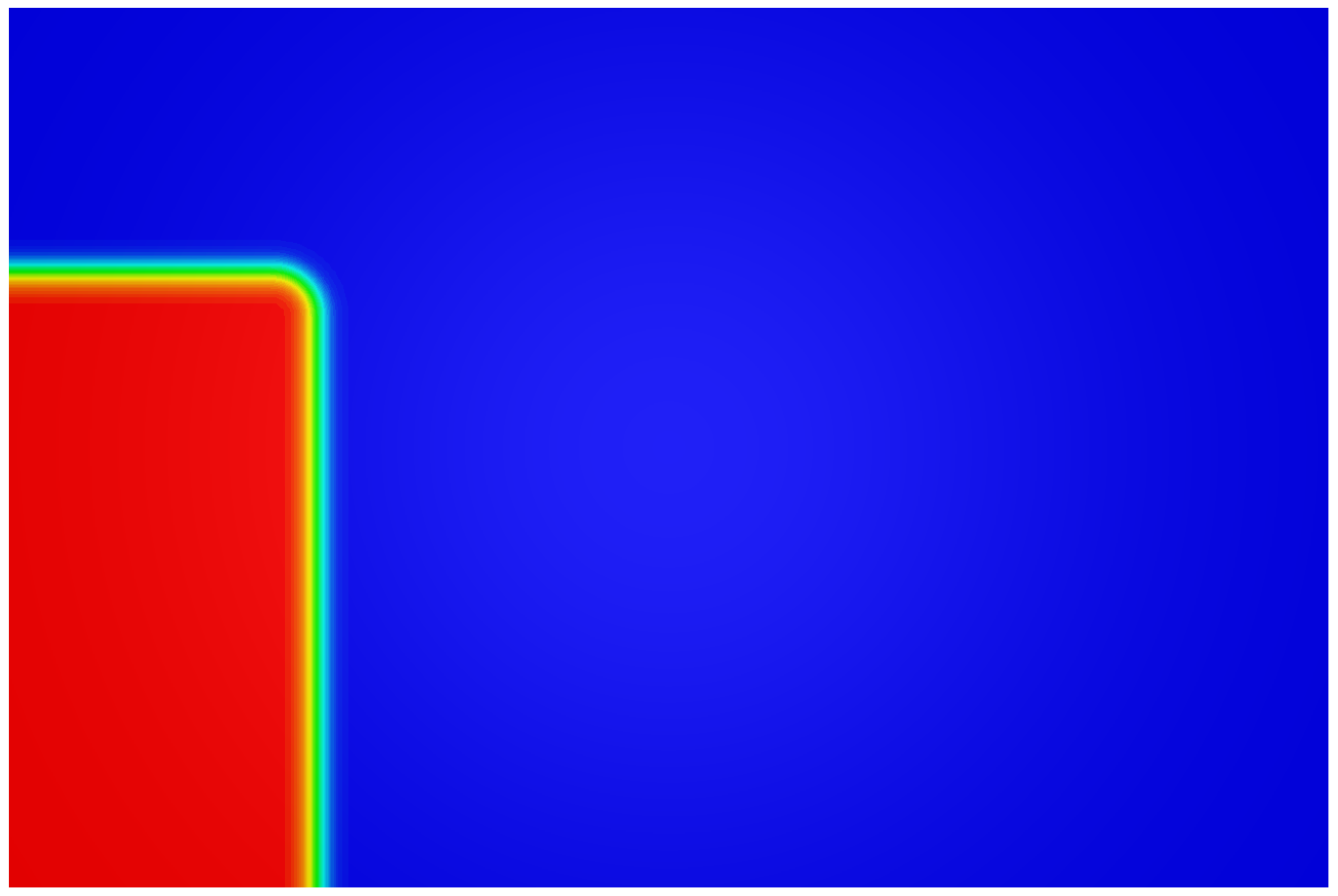}
		\subcaption{$t=0$}
	\end{subfigure}
	\begin{subfigure}{.32\textwidth}
		\centering	
		\includegraphics[width = 0.95\textwidth]{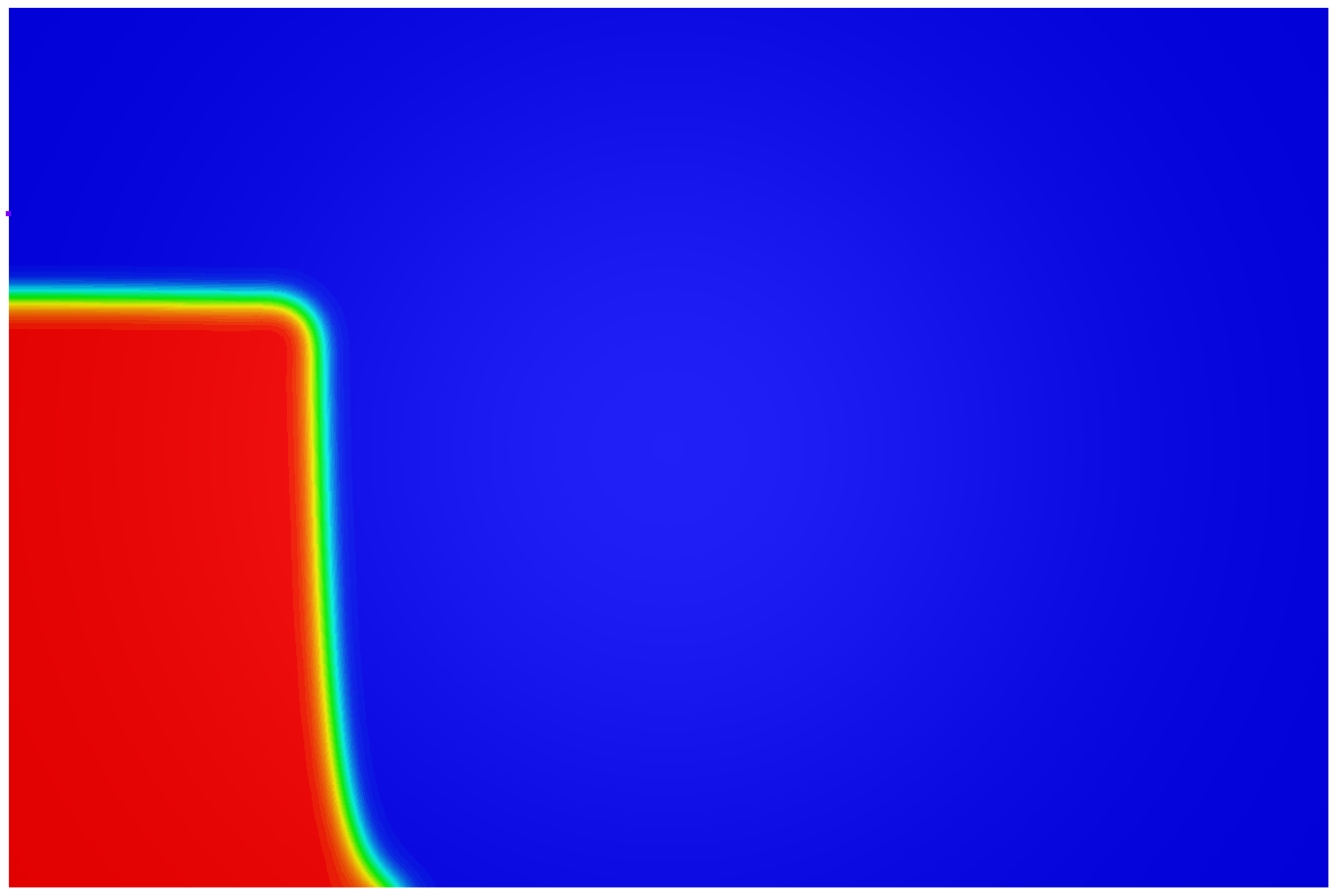}
		\subcaption{$t=0.8$}			
	\end{subfigure}
	\begin{subfigure}{.32\textwidth}
		\centering		
		\includegraphics[width = 0.95\textwidth]{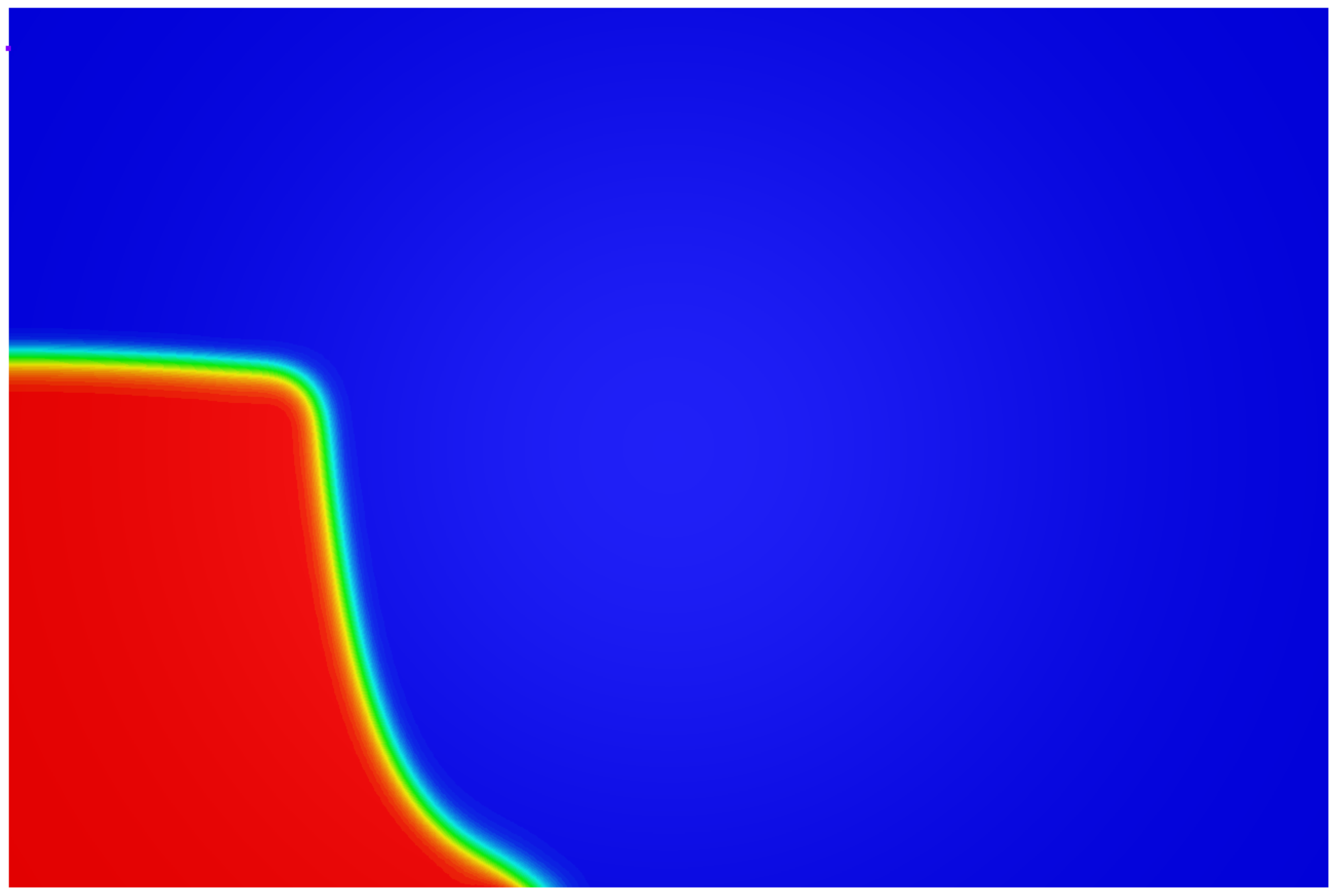}
		\subcaption{$t=1.6$}			
	\end{subfigure}
	\begin{subfigure}{.32\textwidth}
		\centering		
		\includegraphics[width = 0.95\textwidth]{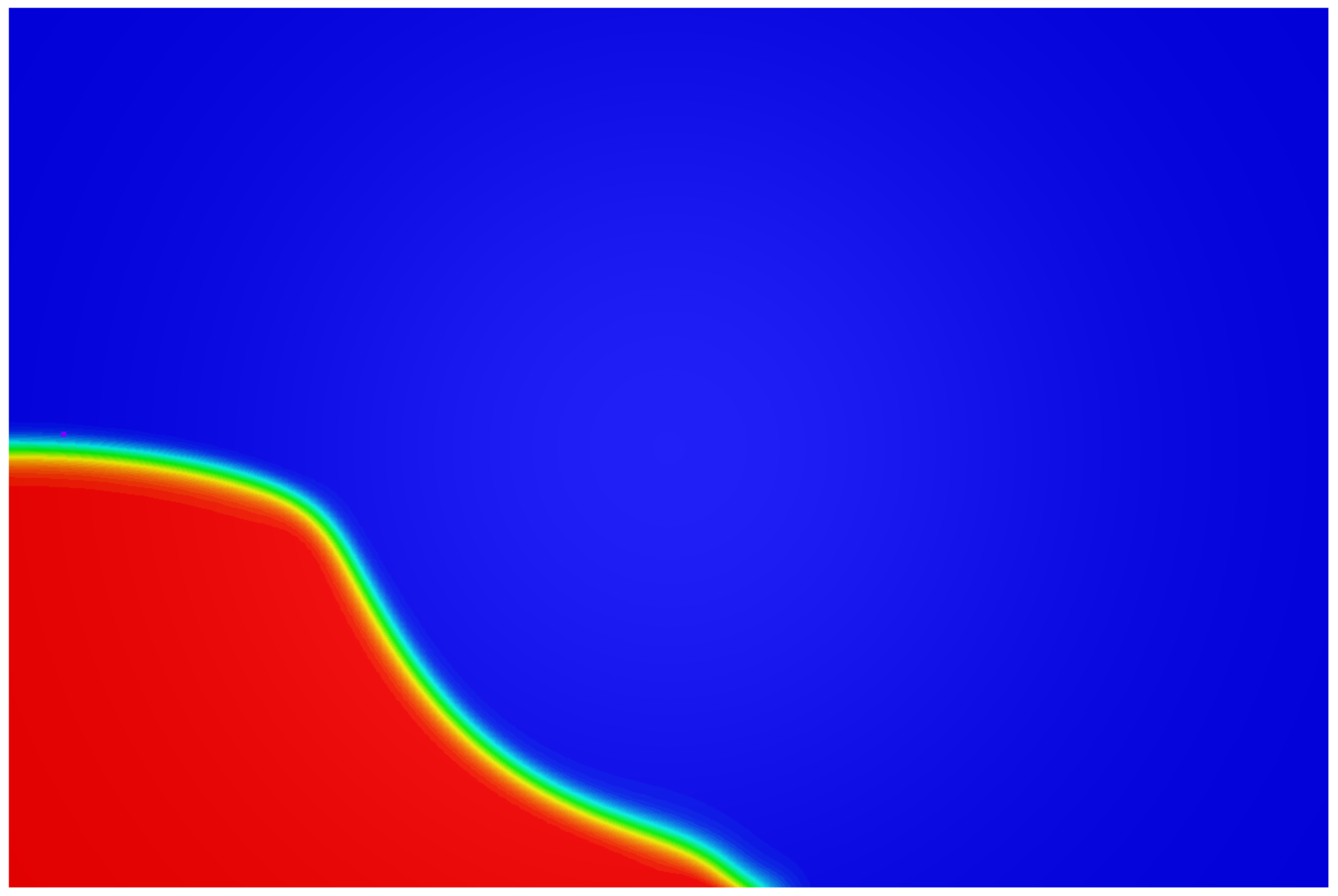}
		\subcaption{$t=2.4$}			
	\end{subfigure}
	\begin{subfigure}{.32\textwidth}
		\centering		
		\includegraphics[width = 0.95\textwidth]{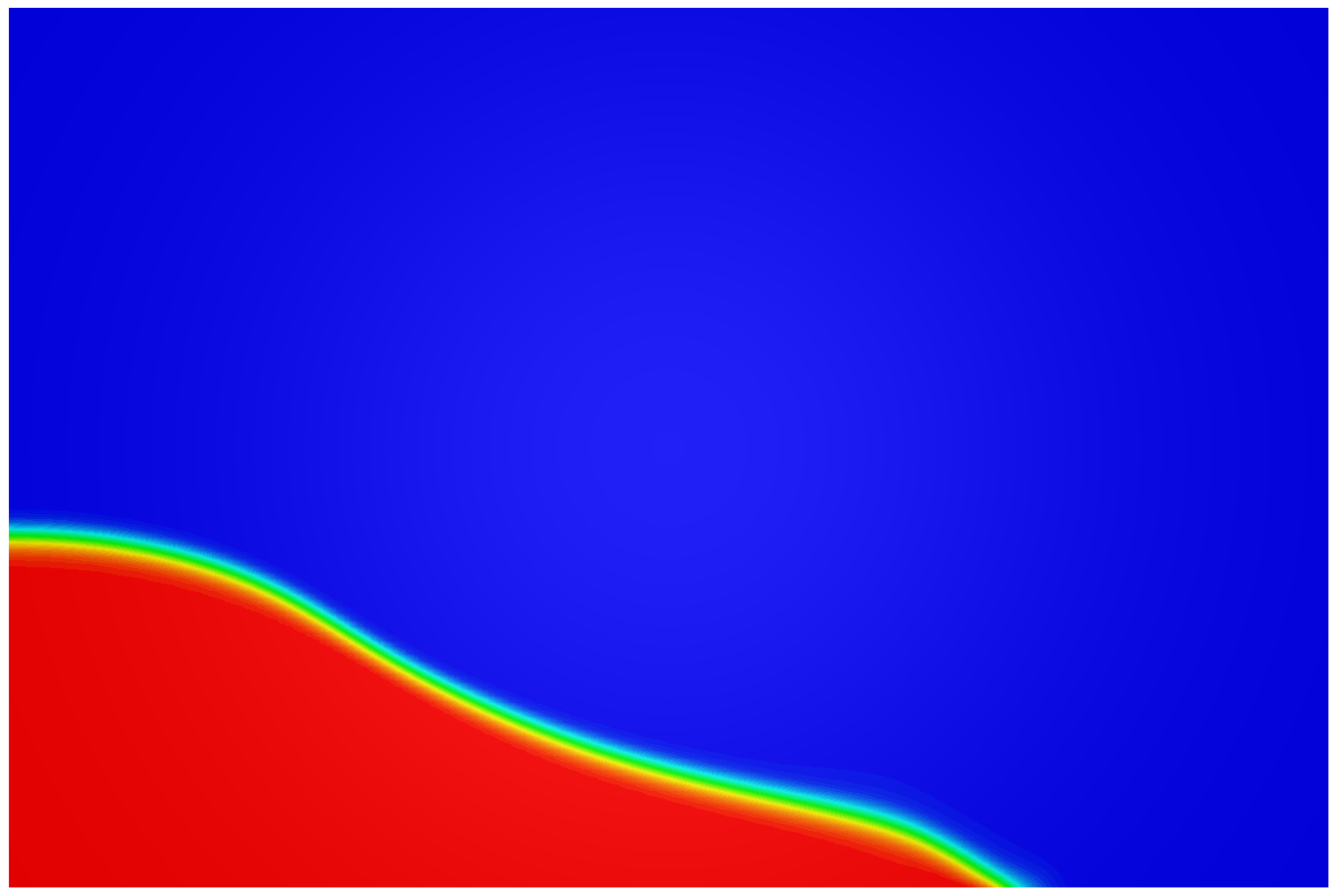}
		\subcaption{$t=3.2$}			
	\end{subfigure}
	\begin{subfigure}{.32\textwidth}
		\centering		
		\includegraphics[width = 0.95\textwidth]{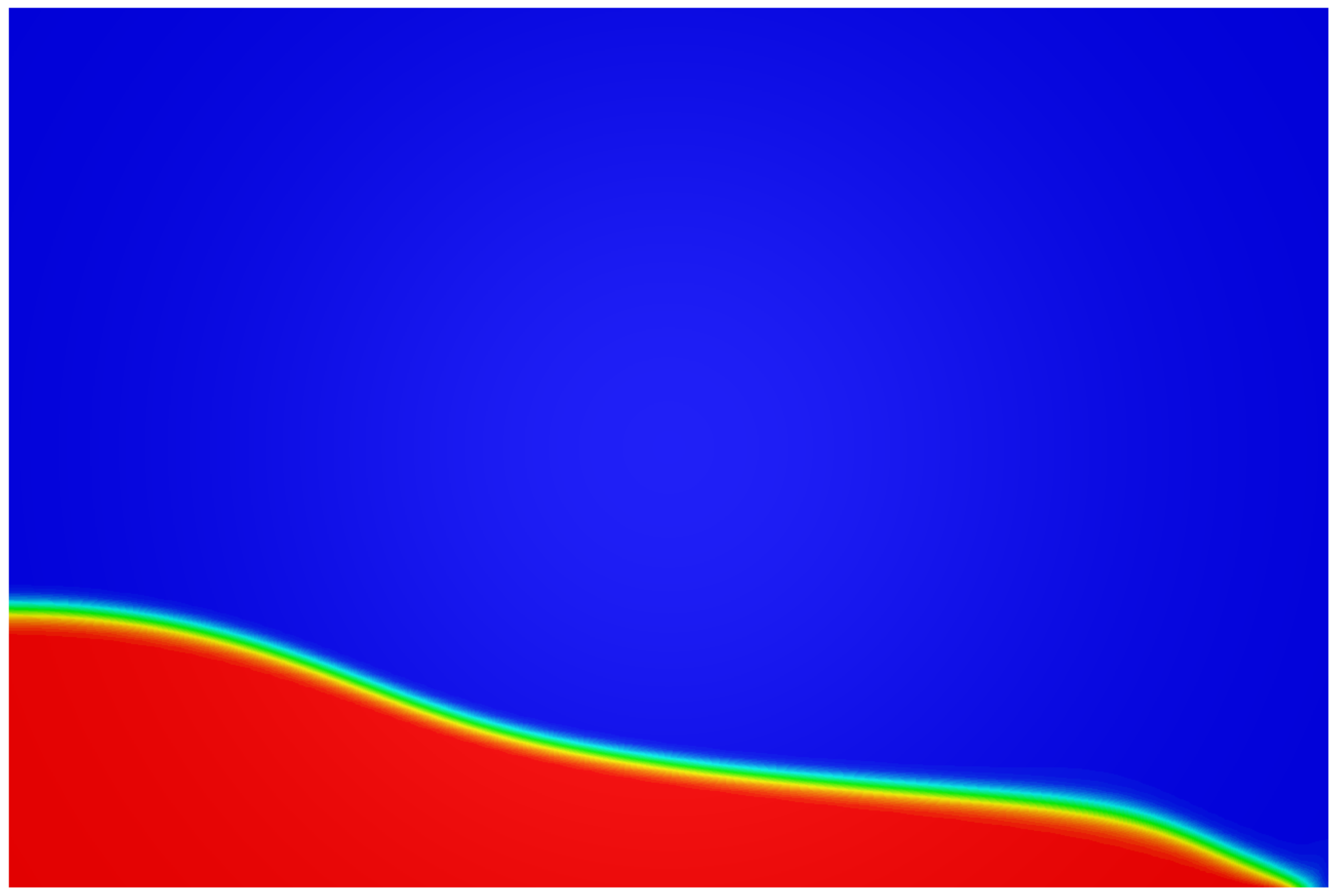}
		\subcaption{$t=4$}			
	\end{subfigure}		
	\caption{broken dam: $\varphi$ evolution using $192\times128$ elements.}
	\label{fig:broken_dam_evolution}	
\end{figure}

 The mesh consists of $192\times128$ linear elements, and the time step size is set to $\Delta t = 0.01$. The geometry of the problem and the tip displacement evolution are shown in Figure \ref{fig:broken_dam_geom}. 
The tip displacement evolutions agrees well with experimental results taken from \cite{Ramaswamy1987}.
The evolution of $\varphi$ is shown at several time instances in Figure \ref{fig:broken_dam_evolution}.
%
\subsection{Sloshing tank} \label{sec: numerical examples: sloshing tank}
\begin{figure}[bt!]
	\centering
	\begin{subfigure}{.40\textwidth}
		\centering
		\includegraphics[width=0.85\textwidth]{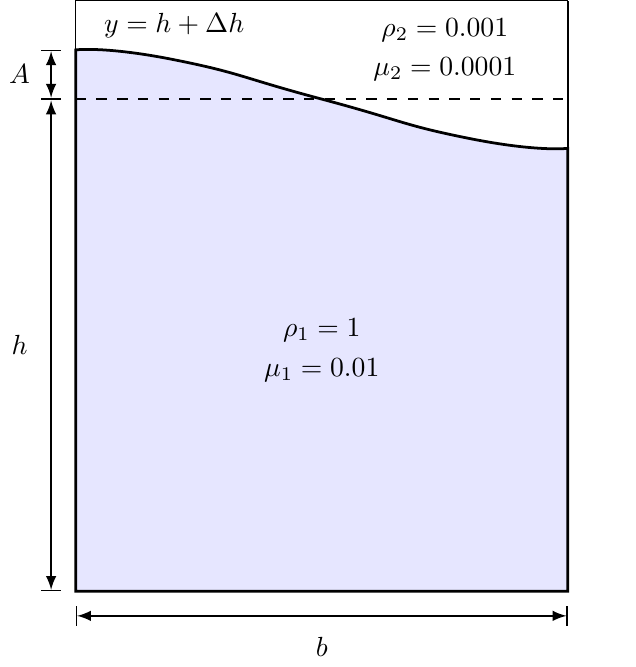}
	\end{subfigure}
	\begin{subfigure}{.40\textwidth}
		\centering
		\includegraphics[width=0.95\textwidth]{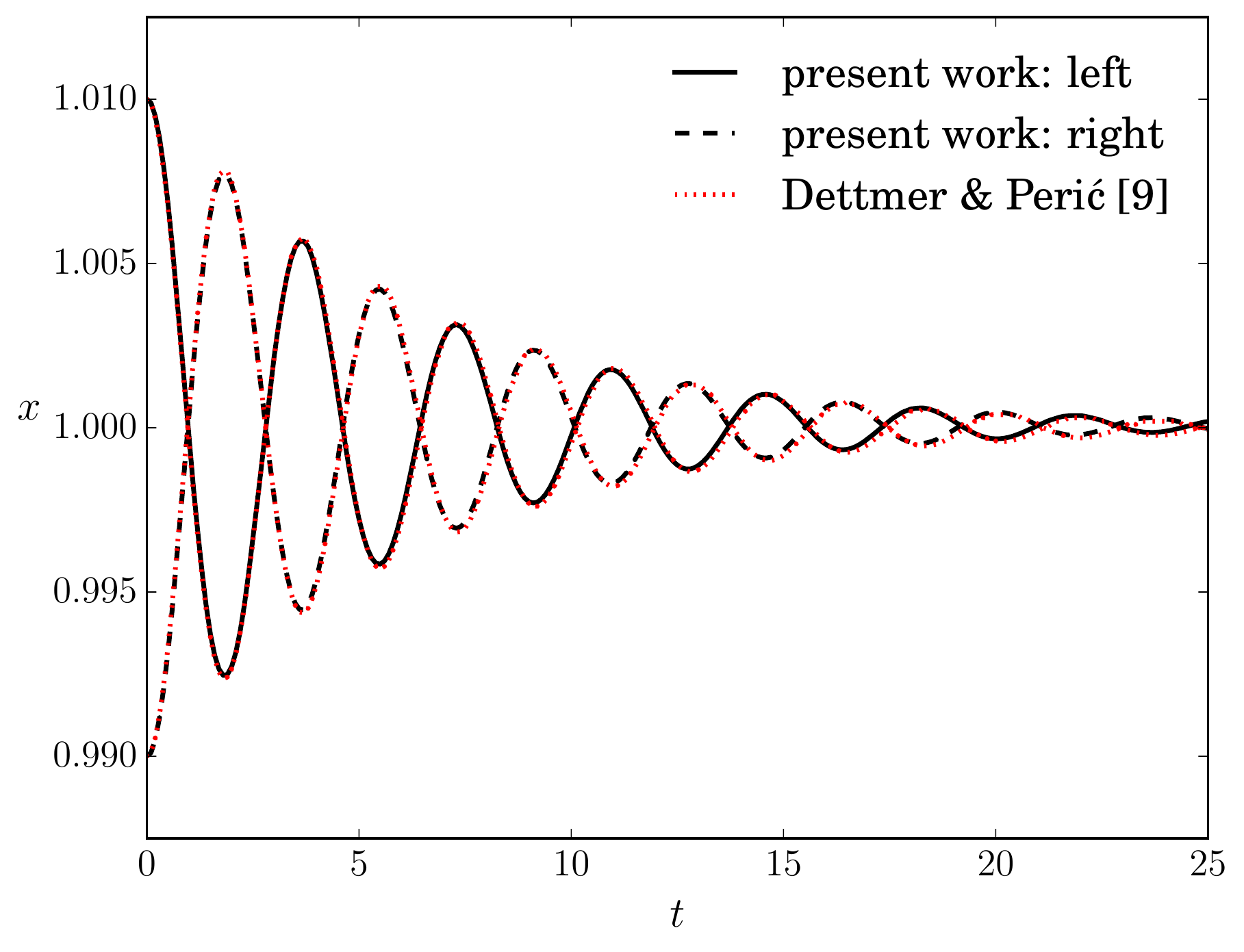}
	\end{subfigure}	
	\caption{Sloshing tank: Geometry (left) and amplitude frequency (right), with 48,000 linear quadrilateral elements.}
	\label{fig:sloshing_Tank_Geom}	
\end{figure}

%
Following \cite{Dettmer2006}, a sloshing tank of size $[0,1] \times [0,1.2]$ is set up with the following parameters $\rho_1 = 1$, $\rho_2 = 0.001$, $\mu_1 = 0.01$, $\mu_2 = 0.0001$ , $\bm{b} = [0,-1]^T$. The surface tension effects are neglected. A uniform mesh with 48,000 linear quadrilateral elements is selected, and a time step size of $\Delta t = 0.1$ is chosen. The fluid at the edges is allowed to adopt any contact angle by using \eqref{eq:free boundary}. Figure \ref{fig:sloshing_Tank_Geom} shows that the frequency at the left and right edges agrees excellently with \cite{Dettmer2006}.
%
%
\subsection{Rayleigh-Taylor instability} \label{sec: numerical examples: rayleigh-taylor instability}
\begin{figure}[tb!]
	\captionsetup[subfigure]{labelformat=empty}
	\centering
	\begin{subfigure}{.15\textwidth}
		\centering
		\includegraphics[width = 0.9\textwidth]{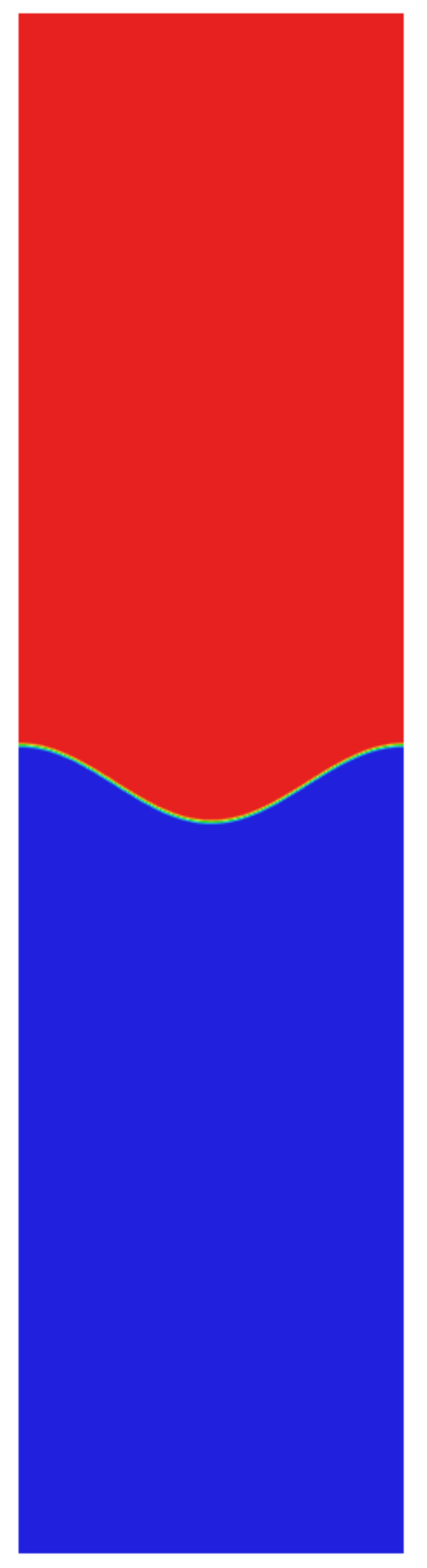}
		\subcaption{$t=0$}
	\end{subfigure}
	\begin{subfigure}{.15\textwidth}
		\centering	
		\includegraphics[width = 0.9\textwidth]{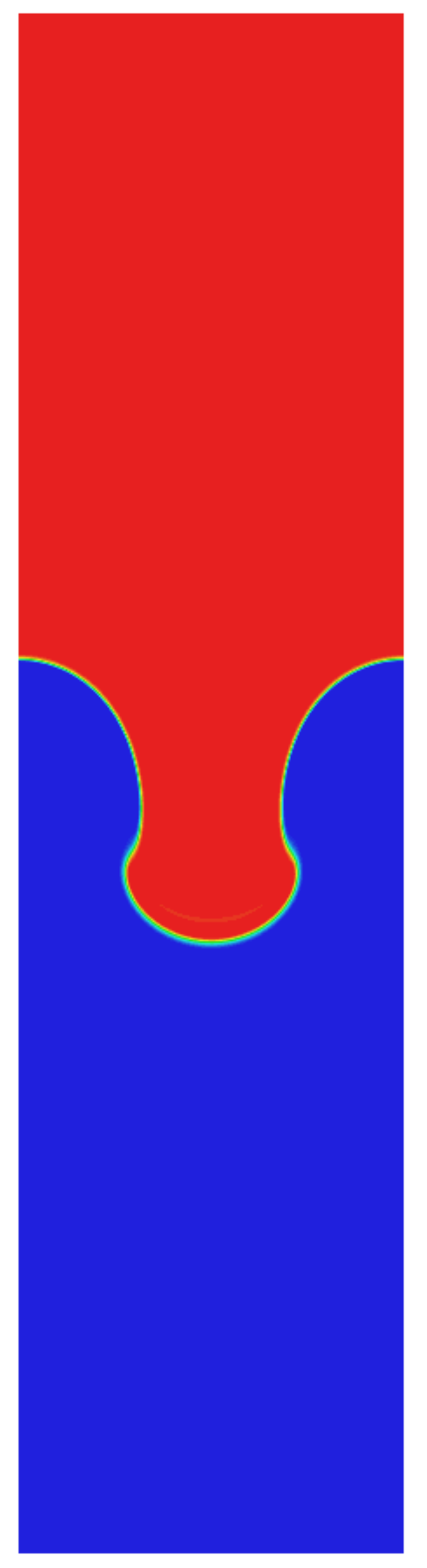}
		\subcaption{$t=0.5$}			
	\end{subfigure}
	\begin{subfigure}{.15\textwidth}
		\centering		
		\includegraphics[width = 0.9\textwidth]{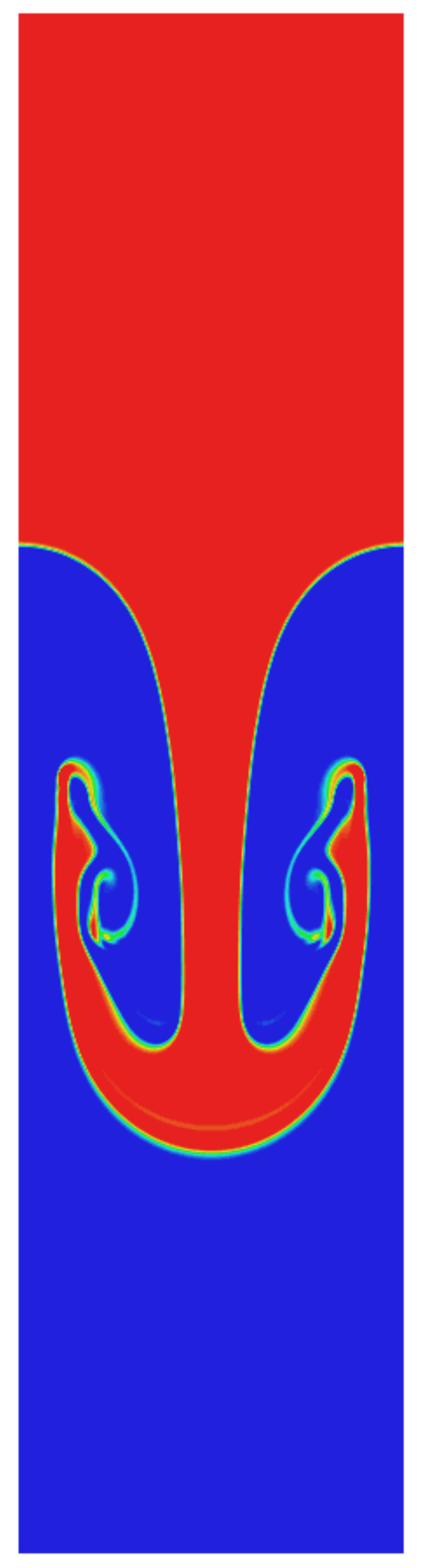}
		\subcaption{$t=1$}			
	\end{subfigure}
	\begin{subfigure}{.15\textwidth}
		\centering		
		\includegraphics[width = 0.9\textwidth]{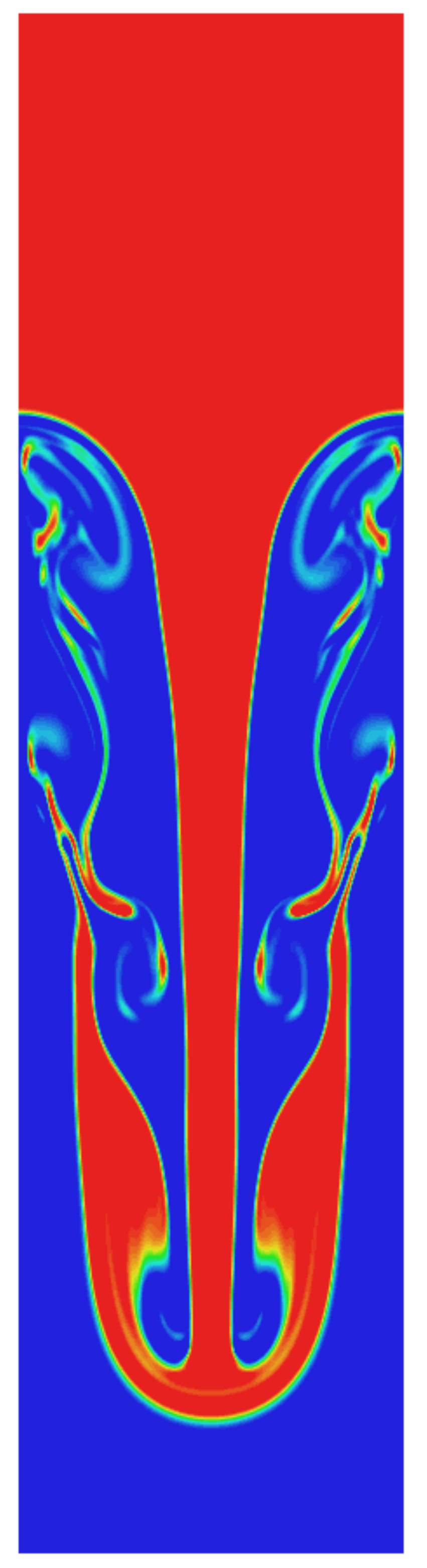}
		\subcaption{$t=1.5$}			
	\end{subfigure}
	\begin{subfigure}{.15\textwidth}
		\centering		
		\includegraphics[width = 0.9\textwidth]{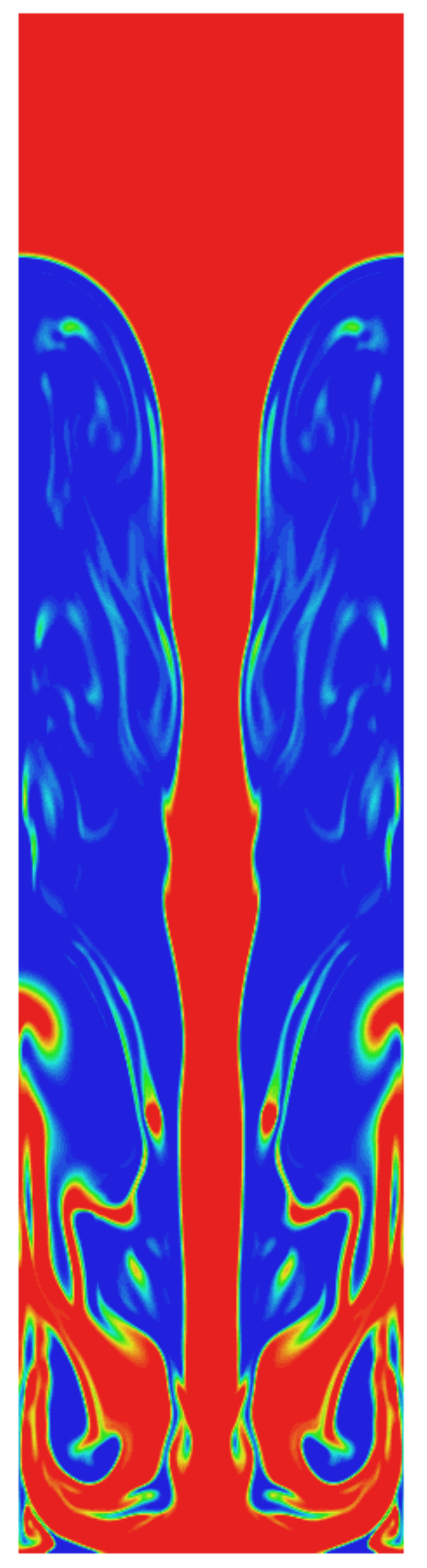}
		\subcaption{$t=2$}			
	\end{subfigure}	
	\caption{Rayleigh-Taylor instability: Evolution of phase field variable $\varphi$, with $256\times1024$ linear elements.}
	\label{fig:rayleigh_taylor_instability}	
\end{figure}

\begin{figure}[tb!]
	\centering
	\includegraphics[width=0.4\textwidth]{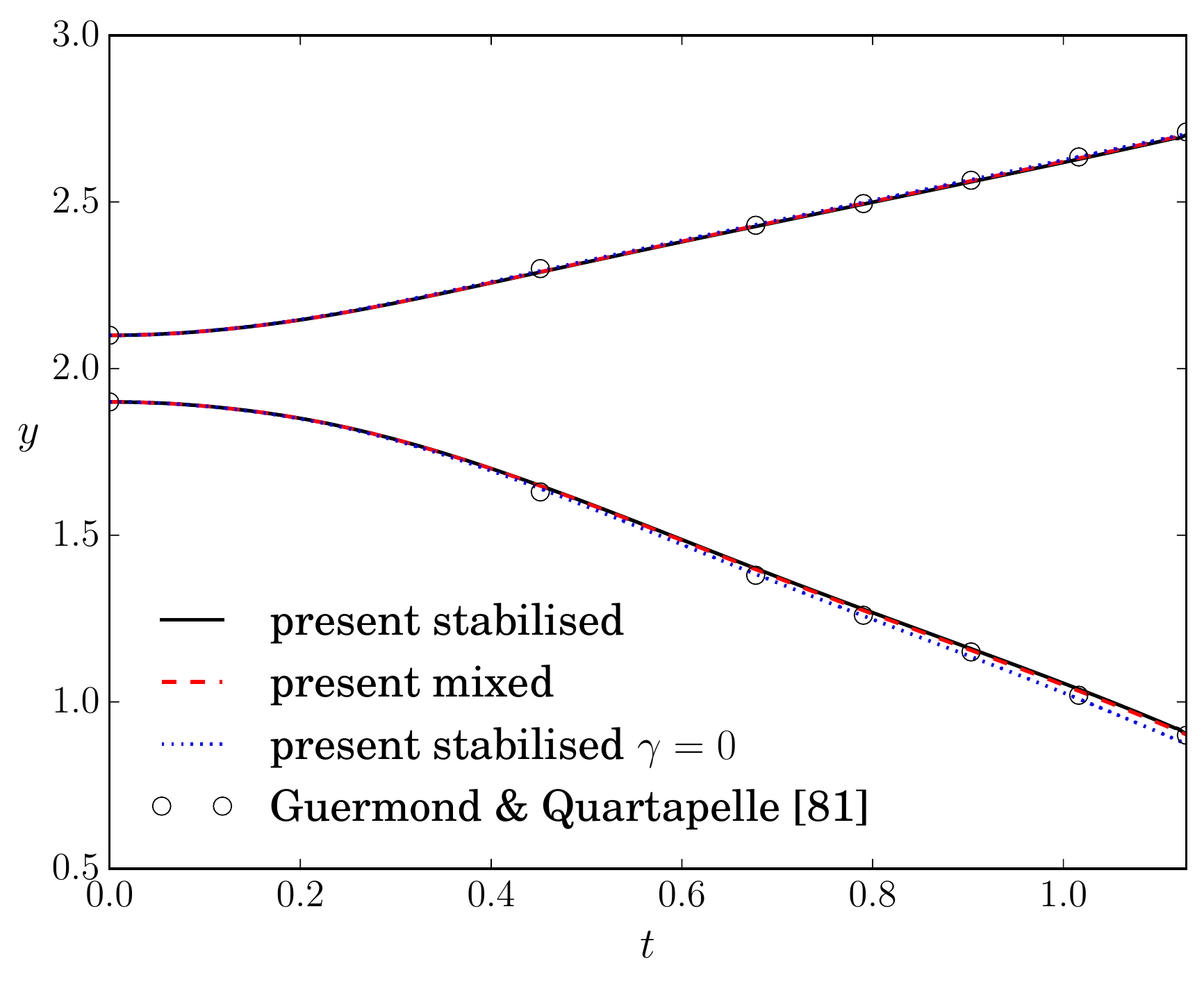}
	\caption{Rayleigh-Taylor instability: $y$ position of interface at the left wall and at the centre. Comparison with \cite{Guermond2000}.}	
	\label{fig:rayleigh_taylor_instability_graph}	
\end{figure}

%
The Rayleigh-Taylor instability problem described in \cite{Hosseini2017} is considered. The conventional Rayleigh-Taylor problem consists of a fluid $A$ sitting on top of a less dense fluid $B$. Any perturbation between the fluid layers in combination with gravitational force, will cause fluid $A$ to drive into fluid $B$ causing the well know mushroom cloud effect synonymous with Rayleigh-Taylor instability. A domain of size $[0,1]\times[0,4]$ is considered with the higher density fluid having $\rho_1=3$ and $\mu_1 = 0.0031316$, and the lower density fluid having $\rho_2=1$, $\mu_2 = 0.0031316$. The gravitational acceleration is taken as $g=9.80665$. The initial perturbation is given by
\begin{align}
	\varphi(x,y) = \tanh\left( \frac{y - 2 - 0.1 \cos(2\pi x)}{\sqrt{2}\epsilon} \right).
\end{align}
The upper and lower boundaries are set to no-slip, and the left and right boundaries are set to slip conditions.
In concurrence with \cite{Hosseini2017}, the other parameters are chosen as follows: $\gamma = 0.01$, $\epsilon = 0.005$, $\alpha = \pi/2$. The mobility function is chosen as $M_2(\varphi)$ with $D=4\cdot 10^{-5}$. Two meshes are compared, a mesh with $256\times1024$ linear stabilised elements and a mesh with $128\times512$ mixed Taylor-Hood elements. For comparison we consider the results from Guermond and Quartapelle \cite{Guermond2000} where surface tension effects are ignored. 
It should be mentioned that in \cite{Hosseini2017}, the author states that the surface tension coefficient is set as small but not zero, in order to avoid the CH equation becoming a pure transport equation, since setting $\gamma=0$ would render $\eta=0$ in the conventional formulation. In this work, we consider $\gamma=0.01$ with the conventional Abels et al. formulation \cite{Abels2011}, as well as $\gamma=0$ with the formulation presented in Section \ref{sec: Numerical formulation}. For the latter formulation the mobility coefficient is increased to $D=10^{-3}$, since it is no longer necessary to set it so small.
The evolution of $\varphi$ is shown in Figure \ref{fig:rayleigh_taylor_instability}. 
Figure \ref{fig:rayleigh_taylor_instability_graph} shows the position along the vertical axis of the rising and falling interface tips for the stabilised and mixed formulations.
The time scaling of the reference solution \cite{Guermond2000} required mapping according to $t = \sqrt{2/g} \ \hat{t}$ to account for the non-dimensionality of the variables. The results for the stabilised and mixed formulations are in agreement with the reference, and nearly indistinguishable from each other. Observing the plot corresponding to $\gamma=0$, it appears that neglecting surface tension effects does not really alter the displacements observed, although it does confirm that $\gamma$ can be set to zero without encountering numerical problems.
	\section{Conclusions} \label{sec: Conclusions}
In this work two novel finite element formulations are presented for modelling the Navier-Stokes-Cahn-Hilliard equations; The first uses mixed Taylor-Hood elements, while the second uses linear equal order stabilised SUPG/PSPG elements. The former formulation is primarily considered for comparative purposes. The models are formulated with two key aspects in mind: 
\begin{enumerate}[label=\roman*.]
	\item The ability to deactivate surface tension effects. This is done by removing the surface tension coefficient from the Cahn-Hilliard equations, and hence any instabilities which would result from setting it to zero. Thus, it is ensured that the Cahn-Hilliard equation exclusively deals with phase dynamics, while the Navier-Stokes equations alone control the physical phenomena.
	\item Computational efficiency, which relates specifically to the stabilised formulation. Here the standard SUPG/PSPG stabilisation is introduced in the momentum equation, which allows for the use of efficient equal order linear elements. 
	The employment of linear elements greatly improves the overall computational efficiency, which is crucial for the simulation of realistic three dimensional problems.
\end{enumerate}
\par A number of benchmark/example problems are solved with the proposed methodologies. The examples in Sections \ref{sec: numerical examples: static bubble}-\ref{sec: numerical examples: filling drop_3D} demonstrate problems dominated by surface tension, while problems without surface tension are investigated in Sections \ref{sec: numerical examples: broken dam}-\ref{sec: numerical examples: rayleigh-taylor instability}. 
In all cases the available reference solutions are reproduced accurately. Particularly noteworthy is the agreement of the results with those obtained from an arbitrary Lagrangian-Eulerian (ALE) based strategy in \cite{Dettmer2006}. It is demonstrated in Sections \ref{sec: numerical examples: large oscillations} and \ref{sec: numerical examples: filling drop_3D} that the volumes of the two fluid phases are conserved with good accuracy. 
%
	
	\FloatBarrier
	\bibsection{\section{References}}
	\bibliographystyle{unsrt}
	\bibliography{library}
	
\end{document}